%% file: htcams-arxiv-main.tex
\documentclass{report}
\usepackage[english]{babel}

\usepackage[utf8]{inputenc}
\usepackage[T1]{fontenc}
\usepackage{amsmath}
\usepackage{amsfonts}
\usepackage{amssymb}
\usepackage{amsthm}
\usepackage[version=4]{mhchem}
\usepackage{stmaryrd}
\usepackage{graphicx}
\usepackage[export]{adjustbox}
\usepackage{mathrsfs}
\usepackage{alltt}
\usepackage{cancel}
\usepackage{tikz}
\usepackage{tikz-cd}
\usetikzlibrary{calc}
\usetikzlibrary{fit}
\usepackage{xcolor}
\usepackage{tocloft}

\graphicspath{ {./images/} }

\usepackage[letterpaper,margin=1.0in]{geometry}

\usepackage{enumitem}
\usepackage{longtable}
\usepackage{mathtools}
\usepackage{array}

\usepackage{url}

\usepackage{subfiles}

\newcommand{\ourcases}[1]{{\left\{\begin{array}{ll} #1 \end{array}\right.}}

\newcommand{\ourcasesclosed}[1]{{\left\{\begin{array}{ll} #1 \end{array}\right]}}

\newcommand{\mat}[1]{\begin{pmatrix} #1 \end{pmatrix}}
\newcommand{\fmat}[1]{{\text{\footnotesize\arraycolsep=0.75\arraycolsep{$\mat{#1}$}}}}
\newcommand{\smat}[1]{{\text{\scriptsize\arraycolsep=0.66\arraycolsep{$\mat{#1}$}}}}
\newcommand{\tmat}[1]{{\text{\tiny\arraycolsep=0.3\arraycolsep{$\mat{#1}$}}}}

\newcommand{\matl}[1]{\begin{pmatrix*}[l] #1 \end{pmatrix*}}
\newcommand{\fmatl}[1]{{\text{\footnotesize\arraycolsep=0.75\arraycolsep{$\matl{#1}$}}}}
\newcommand{\smatl}[1]{{\text{\scriptsize\arraycolsep=0.66\arraycolsep{$\matl{#1}$}}}}
\newcommand{\tmatl}[1]{{\text{\tiny\arraycolsep=0.3\arraycolsep{$\matl{#1}$}}}}

\newcommand{\theset}[1]{{\{#1\}}}
\newcommand{\ceiling}[1]{{\left\lceil{#1}\right\rceil}}
\newcommand{\floor}[1]{{\left\lfloor{#1}\right\rfloor}}
\newcommand{\component}[2]{{{\left[{#2}\right]}_{#1}}}

\newcommand{\littleoh}[1]{{o{\kern-0.75pt}}\left({#1}\right)}

\newcommand{\ourunderbracket}[1]{\underbracket[0.14ex]{#1}}

\newcommand{\nostretch}[1]{{\renewcommand{\arraystretch}{1.0}{#1}}}

\DeclareMathOperator{\add}{add}

\DeclareMathOperator{\case}{case}
\DeclareMathOperator{\coalesce}{coalesce}
\DeclareMathOperator{\cod}{cod}
\DeclareMathOperator{\countfn}{count}
\DeclareMathOperator{\dom}{dom}
\DeclareMathOperator{\Endop}{End}
\DeclareMathOperator{\exponentiate}{exponentiate}
\DeclareMathOperator{\Free}{Free}
\DeclareMathOperator{\Hom}{Hom}
\DeclareMathOperator{\id}{id}
\DeclareMathOperator{\incr}{incr}
\DeclareMathOperator{\Leftop}{Left}
\DeclareMathOperator{\Ob}{Ob}
\DeclareMathOperator{\Set}{\underline{Set}}
\DeclareMathOperator{\UHom}{{\cup}Hom}

\DeclareMathOperator{\undefinedop}{undefined}
\newcommand{\undefined}{{\undefinedop}}

\makeatletter
\DeclareRobustCommand{\sqcdot}{\mathbin{\mathpalette\morphic@sqcdot\relax}}
\newcommand{\morphic@sqcdot}[2]{%
  \sbox\z@{$\m@th#1\centerdot$}%
  \ht\z@=.5\ht\z@
  \vcenter{\box\z@}%
}
\makeatother
\newcommand{\smallbsq}{{\;\textcolor{black}{\raisebox{.45ex}{\rule{.6ex}{.6ex}}}\;}}
\newcommand{\smallblacksquare}{{\textcolor{black}{\raisebox{.45ex}{\rule{.6ex}{.6ex}}}}}

\newcommand{\cC}{\mathscr{C}}
\newcommand{\cD}{\mathscr{D}}
\newcommand{\cF}{\mathscr{F}}

\newcommand{\cM}{\mathscr{M}}
\newcommand{\cN}{\mathscr{N}}
\newcommand{\cS}{\mathscr{S}}

\newcommand{\tikzmark}[1]{\tikz[overlay,remember picture] \node (#1) {};}

\newcounter{tikznodemarkers}
\newcommand\circletext[2]{%
    \tikz[overlay,remember picture] 
        \node (marker-\arabic{tikznodemarkers}-a) at (0,1.5ex) {};%
    #2%
    \tikz[overlay,remember picture]
        \node (marker-\arabic{tikznodemarkers}-b) at (0,0){};%
    \tikz[overlay,remember picture,inner sep=#1]
        \node[draw, rounded corners, rectangle, fit=(marker-\arabic{tikznodemarkers}-a.center) (marker-\arabic{tikznodemarkers}-b.center)] {};%
    \stepcounter{tikznodemarkers}%
}

\theoremstyle{plain}
\newtheorem{theorem}{Theorem}[chapter]
\newtheorem{corollary}[theorem]{Corollary}
\newtheorem{lemma}[theorem]{Lemma}

\newtheorem{result}[theorem]{Result}

\theoremstyle{definition}
\newtheorem{algorithm}[theorem]{Algorithm}
\newtheorem{algorithm-idea}[theorem]{Algorithm Idea}

\newtheorem{definition}[theorem]{Definition}
\newtheorem{example}[theorem]{Example}

\theoremstyle{remark}
\newtheorem{remark}[theorem]{Remark} 
\newtheorem{remarks}[theorem]{Remarks}

\title{
How to Compute a Moving Sum\\[1ex] 
\Large Windowed Recurrences -- A Monograph
}
\author{David K. Maslen and Daniel N. Rockmore}
\date{July 19, 2025\\[0.5ex] (Corrected February 8, 2026)}

\begin{document}

\maketitle

\pagenumbering{roman}

\setcounter{tocdepth}{1} 
\tableofcontents

\clearpage

\chapter*{Disclaimer}

\vspace*{-5ex}
The information, views, and opinions expressed herein are solely those of the authors and do not necessarily represent the views of Point72 or its affiliates. Point72 and its affiliates are not responsible   for, and did not verify for accuracy, any of the information contained herein.

\chapter*{Note on Corrections}

\vspace*{-5ex}
In this revision of the monograph we have corrected several hundred errors. These are almost entirely typos, copy-paste errors, and errors introduced through the manual and automated processes by which the document was transcribed from notes. We have re-validated and corrected all the pseudo-code, this time by testing pseudo-code that was copied from the monograph, rather than our previous reverse approach of translating working and tested code into pseudo-code. We have clarified and fixed the proof of Theorem~\ref{theorem:slick-deque-correctness}, and replaced the one-line proof of Theorem~\ref{theorem:semi-associativity-right-identity} with a correct one-line proof. Example~\ref{example:vector-function-action-from-set-automorphism} was incorrect, and has been replaced.

\clearpage

\pagenumbering{arabic}

\subfile{htcams-arxiv-ch01-introduction.tex}

\subfile{htcams-arxiv-ch02-moving-sums.tex}

\subfile{htcams-arxiv-ch03-two-stacks.tex}

\subfile{htcams-arxiv-ch04-dew.tex}

\subfile{htcams-arxiv-ch05-other-sequential-algorithms.tex}

\subfile{htcams-arxiv-ch06-windowed-recurrences.tex}

\subfile{htcams-arxiv-ch07-semi-associativity.tex}

\subfile{htcams-arxiv-ch08-algorithms-for-windowed-recurrences.tex}

\subfile{htcams-arxiv-ch09-categories-and-magmoids.tex}

\subfile{htcams-arxiv-ch10-introduction-to-vector-algorithms.tex}

\subfile{htcams-arxiv-ch11-vector-sliding-window-star-products.tex}

\subfile{htcams-arxiv-ch12-exponentiation-in-semigroups.tex}

\subfile{htcams-arxiv-ch13-vector-sliding-window-star-products-algorithms.tex}

\subfile{htcams-arxiv-ch14-vector-windowed-recurrences.tex}

\subfile{htcams-arxiv-ch15-pseudo-code.tex}

\subfile{htcams-arxiv-ch16-examples-and-constructions.tex}


\bibliographystyle{abbrv}
\bibliography{htcams-bibliography}

\end{document}

%% file: htcams-arxiv-ch01-introduction.tex
\chapter{Introduction}

\noindent
This monograph is motivated by a deceptively simple computational problem: The efficient  computation of {\em windowed recurrences}, quantities that depend on a moving window of data.  At its core, a {\em windowed recurrence} is a calculation applied iteratively to a sliding window over a data stream. The canonical example is the moving sum, where each output is the sum of the previous $n$ data points. But the concept generalizes far beyond addition, and we can consider windowed products, minima, function applications, and compositions under arbitrary (even non-associative) binary operations. 

While these computations have always been fundamental in the applied sciences, in a world of streaming and distributed data, they are an analytical linchpin.  Whether in low latency real time systems, analysis of DNA sequences, econometric time series analysis, industrial control, or natural language processing, the need to compute over a local sequential context is ubiquitous. 

Our work aims to provide a unifying theoretical framework to this important family of computations, and thus provide the foundation for practical implementation. In particular, we present three important innovations: 

\vspace{1em}
\begin{itemize}
    \item A general algebraic framework for windowed recurrences using the concepts of semi-associativity, semidirect products, and set actions. 
    
    \item New sequential and streaming algorithms for computing sliding window operations, including the {\em Double-Ended Window (DEW)} algorithm with low-latency guarantees.

  \item Compact and efficient parallel and vectorized algorithms derived through a new connection to the theory of semigroup exponentiation and addition chains.
\end{itemize}

\noindent
A guiding principle is the power of the algebraic framework to produce a clean abstract formulation of an important arithmetic process. Historically, this has proved to be hugely important in the development of efficient algorithms, and the windowed recurrence is no exception. Here too we find a transparent translation from algebra to code, with the algebraic structure once again proving to be source of efficiency and simplicity.

\section*{Guide to the Reader}

This monograph is structured into four main parts, each exploring a different facet of the windowed recurrence problem.

\paragraph{Chapters~\ref{chapter:moving-sums}--\ref{chapter:other-sequential-algorithms}: Sequential Algorithms.}
We begin with efficient sequential algorithms for sliding window $\ast$-products, where $\ast$ is associative. Highlights include:
\begin{itemize}
  \item A graphical approach to sequential computation of sliding window aggregates via stacked staggered sequence diagrams.
  \item An analysis of the Two Stacks and DABA algorithms using ideas from majorization theory.
  \item The new {\em Double-Ended Window (DEW)} algorithm with $3N$ complexity and bounded latency.
  \item A theory of {\em selection operators}, enabling a precise analysis of the SlickDeque algorithm.
\end{itemize}

\paragraph{Chapters~\ref{chapter:windowed-recurrences}--\ref{chapter:categories}: Windowed Recurrences and Semi-Associativity.}
These chapters develop the algebraic theory of windowed recurrences:
\begin{itemize}
  \item Definitions for windowed recurrences over functions, set actions, and nonassociative operations.
  \item The development of the theory of semi-associativity for computing with windowed recurrences. This theory describes the algebraic properties that must be obeyed by any data that represents functions and their compositions, and produces the conditions for parallel algorithms for reductions and recurrences, as well as for windowed recurrences.
  
  \item Generalizations to categories and magmoids, enabling recurrences over heterogeneous domains.
\end{itemize}

\paragraph{Chapters~\ref{chapter:vector-algorithms-guide}--\ref{chapter:vector-pseudo-code}: Vector and Parallel Algorithms.}
This part provides high-performance, scalable algorithms:
\begin{itemize}
  \item The reduction of windowed recurrences to semigroup exponentiation in semidirect products.
  \item Efficient implementations of Brauer’s and Thurber’s algorithms for exponentiation.
  \item Compact pseudo-code for vectorized windowed recurrence and non-windowed recurrence computation, including multi-query cases.
  \item Extensions of semi-associativity to vectorized settings.
\end{itemize}

\paragraph{Chapter~\ref{chapter:examples}: A Gallery of Examples.}
This final chapter offers concrete examples, use cases, and algebraic constructions:
\begin{itemize}
  \item Examples which have wide-spread applications in fields, including bioinformatics, natural language processing, and signal processing. 
  \item Techniques to build new recurrence structures from existing ones.
\end{itemize}

\vspace{1em}
\noindent
We hope that readers from a variety of domains---from algebraists to algorithm designers to applied scientists---will find useful ideas and surprising connections in what follows.

%% file: htcams-arxiv-ch02-moving-sums.tex
\chapter{Moving Sums}
\label{chapter:moving-sums}

\section{Definition of Moving Sums}
\label{sec:definition-of-moving-sums}

Assume we are given a sequence of numbers $a_1, a_2, \dots$ and a positive integer $n$, then {\em the moving sum of window length $n$} is the sequence of numbers
\begin{equation}
    y_i = \overbrace{a_i+\dots+a_{i-n+1}}^{n \text { terms }} \label{equation:moving-sum}
\end{equation}
obtained by summing the numbers in a sliding window $a_{i-n+1}, \dots, a_i$ of length $n$. Other names for this are {\em sliding window sum, window sum, sliding sum, rolling sum, or rolling window sum. }
%
%
As written above, the moving sum is defined for $i \geq n$ but the definition is easily extended to $i<n$, and there are several ways to do this. For definiteness we choose the convention that we drop terms from the sum for $i < 1$, so that
\begin{equation*}
y_i = \ourcases{
    a_i + \dots + a_1       & \quad \text{ for $1 \leq i < n$} \\
    a_i + \dots + a_{i-n+1} & \quad \text{ for $i \geq n$}
}
\end{equation*}
%
%
%
%
%
%
Other conventions are possible, and when we use these other conventions we will state so clearly. 

%
%


\section{Notes on Conventions}
\label{sec:notes-on-conventions}

\subsection{Boundary effects and domain of definition}

There are several ways to extend the definition of equation \eqref{equation:moving-sum} to $i<n$.
\begin{enumerate}

\item Drop terms $a_{j}$ from the sum when $j<n$. This is the convention we will mostly use.

\item Define $a_i = 0$ for $i<n$. This is equivalent to 1.

\item Choose not to define $y_i$ for $i < n$, so the calculation of $y_i$ for $i<n$ does not concern us.

\item Extend, if necessary, the numbers you are using with an `undefined' value. Define 
$a_i =  \undefined$ for $i<1$ together with the rule that 
\begin{equation*}
 x + \undefined = \undefined + x =  \undefined
\end{equation*}
so that $y_i = \undefined$ for $i<n$.

\item Extend the sequence $a_1, \dots$ backwards to negative indexes with values of your choosing, so that the sequence is $a_{2-n}, a_{1-n}, \dots, a_{-1}, a_{0}, a_1, a_2, \dots$. Then the definition of $y_i$ via equation \eqref{equation:moving-sum} applies directly.

\item It is of course possible to extend the definition of moving sum to sequences defined on all integer indices, including negative indices, and to finitely supported sequences (i.e., non-zero on finitely many indices) or finitely defined sequences (i.e., defined on finitely many indices) sequences.
\end{enumerate}

\noindent However this is done, we can either assume that the $a_i$ is extended somehow to indices $i<1$, or alternatively that we simply drop these terms from the definition. We will, however, be primarily interested in how to compute the moving sums $y_1, \dots, y_N$ for some finite value $N.$ 

\subsection{Associativity}

We are all familiar with the associativity of the operation $+$.%
\footnote{Note that for floating point arithmetic on commonly used computing hardware, at the time of this writing, $+$ is not associative.} Nevertheless, for definitiveness, and because we will be discussing algorithms for computing moving sums, let us specify the order of operations in the definition as associative from right to left. In other words,
\begin{equation*}
y_i = a_i + (a_{i-1} + (\dots + (a_{i-n+2} + a_{i-n+1}) \dots ))
\end{equation*}

\subsection{Left versus Right}

We have chosen to add new terms to the left of the sum and remove them from the right of the sum. 
\begin{equation*}
y_i = \underset{\text{new}}{a_i} + \ldots + \underset{\text{old}}{a_{i-n+1}}
\end{equation*}
Many authors, however, follow the convention that they add terms to the right and remove from the left. I.e., 
\begin{equation*}
y_i = \underset{\text{old}}{a_{i-n+1}} + \ldots + \underset{\text{new}}{a_i}
\end{equation*}
The reason for our choice will be seen in Chapter~\ref{chapter:windowed-recurrences} when we generalize to arbitrary windowed calculations, and stems from the standard notation for function application: If $\add_x$ denotes the function `addition of $x$', 
\begin{equation*}
    \add_x(y) = x + y
\end{equation*}
then the window sum is 
\begin{align*}
& \add_{a_i}(\add_{a_{i-1}}(\  \dots\ \add_{a_{i-n+2}}( a_{i-n+1} )  \quad \dots \quad)) &  & 
\hspace{-2em} \text{ for } i \geq n, \text{ and } \\
& \add_{a_i}(\add_{a_{i-1}}(\  \dots\qquad \qquad \ \hspace{-0.22em} \add_{a_2} (a_1) \quad \dots \quad)) & & 
\hspace{-2em} \text{ for } i < n 
\end{align*}
so in this notation we apply new functions $\add_{a_i}$ on the left. 
The difference between adding new terms on the left and on the right is cosmetic, and results and algorithms are easily translated from either convention to the other by flipping the order of addends in each addition, or equivalently by using the `opposite' operation defined as $x +_{\text{op }} y=y+x$. As $+$ is commutative $+_\text{op}=+$, but when we come to generalize to noncommutative operations the distinction between $+$ and $+_\text{op}$ matters.

\section{Prefix Sums}
\label{sec:prefix-sums}

The {\em prefix sum} of a sequence of numbers $a_1, a_2, \dots$ is the sequence of numbers

\begin{align*}
z_1    & = a_1 \\
z_2    & = a_2 + a_1 \\
z_3    & = a_3 + a_2 + a_1 \\
\vdots & \\
z_i    & = a_i + \quad \cdots \quad + a_1, \text{ for $i \geq 1$}
\end{align*}
obtained by summing the numbers over an expanding window $a_1, \dots, a_i$. Other names for prefix sums are {\em cumulative sums, partial sums, running sums, running totals}, or {\em a scan}. The same comments about associativity and left versus right apply to prefix sums in the same way as they do to moving sums. In particular, we define the sums by adding new terms on the left of the sum. I.e.,

\begin{equation*}
z_i=a_i+\left(a_{i-1}+\left(\dots+\left(a_2+a_1\right) \cdots\right)\right)
\end{equation*}
As before, the literature commonly defines these by adding terms to the right of the sum, but the difference between the two conventions is cosmetic, and the translation trivial.

%
%

There are efficient parallel algorithms for computing prefix sums. The most straightforward is due to Kogge and Stone \cite{KoggeStone1973}, Ladner and Fischer \cite{LadnerFischer1980}, and Hillis and Steele \cite{HillisSteele1986}, with precursor work by Ofman \cite{Ofman1962}. There is also a related algorithm due to Blelloch \cite{Blelloch1990} \cite{Blelloch1993} \cite{Blelloch1996} which performs less total work but has twice the depth (either $2 \left\lceil \log_2(N + 1)\right\rceil$ or $2 \left\lceil \log_2 N\right\rceil + 1$ vs $\left\lceil\log_2 N\right\rceil$). We won't describe the Kogge-Stone algorithm yet, as it relates closely to the work in Chapters \ref{chapter:vector-sliding-window-*-products}--\ref{chapter:vector-sliding-window-*-product-algorithms}. The work in Chapters~\ref{chapter:vector-sliding-window-*-products}--\ref{chapter:vector-sliding-window-*-product-algorithms} gives a new and simple derivation of that algorithm by relating it to exponentiation in semidirect products, and also presents a systematic approach to deriving variant algorithms and new algorithms for the same problem. The Kogge-Stone algorithm does, however, allow a vectorized description, and under a PRAM (Parallel Random Access Machine) model%
\footnote{See Blelloch \cite{Blelloch1996}.}
it has depth $\left\lceil\log_2 N\right\rceil$ and performs total work $N\left\lceil\log_2 N\right\rceil - 2^{\left\lceil\log_2 N \right\rceil} + 1$.
Note that
%
\begin{align*}
N\left(\left\lceil \log_2 N\right\rceil -2 \right) - 1 & \leq N\left\lceil \log _2 N\right\rceil -2^{\left\lceil \log_2 N \right\rceil} + 1 \\
& \leq N\left(\left\lceil\log_2 N\right\rceil -1\right) + 1
 = {N \left\lceil\log _2 N\right\rceil} - \left( N - 1 \right)
\end{align*}
so the complexity of the algorithm is bounded above by $N\left\lceil\log _2 N\right\rceil$. 


\section{What We Look For in an Algorithm}
\label{sec:algorithm-properties}

In the next sections we will look at some basic methods for computing moving sums, and consider the advantages and pitfalls of the different approaches. Here are some properties we will watch for.

\begin{description}
  \item[Correctness]\ \\ Does the algorithm correctly compute the moving sum?
  
  \item[Accuracy] \ \\ Does the calculation maintain numerical accuracy?
  
  \item[Efficiency] \ \\ How many operations does the algorithm use?
  
  \item[Simplicity] \ \\ Does the algorithm have a lot of special cases? Does it require complicated data structures or indexing? These considerations come to bear when considering implementation.
  
  \item[Parallelizability] \ \\ Can the calculation be (efficiently) distributed across multiple processors?
  
  \item[Vectorizability] \ \\ Can the algorithm be expressed in terms of operations on entire sequences?
  
  \item[Freedom from extraneous choice or data] \ \\ Does the algorithm involve choices or data in the computation of a value that do not appear in the definition of that value? 
  
  \item[Memory] \ \\ How much working space does the algorithm need? 
  
  \item[Streaming] \ \\ Are there online or streaming versions of the algorithm that return one new window sum per value submitted?
  
  \item[Latency] \ \\ For a streaming algorithm, how many operations are performed to produce one new window sum from a newly submitted value?
  
  \item[Generalizability] \ \\ Can the algorithm be generalized to other operations or situations?
  
\end{description}

\noindent In addition to these properties there is an extensive literature considering more advanced features such as out of order processing, variable size windows, multi-query processing, and bulk eviction and insertions. Verwiebe et al.\ \cite{Verweibe2023} give a survey of different types of window aggregation problems. 

\section{The Naive Algorithm}
\label{sec:naive-algorithm}

The most straightforward way to compute a moving sum is to use the definition directly as per the following algorithm.

\begin{algorithm}
\label{algorithm:naive-algorithm}
\ \\
Assume we are given input data $a_1, \ldots, a_N$, and a window length $n$.

\begin{description}
\item[Step 1] Compute
\begin{align*}
y_1 & = a_1 \\
y_2 & = a_2 + a_1 \\
\vdots & \\
y_n & = a_n + a_{n-1} + \dots + a_1 \\
y_{n+1} & = a_{n+1} + a_n + \dots + a_2 \\
\vdots & \\
y_N & = a_N  + a_{N-1} + \dots + a_{N-n+1}.
\end{align*}
\end{description}
\end{algorithm}
\noindent This is a good algorithm from most aspects as it is clearly correct and accurate, though the accuracy depends on the exact method used to compute the sums, e.g., summing successively by pairs in a tree-like fashion keeps numerical errors low. It also is an algorithm that directly corresponds to the definition and does not involve extraneous choices or data in the computation of the $y_i$. 

From the point of view of simplicity, the naive algorithm is simple to describe and implement, involving only a small number of index variables to keep track of which sum is being worked on and which term is being added. A `batch' version of the algorithm working on arrays only requires one item of working space to keep the value of the sum currently being worked on. A streaming version requires $n$ items of working space, to keep the values in the window prior to summation, and one  additional item of working space to keep the sum.%
\footnote{The $n$ items are a fixed size array of length $n$ effectively used as a circular buffer.}

The main issue with the naive algorithm is performance, as it requires $(n-1)\left(N-\frac{n}{2}\right)$ operations. Note that we are only counting the operations that compute the sums, and not the indexing logic, which in practice may be as or more expensive than the operation for the sum itself.
For small $n$ the naive algorithm is performant, but for large $n$ the linear dependence of performance on $n$ is a serious drawback. E.g., for a one-year moving sum of daily data we have $n = 365$, or $n \simeq 250$ for business days. For sliding window analysis of DNA sequences, window lengths of hundreds or thousands of base-pairs may be used \cite{Boyer2012} \cite{Wang2015}. Two dimensional image processing also uses sliding window analysis, with both large window and data sizes.%
\footnote{
Two dimensional sliding windows may easily be broken down into one dimensional moving sum calculations, by first computing moving sums along one dimension with data conditioned on the second coordinate, and then repeating that process on the resulting data with the coordinates switched.
}

Parallelization of the naive algorithm can be done straightforwardly by splitting the calculations for the different $y_i$ over the available processors. And if more processors are available the individual $y_i$ calculations can be individually parallelized by summing in the pattern of a binary tree. This algorithm has depth $\lceil\log _2 n\rceil$.

We'll discuss generalizations of moving sums in much more detail later, but note here that generalizing the naive algorithm to calculations where $+$ is replaced by some other binary operation presents no challenge. There are no algebraic requirements on the operation for the algorithm to generalize other than that the operator is defined for the input and intermediate values. If the operation is nonassociative then the $y_i$ must be calculated by summing from right to left, i.e., 
\begin{equation*}
a_i+\left(a_{i-1}+\left(\dots+\left(a_{i-n+2}+a_{i-n+1}\right) \cdots\right)\right)
\end{equation*}

This leaves vectorizability, from our list of considerations, and the naive algorithm may be represented in vector form without difficulty as follows: Let $a$ denote the sequence $a_1, \dots, a_N$, and $y$ denote the sequence $y_1, \dots, y_N$ and let $+$ be defined on sequences component-wise.
Now define the lag operator $L_{j}$ on sequences by
\begin{equation*}
\component{i}{L_j a} = a_{i-j}
\end{equation*}
where the values for indices $i<1$ are filled in with $0$ (or we could use one of the alternative conventions described in Section \ref{sec:notes-on-conventions}).%
\footnote{
We use the notation $\component{i}{\hspace{1ex}}$ to indicate extraction of the $i^\text{th}$ component of a vector, array, or list, so e.g., $\component{i}{x}$ indicates the $i^\text{th}$ component of the vector $x$. Our arrays and vectors start at index $1$.
}
Then the naive algorithm for computing $y_1, \dots y_N$ can be expressed as
\begin{equation*}
y=a + L_1 a + L_2 a + \dots + L_{n-1} a
\end{equation*}
This vector expression corresponds naturally to a parallel algorithm where each index $i$ is associated with some processor and multiple indexes may be assigned to the same processor. The $+$ operator adds data with the same indexes, and so is perfectly or `embarrassingly' parallel. The lag operators $L_i$ perform no computation but simply communicate data between the processors. Thus when we parallelize the naive algorithm we need at most $(n-1)\left\lceil\frac{N}{p}\right\rceil$ additions per processor where $p$ is the number of processors. There is, however, an amount of communication corresponding to the $n-1$ operators $L_1, \dots, L_{n-1}$.
On a Parallel Random Access Machine, or PRAM, (see \cite{Blelloch1996}) we have
\begin{center}
\begin{tabular}{ll}
work & $(n-1) N$\\   
depth & $(n-1)$ \\
time & $(n-1)\left\lceil\frac{N}{p}\right\rceil$ \\
\end{tabular}
\end{center}
for this approach to parallelization,  
though as noted a depth $\left\lceil\log_2 n\right\rceil$ algorithm  
is also possible.

The next algorithm we look at is more efficient, without the factor $n$ in the complexity, but which presents. a host of issues when applied inappropriately.

\section{The Subtract-on-Evict Algorithm}
\label{sec:subtract-on-evict}

This is the algorithm most software developers will come up with if pressed to find a way to compute moving sums efficiently.%
\footnote{As an experiment ask a friend or colleague}
The name comes from an implementation of the streaming version (see \cite{Hirzel2017}). 
The algorithm proceeds as follows.

\begin{algorithm}[Subtract-on-Evict]
\label{algorithm:subtract-on-evict}
\ \\
Assume we are given input data $a_1, \ldots, a_N$, and a window length $n$. 

\begin{description}
\item[Step 1] 
First compute
\begin{align*}
y_1 & = a_1 \\
y_2 & = a_2 + y_1 \\
\vdots & \\
y_n & = a_n+y_{n-1}
\end{align*}
\item[Step 2] Then compute
\begin{align*}
y_{n+1} & = a_{n+1} + y_n - a_1 \\
y_{n+2} & = a_{n+2} + y_{n+1} - a_2 \\
\vdots \\
y_N & = a_N + y_{N-1} - a_{N-n}
\end{align*}
\end{description}
\end{algorithm}
\noindent There are, of course, two versions of this algorithm according to whether $y_i$ is computed as $a_i + (y_{i-1} - a_{i-n})$ or $(a_i + y_{i-1}) - a_{i-n}$, and a third version, using commutativity, computes $y_i$ as $y_{i-1} + (a_i - a_{i-n})$. The primary advantage of this algorithm is efficiency, as it requires only $N-1$ additions and $\max (N-n, 0)$ subtractions, for a total of $2 N-n-1$ operations when $N>n$ and $N-1$ operations when $N \leq n$.

In many settings this is an efficient and simple algorithm with no drawbacks. It has both `batch' and streaming versions, and the streaming version is low latency, requiring only two operations to produce each new window sum after the initial startup phase. The batch version requires only 1 item of working space, and the streaming version $n+1$ items. As with the naive algorithm the streaming version can be implemented using a fixed size array of length $n$ (effectively a circular buffer) and one additional item to keep the sum.

However, in some settings, drawbacks to the Subtract-on-Evict algorithm are evident. The main problem is correctness. Many applications use numbers which may have undefined, not-a-number (NaN), missing, or infinite values, and these values do not have an additive inverse (i.e., negative) that cancels them to give 0. Consider, for example, the situation where $a_i$ is undefined at index $i$. Under the Subtract-on-Evict algorithm the window sum $y_i$, and all subsequent $j$ with $j>i$ will be undefined, and this problem will persist even for $j$ where the undefined value $a_i$ has passed out of the window. Thus, the Subtract-on-Evict algorithm does not correctly compute the moving sum when undefined or infinite values occur. As an illustration, consider the situation in the table below with window length $n=3$ and the sliding window sum $y_i$ computed by Subtract-on-Evict.
\begin{center}
\begin{tabular}{l|cccccccc}
$i$ & 1 & 2 & 3 & 4 & 5 & 6 & 7 & 8 \\
\hline
$a_i$ & 0 & -1 & 5 & \text{undef} & 7 & 5 & 1 & -3 \\
$y_i$  & 0 & -1 & 4 & \text{undef}  & \text{undef} & \text{undef} & \text{undef} &  \text{undef}\\
\end{tabular}
\end{center}

Accuracy is also a concern for the Subtract-on-Evict algorithm when using floating point arithmetic, and there are two separate mechanisms by which the numerical precision can deteriorate.
\begin{enumerate}
    \item The computation of $y_i$ for $i \gg n$ involves many more operations than the definition. For $i>n$ it involves $(i-1)$ additions and $i-n$ subtractions, whereas the definition involves only $n-1$ additions. In situations where $i \gg n$ this can cause the value of $y_i$ computed using Subtract-on-Evict to be much less accurate.
    \item A second accuracy problem occurs when a value $a_i$ enters the sum that is much larger than others. This causes $y_i$ to be large, and also $y_{i+1}, \dots, y_{i+n-1}$ will be large. The trouble comes when $a_i$ drops out of the window and we compute
    $y_{i+n} = a_{i+n} + (y_{i+n-1}-a_i)$.
    This involves a difference of two large numbers leading to reduced accuracy for $y_{i+n}$ and subsequent moving sums. A simple example with IEEE 754 double precision arithmetic illustrates the point, where again $n=3$ and the $y_i$ are computed by Subtract-on-Evict.
\begin{center}
\begin{tabular}{l|cccccccc}
$i$ & 1 & 2 & 3 & 4 & 5 & 6 & 7 & 8 \\
\hline
$a_i$ & $0.1$ & $0.1$ & $1\mathrm{e}20$ & $0.1$ & $0.1$ & $0.1$ & $0.1$ & $0.1$ \\
$y_i$  & $0.1$ & $0.2$ & $1\mathrm{e}20$ & $1\mathrm{e}20$  & $1\mathrm{e}20$ & $0.1$ & $0.1$ & $0.1$\\
\end{tabular}
\end{center}
Notice how the value of the window sum computed by the Subtract-on-Evict algorithm has drifted and this error persists in all window sums from index 6 onward. In particular, index 6 should be close to $0.3$ rather than $0.1$. If we had instead evaluated $y_i$ using $\left(a_i + y_{i-1}\right)-  a_{i-n}$ then $y_6$, $y_7$, $y_8$ would have been $0.0$ instead of $0.1$. The same phenomenon can also occur when a group of the $a_i$ are larger than others.
\end{enumerate}
The correctness and accuracy problems of Subtract-on-Evict stem from the use of extraneous data in the calculation, together with the requirement for exact inverses. The calculation of $y_i$ in Subtract-on-Evict depends on $a_1, \dots, a_i$, whereas it should only depend on $a_{i-n+1}, \dots, a_i$.

Regarding generalizability, the Subtract-on-Evict algorithm can be generalized to any associative operator with inverses, i.e., to group settings (in the sense of group theory in abstract algebra), but we see that it is too limited to handle even common situations such as missing data and floating point arithmetic. It also will not work directly with types that do not have an inverse or difference operation. The algorithm is inherently sequential so that it does not lend itself to parallelization or vectorization. We shall now turn, however, to a variant of the Subtract-on-Evict algorithm that is vectorizable and hence parallelizable.

\section{The Difference of Prefix Sums Algorithm}
\label{sec:difference-of-prefix-sums}

\begin{algorithm}
\label{algorithm:difference-of-prefix-sums}
\ \\
Assume we are given input data $a_1, \ldots, a_N$, and a window length $n$. 
\begin{description}
    \item[Step 1] 
    First compute the prefix sums
\begin{align*}
& z_1=a_1 \\
& z_2=a_2+a_1 \\
& \vdots \\
& z_N=a_N+\dots+a_1
\end{align*}

\item[Step 2]
Then compute the moving sums as
\begin{equation*}
y_i = \ourcases{
    z_i,         & \text{if $i \leq n$} \\
    z_i-z_{i-n}, & \text{if $i > n$}
}
\end{equation*}   
\end{description}
\end{algorithm}

\noindent There are several variants of this algorithm depending on how the prefix sums are computed.

\begin{description}
    \item[Variant 1]
Compute the prefix sums using the recurrence
\begin{equation*}
z_i = a_i + z_{i-1}
\end{equation*}
starting from $z_1=a_1$. This variant is a rearrangement of the Subtract-on-Evict algorithm, and has the same overall operation counts.

\item[Variant 2]
Compute the prefix sums using the Kogge-Stone algorithm \cite{KoggeStone1973}, or the algorithm of Blelloch \cite{Blelloch1993}. If we use Kogge-Stone then this will perform total work of $N\left\lceil \log_2 N\right\rceil - 2^{\left\lceil\log _2 N\right\rceil}+1$ additions and $N-n$ subtractions. As we shall see later in these notes,%
\footnote{
See Section~\ref{sec:parallel-prefix-sum-algorithms}.
}
the Kogge-Stone algorithm can be written in vector form, and this leads to a vectorization of the second variant of the Difference of Prefix Sums. We have
\begin{align*}
z & = \text{Vectorized Kogge-Stone Prefix-Sum}(a) \\
y & = z-L_n z
\end{align*}
where $L_n$ is the lag operator from Section~\ref{sec:naive-algorithm}. As a parallel algorithm this has depth $\lceil \log _2 N\rceil +1$ under a PRAM model.
\end{description}
\noindent Difference of Prefix Sums suffers from the same correctness and accuracy issues as Subtract-on-Evict, and the generalizability is the same. It is not a streaming algorithm. It involves more values in the computation of $y_i$ than the definition, and so is not free from extraneous data. Memory-wise it requires $N$ items of working space to store the prefix sums.

\section{Examples of Other Sliding Window Calculations}
\label{sec:other-window-calculations}

Before moving on to more algorithms for moving sums, we first look at some other examples of sliding window calculations. This will also help us start to generalize the theory.

\begin{example}[Moving Sums with Missing Data]
A common way to handle missing data is to extend the operation to support an undefined value. This can be achieved by extending the $+$ operation so that
\begin{equation*}
x + y = \ourcases{
    \undefined & \text{if $x = \undefined$ or $y = \undefined$} \\
    x + y            & \text{otherwise}
}
\end{equation*}
As noted before, the undefined value has no inverse, and the Subtract-on-Evict and Difference of Prefix Sums algorithms will not work for this operation. The extended operation is commutative and associative, assuming the original operation had these same properties, and these properties can be used to develop moving sum algorithms. In the case where associativity is only approximate the extended operation maintains the approximate associativity.
\end{example}

\begin{example}[Moving Products]
Moving products are defined analogously to moving sums.

%
\begin{equation*}
   y_i = \ourcases{
    a_i \ldots a_1       & \quad \text{ for $1 \leq i < n$} \\
    a_i \ldots a_{i-n+1} & \quad \text{ for $i \geq n$}
} 
\end{equation*}
For moving products, $0$ is not invertible, so the Subtract-on-Evict and Difference of Prefix Sums algorithms will not work unless all the $a_i$ are non-zero, and the same issues with missing data and undefined values arise. Furthermore, with finite precision arithmetic and floating point numbers, two new problems arise, which are overflow and underflow, and these impact the Difference of Prefix Sums algorithm. 
\smallskip

To see how overflow can occur, consider a moving product, with data length $N=2000$, and window length $n=3$, and $a_i = 2.0$ for $i=1, \dots, 2000$. Assume we are using IEEE 754 double precision arithmetic. According to the definition, we have
\begin{center}
\begin{tabular}{ccccccc}
$i$ & 1 & 2 & 3 & 4 & 5 & $\ldots$ \\
\hline
$a_i$ & 2.0 & 2.0 & 2.0 & 2.0 & 2.0 & $\ldots$\\
$y_i$ & 2.0 & 4.0 & 8.0 & 8.0 & 8.0 & $\ldots$\\
\end{tabular}
\end{center}
So $y_1=2, y_2=4$, and $y_i=8$ for $i \geq 3$. Both the naive algorithm and Subtract-on-Evict compute this correctly. However, Difference of Prefix Sums (applied to products)   overflows on the computation of the prefix product at $i=1024$. Depending on the implementation, this causes either algorithm failure (an error condition or exception) or an undefined or infinite value (incorrectness).
\smallskip

To see how underflow can occur, consider a moving product, with data length $N=2000$, and window length $n=3$, and $a_i=0.5$, for $i=1, \dots, 2000$, and again assume we are using IEEE 754 double precision arithmetic. Then we have
\begin{center} 
\begin{tabular}{lllllll}
$i$  & 1 & 2 & 3 & 4 & 5 & $\ldots$ \\
\hline 
$a_i $ & 0.5 & 0.5 & 0.5 & 0.5 & 0.5 & $\ldots$\\
$ y_i$  & 0.5 & 0.25 & 0.125 & 0.125 &  0.125& $\ldots$ \\
\end{tabular}
\end{center}
As before, the naive algorithm and the Subtract-on-Evict algorithm compute $y_i$ correctly. This time the Difference of Prefix Sums algorithm starts underflowing at the prefix product with $i=1023$, and at $i=1075$ the calculated prefix product is $0.0$. The gradual underflow starting at $i=1023$ causes loss of accuracy, but the zero value at $i=1075$ is not invertible and prevents the algorithm from running correctly from that point onwards.
\end{example}

\begin{example}[Moving Sums and Products with Binary Operations]
\label{example:moving-sums-binary-operations}

This is not so much an example as an obvious generalization. Let $*$ be any binary operation and $a_1, a_2, \dots$ be some objects for which that operation is defined. Then we may define the moving sums or moving products as
%
%
\begin{equation*}
y_i = \ourcases{
    a_i *(a_{i-1} * (\ldots * (a_2 * a_1)             \ldots )) & \quad \text{ for $1 \leq i < n$} \\
    a_i *(a_{i-1} * (\ldots * (a_{i-n+2} * a_{i-n+1}) \ldots )) & \quad \text{ for $i \geq n$}
} 
\end{equation*}
\end{example}
We will give a more formation definition of these in Section~\ref{section:definition-of-windowed-recurrence} Definition~\ref{definition:sliding-window-*-product}, under the name {\em sliding window $*$-products}.

\begin{example}[Moving Max and Min]
The moving $\max$ and $\min$ of a sequence of numbers, $a_i$, satisfy the equations
\begin{align*}
\max \left(a_i, a_{i-1}, \dots, a_{i-n+1}\right) & = 
    \max \left(a_i, \max \left(a_{i-1}, \max \left(\dots, \max \left(a_{i-n+2}, a_{i-n+1}\right) \dots
        \right)\right)\right),\\ 
\min \left(a_i, a_{i-1}, \dots, a_{i-n+1}\right) & =
    \min \left(a_i, \min \left(a_{i-1}, \min\left(\dots, \min \left(a_{i-n+2}, a_{i-n+1}\right), \dots
        \right)\right)\right) 
\end{align*}
Both $\max$ and $\min$ are associative, idempotent, and commutative,%
\footnote{Note that many implementations are neither associative nor commutative, e.g. when there is missing data.}
and binary operations with these operations correspond to meet and join operators of semi-lattices. From an algorithmic standpoint $\max$ and $\min$ have the 
useful property that $\max (x, y) \in \theset{x, y}$, $\min (x, y) \in \theset{x, y}$. 
Binary operations satisfying $x * y \in \theset{x, y}$ are in one to one correspondence with reflexive binary relations, and we call such operations {\em selection operators} or {\em selective}.%
\footnote{We will comment further on this in Section \ref{sec:selection-operators}.}
Both Subtract-on-Evict and Difference of Prefix Sums fail for $\max$ and $\min$ because of the lack of inverses. 
\end{example}

\begin{example}[Fill Forward]
\label{example:fill-forward-I}

The well known operation of filling forward missing data can be represented as a sliding window calculation, where the length of the window is one greater than the maximum number of data points you allow to be filled from any non-missing value. The binary operation in this case is {\em coalesce}, and is defined as
\begin{equation*}
\coalesce(a, b) = \ourcases{
    b & \text{if $a$ is undefined, else} \\
    a
}
\end{equation*}
The windowed fill-forward calculation itself is
\begin{align*}
y_i & = \coalesce(a_i, \dots, a_{i-n+1}) \\
    & = \coalesce(a_i, \coalesce(a_{i-1}, \coalesce(\dots, \coalesce(a_{i-n+2}, a_{i-n+1}) \dots)))
\end{align*}
and as before we drop $a_i$ with $i<1$ from the calculation and so have
%
%
\begin{equation*}
y_i = \ourcases{
    \coalesce(a_i, \ldots, a_1)       & \quad \text{ for $1 \leq i < n$} \\
    \coalesce(a_i, \ldots, a_{n+i-1}) & \quad \text{ for $i \geq n$}
} 
\end{equation*}
Note that for this operation the advantages of ordering from right to left start to become apparent. The operation `$\coalesce$' is associative and shares the property $\coalesce(x, y) \in \theset{x, y}$ that we observed for $\max$ and $\min$, and so it is a {\em selection operator}, which comes from a reflexive binary relation (see Section \ref{sec:selection-operators}). In this case the associated relation is $x R y \Leftrightarrow (x = \undefined \text{ or } y=x)$. As with $\max$ and $\min$ we have no inverses, so Subtract-on-Evict and Difference of Prefix Sums do not apply. `coalesce' is also noncommutative, so this is our first example (in these notes) of a noncommutative sliding window calculation, and is also a practical and commonly used noncommutative window calculation.    
\end{example}

\begin{example}[Sliding Window Continued Fractions]
\label{example:sliding-window-continued-fractions}

Sliding window continued fractions are defined as follows.
{
\allowdisplaybreaks 
\begin{align*}
y_1     & = a_1 \\
y_2     & = a_2+\cfrac{1}{a_1} \\
y_3     & = a_3+\cfrac{1}{a_2+\cfrac{1}{a_1}} \\
\vdots  & \\
y_n     & =  a_n+\cfrac{1}{\ddots+\cfrac{1}{a_2+\cfrac{1}{a_1}}} \\
y_{n+1} & =  a_{n+1}+\cfrac{1}{\ddots+\cfrac{1}{a_3+\cfrac{1}{a_2}}} \\
\vdots  & \\
y_i     & =  a_i+\cfrac{1}{\ddots+\cfrac{1}{a_{i-n+2}+\cfrac{1}{a_{i-n+1}}}}
\end{align*}
}
%
%
The binary operation associated with a sliding window continued fraction is
\begin{equation*}
a * b=a+\frac{1}{b}
\end{equation*}
This operation is nonassociative, so to handle operations like these we will need techniques to mitigate the nonassociativity. 
\end{example}

\begin{example}[Moving Sums with Scale Changes]
\label{example:moving-sums-with-scale-changes}

This situation occurs frequently with financial time series, e.g. securities prices and trading volumes, where scale changes can result from corporate or government actions. In addition to the input data $a_1, a_2, \dots$, we are given a multiplier $m_i$, which can be thought of as a `change of units factor' from one index to the next. The definition of the sliding window calculation in this case is

\begin{equation*}
y_i = a_i + m_i\left(a_{i-1}+m_{i-1}\left(\ldots+m_{i-n+3}\left(a_{i-n+2}+m_{i-n+2} a_{i-n+1}\right) \ldots\right)\right)
\end{equation*}
There are several, ultimately equivalent,  ways to fit these sums into a framework of moving sums with binary operations.  One way is to vary the operations $*$ by defining
\begin{displaymath}
a *_m b=a + m b
\end{displaymath}
and 
\begin{displaymath}
y_i=a_i *_{m_i}\left(a_{i-1} *_{m_{i-1}}\left(\ldots *_{m_{i-n+3}}\left(a_{i-n+2} *_{m_{i-n+2}} a_{i-n+1}\right) \ldots\right)\right)
\end{displaymath}
Alternatively the $a_i$ and $m_i$ can be grouped together to give
\begin{equation*}
\mat{m\\ a} \bullet\, b = a + m b
\end{equation*}
and
\begin{equation*}
y_i=\mat{m_i \\ a_i} \bullet \left(\mat{m_{i-1}\\ a_{i-1}} \bullet \left(\ldots \bullet\left(\mat{m_{i-n+2}\\ a_{i-n+2}} \bullet a_{i-n+1}\right) \ldots\right)\right)
\end{equation*}
Equivalently we can define functions
\begin{equation*}
f_\tmat{m\\ a}(b)=a+m b.
\end{equation*}
Thus, perhaps the most general way to formulate this calculation is to assume we have a sequence of functions $f_i=f_\tmat{m_i\\ a_i}$, and to define a windowed recurrence $y_i$ via
\begin{equation*}
y_i=f_i\left(f_{i-1}\left(\ldots f_{i-n+2}\left(a_{i-n+1}\right) \ldots\right)\right)
\end{equation*}
In this formulation the $a_i$ and $m_i$ are carried in the data describing the functions $f_i$. 
Regardless of the formulation, these operations are noncommutative, nonassociative, and without inverses, so techniques for handling those situations must be used. `Moving sums with scale changes' are equivalent to linear recurrences.
\end{example}

\begin{example}[Moving Sums with Sign Changes]
\label{example:moving-sums-with-sign-changes}

An obvious special case of moving sums with scale changes is if the multiplier $m_i$ is $\pm 1$. (Of course for scale changes we want $m_i>0$, but the formula allows the possibility of negative $m_i$.) This gives
\begin{equation*}
y_i=a_i+\varepsilon_i\left(a_{i-1}+\varepsilon_{i-1}\left(\ldots+\varepsilon_{i-n+3}\left(a_{i-n+2}+\varepsilon_{i-n+2} a_{i-n+1}\right) \ldots\right)\right)
\end{equation*}
These can occur when you want to average a directional quantity where the sign is not well defined, e.g., a `moving average of one-dimensional subspaces'. The considerations for calculating these sums are of course the same as for moving sums with scale changes, or any other linear recurrence.
\end{example}

\section{Sliding Window Sum Algorithms}
\label{sec:sliding-window-sum-algorithms}

We have seen from these examples that to compute sliding window calculations we must handle situations where the operation has no inverses, may be noncommutative, may be nonassociative, or there may even be no binary operation, but instead a `window of functions to apply'. The foundation for handling all these cases, however, will be the case of an associative binary operation, and we now turn our attention to algorithms for that case.

In the next three chapters we look at algorithms that compute moving sums for associative operators and which do not require the existence of inverses. Over the past three decades a large number of algorithms for this have been developed, and we refer the interested reader to the survey articles of Verweibe et al.\ \cite{Verweibe2023}, and Tangwongsan et al.\ \cite{Tangwongsan2022}, as well as the article of Shein et al.\ \cite{Shein2018}. We will give an analysis of the {\em Two Stacks} algorithm \cite{Tangwongsan2015b} then present a new algorithm, which we call the {\em Double-Ended Window} (DEW) algorithm. We then give some comments on related the related {\em DABA and DABA Lite} algorithms of Tangwongsan, Hirzel, and Schneider \cite{Tangwongsan2015b} \cite{Tangwongsan2017} \cite{Tangwongsan2021}, and on the {\em SlickDeque} algorithm of Shein \cite{Shein2019}. 

There are many sequential and parallel sliding window algorithms and implementations that we do not cover here. These include: B-INT, L-INT (Arasu and Widom \cite{ArasuWidom2004}), PANES (Li et al.\ \cite{LiMaier2005}), Pairs, Fragments (Krishnamurthy et al.\ \cite{Krishnamurthy2006}), Flat FAT (Tangwongsan et al.\ \cite{Tangwongsan2015a}), Cutty (Carbone et al.\ \cite{Carbone2016}), SABER (Koliousis et al.\ \cite{Koliousis2016}), Flat FIT (Shein et al.\ \cite{Shein2017}), Scotty (Traub et al.\ \cite{Traub2018} \cite{Traub2021}), Hammer Slide (Theodorakis et al.\ \cite{Theodorakis2018}), FiBA (Tangwongsan et al.\ \cite{Tangwongsan2019}), CBiX (Bou et al.\ \cite{Bou2019}), Slide Side (Theodorakis et al.\ \cite{Theodorakis2020a}), Light SABER (Theodorakis et al.\ \cite{Theodorakis2020b}), and PBA (Zhang et al.\ \cite{Zhang2021}).

%% file: htcams-arxiv-ch03-two-stacks.tex
\chapter{The Two Stacks Algorithm}
\label{chapter:two-stacks}

The Two Stacks algorithm was developed by Tangwongsan, Hirzel, and Schneider \cite{Tangwongsan2015b} and generalizes an idea posted on Stack Overflow \cite{adamax2011} for maintaining the minimum of a queue. The name of the algorithm comes from a particular implementation of the algorithm, which uses two `stack' data structures. There are two distinct aspects to the algorithm. One is the sequence of binary $*$ operations that is performed, i.e., the pattern of usage of $*$. The second is the data structures and bookkeeping used to cause this pattern of operations to be executed. For this algorithm there are many ways to organize the bookkeeping, and these also depend on how the input data is stored, organized, and presented to the algorithm. These are important implementation details and they vary widely from use case to use case. For this reason we focus primarily on the first aspect, i.e., which $*$ operations are performed. 

In this chapter we assume that $*$ is an associative binary operator, and our goal is to compute the moving sums (or moving products)
\begin{equation*}
y_i = \ourcases{
    a_i * \ldots * a_1       & \quad \text{for $1 \leq i < n$} \\
    a_i * \ldots * a_{i-n+1} & \quad \text{for $i \geq n$}
} 
\end{equation*}
where as usual the $a_i$ are considered to drop out of the product when $i<1$. The operator $*$ is not assumed to be commutative, or to have inverses, or other properties, unless otherwise stated.

\section{Evict then Insert}
\label{sec:evict-then-insert}

We start with a version of Two Stacks we call the {\em Evict-then-Insert} version. This computes the moving sums in batches of size $n$, by piecing together the result of prefix-sum and suffix-sum calculations. 

\begin{algorithm}
\label{algorithm:two-stacks}

Our goal is to compute $y_1, \ldots, y_N$. To do this the algorithm proceeds by computing

\begin{center}
\begin{tabular}{ll}
$y_1, \ldots, y_n$            & (Batch 1) \\
$y_{n+1}, \ldots, y_{2 n}$    & (Batch 2) \\
$\vdots$                      &  \\
$y_{k n+1}, \ldots y_{k n+r}$ &  (Last Batch), where $N=kn+r$.
\end{tabular}
\end{center}

\noindent To compute the first batch $y_1, \ldots, y_n$, we use a sequential prefix-sum calculation. I.e.,

\begin{equation*}
\begin{array}{ll}
y_1 = a_1 \\
y_2 = a_2 * y_1 & =a_2 * a_1 \\
\ \vdots & \\
y_n = a_n * y_{n-1} & =a_n * \ldots * a_1
\end{array}
\end{equation*}
The second, and subsequent batches are computed differently from the first batch. To compute the batch $y_{m+1,}, \ldots, y_{m+n}$, where $m$ is a multiple of $n$, proceed as follows. First compute the backwards prefix sums,%
\footnote{Another name for a backward prefix sum is a suffix sum.}  
$u_{m+1}, \ldots, u_{m+n-1}$,
as
\begin{align*}
u_{m+n-1} & = a_{m} \\
u_{m+n-2} & = a_{m} * a_{m-1} \\
\vdots & \\
u_{m+1} & = a_{m} * \ldots * a_{m-n+2}
\end{align*}
using the recursion $u_{m+n-j-1}=u_{m+n-j} * a_{m-j}$ for $j=1, \ldots, n-2$, starting from $u_{m+n-1} = a_m$.
Next compute the prefix sums
\begin{align*}
& v_{m+1}=a_{m+1} \\
& v_{m+2}=a_{m+2} * a_{m+1} \\
& \vdots \\
& v_{m+n}=a_{m+n} * \ldots * a_{m+1}
\end{align*}
using the recursion $v_{i+1}=a_{i+1} * v_i$, starting from $v_{m+1}=a_{m+1}$. Then finally complete the batch by computing the moving sums as
\begin{align*}
& y_{m+1}=v_{m+1} * u_{m+1} \\
& \vdots \\
& y_{m+n-1}=v_{m+n-1} * u_{m+n-1} \\
& y_{m+n}=v_{m+n}
\end{align*}

\end{algorithm}
\begin{remarks} \
\begin{enumerate}
\item 
If the pattern of computation in Algorithm \ref{algorithm:two-stacks} is not immediately clear then the graphical approach to follow may help.
  
\item 
It should be clear that if $N$ is not a multiple of $n$ then the last batch is truncated. In that case all of the suffix sums $u_{kn+1}, \ldots, u_{kn + n - 1}$ must still be computed but only the prefix sums $v_i$ with $i \leq N$ need be computed.

\item Only one of the prefix sums $v_i$ needs to be remembered at a time. Once a prefix sum $v_i$ has been used to compute both $y_i$ and $v_{i+1}$ it may be forgotten. So to avoid unnecessary memory use one should first compute the suffix sums $u_i$ in a batch, and then interleave the computation of the $v_i$ and $y_i$. This variant of the algorithm is called {\em Two Stacks Lite} in Tangwongsan et al.\ \cite{Tangwongsan2021}, or {\em Hammer Slide} in Theodorakis et al.\ \cite{Theodorakis2018}.

\item 
The name Evict-then-Insert comes from a streaming version of the algorithm \cite{Tangwongsan2015b} where new input items $a_i$ are inserted to a data structure, thus increasing window size, and old items $a_{i-n}$ are evicted from the data structure. In steady state the order of insertion vs eviction matters and the algorithm above corresponds to evicting first and then inserting.

\end{enumerate}
\end{remarks}

\section{Graphical Description of Two Stacks}

The Two Stacks algorithm can be visualized by a tabular diagram, which we call a {\em stacked staggered sequence diagram}.

\begin{example}[Stacked Staggered Sequence Diagram]

Here is a stacked staggered sequence diagram for Two Stacks for $n=4$ and $N=10$.

%
\begin{center}
\includegraphics[scale=1.0]{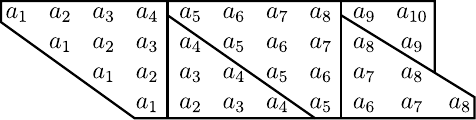}
\end{center}
Let the $i^\text{th}$ column refer to the column with $a_i$ on the top row. In this example it is clear that to compute the sliding window $*$-products $y_i$, we must compute the product of the entries in each column, with lower row entries appearing on the left of the product. We have divided the diagram into 5 regions, which we can label $A$, $B$, $C$, $D$, $E$ as follows.

\begin{center}
\includegraphics[scale=1.0]{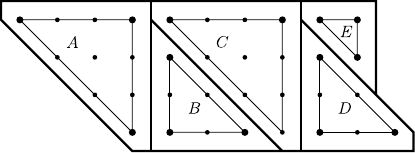}    
\end{center}
We say that a region is {\em vertically connected} if its intersection with any column consists of adjacent entries. Clearly $A$, $B$, $C$, $D$, $E$ are each vertically connected. 

For each region $R$, let $R_i$ denote the $*$-product of the region's entries in the $i^\text{th}$ column, with lower entries in the diagram appearing on the right of the product. Since the regions partition the diagram, and the window sums  $y_i$ are the $*$-products of the entries in each column, it follows that the window products can be formed out of products of the $R_i$ for the regions intersecting each column. Algebraically this is equivalent to the following computation.
\begin{align*}
                            & & A_1 \ &= a_1 \\
                            & & A_2 \ &= a_2 * a_1 \\
                            & & A_3 \ &= a_3 * a_2 * a_1 \\
                            & & A_4 \ &= a_4 * a_3 * a_2 * a_1\\
\\[-2.0ex]
B_5 \  &=  a_4 * a_3 * a_2    & C_5 \ &= a_5\\
B_6 \  &=  a_4 * a_3          & C_6 \ &= a_6 * a_5\\ 
B_7 \  &=  a_4                & C_7 \ &= a_7 * a_6 * a_5\\
       &                      & C_8 \ &= a_8 * a_7 * a_6 * a_5\\
\\[-2.0ex]
D_9 \  &=  a_8 * a_7 * a_6    & E_9 \  &= a_9\\
D_{10} &=  a_7 * a_6          & E_{10} &= a_{10} * a_{9}\\ 
D_{11} &=  a_6 & &
\end{align*}
%
Note that the $A_i$, $C_i$, and $E_i$ are prefix sums and so may be computed efficiently, and the $B_i$ and $D_i$ are suffix-sums (backwards prefix sums). Once these are computed, the window sums $y_i$ may be computed as
\begin{align*}
y_1 &= A_1  & y_5 &= C_5 * B_5 & y_9 \  &= E_9 * D_9 \\
y_2 &= A_2  & y_6 &= C_6 * B_6 & y_{10} &= E_{10} * D_{10} \\
y_3 &= A_3  & y_7 &= C_7 * B_7 & \\
y_4 &= A_4  & y_8 &= C_8 * B_8 &
\end{align*}
%
This approach to computing $y_1, \ldots, y_{10}$ is exactly the Evict-then-Insert version of Two Stacks described in Section~\ref{sec:evict-then-insert}.    
\end{example}

Returning to the general case, the stacked staggered sequence diagram for the sequence $a_1, a_2, \ldots$, and a given $n$, is the following staggered table.
\begin{displaymath}
\begin{array}{cccccccccc}
a_1 & a_2 & a_3 & \dots  & a_n    & a_{n+1} & \dots  & a_{2n}  & a_{2n+1} & \dots  \\
    & a_1 & a_2 & \dots  &        & a_n     & \dots  &         & a_{2n}   & \dots  \\
    &     & a_1 & \dots  &        &         &        &         &          &        \\
    &     &     & \ddots & \vdots & \vdots  &        & \vdots  &          &        \\
    &     &     &        & a_1    &  a_2    & \dots  & a_{n+1} & a_{n+2}  & \ldots  
\end{array}
\end{displaymath}
The general algorithm for using these diagrams to compute sliding window $*$-products follows directly.
\begin{algorithm}[Sliding Window $*$-Products from Diagrams]
\label{algorithm:stacked-staggered-sequence-diagram} \
\begin{description}
\item[Step 1] 
Partition the stacked staggered sequence diagram into regions which are vertically connected.

\item[Step 2] 
For any region $i$, let $R_i$ denote the $*$-product of the entries in the intersection of the region with the $i^\text{th}$ column, with entries that are lower in the diagram appearing on the right of the product.
  
\item[Step 3] 
To compute the $i^\text{th}$ window sum, compute the $*$-product
\begin{equation*}
y_i=\underset{\substack{\text{regions }  R\\ \text { intersecting } \\ \text { column } i}}{\scalebox{3.0}{\raisebox{-1ex}{*}}} R_i
\end{equation*}
where regions appearing lower in the column are on the right of the product.%
\footnote{
Note that if $*$ is commutative then the condition that regions are vertically connected may be dropped, and the products computed in any order.
}
\end{description}
\end{algorithm}
\noindent This procedure does not by itself produce an efficient algorithm, but must be combined with other techniques, in particular the following:
\begin{enumerate}
  \item Right-angled isosceles triangular regions with a downwards sloping hypotenuse correspond to prefix sums and suffix sums and may be computed efficiently by cumulating the sums sequentially.%
\footnote{For a parallel computation, this could also be achieved using a parallel prefix sum algorithm.
}
\begin{align*}
\vcenter{\hbox{\includegraphics[scale=1.0]{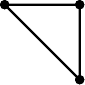}}}  \quad & = \ \text{prefix sums} & \\[\baselineskip]
\vcenter{\hbox{\includegraphics[scale=1.0]{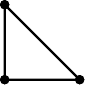}}}  \quad & = \ \text{suffix sums} & \\
\end{align*}
  \item Don't compute a value $R_i$ until it is needed in a $y_i$ calculation, or in another $R_{j}$ value needed by a window product calculation.
\end{enumerate}
We may now describe the Evict-then-Insert version of Two Stacks by the following diagram


\begin{center}
\includegraphics[scale=1.00]{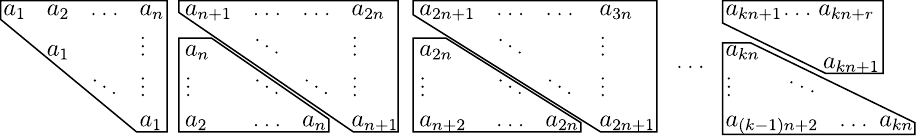} 
\end{center}

\noindent where $N = kn + r$. 
This decomposes the stacked staggered sequence diagram into triangular regions corresponding to prefix and suffix sums, and the Two Stacks algorithm is the algorithm resulting from this decomposition. 

We can also describe the naive algorithm using a stacked staggered sequence diagram. Here is the diagram for $n=4$, $N=10$.


\begin{center}
\includegraphics[scale=1.00]{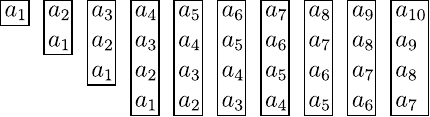}
\end{center}

\noindent In this case the regions do not correspond to prefix or suffix sums.

\section{Two Stacks Variants}
\label{sec:two-stacks-variants}

Graphically we can identify 4 obvious variants of Two Stacks corresponding to differences in the lengths of the triangles, i.e., differences in the lengths of the prefix and suffix sum calculations.

\begin{center}
\begin{tabular}{l m{3.0in} }
Variant & Diagram \\
\\
Evict-then-Insert &
\includegraphics[scale=1.00]{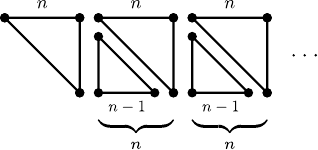} 
\vspace{0.125in}
\\
Combined-Evict-Insert \hspace{0.75in} &
\includegraphics[scale=1.00]{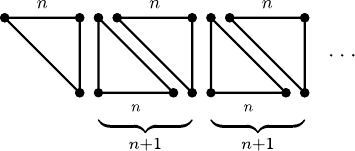} 
\vspace{0.125in}
\\
Variant 3 &
\includegraphics[scale=1.00]{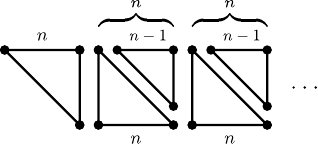}
\vspace{0.125in}
\\
Variant 4 &
\includegraphics[scale=1.00]{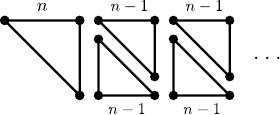}
\vspace{0.125in}
\end{tabular}    
\end{center}
The names of  the Evict-then-Insert and Combined-Insert-Evict variants relate to the development of Two Stacks in the work of Tangwongsan, Hirzel, and Schneider. In Tangwongsan et al.\ \cite{Tangwongsan2015b} \cite{Tangwongsan2017} \cite{Tangwongsan2021} the Two Stacks algorithm is developed through streaming versions that operate by means of three procedure calls named {\tt insert}, {\tt evict}, and {\tt query}. They present several implementation methods, based alternatively on a pair of stack data structures (hence the name Two Stacks), and a double-ended queue. Theodorakis et al.\ \cite{Theodorakis2018} describe an implementation based on a circular buffer data structure, which they call {\em Hammer Slide}. In each of these approaches there is a data structure containing a combination of input values and partial aggregations which is used to compute the window aggregation $a_i * \dots * a_{i-n+1}$ for some window $a_{i-n+1}, \ldots, a_i$, and the {\tt insert}, {\tt evict}, and {\tt query} procedures operate on this data structure. As the {\tt insert} and {\tt evict} procedures are called, items are added or removed from the window and the data structure with input values and aggregates updated accordingly. We describe these procedures in brief. 
\begin{itemize}{}
    \item[]{{\tt{insert($a$)}}:} Insert $a$ to the window and update the values and aggregates in the data structure necessary to compute the window aggregate for the new window. The window length is increased by one. 
    \item[]{{\tt{evict()}}:} Remove the least-recently-inserted item from the window and update the data structure accordingly, including removing values and aggregates no longer required to compute the new window aggregate. The window length is decreased by one.
    \item[]{{\tt{query()}}:} Compute the window aggregate from the values and aggregates in the data structure, and return the result.
\end{itemize}
%
%
We refer the reader to the papers of Tongwangsan et al.\ \cite{Tangwongsan2015b} \cite{Tangwongsan2017} \cite{Tangwongsan2021} for more details on the implementation of these procedures, and for purposes of description of behavior give brief implementation sketch here. We use Peter Landin's off-side rule \cite{Landin1966} to indicate the end of code blocks.

\begin{alltt}
initialization():
    b = 0
    queue = An empty array which allows removal on the left (popleft), and which allows
            appends (pushright) on the right. Items are accessed as queue[1],...,
            where queue[1] is the first item.

insert(a):
    prefix_sum = a if b = length(queue) else a * prefix_sum
    pushright(queue, a)

evict():
    popleft(queue)
    if b = 0 and length(queue) > 0
        for p = length(queue) - 1 to 1 step -1
            queue[p] = queue[p + 1] * queue[p]
    b = length(queue) if b = 0 else b - 1

query():
    return queue[1] if b = length(queue), else 
           prefix_sum if b = 0, else 
           prefix_sum * queue[1]
\end{alltt}

There are two evident approaches to computing window aggregations using {\tt insert}, {\tt evict}, and {\tt query}, which we call the insert-then-evict approach and the evict-then-insert approach. Both start with calls to {\tt insert} for the first $n$ window aggregates, but differ in the order of inserts and evicts for subsequent calculations.

\medskip
{\tt
\begin{center}
\begin{tabular}{lll}
$i$ \hspace{0.75in} & {\normalfont Insert-then-Evict} \hspace{0.5in} & {\normalfont Evict-then-Insert} \\
\hline \\
$1$ & insert($a_1$) & insert($a_1$) \\
 & $y_1=$ query() & $y_1=$ query() \\
 &\\
$2$ & insert($a_2$) & insert($a_2$) \\
& $y_2=$ query() & $y_2=$ query() \\
\vdots & \vdots & \vdots \\ \\
$n$ 
 & insert($a_n$) & insert($a_n$) \\
& $y_n=$ query() & $y_n=$ query() \\
&\\
 $n+1$ & insert($a_{n+1}$) & evict() \\
 & evict() & insert($a_{n+1}$)\\
 & $y_{n+1}=$ query() & $y_{n+1}=$ query() \\
 & \\
$n+2$ & insert($a_{n+2}$) & evict() \\
 & evict() & insert($a_{n+2}$) \\
 & $y_{n+2}=$ query() & $y_{n+2}=$ query() \\
\vdots & \vdots & \vdots \\
\end{tabular}
\end{center}
}
\noindent The order of {\tt evict} and {\tt insert} affects which calculations are performed and leads to algorithms with different batch lengths, i.e., different lengths of the prefix and suffix aggregates that are computed. Evict-then-Insert has the following pattern of computation.

\begin{center}
\includegraphics[scale=1.00]{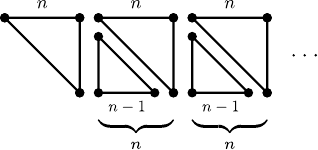}    
\end{center}
Note however that Tangwongsan et al.\ \cite{Tangwongsan2015b} \cite{Tangwongsan2017} \cite{Tangwongsan2021} contain an extra $*$ operation in {\tt evict} that is trivial to remove and we shall ignore this extra operation.%
\footnote{
Indeed, we have removed this operation from our implementation sketch.
}
%
Insert-then-Evict also has an occasional extra operation that is discarded and not used in the computation of a $y_i$ when the window length is fixed.

\begin{center}
\includegraphics[scale=1.00]{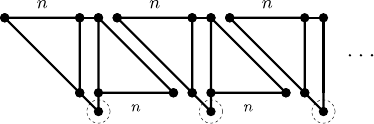}    
\end{center}
This extra operation occurs at $i=n+1, 2(n+1), 3(n+1), \ldots$ and is present because the {\tt insert} procedure is unaware of whether it will be immediately followed by an {\tt evict} operation, and so must prepare for a length $n+1$ window aggregation that may never be queried. This extra operation supports the ability of the algorithm to handle variable window lengths, and so is necessary to the implementations.%
\footnote{
One could of course defer the operation to the next {\tt query} but that would lead to an algorithm with different characteristics.
}
When the window length is fixed (after the first $n$ insertions), the extra $*$ operation in Insert-then-Evict can be avoided by using a fourth procedure, {\tt combined-insert-evict}, which performs insertion and eviction in the same procedure call. Adding {\tt combined-insert-evict} to the supported procedure calls does not hinder the ability to vary window length in these algorithms, though the details obviously depend on which data structures are used to implement Two Stacks. The general approach to {\tt combined-insert-evict} amounts to the following.

\begin{alltt}
combined-insert-evict(a):

    Detect if we are in a situation where an extra operation would occur. 
    
    If we are in such a situation then do not perform the operation but instead use a 
    dummy value, e.g., the input value. Alternatively, do not update the part of the data 
    structure that would have used the unneeded aggregate at all. 
    
    Perform the evict operation as usual.
\end{alltt}
Notice that this inserts {\tt a} and evicts the least recently inserted item from the window and updates the data structure accordingly. The window length is unchanged. In terms of our earlier implementation sketch, {\tt combined-insert-evict} can be implemented using the following pseudo-code.
\begin{alltt}
combined-insert-evict(a):
     if length(queue) > 0
        if b > 0
            insert(a)                                                      The usual case
        else
            pushright(queue, a)                  Insert a without updating the prefix sum
        evict()
\end{alltt}
With a combined insert-evict operation the computation of the $y_i$ proceeds as follows

\medskip
{\tt
\begin{center}
\begin{tabular}{ll}
$i$  \hspace{0.75in} & {\normalfont Combined-Insert-Evict} \\
\hline
\\
$1$ & insert($a_1$) \\
& $y_1=$ query() \\
\vdots & \vdots \\ \\
$n$ & insert($a_n$) \\
& $y_n=$ query() \\
\\
\hline
\\
$n+1$ &  
combined-insert-evict($a_{n+1}$) \\
& $y_{n+1}=$ query() \\
& \\
$n+2$ & 
combined-insert-evict($a_{n+2}$) \\
& $y_{n+2}=$ query()  \\
\vdots & \vdots \\ \\
\end{tabular}
\end{center}
}
\noindent Graphically the Combined-Insert-Evict variant can be represented as follows. 

\begin{center}
\includegraphics[scale=1.00]{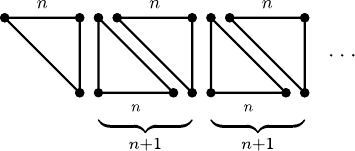}    
\end{center}
In the following when we refer to the Insert-Evict Variant, we will mean the Combined-Insert-Evict Variant with the addition of the extra unused operation at $i=n+1,2(n+1), 3(n+1), \cdots$. 

\section{Two Stacks Complexity}

The operation counts for all five Two Stacks Variants can be easily found from the stacked staggered sequence diagrams by counting the number of $*$ operations required to compute the prefix- and suffix-aggregates and to combine them. Analysis of the operation counts is easier to approach using incremental counts, which is  the number of additional $*$ operations required to compute $y_1, \ldots, y_N$ given that the algorithm has already computed up to $y_1, \ldots, y_{N-1}$. 

Let's first set some notations, and get a few trivial caves out of the way. Let
\begin{equation*}
\countfn_\text{CIE}(N) = \ourcases{\text{\parbox[t]{4.5in}{
The number of $*$ operations required to compute
the window aggregates $y_1,\dots, y_N$  using  
the Combined-Insert-Evict variant of Two Stacks
}}}
\end{equation*}
Similarly define $\countfn_\text{IE}$, $\countfn_\text{EI}$, $\countfn_\text{V3}$, and $\countfn_\text{V4}$  to be the number of $*$ operations required to compute $y_1,\dots, y_N$ using the Insert-Evict, Evict-Insert, Variant 3, and Variant 4 variants of Two Stacks respectively. 
Define the incremental operation counts as
\begin{equation*}
\incr_X(N)=\ourcases{
\countfn_X(N)                    & \text{if $N=1$} \\
\countfn_X(N)-\countfn_X(N-1)  & \text{if $N > 1$}
}
\end{equation*}
where $X$ is one of CIE, IE, EI, V3, or V4. We can now state some trivial cases.\\

\begin{lemma} \label{result:two-stacks-counts-trivial}
For the four variants Combined-Insert-Evict, Evict-Insert, Variant 3, Variant 4, we have
\begin{equation*}
\countfn_X(N)= \ourcases{
0   & \text{if $n=1$ or $N=1$} \\
N-1 & \text{if  $N \leq n$}
}
\end{equation*}
where $X$ is one of CIE, EI, V3, V4. For Insert-Evict
\begin{equation*}
\countfn_\text{IE}(N)=\ourcases{
\left\lfloor\frac{N}{2}\right\rfloor & \text{if $n=1$ or $N=1$} \\
N-1                                  & \text{if $N \leq n$}
}
\end{equation*}
\end{lemma}
\begin{proof}
Trivial.
\end{proof}

\noindent Let's now turn to the increments. These can be read off the stacked staggered sequence diagrams.

\begin{lemma} \label{lemma:two-stacks-increments}
Assume $n \geq 2$. Then the incremental count sequences for Two Stacks variants are as follows. 
\begin{align*}
\text{Combined-Insert-Evict} \hspace{0.5in} & 
\incr_\text{CIE} = 0, \overbrace{1,\dots,1}^{n-1}, \ourunderbracket{n-1, 1, \overbrace{2,\dots,2}^{n-2}, 1}, \ourunderbracket{n-1,1,\overbrace{2,\dots,2}^{n-2},1}, \ldots\\
\\
\text{Insert-Evict} \hspace{0.5in} & 
\incr_\text{IE} = 0, \overbrace{1,\dots,1}^{n-1}, \ourunderbracket{n, 1, \overbrace{2,\dots,2}^{n-2}, 1}, \ourunderbracket{n,1,\overbrace{2,\dots,2}^{n-2},1}, \ldots\\
\\
\text{Evict-Insert} \hspace{0.5in} & 
\incr_\text{EI} = 0, \overbrace{1,\dots,1}^{n-1}, \ourunderbracket{n-1, \overbrace{2,\dots,2}^{n-2}, 1}, \ourunderbracket{n-1,\overbrace{2,\dots,2}^{n-2},1}, \ldots\\
\\
\text{Variant 3} \hspace{0.5in} & 
\incr_\text{V3} = 0, \overbrace{1,\dots,1}^{n-1}, \ourunderbracket{n-1, 1, \overbrace{2,\dots,2}^{n-2}}, \ourunderbracket{n-1,1, \overbrace{2,\dots,2}^{n-2}}, \ldots\\
\\
\text{Variant 4} \hspace{0.5in}
& \incr_\text{V4} = 0, \overbrace{1,\dots,1}^{n-1},   \ourunderbracket{n-1, \overbrace{2,\dots,2}^{n-2}}, \ourunderbracket{n-1,\overbrace{2,\dots,2}^{n-2}}, \ldots\\
\end{align*}
\end{lemma}
\begin{proof}
These all follow from the diagrams. A prefix sum triangle \ $\hbox{\includegraphics[scale=0.75]{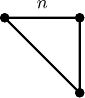}}$ corresponds to $0, \overbrace{1, \dots, 1}^{n-1}$ in the counts for the corresponding indexes. A suffix sum triangle \ $\vcenter{\hbox{\includegraphics[scale=0.75]{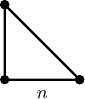}}}$ corresponds to $n-1, \overbrace{0, \dots, 0}^{n-1}$, and whenever there is a boundary between regions in a column we must add 1 to the count for that index. So for instance $\vcenter{\hbox{\includegraphics[scale=0.75]{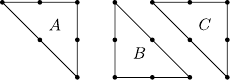}}}$ has incremental counts obtained by adding 3 sequences corresponding to the 3 regions, and adding in an additional sequence for the boundary between $B$ and $C$. I.e.,

\begin{center}
\begin{tabular}{cccccccc}
$A$      & $0$ & $1$ & $1$ & $0$ & $0$ & $0$ & $0$ \\
$B$      & $0$ & $0$ & $0$ & $2$ & $0$ & $0$ & $0$ \\
$C$      & $0$ & $0$ & $0$ & $0$ & $0$ & $1$ & $1$ \\
boundary & $0$ & $0$ & $0$ & $0$ & $1$ & $1$ & $0$ \\
\hline
total    & $0$ & $1$ & $1$ & $2$ & $1$ & $2$ & $1$ 
\end{tabular}
\end{center}

\noindent This gives a general method for calculating incremental operation  counts from diagrams where the regions are formed of triangles of the given shapes. The incremental operation counts can now be read from the diagrams.
\end{proof}

\begin{remark}
We have indicated batches by $\ourunderbracket{\qquad}$ marks on the sequences, and it is clear that the sequences are periodic after the startup batch for $y_1, \ldots, y_n$. The batch lengths and batch operation counts vary between the algorithms after startup. For $n \geq 2$, 
\begin{center}
{ 
\renewcommand{\arraystretch}{1.5}
\begin{tabular}{llllll} 
Algorithm & CIE & IE & EI & V3  & V4 \\ 
\hline 
Batch Length & $n+1$ & $n+1$ & $n$  & $n$ & $n-1$   \\
Batch Op.\ Count & $3 n-3$ & $3 n-2$ & $3 n-4$ &
$3 n-4$ & $3 n-5$   \\
Slope & $\frac{3 n-3}{n+1}$ & $\frac{3 n-2}{n+1}$ & $\frac{3 n-4}{n}$ & $\frac{3 n-4}{n}$& $\frac{3 n-5}{n-1}$\\
\end{tabular} 
} 
\end{center}
where the slope is simply the ratio of the batch operation count to the batch length. What this means is that the operation count function for each variant lies within a band of the given slope, after startup.

\begin{center}
\begin{tikzpicture}[x=0.03in, y=0.015in]
\draw[->] (-2, 0) -- (45,  0); \node[below] at (42, 0) {$N$} ; 
\draw[->] (0, -2) -- (0, 100); \node[above,rotate=90] at (0, 88) {$\text{count}$}; 

\draw ( 1,   0) -- (10,   9);  
\draw (10,   9) -- (11,  18);
\draw (11,  18) -- (12,  19);
\draw (12,  19) -- (20,  35);
\draw (20,  35) -- (21,  36);
\draw (21,  36) -- (22,  45);
\draw (22,  45) -- (23,  46);
\draw (23,  46) -- (31,  62);
\draw (31,  62) -- (32,  63);
\draw (32,  63) -- (33,  72);
\draw (33,  72) -- (34,  73);
\draw (34,  73) -- (42,  89);
\draw (42,  89) -- (43,  90);
\draw (43,  90) -- (44,  99);
\draw (44,  99) -- (45, 100);
\draw[dashed] (4.5,  -4.5) -- (48.5,  103.5); 
\draw[dashed] (8.25,  11.25) -- (46.75,  105.75); 

\end{tikzpicture}
\end{center}

\noindent These functions touch the top and bottom of their bands with period of the batch length. It turns out to be easy to calculate the bands exactly as functions of $n$, and the variant. But to get a flavor, here is a weaker result that holds for all algorithms simultaneously.
\end{remark}

\begin{theorem} \label{theorem:two-stacks-count-bounds}
For each of the 5 variants $X \in \theset{\text{CIE}, \text{IE}, \text{EI}, \text{V3}, \text{V4}}$ we have

\begin{equation*}
\countfn_X(N) = K_X N + c_X(N)
\end{equation*}
where $K_X$ and $c_X$ satisfy the following
\begin{center}
{ 
\renewcommand{\arraystretch}{1.5}
\begin{tabular}{llllll}
$X$ & CIE & IE & EI & V3 & V4 \\
\hline 
$K_X$ & $\frac{3 n-3}{n+1}$ & $\frac{3 n-2}{n+1}$ & $\frac{3 n-4}{n}$ & $\frac{3 n-4}{n}$ & $\frac{3 n-5}{n-1}$ 
\end{tabular}
} 
\end{center}
and
\begin{equation*}
    -2(n-1) \leq c_X(N) \leq -(n-1)
\end{equation*}
for $n \geq 2$ and $n \leq N$. 
\end{theorem}
\begin{proof} 
This will follow from Theorem~\ref{theorem:two-stacks-complexity} by looking at the operation counts at the start and end of each batch.
\end{proof}
\begin{remark}
This result tells us the width of the band is at most $n$. For each variant we can get a sharp bound. E.g., for CIE we have $-2 n+5-\frac{6}{n+1} \leq c_{\text{CIE}}(N) \leq-n+1$ which is sharp for $n \geq 2, n \leq N$. But we will not pursue this further here.
\end{remark}

\section{Cumulative Dominance and the Peter-Paul Lemma}
\label{sec:peter-paul}

To compare the operation counts of the variants, we borrow some ideas from the theory of majorization and stochastic dominance. 

\begin{definition} \label{definition-domination}
For any two finite or infinite sequences of real numbers, $a=a_1, a_2, \ldots$, $b=b_1, b_2, \ldots$, of the same length, We say that $b$ {\em cumulatively dominates} $a$, denoted $a \preccurlyeq b$, if all the partial sums of $a$ are less than the corresponding partial sums of $b$. I.e., 
\[\quad a_1+\ldots+a_i \leq b_1+\ldots+b_i  \text { for all } i\]
\end{definition}

\begin{example}
\[\begin{array}{l} 
{1,1,1,5} \preccurlyeq 1,2,2,3 \preccurlyeq 2,3,2,2  \preccurlyeq  5,1,1,1 \\
1,2,1,1,1\ldots \preccurlyeq  2,1,1,1,1\ldots
\end{array}
\]
\end{example}

\begin{remark} 
The $\preccurlyeq$ order looks similar to majorization, but as the example shows, it is not the same as majorization. In particular $\preccurlyeq$ is a partial order whereas majorization is only a partial order when restricted to increasing sequences. Instead $\preccurlyeq$ is an example of a cone order. (See Marshall and Olkin \cite{Olkin1979}, or Marshall Walkup and Wets \cite{Marshall1967}.) The name {\em cumulatively dominates} is new---Marshall, Walkup and Wets \cite{Marshall1967} call this cone order `Order 1a'.
\end{remark}

There are some obvious transformations of sequences that relate to cumulative domination.

\begin{definition} \label{definition:peter-paul}
Assume $a=a_1, a_2, \ldots$ is a finite or infinite sequence of real numbers, that $x \geq 0$, and that $i, j$, are strictly positive integers. If $a$ is finite then we also assume that $i, j$ are no greater than the length of $a$. Define
\begin{enumerate} 
\item A {\em gift transformation} is a transformation of the form 
\begin{equation*}
    \operatorname{Gift}_{i, x}(a)=a_1, \ldots, a_i+x, a_{i+1},\ldots  
\end{equation*}
which derives its name from the metaphor that the $i^\text{th}$ entry has $x$ `given' to it.

\item A {\em theft transformation} or {\em theft}, is a transformation of the form we

\begin{equation*}
\operatorname{Theft}_{i,x}(a)=a_1, \ldots, a_i-x, a_{i+1}, \ldots .
\end{equation*}
which derives its name from the metaphor that the $i^\text{th}$ entry has $x$ `stolen' from it. 

\item {\em A Peter-Paul transformation} is a transformation of the form
\begin{equation*}
\operatorname{PP}_{i, j, x}(a)=a_1, \ldots, a_i-x, a_{i+1}, \ldots, a_{j}+x, a_{j+1}, \dots
\end{equation*}
where $i<j$, noting that we `rob' $x$ from the earlier entry (Peter) and `give' $x$ to a later entry (Paul).

\item A {\em reverse Peter-Paul transformation} is a transformation of the form

\begin{equation*}
\operatorname{RPP}_{i, j}(a)=a_1, \ldots,  a_i + x, a_{i+1}, \ldots, a_j - x, a_{j+1}, \ldots
\end{equation*}
where $i<j$. In this case we `rob' $x$ from the later entry (Paul) and `give' $x$ to an earlier entry (Peter).

\item An {\em insertion} is a transformation of the form 
\begin{equation*}
\operatorname{Insertion}_{i,x}(a) = a_1, \ldots, a_{i-1}, x, a_i, a_{i+1}, \ldots 
\end{equation*}
in the case where $a$ is an infinite sequence. In the case where $a$ is a finite sequence an insertion has the form
\begin{equation*}
\operatorname{Insertion}_{i,x}(a) = a_1, \ldots, a_{i-1}, x, a_i, \ldots , a_{N-1}
\end{equation*}
where $N$ is the length of $a$.  An insertion is called a {\em low insertion} if $x \leq a_{j}$ for all $j \geq i$, where $j$ is an index of $a$. An insertion is called a {\em high insertion} if $x \geq a_{j}$ for all $j \geq i$. 
\end{enumerate}
\end{definition}

\begin{remark}
Hardy, Littlewood and Polya \cite{HardyLittlewoodPolya1934} call Peter-Paul transformations `transformations T', and Steele \cite{Steele2008} calls them `Robin Hood transformations'. Here we call them Peter-Paul transformations because we are not requiring that $a_i \geq a_{j}$. I.e.,  We are not `stealing from the rich to give to the poor', but rather `robbing Peter to pay Paul' and Peter may or may not be richer than Paul (but he must come first in the sequence).    
\end{remark}

Here is the main result about cumulative domination, which we call the Peter-Paul Lemma.

\begin{lemma}[Peter-Paul] \label{lemma:peter-paul}
Assume $a$ and $b$ are finite or infinite sequences, and in the case they are finite, assume they have the same length.
\begin{enumerate}
\item If $a$ is obtained from $b$ by a theft, or $b$ is obtained from $a$ by a gift, then $a \preccurlyeq b$. 
\item If $a$ is obtained from $b$ by a Peter-Paul transformation or $b$ is obtained from $a$ by a reverse Peter-Paul transformation, then $a \preccurlyeq b$. 
\item If $a$ is obtained from $b$ by $a$ low insertion, then $a \preccurlyeq b$. If $b$ is obtained from $a$ by a high insertion then $a  \preccurlyeq b$.
\item If $a$ and $b$ are finite and of the same length, then
\begin{displaymath}
\begin{array}{rcl}
a \preccurlyeq b & \Leftrightarrow & \text{\parbox[t]{4.5in}{
    $a$ may be obtained from $b$ by a finite number of Peter-Paul transformations and thefts.
}} \\
                 & \Leftrightarrow &  \text{\parbox[t]{4.5in}{
    $b$ may be obtained from $a$ by a finite number of  reverse Peter-Paul transformations and gifts.
}}
\end{array}
\end{displaymath}

\item If $a$ and $b$ are infinite, then 
\begin{displaymath}
\begin{array}{rcl}
 a  \preccurlyeq b &\Leftrightarrow  &  \text{\parbox[t]{4.5in}{
    $a$ may be obtained from $b$ by an infinite sequence of Peter-Paul transformations  $\operatorname{PP}_{i_1, j_1, x_1}, \operatorname{PP}_{i_2, j_2, x_2}, \ldots$ 
    such that the lower indices $i_k \rightarrow +\infty$.
}} \\
& \Leftrightarrow &  \text{\parbox[t]{4.5in}{
    $b$ may be obtained from $a$ by an infinite sequence of reverse Peter-Paul transformations  $\operatorname{RPP}_{i_1, j_1, x_1}, \operatorname{RPP}_{i_2, j_2, x_2}, \ldots$ 
    such that the lower indices $i_k \rightarrow +\infty$.
}}
\end{array}
\end{displaymath}
\end{enumerate}
\end{lemma}

\begin{remark}
Formally, if $a^{(k)}$ denotes the sequence obtained from $a$ after $k$ Peter-Paul transformations then we require that for any $N \geq 1$ then is an $l$, such that for $a_i^{(k)} = a_i^{(l)}$ for all $i \leq N$. I.e., the sequence of sequences $a^{(k)}$ stabilizes at any index $i$ for large enough $k$.
This condition is equivalent to convergence in the discrete product topology on the space of sequences, or equivalently the $I$-adic topology on the sequences viewed as formal power series for the ideal $I = x \mathbb{Z}[[x]]$. Intuitively this is saying that the sequence of transformed sequences obtained from $a$ should converge to $b$ in a `pointwise eventually constant' manner. It sounds technical but the intuition is simple.
\end{remark}
\begin{example}
\begin{displaymath}
0,1,2,3,4,\  \ldots \preccurlyeq 1, 2, 3, 4, 5,\ \ldots
\end{displaymath}
because we can rob from the second position to give to the first, then rob from the third position to give to the second, and so on. Pictorially
\begin{displaymath}
0 \xlongleftarrow{1} 1 \xlongleftarrow{1} 2 \xlongleftarrow{1} 3 \xlongleftarrow{1} 4 \xlongleftarrow{1} \cdots
\end{displaymath}
gives  $\quad 1 \quad 2 \quad 3 \quad 4 \quad 5 \ \ldots$
\end{example}
\begin{proof}[Proof of Lemma~\ref{lemma:peter-paul} (Peter-Paul Lemma)]
Assume $a$ and $b$ are finite or infinite sequences, and in the case they are finite, assume they have the same length.

\begin{enumerate}
\item Suppose $b=a_1, \ldots, a_i+x, a_{i+1}, \ldots$. Then 
\begin{align*}
b_1 + \dots + b_k & = \ourcases{
a_1 + \dots + a_k     & \text{if $k<i$}\\ 
a_1 + \dots + a_k + x & \text{if $k \geq i$}
}\\
& \geq a_1 + \dots + a_k \text{ if  $x \geq 0$} \\
& \leq a_1 + \dots + a_k \text{ if  $x \leq 0$}
\end{align*}

\item Suppose $a=b_1 \ldots, b_i - x, b_{i+1} \ldots, b_j +  x, b_{j+1}, \ldots$. Then 
\begin{align*}
a_1 + \dots + a_k & = \ourcases{
b_1 + \dots + b_k     & \text{if $k<i$ or $k \geq j$}\\ 
b_1 + \dots + b_k - x & \text{if $i \leq k < j$}
}\\
& \leq b_1 + \dots + b_k \text{ if  $x \geq 0$} \\
& \geq b_1 + \dots + b_k \text{ if  $x \leq 0$}
\end{align*}

\item We prove the low insertion case. The high insertion case is similar. Suppose $x \leq b_i, b_{i+1}, \ldots$, and $a=b_1,\ldots, b_{i-1}, x, b_i, b_{i+1}, \ldots$. Then
\begin{align*}
b_1 + \dots + b_k - \left(a_1 + \dots + a_k\right) & = \ourcases{
0       & \text{if $k < i$} \\
b_k - x & \text{if $k \geq i$}
}\\
& \geq  0
\end{align*}

\item The implication from the sequences of transforms to $a \preccurlyeq  b$ follow from 1. and 2. We now show that if $a \preccurlyeq b$ and $a, b$ are of finite length $N$, then $b$ can be obtained from $a$ by $N-1$ reverse Peter-Paul transforms and one gift. 
\begin{displaymath}
\begin{array}{rcl}
a_1, \ldots, a_N & \xrightarrow{\operatorname{RPP}} & a_1+\left(b_1-a_1\right)=b_1, a_2-\left(b_1-a_1\right), a_{3}, \ldots\\
& \xrightarrow{\operatorname{RPP}} &  b_1, a_2-\left(b_1-a_1\right)+\left[\left(b_1+b_2\right)-\left(a_1+a_2\right)\right]=b_2, a_{3}-\left[\left(b_1+b_2\right)-\left(a_1+a_2\right)\right], \ldots \\
& \xrightarrow{\operatorname{RPP}} & b_1, b_2, b_{3}, a_{4}-\left[\left(b_1+b_2+b_3\right)-\left(a_1+a_2+a_3\right)\right], a_5 \ldots \\
& \ldots & \\
& \xrightarrow{\operatorname{RPP}} & b_1, b_2, \ldots, b_{N-1}, b_N -\left[\left(b_1+\ldots+b_N\right)-\left(a_1+\ldots+a_N\right)\right]\\
& \xrightarrow{\operatorname{Gift}\;} & b_1, b_2, \ldots, b_{N-1}, b_N     
\end{array}
\end{displaymath}
A similar argument shows that $a$ may be obtained from $b$ by $N-1$ Peter-Paul transformations which are at indices $(1,2),(2,3), \ldots,(N-1, N)$, followed by a theft at index $N$.\\

\item In the infinite case, if $a \preccurlyeq b$, then to get from $a$ to $b$ we use the same set of reverse Peter-Paul transformations as in 4. but don't stop at index $N-1$, and instead carry on forever. The argument is similar, to get from $b$ to $a$ via an infinite sequence of Peter-Paul transformations which act at indexes $(1,2),(2,3),(3,4), \ldots$.
\end{enumerate}
\end{proof}

\begin{remark}
Note that in the proofs of 4.\ and 5.\ the sequence of transforms used may be explicitly calculated given $a$ and $b$. The Peter-Paul Lemma is essentially due to Hardy, Littlewood, and Polya \cite{HardyLittlewoodPolya1934}.
\end{remark}

\section{Further Two Stacks Complexity Results}

\begin{theorem}[Two Stacks Variant Complexity Comparison]
\label{theorem:two-stacks-count-inequalities}
For $n \geq 1$, $N \geq 1$, we have
\begin{equation*}
\countfn_\text{CIE}(N) \leq \countfn_\text{V3}(N) \leq \countfn_\text{EI}(N) \leq \countfn_\text{V4}(N) \leq 3 N
\end{equation*}
\end{theorem}
\begin{proof} 
We show that
\begin{equation*}
\incr_\text{CIE} \preccurlyeq \incr_\text{V3} \preccurlyeq \incr_\text{EI} \preccurlyeq \incr_\text{V4} \preccurlyeq 3,3,3, \ldots .
\end{equation*}
This follows immediately from the Peter-Paul Lemma. For each cumulative dominance relation we indicate how to get the previous (left-hand) sequence from the subsequent (right-hand) sequence.

\begin{equation*}
\begin{array}{ll}
0, \overbrace{1,\ldots,1}^{n-1}, n-1, 1, \overbrace{2,\ldots,2}^{n-2}, 1, n-1,1, \overbrace{2 \ldots 2}^{n-2}, 1, n-1, \ldots & \text{CIE}\\[0.125in]

\quad \preccurlyeq  0, \overbrace{1,\ldots,1}^{n-1}, n-1, 1, \overbrace{2,\ldots,2}^{n-2}, \ourunderbracket{\quad }, n - 1, 1,\overbrace{2\ldots,2}^{n-2}, \ourunderbracket{\quad }, n-1,  \ldots 
& \text{low insertion (V3)} \\[0.0625in]
\quad \preccurlyeq 0, \overbrace{1, \ldots, 1}^{n-1}, n-1 , 

\tikzmark{a}\overbrace{2,\ldots,2}^{n-2}, 1\tikzmark{b}, n-1, 
\begin{tikzpicture}[overlay,remember picture,out=315,in=225,distance=0.4cm]
    \draw[->,black,shorten >=3pt,shorten <=5pt] (a.center) to (b.center);
\end{tikzpicture}

\tikzmark{a}\overbrace{2,\ldots, 2}^{n-2}, 1\tikzmark{b} , n-1,\ldots
\begin{tikzpicture}[overlay,remember picture,out=315,in=225,distance=0.4cm]
    \draw[->,black,shorten >=3pt,shorten <=5pt] (a.center) to (b.center);
\end{tikzpicture} 

& \text{Peter-Paul (EI)}\\[0.0625in]

\quad \preccurlyeq  0, \overbrace{1,\ldots, 1}^{n-1}, n - 1, \overbrace{2,\ldots,2}^{n-2}, \ourunderbracket{\quad }, n-1, \overbrace{2,\ldots,2}^{n-2}, \ourunderbracket{\quad }, n-1, \ldots\ & \text{low insertion (V4)}\\[0.0625in]

\quad \preccurlyeq 0,\overbrace{2, \tikzmark{a}\ldots, 2}^{n-1}, 0\tikzmark{b}, 
\begin{tikzpicture}[overlay,remember picture,out=315,in=225,distance=0.4cm]
    \draw[->,black,shorten >=3pt,shorten <=6pt] (a.center) to (b.center);
\end{tikzpicture} 

\overbrace{3,\tikzmark{a}\ldots,3}^{n-2}, 1\tikzmark{b},
\begin{tikzpicture}[overlay,remember picture,out=315,in=225,distance=0.4cm]
    \draw[->,black,shorten >=3pt,shorten <=6pt] (a.center) to (b.center);
\end{tikzpicture} 

\overbrace{3,\tikzmark{a}\ldots,3,}^{n-2}  1 \tikzmark{b}, \ldots
\begin{tikzpicture}[overlay,remember picture,out=315,in=225,distance=0.4cm]
    \draw[->,black,shorten >=3pt,shorten <=6pt] (a.center) to (b.center);
\end{tikzpicture} 
& \text{Peter-Paul}\\[0.125in]
\quad \preccurlyeq 3,3,3,3, \ldots & \text{Thefts}
\end{array}
\end{equation*}
%
%
where $\ourunderbracket{\quad }$ indicates the location of insertions, and the curved underlining indicates the positioning of the `Peter-Paul' operations. 
\end{proof}

\noindent The case of Insert-Evict is more complicated.

\begin{theorem}
\ \nopagebreak
\begin{enumerate}
\item $\countfn_\text{CIE}  \leq  \countfn_\text{IE}$ for any window length $n$.
\item $\countfn_\text{V4} \leq  \countfn_\text{IE}$ for window length $n \leq 2$.
\item $\countfn_\text{IE}(N) \leq  \countfn_\text{V4}(N)$ for window length $n \geq 3$ and $N > n+1$.
\item For Evict-Insert and Variant 3, we have, with $X=\textrm{EI}$ or $X=\textrm{V3}$,
\begin{align*}
& \countfn_X         \leq \countfn_\text{IE} & &\text{for $n \leq 3$}\\
& \countfn_X - 1     \leq \countfn_\text{IE} \leq \countfn_X + 2 & &\text{for $n = 4$}\\
& \countfn_\text{IE} \leq \countfn_X + 2 & &\text{for $n \geq 4$}\\
& \countfn_\text{IE} \leq \countfn_X  & & \text{for $n = 5, N > 2n(n+1)$} \\
                               & & & \text{and for  $n \geq 6, N >  n(n+1)$.}
\end{align*}
\end{enumerate}
\end{theorem}

\begin{proof}
\ \noindent
\begin{enumerate}
\item $0, \overbrace{1,\ldots,1}^{n-1}, \ourunderbracket{n-1, 1, \overbrace{2, \ldots,2}^{n-2}, 1, } \ldots \preccurlyeq 0, \overbrace{1,\ldots,1}^{n-1}, \ourunderbracket{n, 1,\overbrace{2,\ldots,2,}^{n-2} 1,} \ldots$ by gifts and the Peter-Paul Lemma. 
\item
For cases $n=1, 2$ we observe that have
\begin{align*}
n=1\colon \quad & 0,0,0,0, \ldots \preccurlyeq 0,1,0,1,0,1,0,1, \ldots \\    
n=2\colon \quad & 0,1,1,1, \ldots \preccurlyeq 0,1,2,1,1,2,1,1,\ldots 
\end{align*}
\item
For the case $n \geq 3$ we observe first that 
\begin{align*}
\incr_\text{IE} & = 0, \overbrace{1,\ldots,1}^{n-1} , n, 1, \overbrace{2,\ldots,2}^{n-2}, 1, n, \ldots\\
\incr_\text{V4} & = 0, \overbrace{1,\ldots,1}^{n-1} , n-1, \overbrace{2, \ldots 2,}^{n-2} n-1, \ldots
\end{align*}
So it is clear that $\countfn_\text{IE}(n+2)=\countfn_\text{V4}(n+2)$, and therefore we only need prove that the sequence of increments for $i \geq n+3$ for V4 cumulatively dominates the sequence of increments for $i \geq n+3$ for IE. In other words we must show that
\begin{equation*}
\ourunderbracket{\overbrace{2,\ldots,2}^{n-2}, 1, n, 1,} \ourunderbracket{\overbrace{2,\ldots,2}^{n-2},1, n, 1,} \ldots \preccurlyeq 
\ourunderbracket{\overbrace{2,\ldots,2}^{n-3},n-1, 2,} \ourunderbracket{\overbrace{2,\ldots,2}^{n-3},n-1, 2,} \ldots 
\end{equation*}
But 
\begin{align*}
\overbrace{2,\ldots,2}^{n-2}, 1, n, 1, \overbrace{2,\ldots,2}^{n-2},1, n, 1, \ldots     
     & \preccurlyeq 
     \overbrace{2,\ldots,2}^{n-2}, 1, n, \ourunderbracket{\quad}, \overbrace{2,\ldots,2}^{n-2}, 1, n, \ourunderbracket{\quad},
     \ldots & & \text{low insertion}\\
      & \preccurlyeq  
     \overbrace{2,\ldots,2}^{n-3}, n - 1, 1, 3, 
     \overbrace{2,\ldots,2}^{n-3}, n - 1, 1, 3, 
     \ldots & & \text{Peter-Paul}\\    
      & \preccurlyeq  
     \overbrace{2,\ldots,2}^{n-3}, n - 1, 2, 2, 
     \overbrace{2,\ldots,2}^{n-3}, n - 1, 2, 2, 
     \ldots & & \text{Peter-Paul}\\   
     & \preccurlyeq  
     \overbrace{2,\ldots,2}^{n-3}, n - 1, 2, \ourunderbracket{\quad}, \overbrace{2,\ldots,2}^{n-3}, n - 1, 2, \ourunderbracket{\quad}, 
     \ldots & & \text{low insertion}
\end{align*}
%
The result follows by the Peter-Paul Lemma. 
\item The case $n \leq 3$ follows easily by the Peter-Paul Lemma.
The remaining cases for $n \geq 4$ follow by looking at the difference $\countfn_\text{IE} - \countfn_\text{EI}$. In general the difference sequence for $n \geq 4$ is
\begin{equation*}
\begin{array}{c}
\overbrace{0,\ldots,0}^{n}, 1,  \overbrace{0,\ldots,0}^{n-2},3-n, 1, 
\overbrace{0,\ldots,0}^{n-3}, 1, 4-n, 3-n, 1,  \overbrace{0,\ldots,0}^{n-4},1, 4-n, 4-n, 3-n, 1, \overbrace{0,\ldots,0}^{n-5} \ldots\\
\ldots, 1, 0, 1, \overbrace{4-n,\ldots,4-n}^{n-3}, 3-n, 1, 1,   \overbrace{4-n,\ldots,4-n}^{n-2},3-n, \underset{\substack{ \uparrow\\ n(n+1)}}{2} ,\
 \overbrace{4-n,\ldots,4-n}^{n }
 \end{array}
\end{equation*}
%
This sequence repeats from position $n(n+1) + 1$ but shifted $4-n$ lower.
The maximum of the sequence is therefore $+2$ which occurs at $N=n(n+1)$. This is the last strictly positive value when $n \geq 6$. When $n=5$ the last strictly positive value is $+1$, which occurs at $N = 2n(n+1)$. The proof for $X = \text{V3}$ is similar.
\end{enumerate}
\end{proof}

\noindent We now give formulae for the operation counts.
\begin{theorem} 
\label{theorem:two-stacks-complexity}

Assume $2 \leq n<N$, then 

\begin{enumerate}
    
\item
$\begin{aligned}[t]
\countfn_{\text{CIE}} & =k(3n - 3) - n + 1 + (r>0)(2 r-1-(r=n)) \\
                       & =3N - 6k - n + 1 - (r>0)(r+1+(r=n))
\end{aligned}$

\noindent where $N=k(n+1)+r$ and $0 \leq r<n+1$. I.e., $k=\left\lfloor\frac{N}{n+1}\right\rfloor, r=N \bmod (n+1)$.

\item
$\begin{aligned}[t]
\countfn_\text{IE}  & = k(3n - 2) - n + 1 + (r>0)(2r - 1 - (r=n)) \\
                    & = 3N - 5k - n + 1   - (r>0)(r + 1 + (r=n))
\end{aligned}$

\noindent where $N=k(n+1)+r$ and $0 \leq r<n+1$. I.e., $k=\left\lfloor\frac{N}{n+1}\right\rfloor, r=N \bmod (n+1)$.

\item
$\begin{aligned}[t]
\countfn_\text{EI}  & = k(3n - 4) - 2n + 3 + (r>0)(n + 2r - 3) \\
                    & = 3N - 4k - 2n + 3   + (r>0)(n - r - 3)
\end{aligned}$

\noindent where $N=kn + r$ and $0 \leq r<n$. I.e., $k=\left\lfloor\frac{N}{n}\right\rfloor, r=N \bmod n$.

\item
$\begin{aligned}[t]
\countfn_\text{V3} & = k(3n - 4) - 2n + 3 + (r>0)(n + 2r - 4 + (r=1)) \\
                   & = 3N - 4k - 2n + 3   + (r>0)(n - r -4 + (r=1))
\end{aligned}$

\noindent where $N=kn+r$ and $0 \leq r<n$. I.e., $k=\left\lfloor\frac{N}{n}\right\rfloor, r=N \bmod n$.

\item
$\begin{aligned}[t]
\countfn_\text{V4} & = k(3n - 5) - 2n + 4 + (r>0)(n + 2r - 3) \\
                   & = 3N - 2k - 2n + 1   + (r>0)(n - r - 3)
\end{aligned}$

\noindent where $N-1 = k(n-1)+r$ and $0 \leq r < n - 1$. I.e., $k=\left\lfloor\frac{N-1}{n-1}\right\rfloor, r=(N-1) \bmod (n-1)$.
\end{enumerate}

\noindent where the relational operators in the formulae take the values $1$ and $0$ for `true' and `false' respectively.
\end{theorem}
\begin{proof}
The 5 cases are similar, so we only give the proof of 1.\ here. It helps to refer to the staggered stacked sequence diagram.

\begin{center}
\includegraphics[scale=1.00]{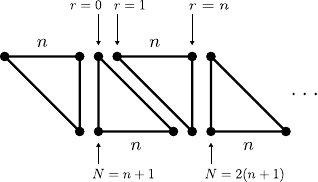}    
\end{center}
Within that diagram, there are two cases to consider.

\begin{itemize} 
\item[Case] $N=k(n+1)$:
This is the case where $r=0$. We know that at $N=n+1$ the operation count is $2(n-1)$, and also that the operation count for each batch of $n+1$ window sums is $3 n-3$. Therefore,
\begin{align*}
\countfn_\text{CIE}(N) & = 2(n-1) +(k-1)(3n - 3) \\
                       & = k(3n - 3) - n + 1
\end{align*}

\item[Case] $N=k(n+1)+r, \quad 0 \leq r<n+1$: 
For this case we must add the extra operations required when $r>0$. When $r>0 $ there are $r-1$ extra operations required to compute the prefix aggregate. When $0<r<n$ there are $r$ operations required to add the prefix aggregate to the already computed suffix aggregate, but when $r=n$ we do not require this operation. Hence, the additional operation count is $(r>0)(2r - 1 + (r=n))$. Hence 
\begin{equation*}
\countfn_\text{CIE}(N) = k(3n - 3) - n + 1 + (r>0)(2r - 1 + (r=n))
\end{equation*}

To get the second form of the formula, substitute $kn = N - k - r$ into this formula.
\end{itemize}

\noindent This proves 1. Cases 2., 3., 4., and 5.\ are similar. 
\end{proof}

\section{Two Stacks Properties}
\label{sec:two-stacks-properties}

We now look at how Two Stacks performs relative to the properties listed in Section~\ref{sec:algorithm-properties}. 

\begin{description}
\item[Correctness] \ \\ \nopagebreak
Two Stacks produces correct results when $*$ is associative. 

\item[Accuracy] \ \\ \nopagebreak
Two Stacks fares well on accuracy. In the case where $*$ is approximately associative, note that Two Stacks performs exactly $n-1$ $*$-operations in the computation of each window sum $y_i$, for $i \geq n$, and there are no cancellations as we saw with Subtract-on-Evict. The error analysis depends on $*$, but if each $*$ operation introduces an error $\varepsilon$, and errors compound linearly, then the error in $y_i$ will be bounded by $(n-1) \varepsilon$. If the prefix and suffix aggregates are computed using parallel prefix and suffix sum algorithms with depth $\left\lceil\log_2 n\right\rceil$, then the error will be bounded by $\left(2 \left\lceil\log_2 n \right\rceil + 1\right) \varepsilon$. So Two Stacks is accurate.

\item[Efficiency] \ \\ \nopagebreak
The number of $*$ operations is bounded by $3N$, and the bookkeeping is linear in $N$.

\item[Simplicity] \ \\ \nopagebreak
The algorithm is simple. There are short implementations in code, and the algorithm has a straightforward description.

\item[Freedom from extraneous choices or data] \ \\ \nopagebreak
The only quantities used in the computation of $y_i$ are
$a_{i-n+1}, \ldots, a_i$.

\item[Streaming] \ \\ \nopagebreak
There are streaming implementations based on the {\tt insert}, {\tt query}, {\tt evict}, and {\tt{combined-insert-evict}} procedures.

\item[Memory] \ \\ \nopagebreak
Both batch and streaming versions can be implemented using $n$ items of working space to store aggregates and window item values, plus possibly an additional item used when combining the prefix and suffix aggregates.

\item[Generalizability] \ \\ \nopagebreak
The algorithm works for any (computable) associative operator, and does not require other properties such as commutativity, invertibility, or being `max-like' (i.e., a `selection operator').

\end{description}
This leaves parallelizability, vectorizability, and latency, which we now address.
\medskip

\begin{description}
\item[Two Stacks Parallelizability] \ \\ \nopagebreak
There are two direct approaches to parallelizing Two Stacks, both of which are discussed in Snytsar and Turakhia \cite{SnytsarTurakhia2019}, and Snytsar \cite{Snytsar2023b}. The first approach is to break up the calculation according to batches. This works in a similar fashion for all variants, and we illustrate this for the Combined-Insert-Evict variant.
 
\begin{center}
\begin{tabular}{  m{0.6875in} |  m{0.75in} |  m{0.75in} | m{0.75in} | m{0.5in} }
batch 1 & batch 2 & batch 3 & batch 4 & $\cdots$ \\
\includegraphics[scale=1.00]{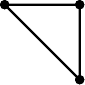} &
\includegraphics[scale=1.00]{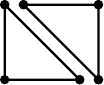} &
\includegraphics[scale=1.00]{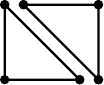} &
\includegraphics[scale=1.00]{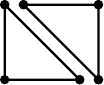} &
$\cdots$ \\
processor 1 & processor 2 & processor 3 & processor 4 & $\cdots$
\end{tabular}    
\end{center}
Although the batches have overlapping input data they do not share intermediate or output computation and so may be computed in parallel. A second parallelization opportunity arises from using parallel algorithms to compute the prefix and suffix aggregates. This parallelizes within each batch

\begin{center}
\newcommand\topstrut{\rule{0pt}{2.6ex}}
\begin{tabular}{  m{0.6875in} |  m{0.6875in} |  m{0.6875in} | m{0.5in} }
batch 1 & batch 2 & batch 3 &  $\cdots$ \\[1ex]
parallel prefix sum & parallel prefix sum & parallel prefix sum & $\cdots$ \\[0.25ex]
 \hspace{0.1875in} $\big\downarrow$ &  
 \hspace{0.3125in} $\big\downarrow$ &  
 \hspace{0.3125in} $\big\downarrow$ & \\[0.25ex]
\includegraphics[scale=1.00]{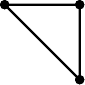} &
\includegraphics[scale=1.00]{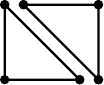} &
\includegraphics[scale=1.00]{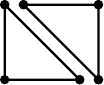} &
$\cdots$ \\
& \hspace{0.1875in} $\big\uparrow$ &  \hspace{0.1875in} $\big\uparrow$ & \\[0.5ex]
& parallel suffix sum & parallel suffix sum & 
$\cdots$
\end{tabular}    
\end{center}
Both these approaches require only associativity to work and do not rely on commutativity or other properties of $*$.

\item[Two Stacks Vectorizability] \ \\ \nopagebreak
The Two Stacks algorithm is vectorizable within each batch, as shown by the second parallelization approach. To vectorize across the entire sequence of inputs $a_1, \ldots, a_N$, and outputs $y_1, \ldots, y_N$ requires vector operations which handle the boundaries between batches. In essence this requires a `shift within batch' operation. 

\item[Two Stacks Latency] \ \\ \nopagebreak
The Two Stacks algorithm requires less than $3N$ $*$-operations to compute the $N$ window aggregates $y_1, \ldots, y_N$, but as highlighted in Tangwongsan et al.\ \cite{Tangwongsan2015b} \cite{Tangwongsan2017}, the number of operations required to compute each additional $y_i$ is not constant, and instead spikes periodically. We saw this in the Lemma~\ref{lemma:two-stacks-increments} in the incremental operator counts. For example, for the Combined-Insert-Evict Variant the incremental counts are

\begin{equation*}
\incr_\text{CIE} = 0, \overbrace{1, \ldots, 1}^{n-1}, {\underbrace{n-1}_{\substack{\uparrow \\ \text{spike}}}},1, \overbrace{2,\dots,2}^{n-2},1, 
{\underbrace{n-1}_{\substack{\uparrow \\ \text{spike}}}}, 1\ldots
\end{equation*}
In streaming applications this causes latency spikes at items $i=n+1,2(n+1), 3(n+1), \ldots$, when $n$ is large. 

\end{description}
\bigskip

In the next section we describe a new algorithm, the Double-Ended Window, or DEW, algorithm that has similar complexity to Two Stacks, with operation counts bounded by $3 N$, but which has all incremental counts $\leq 3$, and hence does not suffer from latency spikes.

%% file: htcams-arxiv-ch04-dew.tex
\chapter{A New Algorithm: The Double-Ended Window (DEW) Algorithm}
\label{chapter:dew}

This is a new algorithm which addresses the latency spike problem of Two Stacks while preserving the efficiency of Two Stacks. In this chapter we again assume that $*$ is an associative binary operator, and our goal is to compute the moving sums (or moving products)
\begin{equation*}
y_i = \ourcases{
    a_i * \ldots * a_1     & \quad \text{ for $1 \leq i < n$} \\
    a_i * \ldots * a_{i-n+1} & \quad \text{ for $i \geq n$}
} 
\end{equation*}
The operator $*$ is not assumed to be commutative, or to have inverses, or other properties, unless otherwise stated.

\section{Stacked Staggered Sequence Diagrams Again}

In order to develop window aggregation algorithms with improved properties, we need to increase our repertoire of regions we can interpret and compute efficiently.

\medskip
\begin{center}
{\bf{An initial repertoire}} 
\nopagebreak

\begin{longtable}
{m{1.0in} m{0.5in} |m{1.25in}|m{1.75in}|c}
Region &  & Interpretation & Column Aggregates &  Complexity \\
       &  &                & (left to right)   &  \\
\endfirsthead
Region &  & Interpretation & Column Aggregates &  Complexity \\
       &  &                & (left to right)   &  \\
\endhead
\parbox[b]{1.0in}{\includegraphics[scale=1.00]{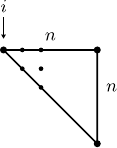}} & &  
prefix sum &  \parbox[b]{1.25in}{
\begin{equation*}
\begin{split}
& a_i \\
& a_{i+1} * a_i\\
& \vdots \\
& a_{i+n-1} * \cdots * a_i
\end{split}
\end{equation*}
} & $n-1$\\
\parbox[b]{1.0in}{\includegraphics[scale=1.00]{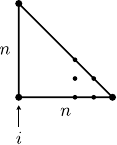}} & &  suffix sum &  \parbox[b]{1.25in}{
\begin{equation*}
\begin{split}
& a_i * \cdots * a_{i-n+1} \\
& a_i * \cdots * a_{i-n+2}\\
& \vdots \\
& a_i
\end{split}
\end{equation*}
}
& $n-1$\\
\parbox[b]{1.0in}{\includegraphics[scale=1.00]{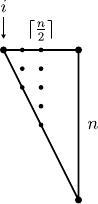}} & $n$ odd &  
double-ended sum starting from a single point &  \parbox[b]{1.75in}{
\begin{equation*}
\begin{split}
& a_i \\
& a_{i+1} * a_i * a_{i-1}\\
& \vdots \\
& a_{i+ \lceil\frac{n}{2}\rceil - 1} * \cdots * a_{i - \lceil\frac{n}{2}\rceil + 1}
\end{split}
\end{equation*}
}
& $n-1$\\
\parbox[b]{1.0in}{\includegraphics[scale=1.00]{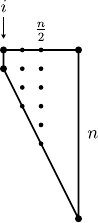}}  & $n$ even &  
double-ended sum starting from two points &  \parbox[b]{1.5in}{
\begin{equation*}
\begin{split}
& a_i * a_{i-1}\\
& a_{i+1} * a_i * a_{i-1} * a_{i-2}\\
& \vdots \\
& a_{i + \frac{n}{2} - 1} * \cdots * a_{i - \frac{n}{2}}
\end{split}
\end{equation*}
}
& $n-1$\\
\end{longtable}
\end{center}

\section{Flips and Slides}

In a stacked staggered sequence diagram there can clearly be regions that represent the same calculations, and there are transformations of regions which preserve the calculations the regions represent. We will use two such transformations, which we call slides and flips. The slide is self explanatory. If we slide any region down and to the right, or up and to the left, along a $45^{\circ}$ sloped line, the calculation it represents is unchanged.

\begin{center}
\includegraphics[scale=1.00]{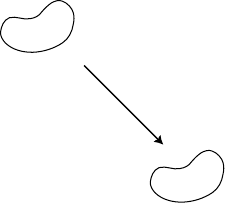}
\hspace{0.5in}
\includegraphics[scale=1.00]{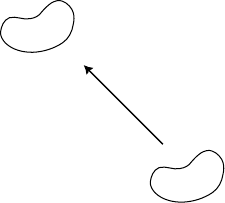}
\end{center}
A single illustration suffices to demonstrate the principle.


\begin{center}
\includegraphics[scale=1.00]{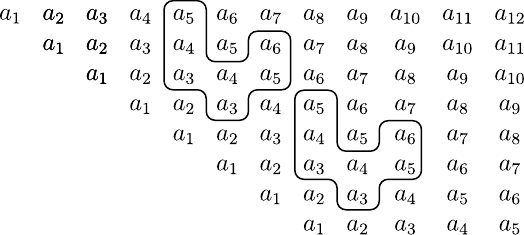}
\end{center}
%
Slides preserve the shape of the region as well as the column aggregates of the region.

Flips are only slightly more complicated than slides. For a flip we slide the individual columns of the region so as to reverse their order.

\begin{center}
\includegraphics[scale=1.00]{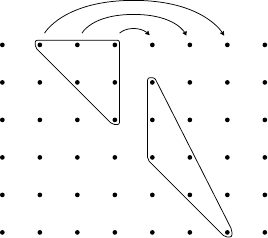}
\end{center}
%
Notice how this skews the shape of the region and preserves the column aggregates while reversing their order.

By combining flips and slides we can expand our repertoire of regions further.

\medskip
\begin{center}
{\bf{An expanded repertoire}}
\nopagebreak

\begin{longtable}{m{1.5in} m{0.5in}|m{1.25in}|c}
Region & & Interpretation & Complexity \\
\endfirsthead
Region & & Interpretation & Complexity \\
\endhead
\parbox[c]{1.5in}{\includegraphics[scale=1.00]{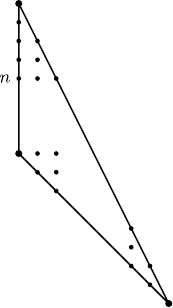}} &  &
flipped prefix sum & $n-1$\\ \vspace{0.125in}
\parbox[c]{1.5in}{\includegraphics[scale=1.00]{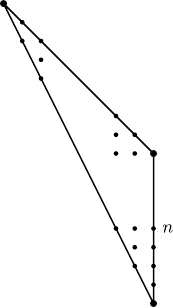}} &  &
flipped suffix sum & $n-1$\\ \vspace{0.125in}
\parbox[c]{1.5in}{\includegraphics[scale=1.00]{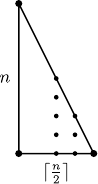}} & $n$ odd  &  
flipped single point double-ended sum & $n-1$\\ \vspace{0.125in}
\parbox[c]{1.5in}{\includegraphics[scale=1.00]{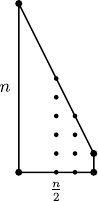}} & $n$ even  &  
flipped two point double-ended sum & $n-1$\\ \vspace{0.125in}
\parbox[c]{1.5in}{\includegraphics[scale=1.00]{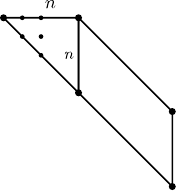}} &  &
prefix sum with slide of final column & $n-1$\\ \vspace{0.125in}
\parbox[c]{1.5in}{\includegraphics[scale=1.00]{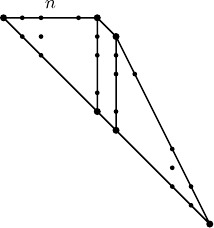}} &  &
prefix sum with flip & $n-1$\\ \vspace{0.125in}
\parbox[c]{1.5in}{\includegraphics[scale=1.00]{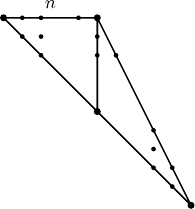}} & &
prefix sum with flip around final column & $n-1$\\ \vspace{0.125in}
\parbox[c]{1.5in}{\includegraphics[scale=1.00]{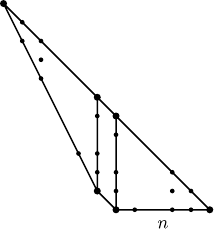}} & &
suffix sum with flip & $n-1$\\ \vspace{0.125in}
\parbox[c]{1.5in}{\includegraphics[scale=1.00]{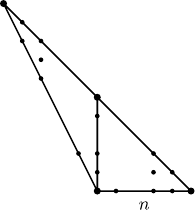}} &  & 
suffix sum with flip around final column & $n-1$\\ \vspace{0.125in}
\parbox[c]{1.5in}{\includegraphics[scale=1.00]{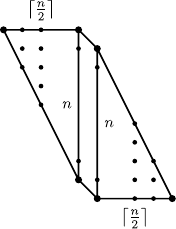}} & $n$ odd  & 
single point double-ended sum with flip & $n-1$\\ \vspace{0.125in}
\parbox[c]{1.5in}{\includegraphics[scale=1.00]{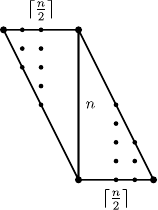}} & $n$ odd & 
single point double-ended sum with flip around final column & $n-1$\\ \vspace{0.125in}
\parbox[c]{1.5in}{\includegraphics[scale=1.00]{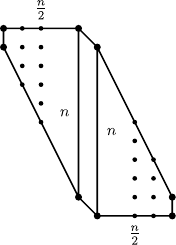}} & $n$ even & 
two point double-ended sum with flip & $n-1$\\ \vspace{0.125in}
\parbox[c]{1.5in}{\includegraphics[scale=1.00]{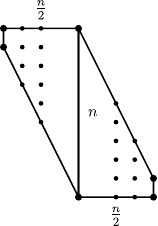}} & $n$ even & 
two point double-ended sum with flip around final column & $n-1$\\ \vspace{0.125in}

\end{longtable}
\end{center}

\section{The DEW Algorithm Graphically}
\label{sec:dew-graphically}

The idea of the DEW algorithm is to use double-ended sums and flipped double-ended sums to fill out the stacked staggered sequence diagram for window aggregates. Roughly the idea is the following. picture

\vspace{0.25in}
\begin{tabular}{m{1in}m{4.3in}}
Idea of DEW & 
\includegraphics[scale=1.00]{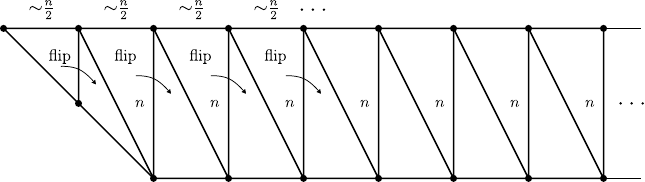}
\end{tabular}
\vspace{0.25in}

\noindent As with Two Stacks the algorithm proceeds by computing the window aggregates in batches, though this time the batches have size ${\sim} \frac{n}{2}$ and each batch sets up the next batch via a flip. The diagram above is not precise, but we can already get a rough complexity estimate. Each batch of length ${\sim}\frac{n}{2}$ (after the first) requires $n-1$ operations to compute the double-ended aggregates and ${\sim} \frac{n}{2}$ operations to combine the double-ended aggregates with the flipped double-ended aggregates from the previous batch. Hence we expect ${\sim} 3 N$ operations to compute the window aggregates $y_{1}, \ldots, y_{N}$. So based on this idea we expect to have an algorithm competitive with Two Stacks. In practice there are slight differences between batches depending on whether $n$ is odd or even---after all if $n$ is odd then $\frac{n}{2}$ is not an integer and cannot be a batch size. We now account for these details.

As with Two Stacks, DEW comes in several variants, but this time the stacked staggered sequence diagrams also depend on whether $n$ is odd or even. We also list the operation count increments in the same table.
\medskip
\begin{center}
{\bf{DEW Variants}}
\nopagebreak \medskip

\begin{tabular}{m{0.55in} m{2.9in} m{2.55in}}
Variant 1 & Diagram & Increments \\
\vspace{0.25in}
$n$ even &
\vspace{0.125in}
\includegraphics[scale=1.00]{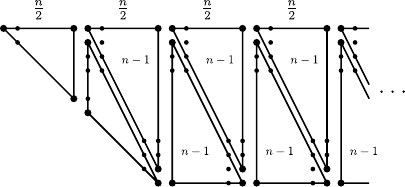} & 
$0, \overbrace{1, \dots, 1}^{\frac{n}{2} - 1}, 
\ourunderbracket{1, \overbrace{3, \dots, 3}^{\frac{n}{2} - 1},}_{\frac{n}{2}}
\ourunderbracket{1, \overbrace{3, \dots, 3}^{\frac{n}{2} - 1},}_{\frac{n}{2}}
\dots$
\\
\vspace{0.25in}
\vspace{0.1in}
$n$ odd &
\vspace{0.1in}
\includegraphics[scale=1.00]{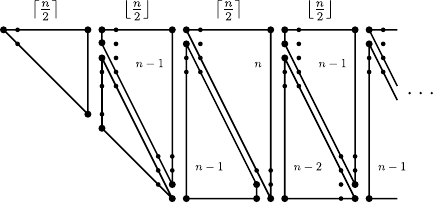} &
\vspace{0.1in}
$0, \overbrace{1, \dots, 1}^{\lceil\frac{n}{2}\rceil - 1}, 
\ourunderbracket{
\ourunderbracket{2, \overbrace{3, \dots, 3}^{\lceil\frac{n}{2}\rceil - 2},}_{\lfloor\frac{n}{2}\rfloor} 
\ourunderbracket{1, \overbrace{3, \dots, 3}^{\lceil\frac{n}{2}\rceil - 2}, 2,}_{\lceil\frac{n}{2}\rceil}
}_n
\dots
$
\end{tabular}
\begin{tabular}{m{0.55in} m{2.9in} m{2.55in}}
Variant 2 & & \\
$n$ even &
\includegraphics[scale=1.00]{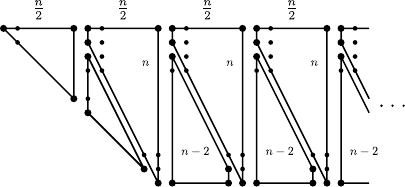} &
$0, \overbrace{1, \dots, 1}^{\frac{n}{2} - 1}, 
\ourunderbracket{2, \overbrace{3, \dots, 3}^{\frac{n}{2} - 2}, 2}_{\frac{n}{2}}
\ourunderbracket{2, \overbrace{3, \dots, 3}^{\frac{n}{2} - 2}, 2}_{\frac{n}{2}}
\dots$

\\
\vspace{0.25in}
\vspace{0.1in}
$n$ odd &
\vspace{0.1in}
\includegraphics[scale=1.00]{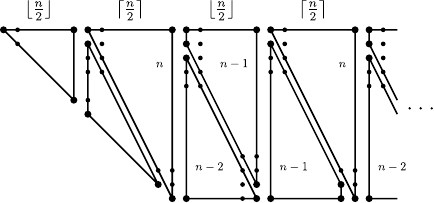} &
\vspace{0.1in}
$0, \overbrace{1, \dots, 1}^{\lceil\frac{n}{2}\rceil - 2}, 
\ourunderbracket{
\ourunderbracket{1, \overbrace{3, \dots, 3}^{\lceil\frac{n}{2}\rceil - 2}, 2,}_{\lceil\frac{n}{2}\rceil}
\ourunderbracket{2, \overbrace{3, \dots, 3}^{\lceil\frac{n}{2}\rceil - 2},}_{\lfloor\frac{n}{2}\rfloor} 
}_n
\dots
$
\end{tabular}
\end{center}
These diagrams above show 4 patterns of calculation, but we have grouped these into 2 variants to correspond to the (circular) algorithms to be discussed in Section~\ref{sec:dew-implementation-sketches}. Another way to understand the grouping is to note that the double-ended aggregates in the algorithms either start with single or double points, and these follow different patterns.

\begin{center}
\begin{tabular}{lllllll}
Variant & parity & \multicolumn{5}{l}{batch double sum type}  \\
\hline \\[-2ex]
Variant 1 & $n$ even & start & single & single & single & \ldots \\
          & $n$ odd  & start & double & single & double & \ldots \\
Variant 2 & $n$ even & start & double & double & double & \ldots \\
          & $n$ odd  & start & single & double & single & \ldots \\
\end{tabular} 
\end{center}
How should the start batches fit into this pattern? For some implementations of DEW it makes most sense to think of these as simply different kinds of batches---they correspond to single-ended prefix sums rather than double-ended prefix sums after all. For other implementations it makes sense to think of the start batches as regular batches with some missing data%
\footnote{
The missing data would correspond to not-yet-filled memory locations.
}
and to extend the pattern backwards. I.e.,

\begin{center}
\begin{tabular}{lllllll}
Variant & parity & \multicolumn{5}{l}{batch double sum type}  \\
\hline \\[-2ex]
Variant 1 & $n$ even & single\hspace{1ex} (start) & single & single & single & \ldots \\
          & $n$ odd  & single\hspace{1ex} (start) & double & single & double & \ldots \\
Variant 2 & $n$ even & double (start) & double & double & double & \ldots \\
          & $n$ odd  & double (start) & single & double & single & \ldots \\
\end{tabular} 
\end{center}
So Variant 1 starts the pattern with `single' and Variant 2 starts with `double'. This is a somewhat arbitrary grouping, and indeed there are implementations of DEW where it is natural to instead group Variant 2, $n$ even with Variant 1, $n$ odd and Variant 1, $n$ even with Variant 2, $n$ odd. We can, however, offer two other observations suggesting that our given grouping is more natural, or at least more convenient. Firstly that with this grouping the lengths of the start batches match the lengths of batch $l$ for all odd $l$ for the same variant (where the start batch is batch 1), and secondly that the Variant 1 increments cumulatively dominate the Variant 2 increments. These make statements about the algorithm's complexity less complicated.

\section{DEW Complexity}

First some notation. Let $\countfn_\text{DEW,V1}$ and $\countfn_\text{DEW,V2 }$ denote the $*$-operation count functions for DEW Variant 1 and DEW Variant 2, and let $\incr_\text{DEW,V1}$ and  $\incr_\text{DEW,V2}$ denote the corresponding incremental count functions. The incremental count functions were given in the DEW Variants table in Section~\ref{sec:dew-graphically}, and can also be read off the diagrams in that section. Let's start with some simple observations.

\begin{theorem}[DEW Variant Complexity Comparison]
\label{theorem:dew-count-inequalities}
\ \nopagebreak

\begin{enumerate} 
\item
$\incr_\text{DEW,V1} \leq 3$, and $\incr_\text{DEW,V2} \leq 3$.
I.e., DEW has no latency spikes.

\item The slope of $\countfn_\text{DEW,V1}$ and $\countfn_\text{DEW,V2}$ after startup is $\frac{3 n-4}{n}$, for $n \geq 2$. I.e.,
\begin{align*}
\countfn_\text{DEW,V1}(N+n)  &=\countfn_\text{DEW,V1}(N) + 3n - 4 \\
\countfn_\text{DEW,V2}(N+n)  &=\countfn_\text{DEW,V2}(N) + 3n - 4
\end{align*}
for $N \geq \left\lceil\frac{n}{2}\right\rceil$, $n\geq 2$.

\item 
$\countfn_\text{DEW,V1} \leq \countfn_\text{DEW,V2} \leq (N \longmapsto 3 N)$

\item 
$\countfn_\text{DEW,V2} \leq \countfn_\text{DEW,V1} + 1$
\end{enumerate}
\end{theorem}
\begin{proof}
1.\ and 2.\ follow directly from the increments. 3.\ follows from the Peter-Paul Lemma. We handle the cases $n$ even and $n$ odd separately.
\begin{align*}
&\text{$n$ even \qquad} &
\overbrace{0, 1\ldots, 1}^{\frac{n}{2}},\overbrace{1, 3\ldots, 3}^{\frac{n}{2}}\overbrace{1, 3\ldots, 3}^{\frac{n}{2}}, \ldots & \preccurlyeq
\overbrace{0, 1\ldots, 1}^{\frac{n}{2}},
\overbrace{\tikzmark{a} 2, 3\ldots, 3, 2\tikzmark{b}}^{\frac{n}{2}},
\begin{tikzpicture}[overlay,remember picture,out=315,in=225,distance=0.4cm]
    \draw[->,black,shorten >=3pt,shorten <=5pt] (a.center) to (b.center);
\end{tikzpicture} 
\overbrace{\tikzmark{a} 2, 3\ldots, 3, 2\tikzmark{b}}^{\frac{n}{2}},
\begin{tikzpicture}[overlay,remember picture,out=315,in=225,distance=0.4cm]
    \draw[->,black,shorten >=3pt,shorten <=5pt] (a.center) to (b.center);
\end{tikzpicture} 
\ldots \\[0.125in]
&\text{$n$ odd \qquad} &
\overbrace{0, 1\ldots, 1}^{\left\lceil\frac{n}{2}\right\rceil},
\ourunderbracket{
\overbrace{2, 3\ldots, 3}^{\left\lfloor\frac{n}{2}\right\rfloor},
\overbrace{1, 3\ldots, 3, 2}^{\left\lceil\frac{n}{2}\right\rceil},
}_{n}
\ldots & \preccurlyeq
\overbrace{0, 1\ldots, 1}^{\left\lceil\frac{n}{2}\right\rceil},
\ourunderbracket{
\overbrace{\tikzmark{a} 3\ldots, 3, 2\tikzmark{b}}^{\left\lfloor\frac{n}{2}\right\rfloor},
\begin{tikzpicture}[overlay,remember picture,out=315,in=225,distance=0.4cm]
    \draw[->,black,shorten >=3pt,shorten <=5pt] (a.center) to (b.center);
\end{tikzpicture} 
\overbrace{\tikzmark{a} 2, 3\ldots, 3, 1\tikzmark{b}}^{\left\lceil\frac{n}{2}\right\rceil},
\begin{tikzpicture}[overlay,remember picture,out=315,in=225,distance=0.4cm]
    \draw[->,black,shorten >=3pt,shorten <=5pt] (a.center) to (b.center);
\end{tikzpicture} 
\vphantom{\Big|}}_{n}
\ldots \\
& & & = 
\overbrace{0, 1\ldots, 1}^{\left\lfloor\frac{n}{2}\right\rfloor},
\ourunderbracket{
\overbrace{1, 3\ldots, 3, 2}^{\left\lceil\frac{n}{2}\right\rceil},
\overbrace{2, 3\ldots, 3}^{\left\lfloor\frac{n}{2}\right\rfloor},
}_{n}
\ldots
\end{align*}
For 4.\ we note that the Peter-Paul transformations used for 3.\ are non-overlapping and each transfer a count of 1.
\end{proof}
\begin{remark}
It is informative to contrast the inequalities in Theorem~\ref{theorem:dew-count-inequalities} Part 1 with those of Theorem~\ref{theorem:two-stacks-count-inequalities}. Both compare the increments of an operation count sequences with the constant sequence $3,3,3,\ldots$, but in the DEW case we have the stronger relation $\incr_X \leq 3$, i.e., $\incr_X$ is {\em dominated} by $3$, whereas in the Two Stacks case we have the weaker relation $\incr_X \preccurlyeq 3$, i.e., $\incr_X$ is {\em cumulatively dominated} by $3$. This reflects that DEW is a more fully de-amortized algorithm than Two Stacks.
\end{remark}

To compare with Two Stacks we denote the count and increment functions of Two Stacks variants by $\countfn_\text{TS,X}$ and $\incr_\text{TS,X}$, where $X$ is the variant.

\begin{theorem}[DEW Complexity Comparison with Two Stacks]
\label{theorem:dew-two-stacks-count-inequalities}
\ \\ \nopagebreak
\begin{enumerate}
\item $\countfn_\text{TS,V3} \leq \countfn_\text{DEW,V1} \leq \countfn_\text{DEW,V2}$

\item $\countfn_\text{DEW,V1} \leq \countfn_\text{DEW,V2} \leq \countfn_\text{TS,V3}\, +\,  n - 2$, for $n \geq 2$.

\item $\countfn_\text{DEW,V2}(N) = \countfn_\text{TS,V3}(N)$, for $N=n+1, 2n+1, 3n+1, \ldots$.
\end{enumerate}
\end{theorem}
\begin{proof}
We start with the observations that $\countfn_\text{DEW,V1}(n+1) = \countfn_\text{TS,V3}(n+1) = 2n-2$, and $\countfn_\text{DEW,V2}(n+1) = 2n-1$. Parts 1.\ and 2.\ then follow from the Peter Paul Lemma, and the periodicity of the increments. For part 3. note that the maximum difference between Two Stacks Variant 3 and the DEW variants occurs at $N=n,\, 2n,\, 3n, \ldots$, and is $n-2$ at these points.
\begin{align*}
\countfn_\text{DEW,V1}(n) & =\countfn_\text{DEW,V2 }(n) = 2n - 3\\
\countfn_\text{TS,V3}(n)  & = n - 1
\end{align*}
Therefore the maximum difference is $n-2$.
\end{proof}
\begin{remark}
In essence Two Stacks gets a head start over DEW because length of the startup batch for Two Stacks is double that of DEW, but DEW catches up once per period of length $n$.
\end{remark}
\begin{result}[DEW Complexity Formulae]
\label{result:dew-complexity}
\  \nopagebreak
\begin{enumerate}
\item 
$\begin{aligned}[t]
\countfn_\text{DEW,V1}(N)
& = \ourcases{
0          & \text{if $n=1$  or $N=1$, else} \\
N-1        & \text{if $n=2$  or  $N \leq\left\lceil\frac{n}{2}\right\rceil$, else}\\
3N - n - 3 & \text{if $\left\lceil\frac{n}{2}\right\rceil < N \leq n$}
}
\end{aligned}$

\noindent Furthermore, if $n>2$ and $N>n$, then
\begin{align*}
\countfn_\text{DEW,V1}(N)
 & = k(3n - 4) - n + 1 + (r>0)\left(
          3r - 2 - (r > \left\lfloor\frac{n}{2}\right\rfloor) 
                 - (r > \left\lceil \frac{n}{2}\right\rceil )
                \right) \\
& = 3N - 4k - n + 1 - \left(2(r>0) + 
                            (r > \left\lfloor\frac{n}{2}\right\rfloor) 
                          + (r > \left\lceil \frac{n}{2}\right\rceil )
                \right)
\end{align*}
where $N=kn + r$ and $0 \leq r < n$. I.e. $k=\left\lfloor\frac{N}{n}\right\rfloor, r = N \bmod n$
\item 
$\begin{aligned}
\countfn_\text{DEW,V2}(N)
& = \ourcases{
0          & \text{if $n=1$  or $N=1$, else} \\
N-1        & \text{if $n=2$  or  $N \leq \left\lceil\frac{n}{2}\right\rceil$, else}\\
3N - n - 2 - \left(N = n\right) & \text{if $\left\lceil\frac{n}{2}\right\rceil < N \leq n$}
}
\end{aligned}$

\noindent Furthermore, if $n>2$ and $N>n$, then
\begin{align*}
\countfn_\text{DEW,V2}(N)
 & = k(3n - 4) - n + 1 + (r>0)\left(
          3r - 1 - (r > \left\lfloor\frac{n-1}{2}\right\rfloor) 
                 - (r > \left\lceil \frac{n-1}{2}\right\rceil )
                \right) \\
& = 3N - 4k - n + 1 - \left((r>0) + 
                            (r > \left\lfloor\frac{n-1}{2}\right\rfloor) 
                          + (r > \left\lceil \frac{n-1}{2}\right\rceil )
                \right)
\end{align*}
where $N=kn + r$ and $0 \leq r < n$. I.e. $k=\left\lfloor\frac{N}{n}\right\rfloor, r = N \bmod n$
\end{enumerate}
\end{result}
\begin{proof}
Similar to the proof of \ref{theorem:two-stacks-complexity}.
\end{proof}

\section{The DEW Algorithm Algebraically}
\label{dew-algebraically}

We now describe the DEW algorithm using algebraic formulae. For reasons of brevity we describe Variant 1, as Variant 2 is similar.
As with Two Stacks, DEW proceeds in batches, but each batch is now of length $\left\lceil\frac{n}{2}\right\rceil$ or $\left\lfloor\frac{n}{2}\right\rfloor$. Variant 1 starts with a `startup batch' of length $m=\left\lceil\frac{n}{2}\right\rceil$, which is simply a prefix sum. In the following descriptions we define $m = \left\lceil\frac{n}{2}\right\rceil$.

\begin{description}

\item[Batch 1]
Compute

\begin{align*}    
y_1 & = v_1 = a_1 \\
y_2 & = v_2 = a_2 * v_1 \\
y_3 & = v_3  = a_3 * v_2 \\
\vdots & \\
y_m & = v_m = a_m * v_{m-1} 
\end{align*}
%
\end{description}

\noindent Batch 2 is different. It has length $n - m = \left\lfloor\frac{n}{2}\right\rfloor$ and depends on whether $n$ is even or odd. 

\begin{description}
\item[Batch 2]
We first compute $v_i$ as follows, depending on whether $n$ is even or odd.

\begin{center}
\begin{tabular}{c|c}
$n$ even case & $n$ odd case\\
\parbox[t]{2in}{
\begin{align*}
v_{m+1} & = a_{m+1} \\
v_{m+2} & = a_{m+2} * v_{m+1} * a_m &\\
v_{m+3} & = a_{m+3} * v_{m+2} * a_{m-1} &\\
\vdots & \\
v_n & = a_n * v_{n-1} * a_2
\end{align*}
}
&
\parbox[t]{2in}{
\begin{align*}
v_{m+1} & =a_{m+1} * a_m \\
v_{m+2} & =a_{m+2} * v_{m+1} * a_{m-1} \\
v_{m+3} & =a_{m+3} * v_{m+2} * a_{m-2} \\
\vdots & \\
v_n    & =a_n * v_{n-1} * a_2
\end{align*}
}
\end{tabular}
\end{center}
In other words

\begin{equation*}
v_{m+1} =\ourcases{
a_{m+1}       & \text{if $n$ is even}\\
a_{m+1} * a_m & \text{if $n$  is odd}
}
\end{equation*}
and 
\begin{equation*}
v_{m+i} = \ourcases{
a_{m+i} * v_{m+i-1} * a_{m+2-i} & \text{if $n$ is even} \\ 
a_{m+i} * v_{m+i-1} * a_{m+1-i} & \text{if $n$ is odd}
}
\end{equation*}
for $1 < i \leq n-m$. 
As the $v_{m+i}$ are computed in Batch 2, the $y_{m+i}$ can be computed in turn follows

\begin{center}
\begin{tabular}{c|c}
$n$ even case & $n$ odd case \\
\parbox[t]{2in}{
\begin{align*}
y_{m+1} & =v_{m+1} * v_m \\
y_{m+2} & =v_{m+2} * v_{m-1} \\
\vdots  & \\
y_n     & = v_n * v_1
\end{align*}
}
& 
\parbox[t]{2in}{
\begin{align*}
y_{m+1} & =v_{m+1} * v_{m-1} \\
y_{m+2} & =v_{m+2} * v_{m-2} \\
\vdots &  \\
y_n & = v_n * v_1
\end{align*}
}
\end{tabular}
\end{center}
which is to say

\begin{equation*}
y_{m+i}= \ourcases{
    v_{m+i} * v_{m+1-i} & \text{ if $n$ is even} \\
    v_{m+i} * v_{m-i}   & \text{ if $n$ is odd}
}
\end{equation*}
$y_{m+i}$ can be computed as soon as $v_{m+i}$ has been computed, so the order of computation is $v_{i+1}, y_{i+1}$, $v_{i+2}, y_{i+2}, v_{i+3}, y_{i+3}, \ldots$.
Also note that in the case where $n$ is even Batch 2 has length $m$, and in the odd case Batch 2 has length $m-1$.

\item[Batch 3]
For Batch 3 the double-ended sums $v_{n+i}$ start from a single point in both the $n$ even and $n$ odd cases, and the batch has length $m=\left\lceil\frac{n}{2}\right\rceil$ in both cases.

\begin{align*}
v_{n+1} & = a_{n+1} \\
v_{n+2} & = a_{n+2} * v_{n+1} * a_n \\
\vdots &  \\
v_{n+m} & = a_{n+m} * v_{n+m-1} * a_{n+2-m}
\end{align*}
and 
\begin{equation*}
v_{n+i}=a_{n+i} * v_{n+i-1} * a_{n+2-i}
\end{equation*}
for $1 < i \leq m$. For $1 \leq i < m$ we then compute $y_{n+i}$ as

\begin{equation*}
y_{n+i}=v_{n+i} * v_{n+1-i}
\end{equation*}
but there is again a difference between $n$ even and $n$ odd when we reach $y_{n+m}$, as
\begin{equation*}
y_{n+m}=\ourcases{
v_{n+m} * v_{m+1} & \text{ if $n$ is even}\\
v_{n+m}           & \text{ if $n$ is odd}
}
\end{equation*}
\end{description}

\noindent After Batch 3 the pattern starts repeating, with Batch 4 similar to Batch 2, Batch 5 similar to Batch 3, and so on. The entire algorithm for Variant 1 may therefore be summarized in the following table

\begin{center}
\begin{tabular}{l|l}
Batch $l$ & Description \\
\hline \\[-2ex]
\parbox[t]{0.75in}{Batch $1$} 
&
\parbox[t]{4.4in}{
Same for $n$ even and $n$ odd.
\begin{flalign*}
& m    = \left\lceil\frac{n}{2}\right\rceil & \\
& v_1  = a_1 & \\
& v_i  = a_i * v_{i-1} & \text{for $1 < i \leq m$} \\
& y_i  = v_i           & \text{for $1 \leq i \leq m$}
\end{flalign*}
}\\[1ex]
\hline \\[-2ex]
\parbox[t]{0.75in}{Batch $l$\\$l$ even}
&
\parbox[t]{4.4in}{
\vspace{-4.0ex}
\begin{flalign*}
& m = \left\lceil\frac{n}{2}\right\rceil & \\
& l = 2 k\\
& M = (k-1)n + m\\
& v_{M+1} =\ourcases{
    a_{M+1}         & \text{if $n$ is even} \\
    a_{M+1} * a_{M} & \text{if $n$ is odd}
} & \ \\
& v_{M+i} = \ourcasesclosed{
    a_{M+i} * v_{M+i-1} * a_{M+2-i} & \text{if $n$ is even } \\
    a_{M+i} * v_{M+i-1} * a_{M+1-i} & \text{if $n$ is odd }
}
& \text{for $1 < i \leq n - m$}\\
& y_{M+i}  =\ourcasesclosed{
v_{M+i} * v_{M+1-i}  & \text{if $n$ is even} \\
v_{M+i} * v_{M-i}    & \text{if $n$ is odd }
}
& \text{for $1 \leq i \leq n - m$}
\end{flalign*}
} 
\\[1ex]
\hline \\[-2ex]
\parbox[t]{0.75in}{Batch $l$\\$l$ odd}
&
\parbox[t]{4.4in}{
\vspace{-4.0ex}
\begin{flalign*}
& m = \left\lceil\frac{n}{2}\right\rceil & \\
& l       = 2k + 1\\
& M       = kn\\
& v_{M+1} = a_{M+1} \\
& v_{M+i} = a_{M+i} * v_{M+i-1} * a_{M+2-i} & \text{for $1 < i \leq m$}\\
& y_{M+i} = v_{M+i} * v_{M+1-i}             & \text{for $1 \leq i < m$} \\
& y_{M+m} = \ourcases{
v_{M + m} * v_{M+1-m}    & \text{if $n$ is even}\\
v_{M + m}                & \text{if $n$ is odd}\\
}
\end{flalign*}
} 
\end{tabular}
\end{center}

\section{Three Implementation Sketches for DEW}
\label{sec:dew-implementation-sketches}

We now give an algorithmic description of DEW on a sequential random access machine. There are many ways to do this corresponding to different approaches to organizing the bookkeeping of the algorithm. The approach we will take uses a fixed length away of length $n$ to store input values and double-ended aggregates, and treats that array, effectively, as a circular buffer. In keeping with the conventions so far we shall start counting array indexes at $1$, and denote the contents of this array {\tt{arr[$1$]}}, {\tt{arr[$2$]}}, $\ldots$, {\tt{arr[$n$]}}.

\begin{center}
\begin{tabular}{m{3in} m{3in}}
\includegraphics[scale=1.00]{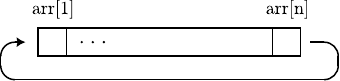}
& 
\includegraphics[scale=1.00]{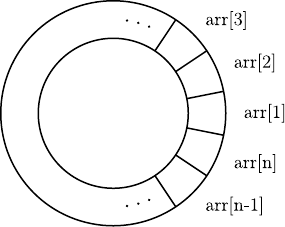}
\end{tabular}    
\end{center}
The circular algorithm for DEW uses two indexes (or pointers) $p$, $q$, which move around the array in opposite directions. At each step of the algorithm we do some computation using array contents and new data, then store results in the array cells pointed to by $p$ and $q$. Next we update $p$ and $q$, wrapping around the array if necessary, and finally (for that step) return a new window aggregate.

\begin{center}
\includegraphics[scale=1.00]{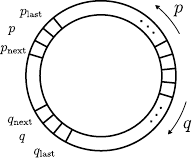}    
\end{center}
The two DEW Variants are distinguished by differing starting positions for $p$ and $q$. Variant 1 starts with $p=q=1$, and Variant 2 start with $p=1, q=n$. After the start, the rules for the two variants are the same, but for both variants there are special cases for handling empty array cells during the startup phase (these are dropped from any $*$-products being computed), and also special cases for when $|p-q|_\text{circular} = \min\left((p-q) \bmod n, (q-p) \bmod n\right) \leq 1$. 

We now describe the circular DEW algorithm using pseudo-code, but without any consideration for the efficiency of the bookkeeping. We shall indicate afterwards how to make this bookkeeping more efficient. We again use Landin's off-side rule \cite{Landin1966} to indicate the end of code blocks.

\begin{algorithm}[Circular DEW---Basic Version] 
\ \nopagebreak \\[-3ex] \nopagebreak
\begin{alltt}
initialization(n):
    p = q = 1        for Variant 1
    p = 1, q = n     for Variant 2
    arr = empty array of length n starting at arr[1]

insert(x): This also `evicts' to keep the window length \(\leq\) n
    p_last, q_last = ((p - 2) mod n) + 1, (q mod n) + 1
    p_next, q_next = (p mod n) + 1, ((q - 2) mod n) + 1
    dea = x if p = q or (p = q_last and arr[p] is empty), else
          x * arr[p] if p = q_last, else
          x * arr[p_last] if arr[p] is empty, else
          x * (arr[p_last] * arr[p])            
    agg = dea if q_next = p or arr[q_next] is empty, else
          dea * arr[q_next]                
    arr[q] = x
    if p \(\neq\) q then
        arr[p] = dea
    p, q = p_next, q_next
    return agg   
\end{alltt}
\end{algorithm}

Here are diagrams indicating the first several iterations of DEW for $n=8$ and $n=9$, and for both variants. In these diagrams, an integer $i$ outside the circle is used to indicate $a_i$ stored at the corresponding location, and a pair of integers $(i\ j)$ indicates the $*$-product $a_i * a_{i - 1} * \ldots * a_j$ stored at the corresponding location.

\begin{center}
\begin{tabular}{m{0.5in} m{2.5in} m{2.5in}}
& Variant 1 & Variant 2 \\
\vspace{0.875in} $n = 8$ & 
\parbox[c]{2.5in}{
\includegraphics[scale=1.00]{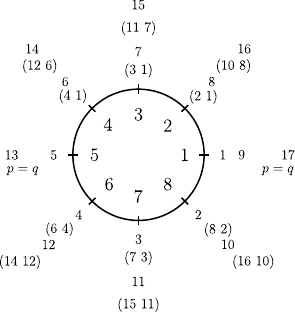}}
&
\parbox[c]{2.5in}{
\includegraphics[scale=1.00]{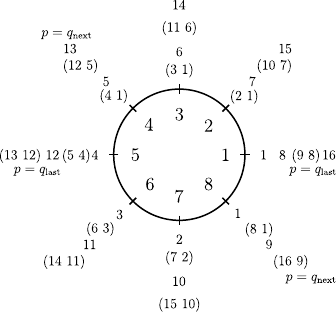}}
\\
\vspace{0.25in}\vspace{0.875in} $n = 9$ & 
\vspace{0.25in}
\parbox[c]{2.5in}{
\includegraphics[scale=1.00]{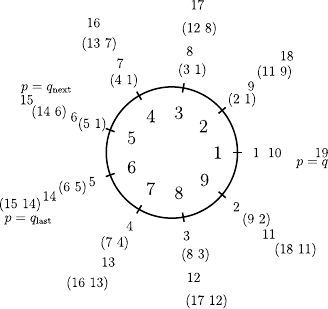}}
&
\vspace{0.25in}
\parbox[c]{2.5in}{
\includegraphics[scale=1.00]{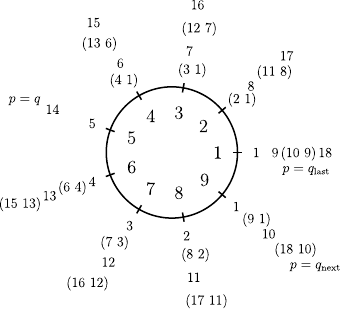}}
\end{tabular}
\end{center}

This algorithm gives a compact description of circular DEW, but it does have short-comings. Firstly, it requires that the array cells {\tt{arr[$i$]}} can be empty or non-empty, and this may require extra overhead to implement. Secondly, the bookkeeping is inefficient, with extra quantities being updated, and many conditionals required before the common cases are reached. Both of these issues are easily remedied. The empty cells can be handled with additional bookkeeping to detect the empty cells without needing to access them, so they can then be implemented as uninitialized cells rather than empty cells. The efficiency of the bookkeeping can be addressed using a `sentinel index' that is updated whenever special cases are executed, and whose purpose is to allow the execution path to reach the common case with a single conditional $p \neq$ sentinel. With this sentinel logic the DEW implementation becomes competitive with Two Stacks (also implemented with a sentinel for bookkeeping efficiency).

\begin{remark}
The {\tt insert} procedure we have described for DEW does not satisfy the same properties as the {\tt insert} procedure described for Two Stacks in Section~\ref{sec:two-stacks-variants}, as it also performs an `evict' if adding the item to the window would increase the window length beyond $n$. Thus it behaves like {\tt insert} (of Section~\ref{sec:two-stacks-variants}) when fewer than $n$ items have been inserted so far, and behaves like {\tt combined-insert-evict} thereafter. In terms of the procedures in Section~\ref{sec:two-stacks-variants} its behavior is thus `{\tt insert} if less than $n$ items inserted else {\tt combined-insert-evict}'. A second difference with Section~\ref{sec:two-stacks-variants} is that it returns the window aggregate directly rather than storing results and relying on query to return the result.
\end{remark} 

\begin{algorithm}[Circular DEW---Sentinel Version]
\ \\ 
For simplicity, we only record the algorithm for Variant 1 here. Variant 2 is similar.

%
%
\begin{alltt}
initialization(n):
    p = q = 1
    sentinel = 1
    mode = ONE if n = 1 else TWO if n = 2 else START
    arr = uninitialized array of length n starting at arr[1]

insert(x): This also evicts to keep the window length \(\leq\) n
    if p \(\neq\) sentinel
        dea = x * (arr[p - 1] * arr[p])
        agg = dea * arr[q - 1]
        arr[p] = dea
        arr[q] = x
        p = p + 1
        q = q - 1
        return agg
    else if mode = REGULAR
        if p = n + 1
            p = 1
            q_next = n
            agg = x * arr[q_next]
            sentinel = \(\lfloor\)n/2\(\rfloor\) + 1    use integer floor division by 2 or binary right shift
        else if q = p
            q_next = q - 1
            agg = x * arr[q_next]
            sentinel = n + 1
        else if q = p - 1
            q_next = q - 1
            dea = x * arr[p]
            agg = dea * arr[q_next]
            arr[p] = dea
            sentinel = n + 1
        else
            q_next = q - 1
            agg = x * (arr[p-1] * arr[p])
            arr[p] = agg
            sentinel = p + 1
        arr[q] = x
        p = p + 1
        q = q_next
        return agg
    else if mode = START
        if p = 1
            q_next = n
            agg = x
            sentinel = 2
        else if p = q
            q_next = q - 1
            agg = x * arr[q_next]
            mode = REGULAR
            sentinel = n + 1
        else if p = q - 1 
            q_next = q - 1
            agg = x * arr[p - 1]
            mode = REGULAR
            arr[p] = agg
            sentinel = p + 1
        else
            q_next = q - 1
            agg = x * arr[p-1]
            arr[p] = agg
            sentinel = p + 1
        arr[q] = x
        p = p + 1
        q = q_next
        return agg
    else if mode = TWO 
        p_next = 2 if p = 1 else 1
        if q = 0
            agg = x * arr[p_next]
        else
            agg = x
            q = 0
        arr[p] = x
        sentinel = p_next
        p = p_next
        return agg
    else
        return x
\end{alltt}
\end{algorithm}

There are alternative approaches to implementing DEW, corresponding to different ways of organizing the bookkeeping, and we briefly describe another such approach.
Instead of storing the input values and double-ended aggregates in the same array, we can store them in two separate arrays, and use a single index variable $p$. We call the two arrays {\tt values}, and {\tt aggregates}, and each has length $\lceil\frac{n+1}{2}\rceil$.

\begin{displaymath}
\begin{array}{lr}
\begin{array}{l}
\text{\tt{values}}\\
\text{{\tt{aggregates}}\quad}
\end{array}
&
\overbrace{
\begin{array}{|c|c|c|c|c|}
\cline{1-5}
&  & \cdots \text{\hspace{1.25in}} & &\\
\cline{1-5}
&  & \cdots \text{\hspace{1.25in}} & &\\
\cline{1-5}
\end{array}
}^{\lceil\frac{n+1}{2}\rceil}
\end{array}
\end{displaymath}
\vspace{0.125in}
\begin{algorithm}[Alternative DEW Implementation]
\ \\
We describe the algorithm for Variant 1. Variant 2 is similar.

\begin{alltt}
initialization(n):
    p = 1
    step = 1 if n > 1 else 0
    values = empty array of length \(\lceil\)(n+1)/2)\(\rceil\) starting at values[1]
    aggregates = empty array of length \(\lceil\)(n+1)/2)\(\rceil\) starting at aggregates[1]

insert(x): This also evicts to keep the window length \(\leq\) n
    dea = x if p = 1 or (p = \(\lceil\)(n+1)/2)\(\rceil\) and n is even), else
          x * values[p] if step = 0, else
          x * aggregates[p - step] if values[p] is empty, else
          x * (aggregates[p - step] * values[p])
    step = 0  if (p = \(\lceil\)(n+1)/2)\(\rceil\) and step = 1 and n is odd) or n = 1, else
           -1 if p = \(\lceil\)(n+1)/2)\(\rceil\), else
           1  if p = 1 and step = -1, else
           step
    agg = dea if step = 0 or aggregates[p + step] is empty, else
          dea * aggregates[p + step]
    values[p] = x
    aggregates[p] = dea
    p = p + step
    return agg
    
\end{alltt}

\end{algorithm}
\noindent As with the circular version of DEW this implementation can easily be rewritten to use a sentinel and to avoid explicit checks for empty cell contents. When this is done its performance is essentially the same as the circular version with a sentinel.

\section{DEW Properties}
\label{sec:dew-properties}

\begin{description}

\item[Correctness] \ \\ 
DEW produces correct results provided $*$ is associative.

\item[Accuracy] \ \\ 
DEW fares well on accuracy. Its accuracy is similar to sequential versions of Two Stacks with error in $y_i$ roughly bounded by $(n-1) \varepsilon$, where $\varepsilon$ is a bound on the error introduced by each $*$ operation.

\item[Efficiency] \ \\ 
The number of $*$ operations is bounded by $3N$, and the bookkeeping is linear in $N$.

\item[Simplicity] \ \\
The algorithm is simple. There are short implementations in code, and the algorithm can be explained to, and understood by, a high school freshman.

\item[Memory] \ \\
The circular DEW implementation uses $n$ items of working space, plus two items to combine double-ended aggregates and form the window aggregate.

\item[Freedom from extraneous choice or data] \ \\
The only quantities used in the computation of $y_i$ are $a_{i-n+1}, \ldots, a_{i}$.

\item[Streaming] \ \\
There are streaming implementations based on {\tt insert} with auto-evict after the first $n$ items. 

\item[Latency] \ \\
DEW requires at most $3$ $*$-operations to compute each new window aggregate.

\item[Parellelizability] \ \\
Not obviously parallelizable, other than breaking into sections with overlap.

\item[Vectorizability] \ \\
Not obviously vectorizable.

\item[Generalizability] \ \\
DEW works for any (computable) associative operator, and does not require other properties such as commutativity, invertibility, or being a `selection operator'.

\end{description}

%% file: htcams-arxiv-ch05-other-sequential-algorithms.tex
\chapter{Other Sequential Sliding Window Algorithms}
\label{chapter:other-sequential-algorithms}

In this chapter we assume that $*$ is an binary operator, and our goal is to compute the moving sums (or moving products)
\begin{equation*}
y_i = \ourcases{
    a_i * \ldots * a_1     & \quad \text{ for $1 \leq i < n$} \\
    a_i * \ldots * a_{i-n+1} & \quad \text{ for $i \geq n$}
} 
\end{equation*}
For the first part of the chapter, which discusses the DABA algorithm and its variants, we again assume that $*$ is associative. For the discussion of the SlickDeque algorithm, however, we drop the associativity assumption, as that algorithm requires a different set of assumptions for correctness. A summary of the properties assumed by the different algorithms is given in Section~\ref{sec:sliding-window-summary}.

\section{DABA and Variants}
\label{sec:daba-and-variants}

The DABA and DABA Lite algorithms are window aggregation algorithms developed by Tangwongsan, Hirzel, and Schneider in Tangwongsan et al.\ \cite{Tangwongsan2015b} \cite{Tangwongsan2017} \cite{Tangwongsan2021}. These were the first such algorithms to be developed that had linear complexity while avoiding the latency spike problem. DABA has operation count bounded by $5N$ and DABA Lite has operation count bounded by $4N$. In comparison with DEW, both DABA and DABA Lite have increased $*$-operation count, and more complicated bookkeeping. But they also have an important additional property, not shared by DEW, in that they support variable size windows through an {\tt insert}, {\tt evict}, {\tt query} interface. An example where this is crucial is the implementation of time-based sliding window algorithms. Suppose each data item to be aggregated is paired with a time stamp, and we wish to aggregate all observations within a time $T$ of the latest observation. Using the {\tt evict}, {\tt insert}, and {\tt query} procedures of DABA, or DABA Lite, or Two Stacks, this can be achieved easily as follows. Here {\tt insert} refers to the insert procedure {\em without} eviction, and increases the length of the data in the aggregator by $1$.

\begin{algorithm}[Time-based Sliding Window Aggregation]
\ \\ 
To add a new data point $x$ and compute a time-based sliding window aggregate with window length $T$, perform the following operations.
\begin{alltt}
insert-and-compute-aggregate(x):
    call insert(x) for the aggregator
    insert x into a FIFO queue
    iterate from the front of the queue removing items with time stamp < timestamp(x) - T
        and call evict() on the aggregator for each such item found
    stop when an item is found with time stamp \(\geq\) timestamp(x) - T
    call query() on the aggregator and return the result
\end{alltt}
    
\end{algorithm}

\section{DABA Diagrams}

The DABA and DABA Lite algorithms have a complex startup behavior, though the algorithms themselves are simply described. Once steady state is reached, however, their stacked staggered sequence diagrams are easy to draw. As with Two Stacks, there are different diagrams depending on whether we call {\tt insert} and {\tt evict} in the order {\tt insert-evict}, or {\tt evict-insert} in the steady state. There are also different diagrams for $n$ even versus $n$ odd, as we saw with DEW. DABA also has an occasional pair of unnecessary $*$ operations that occur immediately after a `flip' in the {\tt insert-evict} $n$ odd case. These wasted $*$ operations can easily be avoided by using a combined {\tt insert-evict} operation that detects when the flip happens and adjusts accordingly. 

Here are the steady state stacked staggered sequence diagrams for DABA for $n=9$, $n=10$, and in general. The braces, where present, indicate batches.


\begin{center}
{\bf{DABA Steady State, n = 9, 10}}  
\nopagebreak
\medskip

\begin{tabular}{m{1.0in} m{1.0in} m{1.0in} m{1.25in}}
insert-evict & insert-evict & evict-insert & \hspace{0.0625in} evict-insert \\
& (combined) & & \\
$n = 10$ & $n = 9$ & $n = 10$ & \hspace{0.0625in} $n = 9$ \\
\parbox[t]{1.0in}{
\vspace{0.13786458333in} 
\includegraphics[scale=1.00]{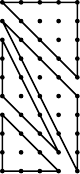} 
} &
\parbox[t]{1.0in}{
\vspace{0.13786458333in} 
\vspace{0.125in}
\includegraphics[scale=1.00]{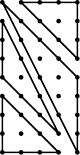} 
} &
\parbox[t]{1.0in}{
\vspace{0.13786458333in} 
\includegraphics[scale=1.00]{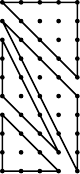} 
} &
\parbox[t]{1.25in}{
\vspace{0.0000in}  
\vspace{0.125in}
\includegraphics[scale=1.00]{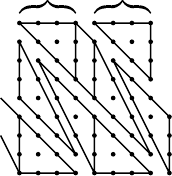} 
}
\end{tabular}
\end{center}

\medskip

\begin{center}
{\bf{DABA Steady State}}
\nopagebreak
\medskip

\begin{tabular}{m{1.0in} m{1.0in} m{1.0in} m{1.45in}}
\hspace{-0.043625in}   \hspace{0.25in}  insert-evict & 
\hspace{-0.04209375in} \hspace{0.125in} insert-evict & 
\hspace{-0.043625in}   \hspace{0.25in}  evict-insert & 
\hspace{-0.04209375in} \hspace{0.125in} evict-insert   
\\
& \hspace{-0.04209375in} \hspace{0.125in} (combined) & & \\
\hspace{-0.043625in}   \hspace{0.25in}  $n$ even & 
\hspace{-0.04209375in} \hspace{0.125in} $n$ odd  & 
\hspace{-0.043625in}   \hspace{0.25in}  $n$ even & 
\hspace{-0.04209375in} \hspace{0.125in} $n$ odd    
\\
\parbox[t]{1.0in}{
\vspace{0.00775in} 
\includegraphics[scale=1.00]{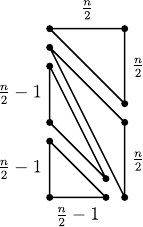} 
} &
\parbox[t]{1.0in}{
\vspace{0.0000in}  
\vspace{0.125in}
\includegraphics[scale=1.00]{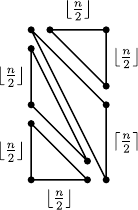}
} &
\parbox[t]{1.0in}{
\vspace{0.00775in} 
\includegraphics[scale=1.00]{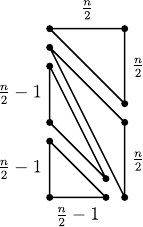} 
} &
\parbox[t]{1.45in}{
\vspace{0.0000in}  
\vspace{0.125in}
\includegraphics[scale=1.00]{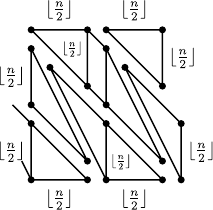} 
}
\\
\end{tabular}
\end{center}

\noindent The increments for DABA can be read off the diagram. For example, the steady state $*$-operation count increments for insert-evict (or evict-insert) $n$ even are $3,\underbrace{5, \ldots, 5}_{\frac{n}{2}-2},3$. This, however, differs from the operation counts of the DABA algorithms presented in Tangwongsan et al.\ \cite{Tangwongsan2015b} \cite{Tangwongsan2017} \cite{Tangwongsan2021}. The reason for the discrepancy is that the stacked staggered sequence diagrams represent the operations the algorithm uses to compute each window sum, but not the order of operations. The algorithms we have been reading off the diagrams compute each quantity only when needed, and as soon as it is needed, but not sooner, and are more fully deamortized than the DABA algorithms in Tangwongsan et al.\ \cite{Tangwongsan2015b} \cite{Tangwongsan2017} \cite{Tangwongsan2021}. The DABA algorithms in Tangwongsan et al.\ \cite{Tangwongsan2015b} \cite{Tangwongsan2017} \cite{Tangwongsan2021} perform some calculations ahead of time and before they are needed, and this increases their latency. Their stacked staggered sequence diagrams show that they can be deamortized further. To avoid confusion, we refer to the more fully deamortized versions, corresponding to the diagrams, as DDABA and DDABA Lite, though the actual sequence of $*$ operations to compute each window sum is identical to DABA and DABA Lite, respectively. 

The startup for DDABA and DDABA Lite can either be the regular DABA or DABA Lite startup, or, in the case where $n$ is known in advance, a simplified startup may be obtained by applying the steady state diagram algorithms to the case with missing data (i.e., missing data for $a_i$ with $i < 0$). Here are the DDABA cases to illustrate.


\begin{center}
{\bf{Simplified (D)DABA Startup}}
\nopagebreak
\medskip

\begin{tabular}{m{1.3125in} m{1.25in} m{1.3125in} m{1.75in}}
\hspace{0.125in} insert-evict & \hspace{0.125in} insert-evict & \hspace{0.125in} evict-insert & \hspace{0.125in} evict-insert  
\\
& \hspace{0.125in} (combined) & & 
\\
\hspace{0.125in} $n$ even & \hspace{0.125in} $n$ odd  & \hspace{0.125in} $n$ even & \hspace{0.125in} $n$ odd 
\\
\parbox[t]{1.3125in}{
\vspace{0.00775in} 
\includegraphics[scale=1.00]{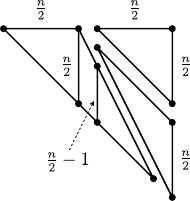}
} &
\parbox[t]{1.25in}{
\vspace{0.0000in}  
\includegraphics[scale=1.00]{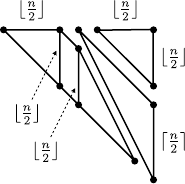}
} &
\parbox[t]{1.3125in}{
\vspace{0.00775in} 
\includegraphics[scale=1.00]{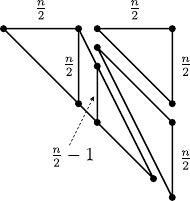}
} &
\parbox[t]{1.75in}{
\vspace{0.0000in}  
\includegraphics[scale=1.00]{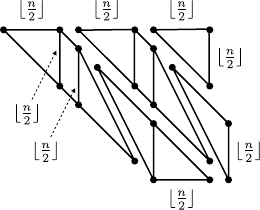} 
}
\\
\end{tabular}
\end{center}

We now show the steady state stacked staggered sequence diagrams for (D)DABA Lite. The diagrams for simplified startup are easily obtained by `removing triangles' for missing data and `adding a triangle in front'. In the (D)DABA Lite diagrams we have shaded regions corresponding to suffix $*$-aggregates that are accumulated before they are required---the algorithms can also be implemented without this `eager accumulation', but they would then suffer from latency spikes. Thus we see that eager evaluation of $*$-aggregates can either improve or worsen latency depending on the context. The shaded areas precompute the triangles marked $L$ in the next batch.


\begin{center}
{\bf{(D)DABA Lite Steady State}}
\nopagebreak
\medskip

\begin{tabular}{m{1.25in} m{1.25in} m{1.25in} m{1.5in}}
\hspace{-0.043625in}   \hspace{0.25in}  insert-evict & 
\hspace{-0.04209375in} \hspace{0.125in} insert-evict & 
\hspace{-0.043625in}   \hspace{0.25in}  evict-insert & 
\hspace{-0.04209375in} \hspace{0.125in} evict-insert   
\\
& \hspace{-0.04209375in} \hspace{0.125in} (combined) & & \\
\hspace{-0.043625in}   \hspace{0.25in}  $n$ even & 
\hspace{-0.04209375in} \hspace{0.125in} $n$ odd  & 
\hspace{-0.043625in}   \hspace{0.25in}  $n$ even & 
\hspace{-0.04209375in} \hspace{0.125in} $n$ odd    
\\
\parbox[t]{1.25in}{
\vspace{0.00775in} 
\includegraphics[scale=1.00]{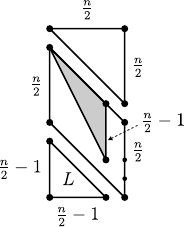} 
} &
\parbox[t]{1.25in}{
\vspace{0.0000in}  
\vspace{0.125in}
\includegraphics[scale=1.00]{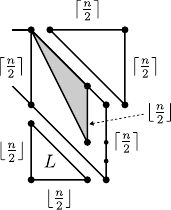}
} &
\parbox[t]{1.25in}{
\vspace{0.00775in} 
\includegraphics[scale=1.00]{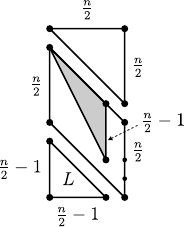} 
} &
\parbox[t]{1.5in}{
\vspace{0.0000in}  
\vspace{0.125in}
\includegraphics[scale=1.00]{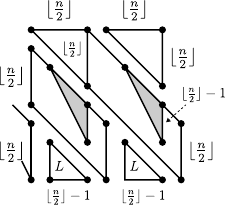} 
}
\\
\end{tabular}
\end{center}

\begin{remark}
Note that the versions of DABA Lite in Tangwongsan et al.\ \cite{Tangwongsan2021} have an extra $*$ operation due to accumulating the suffix $*$-aggregate one more step than necessary---essentially the shaded area is extended by an extra column. In the {\tt insert-evict} case there is now only one flip-related wasted operation rather than two, though the number of wasted operators per batch in this case remains at $2$ (because of the suffix aggregate). We remove all of these wasted operations from our DABA Lite diagrams and operation counts.
\end{remark}

We may now give the steady state operation count increments for DDABA and DDABA Lite, as usual reading them off the diagrams. We account for the operation counts of the shaded regions on the columns where they occur, and do not count the $*$ operations for the regions marked $L$, as these have already been computed.

\begin{theorem}[DABA increments]
The steady state increments for DDABA and DDABA Lite are as follows.

\begin{tabular}{lcccc}
& insert-evict & (combined) insert-evict & evict-insert & evict-insert\\
& $n$ even & $n$ odd & $n$ even & $n$ odd \\
DDABA & 
$3, \overbrace{5,\ldots 5}^{\frac{n}{2}-2} ,3$ & 
$2,4, \overbrace{5, \ldots, 5}^{\lceil\frac{n}{2}\rceil-3},3$  &
$3, \overbrace{5, \ldots, 5}^{\frac{n}{2}-2},3$  &
$3,4, \overbrace{5, \ldots, 5}^{\lfloor\frac{n}{2}\rfloor-2}$\\
\\
DDABA Lite &
$2, \overbrace{4,\ldots 4}^{\frac{n}{2}-2} ,2$ & 
$2,3, \overbrace{4, \ldots, 4}^{\lceil\frac{n}{2}\rceil-3},2$  &
$2, \overbrace{4, \ldots, 4}^{\frac{n}{2}-2},2$  &
$2,3, \overbrace{4, \ldots, 4}^{\lfloor\frac{n}{2}\rfloor-2}$
\end{tabular}
\end{theorem}

\begin{theorem}[DABA Batch Operation Counts and Slopes]
The steady state batch operation counts and slopes for DDABA and DDABA Lite are as follows.

\begin{center}
\begin{tabular}{lcccc}
& insert-evict & (combined) insert-evict & evict-insert & evict-insert\\
& $n$ even & $n$ odd & $n$ even & $n$ odd \\
\hline\\[-2ex]
DDABA        & & & &\\[1ex]
Batch length & $\frac{n}{2}$ & $\left\lceil\frac{n}{2}\right\rceil$ & $\frac{n}{2}$ & $\left\lfloor\frac{n}{2}\right\rfloor$ \\[1ex]
Batch operation count & $5\frac{n}{2} - 4$ & $5\left\lceil\frac{n}{2}\right\rceil - 6$ & $5\frac{n}{2} - 4$ & $5\left\lfloor\frac{n}{2}\right\rfloor - 3$\\[1ex]
Slope & $\frac{5n-8}{n}$ & $\frac{5n-7}{n+1}$ & $\frac{5n-8}{n}$ & $\frac{5n-11}{n-1}$\\[1ex]
\hline\\[-2ex]
DDABA Lite       & & & &\\[1ex]
Batch length & $\frac{n}{2}$ & $\left\lceil\frac{n}{2}\right\rceil$ & $\frac{n}{2}$ & $\left\lfloor\frac{n}{2}\right\rfloor$ \\[1ex]
Batch operation count & $4\frac{n}{2} - 4$ & $4\left\lceil\frac{n}{2}\right\rceil - 5$ & $4\frac{n}{2} - 4$ & $4\left\lfloor\frac{n}{2}\right\rfloor - 3$\\[1ex]
Slope & $\frac{4n-8}{n}$ & $\frac{4n-6}{n+1}$ & $\frac{4n-8}{n}$ & $\frac{4n-10}{n-1}$\\
\end{tabular}    
\end{center}
\end{theorem}


\section{SlickDeque}
\label{sec:slick-deque}

%
%

The SlickDeque algorithm of Shein \cite{Shein2019} \cite{Shein2018} is a multi-query%
\footnote{
See Shein \cite{Shein2019} for further discussion of multi-query algorithms.
}
algorithm capable of computing window aggregates of different window lengths simultaneously. Shein describes two variants, one for invertible operations, which is equivalent to Subtract-on-Evict in the case of a single window length, and a second variant to be used for non-invertible operations, which is the variant we explore here. For non-invertible operations the SlickDeque algorithm makes an additional assumption which is that $x * y \in \theset{x, y}$ for all $x, y$. Such operations are in one to one correspondence with reflexive binary relations, as we will describe in the following section. This, together with the transitivity of the corresponding binary relation, characterizes the operations to which SlickDeque applies. It is interesting to note that associativity of $*$ is not a necessary condition for the correctness of the SlickDeque algorithm, and we shall see this in Examples~\ref{example:selection-operators-6} and \ref{example:selection-operators-7}, and Theorem~\ref{theorem:slick-deque-correctness}. In particular, Theorem~\ref{theorem:slick-deque-correctness} and Theorem~\ref{theorem:slick-deque-transitivity-necessary} show that for an operation satisfying $x * y \in \theset{x, y}$, SlickDeque produces correct results on all inputs if and only if the corresponding reflexive relation is transitive.

In the following sections on selection operators and on SlickDeque we no longer assume that the binary operation $*$ is associative.

\section{Selection Operators}
\label{sec:selection-operators}

\begin{definition}
Let $X$ be a set and $*\colon X \times X \rightarrow X$ be a binary operation on $X$. Then $*$ is a {\em selection operator}, or {\em selective}, if and only if for all $x, y \in X$ we have $x * y \in \theset{x, y}$.
\end{definition}

\begin{definition}
Let $X$ be a set and $*\colon X \times X \rightarrow X$ be a binary operation on $X$. Define the binary relation $R_* \subseteq X \times X$ as follows
\begin{equation*}
x R_* y \Leftrightarrow x * y = y, \quad \text{for all } x, y \in X
\end{equation*}
\end{definition}

\begin{definition}
Let $X$ be a set and let $R \subseteq X \times X$ be a binary relation on $X$. Define the binary operation $*_R \colon X \times X \longrightarrow X$ by
\begin{equation*}
x *_{R} y= \ourcases{
    y & \text{if $x R y$, else}\\ 
    x &
}
\end{equation*}
\end{definition}
\noindent 
Thus $* \longmapsto R_*$ constructs a binary relation from a binary operator, and $R \longmapsto *_R$ constructs a binary operator from a binary relation. The following theorem is a slight generalization of standard results from the theory of semi-lattices, or equivalently commutative bands (see e.g.\ \cite{Clifford1961}). Note however that this result also applies to situations where the operator $*$ is noncommutative.

\begin{theorem}
\label{theorem:selection-operators-and-reflexive-relations}
Assume $X$ is a set, $*\colon X \times X \rightarrow X$ is a binary operation on $X$, and $R \subseteq X \times X$ is a binary relation on $X$. Then
\begin{enumerate}
\item If $*$ is a selection operator then $*$ is idempotent
\item $*_{R}$ is a selection operator
\item If $*$ is idempotent then $R_*$ is reflexive
\item $R_{*_{R}}$ is the reflexive closure of $R$
\item If $R$ is reflexive then $R_{*_R} = R$
\item If $*$ is a selection operator then $*=*_{R_*}$
\item The following are equivalent for a binary operator $*$
    \begin{enumerate}
    \item $*$ is a selection operator
    \item $* = *_R$ for some binary relation $R$
    \item $* = *_R$ for some reflexive binary relation $R$
    \item $* = *_{R_*}$
    \end{enumerate}
\item The following are equivalent for a binary relation $R$
    \begin{enumerate}
    \item $R$ is reflexive.
    \item $R=R_*$ for some idempotent binary operation $*$
    \item $R=R_*$ for some selection operator $*$
    \item $R=R_{*_R}$
    \end{enumerate}
\item Suppose $*$ is a selection operator, and $R$ is reflexive, and $*=*_{R}$ (and hence $R=R_{*}$), then
    \begin{enumerate}
    \item $*$ is commutative $\Leftrightarrow R$ is connected and antisymmetric
    \item $*$ is associative $\Rightarrow R$ is transitive
    \item If $R$ is connected, then $*$ is associative $\Leftrightarrow$ $R$ is transitive
    \end{enumerate}
\end{enumerate}
\end{theorem}
\begin{proof}
It is interesting to note that Theorem~\ref{theorem:selection-operators-and-reflexive-relations} makes statements about properties involving at most 3 elements, and if $*$ is a selection operator then $\theset{x, y, z}$ is closed under $*$ for any $x, y, z$, and therefore Theorem~\ref{theorem:selection-operators-and-reflexive-relations} can be proven by looking exhaustively at all selection operators on 3 element sets. This verification can be automated, thus providing a machine-assisted proof of the theorem. We, however, give a more traditional proof.

For 1.\ note that if $*$ is a selection operation then for any $x\in X$ we have $x * x \in \theset{x, x}$ and hence $x*x = x$. Hence $*$ is idempotent. Part 2.\ follows directly from the definition of $*_R$. For 3.\ note that if $*$ is idempotent then $x R_* x \Leftrightarrow x * x = x$ which is true for any $x\in X$. Hence $R_*$ is reflexive. For 4.\ note that
\begin{equation*}
x R_{*_R} y \Leftrightarrow x *_R y = y 
            \Leftrightarrow y = \ourcasesclosed{
                                    y & \text{if $x R y$ , else}\\
                                    x & }
            \Leftrightarrow \left( x R y \text{ or } x = y \right)
\end{equation*}
For 5.\ note that if $R$ is reflexive then
\begin{equation*}
x R_{*_R} y \Leftrightarrow \left( x *_R y = y \right)
            \Leftrightarrow \left((\text{$y$ if $x R y$ else $x$}) = y \right)
            \Leftrightarrow \left( x R y \text{ or } x = y \right)
            \Leftrightarrow x R y
\end{equation*}
For 6.\ note that 
\begin{equation*}
x *_{R_*} y = \ourcasesclosed{y & \text{if $x R_* y$, else}  \\ x &}
            = \ourcasesclosed{y & \text{if $x * y = y$, else}\\ x &}
\end{equation*}
but if $*$ is a selection operator then either $x * y = y$ or $x * y = x$, and hence
\begin{equation*}
x *_{R_*} y = \ourcasesclosed{y & \text{if $x * y = y$, else}\\ x & \text{if $x * y = x$}}
= x * y
\end{equation*}
For 7.\ we have (c) $\Rightarrow$ (b) trivially, and (b) $\Rightarrow$ (a) by 2., and (a) $\Rightarrow$ (d) by 6. For (d) $\Rightarrow$ (c) assume that $* = *_{R_*}$. Then $*$ is a selection operator by 2.\ and hence idempotent by 1.\ and hence $R_*$ is reflexive by 3.\ Thus (c) holds. For 8.\ we have (a) $\Rightarrow$ (d) by 5., and (d) $\Rightarrow$ (c) by 2.\, and (c) $\Rightarrow$ (b) by 1., and (b) $\Rightarrow$ (a) by 3.

For 9.(a) we first assume that $*$ is commutative and show that $R$ is connected and antisymmetric. For antisymmetry, if $x R y$ and $y R x$ then $x*y = y$ and $y * x = x$, but then by commutativity $x = y*x=x*y=y$. For connectedness, suppose that $x,y\in X$. Since $*$ is a selection operator we must have $x * y = x$ or $x * y = y$. But by commutativity we then have $y * x = x$ or $x * y = y$. Therefore $y R x$ or $x R y$. 

Now we must prove the converse and so assume that $R$ is connected and antisymmetric. Since $R$ is reflexive we also know that $R$ is strongly connected. Since $*$ is a selection operator we know that $x*y\in \theset{x, y}$, and $y*x\in \theset{x, y}$. This give us 4 cases to consider and in each of them we must show that $x*y = y*x$. For the case $x * y = x$ and $y * x = x$ clearly $x * y = y*x$, and similarly for the case where $x*y = y$ and $y*x = y$. For the case where $x * y = y$ and $y * x = x$ we have $x R y$ and $y R x$ and hence by antisymmetry it follows that $x = y$ and hence $x*y = y*x$. The remaining case for 9.(a) is where $x * y = x$ and $y * x = y$. By connectivity and reflexivity we know that $R$ is strongly connected and hence $x R y$ or $y R x$, and therefore $x * y = y$ or $y * x = x$ which reduces this case to the already handled cases.

For 9.(b) Assume that $*$ is associative and $x R y$ and $y R z$. Then $x*y=y$ and $y*z=z$. Therefore $x*z = x*(y*z) = (x*y)*z = y*z = z$, and hence $x R z$.

For 9.(c) We again have a multi-case analysis. By 9.(b) we know that associativity of $*$ implies transitivity of $R$ so it remains to show that if $R$ is connected and transitive then $*$ must be associative. So we suppose $R$ is connected and transitive and that $x,y,z\in X$. Since $*$ is a selection operator we have $x*y\in \theset{x, y}$ and $y * z\in \theset{y, z}$ and there are four cases to consider. For the case $x * y = x$ and $y * z = z$ we have $x * (y * z) = x * z = (x * y) * z$. For the case $x * y = y$ and $y * z = y$ we have $x *(y * z) = x * y = y = y * z = (x * y) * z$. In the case $x * y = y$ and $y * z = z$ we have $x R y$ and $y R z$ and hence $xRz$ by transitivity. But then it follows that $x * z = z$ and hence $x * (y * z) = x * z = z = y * z = (x * y) * z$. 

The remaining case for 9.(c) is where $x * y = x$ and $y * z = y$. By connectivity and reflexivity we know that $z R y$ or $y R z$. If $y R z$ then $y * z = z$ and thus we have $x * y = x$ and $y * z = z$ which is an already handled case. So we assume that $z R y$ and thus $z * y = y$. Now recalling again that $*$ is a selection operator we must have $x * z = x$ or $x * z = z$. If $x * z = x$ then $x * (y * z) = x * y = x = x * z = (x * y) * z$. This leaves the final case where $x * z = z$. Hence we have $x R z$, and we already assumed $z R y$, so by transitivity we have $x R y$ and hence $x * y=y$. So in this final case $x * y = y$ and $y * z = y$ which is an already handled case.
\end{proof}

\noindent We now give several examples of selection operators.

\begin{example} \label{example:selection-operators-1}
The binary operation $x * y = \max(x, y)$, where $x$, $y$ are integers, has corresponding relation $R = R_{\max} = \ \leq$. I.e., $xRy \Leftrightarrow x \leq y$.
$R$ is reflexive, antisymmetric, connected, transitive. $*$ is idempotent, selective, commutative, associative.
\end{example}

\begin{example} \label{example:selection-operators-2}
The binary operation $x * y=\operatorname{first}(x, y) = x$, has corresponding relation
$R = R_{\operatorname{first}} = \text{equality}$. I.e., $x R_{\operatorname{first}} y \Leftrightarrow x = y$.
$R$ is reflexive, antisymmetric, non-connected, transitive. $*$ is idempotent, selective, noncommutative, associative.
\end{example}

\begin{example} \label{example:selection-operators-3}
The binary operation $x * y = \coalesce(x, y) = (y \text{ if $x$ is undefined else } x)$ has the corresponding relation given as follows.
\begin{equation*}
x R y \Leftrightarrow x R_{\coalesce} y 
      \Leftrightarrow \left(\coalesce(x, y) = y \right)
      \Leftrightarrow \left(\text{$x$ is $\undefined$ or $y=x$} \right)
\end{equation*}
Here `$\undefined$' is a value. $R$ is reflexive, antisymmetric, non-connected, transitive. $*$ is idempotent, selective, noncommutative, associative.
\end{example}

\begin{example} \label{example:selection-operators-4}
The binary operator $x * y = \operatorname{first}(x, y)$ restricted to the 3 element set $\theset{a, b, c}$ gives rise to the following operator and relation tables for $R$ and $*$. Here $T$ and $F$ represent true and false values for the binary relation.

\begin{center}
\begin{tabular}{l|lll}
$R$ & $a$ & $b$ & $c$ \\
\hline
$a$ & $T$ & $F$ & $F$ \\
$b$ & $F$ & $T$ & $F$ \\
$c$ & $F$ & $F$ & $T$ \\
\end{tabular}
\qquad \qquad
\begin{tabular}{l|lll}
$*$ & $a$ & $b$ & $c$ \\
\hline
$a$ & $a$ & $a$ & $a$ \\
$b$ & $b$ & $b$ & $b$ \\
$c$ & $c$ & $c$ & $c$ 
\end{tabular}    
\end{center}
\end{example}

\begin{example} \label{example:selection-operators-5}
The binary operator $* = \coalesce$ restricted to the 3 element set $\theset{a, b, c}$, where $a = \undefined$, gives rise to the following operator and relation tables for $R$ and $*$.

\begin{center}
\begin{tabular}{c|ccc}
$R$ & $a$ & $b$ & $c$ \\
\hline
$a$ & $T$ & $T$ & $T$ \\
$b$ & $F$ & $T$ & $F$ \\
$c$ & $F$ & $F$ & $T$ 
\end{tabular}
\qquad \qquad
\begin{tabular}{l|lll}
$*$ & $a$ & $b$ & $c$ \\
\hline
$a$ & $a$ & $b$ & $c$ \\
$b$ & $b$ & $b$ & $b$ \\
$c$ & $c$ & $c$ & $c$ 
\end{tabular}    
\end{center}
\end{example}

\begin{example} \label{example:selection-operators-6}
Consider the following binary relation and corresponding selection operator.

\begin{center}
\begin{tabular}{l|lll}
$R$ & $a$ & $b$ & $c$ \\
\hline
$a$ & $T$ & $T$ & $F$ \\
$b$ & $T$ & $T$ & $F$ \\
$c$ & $F$ & $F$ & $T$ 
\end{tabular}
\qquad \qquad
\begin{tabular}{l|lll}
$*$ & $a$ & $b$ & $c$ \\
\hline
$a$ & $a$ & $b$ & $a$ \\
$b$ & $a$ & $b$ & $b$ \\
$c$ & $c$ & $c$ & $c$ 
\end{tabular}    
\end{center}

\noindent Note that $a * (c * b) = a$ and $(a * c) * b = b$.
$R$ is reflexive, non-antisymmetric, non-connected, transitive.
$*$ is idempotent, selective, noncommutative, nonassociative.
This is an example of an operation for which the corresponding relation is transitive. SlickDeque therefore produces correct results for this operation even though the operation is nonassociative (see Theorem~\ref{theorem:slick-deque-correctness}).
\end{example}

\begin{example} \label{example:selection-operators-7}
Consider the following binary relation and corresponding selection operator.

\begin{center}
\begin{tabular}{l|lll}
$R$ & $a$ & $b$ & $c$ \\
\hline
$a$ & $T$ & $T$ & $F$ \\
$b$ & $F$ & $T$ & $F$ \\
$c$ & $F$ & $F$ & $T$ \\
\end{tabular}
\qquad \qquad
\begin{tabular}{l|lll}
$*$ & $a$ & $b$ & $c$ \\
\hline
$a$ & $a$ & $b$ & $a$ \\
$b$ & $b$ & $b$ & $b$ \\
$c$ & $c$ & $c$ & $c$ 
\end{tabular}    
\end{center}

\noindent Note that $a *(c * b)=a$ and $(a*c)*b=b$.
$R$ is reflexive, antisymmetric, non-connected, transitive.
$*$ is idempotent, selective, noncommutative, nonassociative.
This is another example of an operation where the corresponding relation is transitive, and for which SlickDeque produces correct results even though the operation is nonassociative (see Theorem~\ref{theorem:slick-deque-correctness}).
\end{example}

\begin{example} \label{example:selection-operators-8}
Consider the following binary relation and corresponding selection operator.

\begin{center}
\begin{tabular}{l|lll}
$R$ & $a$ & $b$ & $c$ \\
\hline
$a$ & $T$ & $T$ & $F$ \\
$b$ & $F$ & $T$ & $T$ \\
$c$ & $T$ & $F$ & $T$ \\
\end{tabular}
\qquad \qquad
\begin{tabular}{l|lll}
$*$ & $a$ & $b$ & $c$ \\
\hline
$a$ & $a$ & $b$ & $a$ \\
$b$ & $b$ & $b$ & $c$ \\
$c$ & $a$ & $c$ & $c$ 
\end{tabular}    
\end{center}

\noindent Note that $a *(b * c) = a$ and $(a * b) * c = c$.
$R$ is reflexive, antisymmetric, connected, intransitive.
$*$ is idempotent, selective, commutative, nonassociative.
This is an example of an operation where the corresponding relation is intransitive and for which SlickDeque does not produce correct results even though the operation is a selection operator (see Theorems~\ref{theorem:slick-deque-correctness} and \ref{theorem:slick-deque-transitivity-necessary}).
    
\end{example}

\begin{example} \label{example:selection-operators-9}
Consider the following binary relation and corresponding selection operator.

\begin{center}
\begin{tabular}{l|lll}
$R$ & $a$ & $b$ & $c$ \\
\hline
$a$ & $T$ & $T$ & $T$ \\
$b$ & $T$ & $T$ & $F$ \\
$c$ & $T$ & $F$ & $T$ \\
\end{tabular}
\qquad \qquad
\begin{tabular}{l|lll}
$*$ & $a$ & $b$ & $c$ \\
\hline
$a$ & $a$ & $b$ & $c$ \\
$b$ & $a$ & $b$ & $b$ \\
$c$ & $a$ & $c$ & $c$ 
\end{tabular}    
\end{center}

\noindent Note that $c *(a * b) = c$ and $(c * a) * b = b$.
$R$ is reflexive, non-antisymmetric, non-connected, intransitive.
$*$ is idempotent, selective, noncommutative, nonassociative.
This is another example of an operation where the corresponding relation is intransitive, and for which SlickDeque does not produce correct results even though the operation is a selection operator (see Theorems~\ref{theorem:slick-deque-correctness} and \ref{theorem:slick-deque-transitivity-necessary}).

\end{example}

The correspondence $* \longmapsto R_{*}$ described in Theorem~\ref{theorem:selection-operators-and-reflexive-relations} may be formulated in alternative ways. Instead of the definition $x R_{*} y \Leftrightarrow x * y=y$, we could have used $y * x=y$, or $y * x=x$, or $x * y=x$. These are all simply related by the opposite operation and opposite relation defined by

\begin{align*}
& x *_\text{op} y = y * x \\
& x R_\text{op} y = y R x
\end{align*}
Thus corresponding to $*$ we have 4 relations $R_*$, $R_{*_\text{op}}$, $\left(R_*\right)_\text{op}$, and $\left(R_{*_\text{op}}\right)_\text{op}$, and corresponding to $R$ we have 4 operations $*_{R}, *_{R_\text{op}}, \left(*_{R}\right)_\text{op}, \left(*_{R_\text{op}}\right)_\text{op}$. To translate Theorem~\ref{theorem:selection-operators-and-reflexive-relations} to these other correspondences we can simply note that

\begin{align*}
& \left(* \longmapsto R_{*}\right)^{-1} = 
R \longmapsto *_{R} \\
& \left(* \longmapsto R_{*_\text{op}}\right)^{-1} = 
R \longmapsto\left(*_{R}\right)_\text{op} \\
& \left(* \longmapsto\left(R_{*}\right)_\text{op}\right)^{-1} = 
R \longmapsto *_{R_\text{op}} \\
& \left(* \longmapsto\left(R_{*_\text{op}}\right)_\text{op}\right)^{-1} =
R \longmapsto\left(*_{R_\text{op}}\right)_\text{op}
\end{align*}
for any selection operator $*$ and reflexive binary relation $R$,
and further note that for each of the properties reflexivity, connectedness, antisymmetry, and transitivity, $R$ has the property if and only if $R_\text{op}$ has the same property, and for each of the properties idempotency, commutativity, associativity, and the property of being a selection operator, $*$ has the property if and only if $*_\text{op}$ has the same property.
%

\section{Introduction to SlickDeque}
\label{sec:introduction-to-slick-deque}

%
%

It is instructive now to consider SlickDeque in the case where there is a single window length.%
\footnote{Shein \cite{Shein2019} describes a multi-query algorithm that simultaneously computes window aggregates for multiple window lengths.
}
To motivate the algorithm let us consider the computation of a window aggregate of integers with $* = \max$, and $R = \leq$. Suppose the window length is $n=7$, and the items in the window are

\begin{displaymath}
    \begin{array}{|l|l|l|l|l|l|l|}
\hline 1 & 3 & 6 & 2 & 5 & 1 & 4 \\
\hline
\end{array}
\end{displaymath}
where new items are inserted on the right and items are evicted from the left.
It is easy to see that the first item, 1, cannot be the max in this window because of the 3 which is one item to the right, and also this initial item cannot result in the max for any window obtained from this window by evictions (on the left) or insertions (on the right). Any eviction will remove this 1, and any insertions will result in a window still containing the 3. So we can remove this initial 1 from consideration. A similar argument applies to the 3 however, because of the 6. Also a similar argument applies to the 2 before the 5, and the 1 before the 4. None of these can result in the max of a window obtained from this one by evictions (on the left) or insertions (on the right). So we can delete all these elements from the window without affecting the current aggregation or future aggregations.
This gives the following array.

\begin{displaymath}
    \begin{array}{|l|l|l|l|l|l|l|}
\hline \xcancel{1} & \xcancel{3} & 6 & \xcancel{2} & 5 & \xcancel{1} & 4 \\
\hline
\end{array}
\end{displaymath}
But we still need enough information to know when to evict items from the window, and if we simply delete the items that will never yield the max, we will lose this bookkeeping information. The solution is to record the item number (i.e., the original position in the input data) as well. This gives us the following data structure.

\begin{displaymath}
\begin{array}{r}
\begin{array}{l l l}
3 & 5& 7
\end{array}\\
\begin{array}{|l|l|l|}
\hline 6 & 5 & 4 \\
\hline
\end{array}
\end{array}
\end{displaymath}
The first item in this new data structure is the window-max, and now eviction and insertion are easy to understand. Our current latest item has item number 7, so an eviction should remove the item with item number 1 . However the item with item number 1 is not in the array (or deque), so instead of evicting we should simply increment an index (or pointer/counter) indicating where the start of the window should be. When this index reaches 3 (the item number of the 6), we should then evict the 6. For insertions we should add an item, together with its item number to the end of the array, and then remove any `dead' items to its left that will never yield the max. For example, if we add a 1 we will not create any new dead items and we will get

\begin{displaymath}
\begin{array}{r}
\begin{array}{l l l l}
3 & 5 & 7 & 8 
\end{array}\\
\begin{array}{|l|l|l|l|}
\hline 6 & 5 & 4 & 1 \\
\hline
\end{array}
\end{array}
\end{displaymath}
But if we subsequently add a 5, this will clear out the 1 and the 4, as well as the 5 that was already there. I.e. we clear out everything everything back to the 6. This gives the following data structure.

\begin{displaymath}
\begin{array}{r}
\begin{array}{l l}
3 & 9 
\end{array}\\
\begin{array}{|l|l|}
\hline 6 & 5\\
\hline
\end{array}
\end{array}
\end{displaymath}

The SlickDeque algorithm applies these ideas, with the only difference being that they apply to a general selection operator, whose corresponding relation is transitive. We can describe the algorithm using the {\tt insert}, {\tt evict}, and {\tt query} procedure interface used for Two Stacks and DABA.

\begin{algorithm}[SlickDeque Implementation Sketch]
\
%
%
\begin{alltt}
initialization():
    i = 0                                i is the index of the last element in the window
    j = 0                               j is the index one before the start of the window
    arr = An empty array of pairs which allows removal on the left (popleft)
          and the right (popright), and which allows appends (pushright) on the 
          right. (This could be implemented as a deque using chained arrays, circular 
          buffers, or a link list). Items are accessed as arr[1], ..., arr[last], 
          with arr[last] as the most recently pushed item.

insert(x):
    while length(arr) > 0 and x * arr[last][1] = x
        popright(arr)                                This removes the last item arr[last]
    i = i + 1
    pushright(arr, (x, i))               Pushes the pair (x, i) onto the end of the array

evict():                                             This will fail if the array is empty
    j = j + 1
    if arr[1][2] = j                Compares the second item of the pair in arr[1] with j
        popleft(arr)                         Removes the least recently added item arr[1]

query():
    return arr[1][1]                       Return the first element of the pair in arr[1]
\end{alltt}
\end{algorithm}
\medskip

\noindent In order for SlickDeque to work correctly it is not necessary to assume that $*$ is associative, but instead that $*$ is a selection operator and that $R_*$ is transitive. Of course if $*$ is an associative selection operator then these conditions are fulfilled.

\begin{theorem}[SlickDeque Correctness]
\label{theorem:slick-deque-correctness}
Assume $*$ is a selection operator and $R_*$ is transitive, then SlickDeque computes the window aggregates
\begin{equation*}
    a_i * (a_{i-1} *(\ldots *(a_{j+2} * a_{j+1}) \ldots ))
\end{equation*}
In particular, if $*$ is an associative selection operator then SlickDeque computes these aggregates.
\end{theorem}
\begin{proof}[Proof of SlickDeque Correctness]
Consider the window $a_{j+1}, \ldots, a_i$, and let $y_k = a_k * (\ldots * (a_{j+2} * a_{j+1})\ldots)$. By construction
\begin{equation*}
  y_{k_l} = a_{k_l} * (\ldots * (a_{k_{l-1}} * (\ldots * (a_{k_1} * (\ldots * (a_{j+2} * a_{j+1}) 
                 \ldots)) \ldots )) \ldots)
\end{equation*}
where $k_1, \ldots, k_l$ are the item numbers (second component) of the pairs in the SlickDeque array. Note that $k_l = i$. To prove the theorem we must show that $y_{k_l} = a_{k_1}$. Let $R = R_*$.

First we prove that $y_{k_1} = a_{k_1}$. Note that $y_{k_1} \in \theset{a_{j+1}, \ldots, a_{k_1}}$ as $*$ is a selection operator. So there is an integer $p$ for which $y_{k_1} = a_p$ and $j+1 \leq p \leq k_1$, and we may choose $p$ to be the largest such integer. Clearly $a_p \notin \theset{a_{p+1}, \ldots, a_{k_1}}$. Therefore $y_m = a_p$ for $p \leq m \leq k_1$ as for any $m$ with $p \leq m \leq k_1$ we have $a_p = y_{k_1} \in \theset{y_m, a_{m+1}, \ldots, a_{k_1}}$ and hence $a_p = y_m$. Now suppose $p \neq k_1$. Then, as $(a_{k_1}, k_1)$ is in first position in the array, the entry $(a_p, p)$ must have been removed, so there is some $m$ with $p < m \leq k_1$ such that $a_m * a_p = a_m$. But then we must have $y_m = a_m * y_{m-1} = a_m * a_p = a_m$. Thus $a_m = y_m = a_p = y_{k_1}$, which contradicts the choice of $p$ as the largest $p$ with $a_p = y_{k_1}$ and $j+1 \leq p \leq k_1$. Hence $p = k_1$ and $y_{k_1} = a_{k_1}$.

We now proceed by induction, proving that $a_k * a_{k_1} = a_{k_1}$ for all $k$ with $k_1 \leq k \leq k_l$. Clearly the induction hypothesis holds for $k = k_1$. Consider $k$ with $k_1 \leq k < k_l$ and assume that $a_p * a_{k_1} = a_{k_1}$ for all $p$ with $k_1 \leq p \leq k$. Suppose that there is no $p$ with $k_1 \leq p \leq k$ such that $a_{k+1} * a_p = a_p$. Then $a_{k+1} * a_p = a_{k+1}$ for all $p$ with $k_1 \leq p \leq k$. But then when $(a_{k+1}, k+1)$ was inserted into the array the algorithm must have removed all entries, including $(a_{k_1}, k_1)$, and put $(a_{k+1}, k+1)$ in first place, and this cannot have happened as $k_1 < k+1 \leq k_l$ and therefore $(a_{k_1}, k_1)$ is in first place. Therefore there must be an $p$ with $k_1 \leq p \leq k$ such that $a_{k+1} * a_p = a_p$. Then $a_{k+1} R a_p$ and also $a_p R a_{k_1}$, so by transitivity $a_{k+1} R a_{k_1}$, and hence $a_{k+1} * a_{k_1} = a_{k_1}$. This completes the induction step.
Therefore $a_k * a_{k_1} = a_{k_1}$ for all $k$ with $k_1 \leq k \leq k_l$, and hence $a_k * y_{k_1} = y_{k_1} = a_{k_1}$ for all $k$ with $k_1 \leq k \leq k_l$. Therefore $y_k = y_{k_1} = a_{k_1}$ for all $k$ with $k_1 \leq k \leq k_l$. Hence $y_{k_l} = a_{k_1}$.
\end{proof}

\begin{remarks} \ \nopagebreak
\begin{enumerate}

\item 
Note that the SlickDeque algorithm, as given here, uses the condition {\tt{x*arr[last][1] = x}}, as the removal condition on insertion, whereas Shein et al.\ \cite{Shein2018} use the condition {\tt{arr[last][1]*x = x}}. This is because we are computing the window aggregates $a_i * (a_{i-1}*(\ldots *(a_{i-n+2}*a_{i-n+1})\ldots))$, whereas Shein et al.\ \cite{Shein2018}  computes $((\ldots(a_{i-n+1} * a_{i-n+2}) * \ldots) * a_{i-1}) * a_i$. These are simply related by the opposite operator $*_\text{op}$. The condition we use in the algorithm is $\text{\tt{arr[last][1]}} R_{*_\text{op}} \text{\tt{x}}$, and Shein et al.\ \cite{Shein2018} uses 
$\text{\tt{arr[last][1]}} R_* \text{\tt{x}}$, as the order of our aggregations is opposite to theirs. Note that while $R_{*_\text{op}}$ appears in our formulation of the algorithm, the operator $R_*$ appears directly in the conditions for, and proof of, correctness.

\item
It is easy to see that to compute $N$ window aggregates of length $n$ (after startup), SlickDeque uses at most $2 N$ $*$-operations, and at most $2 N$ equality comparisons on items being aggregated. 

\item 
The transitivity of $R_*$ is required for SlickDeque to work, as can be seen by Example \ref{example:selection-operators-8}, and note that in this case the operator of the counter-example is commutative in addition to being a selection operator. On the other hand associativity is not required, as can be seen by Examples~\ref{example:selection-operators-6} and \ref{example:selection-operators-7}.

\item
SlickDeque does apply to noncommutative operations, and some noncommutative selection operations are important in practical applications. In particular, coalesce is an associative noncommutative selection operator, corresponding to a non-connected relation. The window aggregate in this case is a fill-forward operation. Thus SlickDeque gives an efficient algorithm for fill-forward operations.    

\item
As with Two Stacks and DABA, SlickDeque can be used to compute fixed length window aggregates in steady state by an {\tt insert} followed by an {\tt evict} or an {\tt evict} followed by an {\tt insert}. The {\tt insert-evict} version can result in an extra $*$ operation in cases where a new item $x$ is inserted satisfying $x * z = x$ for all items in the array. In this case the extra operation may be avoided by a {\tt combined-insert-evict} procedure, which is simply {\tt evict} followed by {\tt insert} together with an emptiness check.

\begin{alltt}
combined-insert-evict(x):
    if length(arr) > 0
        evict()
        insert(x)
\end{alltt}
\end{enumerate}
\end{remarks}

Theorem~\ref{theorem:slick-deque-correctness} has a converse, which states that for selection operators $*$ the transitivity of $R_*$ is not only a sufficient condition for correctness of SlickDeque, but is also a necessary condition.

\begin{theorem}[Transitivity Necessary for SlickDeque]
\label{theorem:slick-deque-transitivity-necessary}
Assume $*$ is a selection operator and $R_*$ is intransitive, then there is a window length, and a sequence of input data, for which SlickDeque computes at least one of the corresponding window aggregates incorrectly.
\end{theorem}
\begin{proof}
Assume that $*$ is a selection operator and $R_*$ is intransitive. Then there must be distinct $a$, $b$, $c$, with $a R_* b$ and $b R_* c$ and not $a R_* c$. Then $a * b = b$ and $b * c = c$ and $a * c = a$. Now consider calling the {\tt insert} procedure of SlickDeque on the elements $c$, $b$, $a$ in turn. After this, the array {\tt arr} of the SlickDeque algorithm will contain $[(c, 1), (b, 2), (a, 3)]$. However $a*(b*c)=a*c=a\neq c$, so the product of the elements in the window $c, b, a$ is not equal to the first entry in the pair at the start of the array. Thus SlickDeque computes the sliding window *-product of length 3 incorrectly for the input sequence $c, b, a$. (For an alternative proof, note that the result follows from verification on each of the 35 intransitive reflexive relations on any fixed three element set, and this may be easily automated.)
\end{proof}

\subsection{SlickDeque Latency}


SlickDeque can suffer from latency spikes of up to $n$ (or $n-1$ for {\tt evict-insert}) $*$ operations, and $n$ equality comparisons on some input sequences. Shein et al.\ \cite{Shein2018} note that for data arriving in random $R_*$ (or $R_{*_\text{op}}$) order, the worst case spike of length $n$ will occur infrequently. Whether this is frequent or infrequent or even of concern when it does occur depends, of course, on the operation $*$, the nature of the input data, the window length $n$, and the application for which the calculation is performed.
For concreteness, however, let us consider the case where $*= \max$, $R_*=\leq$, and SlickDeque is operated using the {\tt insert-evict} version. The worst case latency spike for SlickDeque will then occurs when data arrives in descending order for $n$ or more items, and then this is followed by an item that is greater than the previous $n$ items. I.e. $a_{i-n} > a_{i-n+1} > \ldots > a_{i-1}$, and then $a_i \geq a_{i-n+1}$. In terms of the $*$ operator, in general, this condition can be written as $a_j * a_{j-1} \neq a_j$ for $j=i-1, i-2, \ldots, i-n+1$ and then $a_i * a_j=a_i$ for $j=i-1, i-2, \ldots, i-n+1$. This kind of situation can arise whenever the data correspond to decaying or transient responses to sudden changes. Examples of such systems abound.

\begin{itemize}
  \item Decay of temperature after sudden heat pulses.
  \item Damped mechanical or electrical systems after sudden impulses.
  \item Message traffic with bursts of activity.
  \item Response to user control. E.g., a mechanical or industrial control system may move to a new steady state after a user input. This may be followed by another user input
  \item Responses to news.
  \item Responses to crises, or emergencies, or system critical events.
\end{itemize}
In many applications latency is not a concern. However in applications where it is a concern it is not uncommon for low latency to be most important in response to some kind of event.

\section{Summary of Sliding Window Algorithms}
\label{sec:sliding-window-summary}

For the algorithms that we have covered in these notes, we now summarize their performance and properties.

\medskip

\begin{samepage}
\begin{center}
{\bf{Sliding Window Algorithm Performance Characteristics}} \nopagebreak 
\vspace*{6pt} 
{ 
\renewcommand{\arraystretch}{1.5} \nopagebreak 
\begin{tabular}{|l|l|l|l|}
\hline
Algorithm & Complexity Bound & Max.\ Latency & Requirements\\
\hline 
Subtract-on-Evict & \parbox[t]{1.45in}{$2N$ $*$-operations\\ $N$ inversions\\[-1.5ex]} 
                  & \parbox[t]{1.125in}{$2$ $*$-operations\\ $1$ inversion} 
                  & \parbox[t]{1.325in}{Associativity\\ Invertibility} \\
\hline
SlickDeque & \parbox[t]{1.45in}{$2N$ $*$-operations\\ $N$ equality comparisons\\[-1.5ex]} 
            & \parbox[t]{1.125in}{$n-1$ $*$-operations\\ $n-1$ comparisons} 
            & \parbox[t]{1.325in}{$*$ a selection operator\\ $R_*$ is transitive}\\
\hline
Two Stacks &
$3N$ $*$-operations  &
$n-1$ $*$-operations &
Associativity\\
\hline
DEW & $3N$ $*$-operations  & $3$ $*$-operations & Associativity\\
\hline
DABA Lite & $4N$ $*$-operations & $6$ $*$-operations & Associativity\\
\cline{1-1} \cline{3-3}
DDABA Lite & & $4$ $*$-operations & \\
\hline
DABA & $5N$ $*$-operations & $8$ $*$-operations & Associativity\\
\cline{1-1} \cline{3-3}
DDABA & & $5$ $*$-operations & \\
\hline
\end{tabular}
} 
\end{center}
\end{samepage}
\medskip

For a general associative $*$ operator, Two Stacks is simple and has the lowest operation count, followed closely by DEW. If latency is a concern, then DEW combines low operation count with the lowest latency. If variable size windows are important, then Two Stacks is preferable to DEW, or if latency is an also issue, then DABA Lite is preferable. However all of these algorithms are likely to be efficient for high quality implementations when the $*$ operations are cheap. Only when the $*$ operation is expensive does it become more important to choose the algorithm the with the absolute lowest operation count.

If operation count is critical and other properties, such as invertibility or being a selection operator are available then using a more targeted situation-specific algorithm, such as Subtract-on-Evict or SlickDeque may be beneficial. Note however, that Subtract-on-Evict may degrade accuracy or even give incorrect results in situations where the $*$ operation is approximate or not fully invertible.

\section{What Is Next and Why}
\label{sec:what-is-next}

We conclude this chapter with a brief overview of the questions raised which we will address in the following chapters.

\subsubsection*{How to handle nonassociative operations and set actions}
All of the algorithms described in Chapters \ref{chapter:moving-sums}--\ref{chapter:other-sequential-algorithms} work with associative operators, with the exception of SlickDeque which works for selection operators associated with reflexive transitive relations. We have already seen examples of nonassociative operators in Section~\ref{sec:other-window-calculations}, including examples that arise under practical circumstances. These are common in practice, so we must find techniques to handle these. There are general mathematical techniques to relate nonassociative operations to associative operations, and in particular to function composition, which is always associative. We start to explore this in Chapters \ref{chapter:windowed-recurrences}--\ref{chapter:algorithms-for-windowed-recurrences}, and thereby replace the question {\em `Can I relate my nonassociative operation to an associative operation I can compute with?'} with the question {\em `Can I effectively compute a function composition?'}. This replaces algorithmic questions with questions of a mathematical and algebraic nature (though still inherently algorithmic), and leads to {\tt lift}, {\tt compose}, {\tt apply} interfaces to sliding window algorithms, as well as parallel reductions and scans. This is similar to, and explains, the {\tt lift}, {\tt combine}, {\tt lower} interface of Hirzel et al.\ \cite{Hirzel2017}.

An important concept that we explore here is that of semi-associativity, which abstracts the notion of function application in a manner similar to the way that associativity abstracts the notion of function composition. Our definition of semi-associativity has fewer conditions (i.e. is logically weaker) than that of other authors, but is is sufficiently strong to derive parallel algorithms for reductions, prefix sums/scans, and windowed recurrences, as well as to derive the main sequential algorithms for windowed recurrences. The natural object of study here is a set action of a set on another set, rather than a binary operation on a set.

\subsubsection*{What is a good general definition of windowed recurrence?}
A good definition should cover known examples without undue complexity, but also abstract away unnecessary details that can clutter up proofs and algorithms. In Chapter~\ref{chapter:windowed-recurrences} we propose a definition which covers associative and nonassociative cases, and set actions, and has a simple mathematical structure.

\subsubsection*{Are there vectorized algorithms?}
The algorithms of Chapters \ref{chapter:moving-sums}--\ref{chapter:other-sequential-algorithms} provide a selection of properties for users including trade-offs in performance, latency, window length variability, and parallelizability. One missing property of the current offerings is full vectorizability.%
\footnote{
Though also see Snytsar and Turakhia \cite{SnytsarTurakhia2019}.
}

A fully vectorized algorithm is obviously important for use on vector processors, and for GPU computation, as it allows the details of how the vector operations are implemented in software or hardware to be abstracted away from the algorithm itself. The importance of vectorized algorithms does not end at vector or parallel computation, however, and there are other, sometimes more important, reasons why vector algorithms are necessary. These have to do with what operations or interfaces are exposed to algorithm users in real world systems. A data scientist using a table processing system or statistical package may only have access to operations that work on columnar data, or if there are available operations acting on individual numbers these may be vastly less efficient than vectorized versions 
because of hardware or software constraints. A simple vector algorithm allows such users to implement windowed recurrences using these efficient vector operations without having to rewrite the system they are using---something which may not be a technical, or a legal, option for them.

But suppose now that the system already came with efficient and well implemented window aggregation procedures. Would our hypothetical user still have a need for vectorized algorithms? The answer is `Yes, if the user wants to define their own aggregation operations.' Vectorized algorithms, in addition to abstracting away and leveraging vector operations, can also present an interface where the user defined $*$ operation passed in to the system is itself a vector operation. This allows users of such systems to define their own aggregations using efficient vector operations. There are, of course, several ways the system designer could meet the need for user defined vector operations, without using a vectorized algorithm for windowed recurrences. One approach would be to build in a de-vectorization procedure that analyses the user's vector code and compiles it to a fast procedure usable by non-vectorized aggregation algorithms. Another approach is to have built the system so there is no particular speed benefit to vector operations over scalar operations (though that itself may indicate a missed opportunity to fully utilize the available hardware). Neither of these options are available, however, to the user working with an already built and deployed system, if the designers have not included them. For such a user, a vectorized algorithm that works with the system they have, rather than the system they wish they had, is of great benefit.
In Chapters \ref{chapter:vector-algorithms-guide}--\ref{chapter:vector-pseudo-code} we develop fully vectorized algorithms for windowed recurrences. These algorithms have connections to well known constructions in abstract algebra, and also provide new algorithms, and new variants of known algorithms, for computing prefix sums.

\subsubsection*{What are practical examples of windowed recurrences beyond those we have seen?}
The variety of calculations for which function composition can be efficiently computed (or equivalently, lifted to a semigroup or magma with efficiently computable operation) is surprisingly large. Blelloch \cite{Blelloch1993}, Fisher and Ghuloum \cite{FisherGhuloum1994}, and Chin, Takano and Hu \cite{ChinTakano1998} give many examples in the context of parallelization of prefix sums/scans and reductions.

For the practitioner, it is helpful to see commonly used cases. In Chapter~\ref{chapter:examples} we provide a gallery of examples arising from practical applications in a variety of fields, and give the recurrence functions, as well as the corresponding associative and semi-associative operators, semigroups, and function composition formulae, needed to efficiently solve the corresponding window aggregation problem (or prefix sum/scan or parallel reduction problem).

\subsubsection*{How can I compute the windowed recurrence I am interested in?}
When the recurrence you are interested in is not one of the known examples, you need to also have techniques for deriving new function composition formulae. In Chapter~\ref{chapter:examples} we also give examples of general constructions for building new function composition formulae from existing ones. These correspond to algebraic constructions for constructing semigroups and magmas from other semigroups and magmas, and can be used by practitioners to solve concrete windowed recurrence problems.


%% file: htcams-arxiv-ch06-windowed-recurrences.tex
\chapter{Windowed Recurrences}
\label{chapter:windowed-recurrences}

\section{Definition of Windowed Recurrence}
\label{section:definition-of-windowed-recurrence}

In this section we define windowed recurrences, and explore the definition. In Chapter~\ref{chapter:categories} we extend these definitions to categories and magmoids. 

\begin{definition}[Windowed Recurrence]
\label{definition:windowed-recurrence-functions}

Let $X$ be a set, and assume $x_0, x_1, \ldots$ is a sequence of elements of $X$, and that $f_1, f_2, \ldots$, is a sequence of functions $f_i\colon X \rightarrow X$. Let $n$ be a strictly positive integer. Then the {\em windowed recurrence of length $n$}, corresponding to the sequences $\theset{x_i}, \theset{f_i}$, is the sequence

\begin{equation*}
y_i = \ourcases{
f_i(f_{i-1}(\ldots f_1(x_0) \ldots))           &  \text{for $1 \leq i < n$}\\
f_i(f_{i-1}(\ldots f_{i-n+1}(x_{i-n}) \ldots)) &  \text{for $i \geq n$}
}
\end{equation*}
obtained by applying the functions to the data, $\theset{x_i}$, $n$ times.
In other words $y_1 = f_1(x_0), y_2 = f_2(f_1(x_0)), \ldots$, $y_n = f_n(\ldots f_1(x_0) \ldots), 
y_{n+1} = f_{n+1}(\ldots f_2(x_1) \ldots), \ldots, y_i = f_i(\ldots f_{i-n+1}(x_{i-n}) \ldots), \ldots$.
\end{definition}

\noindent Given the data of a set $X$ and a sequence of functions $f_1, f_2, \ldots$ on $X$ we may also define non-windowed recurrences, or simply recurrences, and reductions.

\begin{definition}[Recurrence, Reduction]
\label{definition:recurrence-reduction}

Let $X$ be a set, and assume $x_0 \in X$, and that $f_1, f_2, \ldots$, is a sequence of functions $f_i\colon X \rightarrow X$. 
\begin{enumerate}
\item 
The {\em recurrence} corresponding to the element $x_0$ and sequence $\theset{f_i}$ is the sequence defined by
\begin{equation*}
z_i = \ourcases{
    x_0 & \text{for $i=0$} \\
    f_i(z_{i-1}) & \text{for $i \geq 1$}}
\end{equation*}
In other words, $z_0 = x_0, z_1 = f_1(x_0), z_2 = f_2(f_1(x_0)), z_3 = f_3(f_2(f_1(x_0))), \ldots$.

\item 
Let $N$ be a strictly positive integer. Then the {\em reduction} corresponding to the element $x_0\in X$ and the sequence $\theset{f_i}$ is the element of $X$ given by
\begin{equation*}
    f_N(f_{N-1}(\ldots f_1(x_0)\ldots))
\end{equation*}
\end{enumerate}
\end{definition}

\noindent Thus corresponding to a recurrence relation 
\begin{equation*}
z_i = f_i(z_{i-1})
\end{equation*}
we may define a {\em reduction} given an initial element $z_0 = x_0$ and a strictly positive integer $N$, we may define a {\em recurrence} given just an initial element $z_0 = x_0$, and we may define a {\em windowed recurrence} given a sequence of initial elements $x_0, x_1, \ldots$.

\begin{remarks}[On the Definition of Windowed Recurrence] \begin{samepage} \
\begin{enumerate}
\item 
As with the definition of moving sums, we have chosen a convention for the $i < n$ case corresponding to a choice of how to fill in for the missing functions $f_j$ when $j < 1$. The convention we use corresponds to defining $f_j = \operatorname{id}_X\colon X \rightarrow X\colon x \mapsto x$, for $j < 1$, and $x_j = x_0$, for $j < 1$.
\end{enumerate}
\end{samepage}
\begin{enumerate}
\setcounter{enumi}{1}
\item 
Other conventions are possible, e.g., by extending $f_i$ to $i < 1$ and $x_i$ to $i < 0$ in any way one chooses. Another choice is to choose not to define $y_i$ for $i < n$. Alternatively one can choose to extend $X$ with a value representing an undefined result, and set $y_i = \undefined$ for $i < n$ (and $x_i=\undefined$ for $i < 0$). The possibilities are analogous to those discussed in Section~\ref{sec:notes-on-conventions}.

\item
Windowed recurrences are related to (non-windowed) recurrences as follows. A windowed recurrence corresponds to running a recurrence for a fixed number of steps, $n$, starting from different initial points, $x_{i-n}$, and different positions $f_{i-n+1}$ in the function sequence. More explicitly, if we define recurrences $z_{i, j}$ for each $i \geq 0$, by
\begin{align*}
z_{i, i} & = x_i             & \text{for $i \geq 0$}\\
z_{i, j} & = f_j(z_{i, j-1}) & \text{for $j > i$}
\end{align*}
Then
\begin{equation*}
y_i = \ourcases{
    z_{0,i}   & \text{for $1 \leq i < n$} \\
    z_{i-n,i} & \text{for $i \geq n$}
}
\end{equation*}
The sequence $y_i$ is indicated on the following table, in which each row represents a recurrence starting from a different initial point and with a truncated sequence of functions.
{ 
\renewcommand{\arraystretch}{1.5}
\begin{center}
\begin{tabular}{c|ccccccc}
$i$ & $0$ & $1$ & $2$ & \ldots & $n$ & $n+1$ & \ldots \\
\hline \rule{0pt}{1.2\normalbaselineskip}
$z_{0,i}$ & $x_0$ & \circletext{4pt}{$f_1(x_0)$} & \circletext{4pt}{$f_2(f_1(x_0))$} & \ldots 
& \!\circletext{4pt}{$f_n(\ldots f_1(x_0)\ldots)$} & $f_{n+1}(\ldots f_1(x_0)\ldots)$ & $f_{n+2}(\ldots f_1(x_0)\ldots)$\\
$z_{1,i}$ &       & $x_1$      & $f_2(x_1)$      & \ldots 
& \!$f_n(\ldots f_2(x_1)\ldots)$ & \circletext{4pt}{$f_{n+1}(\ldots f_2(x_1)\ldots)$} & $f_{n+2}(\ldots f_2(x_1)\ldots)$\\
$z_{2,i}$ &       &            & $x_2$           & \ldots 
& \!$f_n(\ldots f_3(x_2)\ldots)$ & $f_{n+1}(\ldots f_3(x_2)\ldots)$ & \circletext{4pt}{$f_{n+2}(\ldots f_3(x_2)\ldots)$}\\
          &       &            &         & $\ddots$ \!& & 
\end{tabular}
\end{center}
} 

\item 
Our definition only considers single term windowed recurrences. This is because any multi-term recurrence may be easily reduced to a single term recurrence as follows.
Suppose $z_i = f_i(z_{i-1}, \ldots, z_{i-p})$ is a multi-term recurrence with initial conditions $z_0 = x_0, \ldots, z_{p-1} = x_{p-1}$. Then we can define an equivalent single term recurrence by

\begin{equation*}
u_i = \smatl{z_{i-p+1}\\ \vdots\\ z_i}, \quad \text{ and } u_{p-1} = \smatl{x_0\\ \vdots\\ x_{p-1}}
\end{equation*}
so that

\begin{equation*}
u_i = \smatl{
\component{2}{u_{i-1}} \\
\vdots \\
\component{p}{u_{i-1}} \\
f_i(u_{i-1})} = g_i(u_{i-1})
\end{equation*}
%
%
where $\component{k}{u}$ denotes the $k$-th component of $u$ and $f_i(u_{i-1}) = f_i(\component{p}{u_{i-1}}, \ldots, \component{1}{u_{i-1}})$.
%
Given a multi-term recurrence, and initial conditions $x_0, x_1, \ldots$, the corresponding multi-term windowed recurrence is the windowed recurrence for the functions $g_i$, and initial conditions $v_i = \smat{x_{i-p+1}\\ \vdots\\ x_i}$. Of course the initial conditions $v_i$ could alternatively be chosen to be arbitrary vectors of length $p$ instead.

\end{enumerate}
\end{remarks}

Before commenting further on the definition of windowed recurrence, let us also formalize the definition of sliding window $*$-product that we used in the preceding chapters. This definition appeared informally in Example~\ref{example:moving-sums-binary-operations}.

\begin{definition}(Sliding Window $*$-Product) \label{definition:sliding-window-*-product}
Let $A$ be a set, and let $*\colon A \times A \rightarrow A$ be a binary operation on $A$. Let $a_1, a_2, \ldots$ be a sequence of elements of $A$, and let $n$ be a strictly positive integer. Then the {\em sliding window $*$-product} of length $n$ corresponding to the binary operation $*$ and the sequence $\theset{a_i}$, is the sequence
\begin{equation*}
y_i = \ourcases{
    a_i *(a_{i-1} * (\ldots * (a_2 * a_1)             \ldots )) & \quad \text{ for $1 \leq i < n$} \\
    a_i *(a_{i-1} * (\ldots * (a_{i-n+2} * a_{i-n+1}) \ldots )) & \quad \text{ for $i \geq n$}
} 
\end{equation*}
\end{definition}

\noindent The definition of sliding window $*$-product has a close relation to the definition of windowed recurrence as the following remarks show.

\begin{remarks}[Windowed Recurrences and Sliding Window $*$-Products] \
\begin{enumerate} \label{remarks:windowed-recurrences-and-sliding-window-*-products}
\item 
If $y_i$ is the windowed recurrence of length $n$ for the sequence of initial values $\theset{x_i}$ and the function sequence $\theset{f_i}$, then $y_i = (f_i \circ \ldots \circ f_{i-n+1})(x_{i-n})$ for $i \geq n$, so $y_i$ is is equivalent to a sliding window $\circ$-product applied to the sequence $x_{i-n}$. This observation is the basis for algorithms for computing windowed recurrences.

\item
If $A$ is a set with a binary operation $*$ and $a_0, a_1, a_2, \ldots$ is a sequence with $a_i \in A$, then we may set $X = A$ and define $f_i(x) = a_i * x$, for $i=1, 2, \ldots$, $x \in X$. Also let $x_i=a_i$, for $i=0, 1, \ldots$. Then the windowed recurrence of length $n$ corresponding to the sequences $\theset{f_1, f_2, \ldots}$, $\theset{x_0, x_1, \ldots}$ is equal to the sequence of sliding window $*$-products of length $n+1$ corresponding to $\theset{a_0, a_1, \ldots}$ with the first element removed. I.e., it is 
\begin{displaymath}
a_1 * a_0, \ldots, a_n*(\ldots * (a_1 * a_0)\ldots), a_{n+1}*(\ldots * (a_2 * a_1)\ldots), \ldots
a_i*(\ldots * (a_{i-n+1} * a_{i-n})\ldots), \ldots
\end{displaymath}
This gives an approach to computing sliding window $*$-products of nonassociative operations. See Section~\ref{sec:nonassociative-operations}.

\item
An alternative relationship between sliding window $*$-products and windowed recurrences is available if $*$ is a binary operation on a set $A$ and there is an element $1\in A$ that is a right identity for $*$, i.e., $a * 1 = a$ for all $a \in A$.%
\footnote{
E.g., this happens when $(A, *)$ is a monoid.
}
Let $a_1, a_2, \ldots$ we a sequence of elements of $A$, and set $X = A$, and $f_i(x) = a_i * x$ for $i \geq 1$, $x \in X$ as before. This time, however, we set $x_i = 1$ for $i \geq 0$. Then the $y_i$ of the windowed recurrence of length $n$ corresponding to $\theset{f_1, f_2, \ldots}$, $\theset{x_0, x_1, \ldots}$ exactly match the $y_i$ of the sliding window $*$-product of length $n$ corresponding to $\theset{a_1, a_2, \ldots}$.

\item Another relationship between sliding window $*$-products and windowed recurrences is as follows. Again we assume $*$ is a binary operation on $A$ and that $*$ has a right identity. Let $a_1, a_2, \ldots$ be a sequence of elements of $A$. Let $X = A$ and define $f_i(x) = a_i * x$, for $i=1, 2, \ldots$, $x \in X$. We now let $\theset{x_i}$ be the sequence $1, a_1, a_2, \ldots$. I.e., $x_0 = 1$ and $x_i = a_i$ for $i \geq 1$. Then the windowed recurrence of length $n-1$ corresponding to sequences $\theset{f_1, f_2, \ldots}$, $\theset{x_0, x_1, \ldots}$ is equal to the sequence of sliding window $*$-products of length $n$ of $\theset{a_1, a_2, \ldots}$. Despite the similarity to the construction in 2.\ this is a different relationship between sliding window $*$-products and windowed recurrences. In Chapter~\ref{chapter:vector-windowed-recurrences} both of these examples are subsumed under the same relationship between vector sliding window $*$-products and vector windowed recurrences, but the shift operators used differ between the two. See Example~\ref{example:vector-set-action-from-vector-*-product} part 2, and also Examples~\ref{example:vector-set-action-variable-length-sequences}, \ref{example:vector-set-action-fixed-length-sequences}, and \ref{example:vector-set-action-unbounded-length-sequences}.

\item 
The constructions of 2.\ and 4.\ may, at first sight, make it appear that the window length conventions for the definitions of {\em windowed recurrence} and {\em sliding window $*$-product} are mismatched. The reasons for the choice of length convention in the definitions are two-fold. Firstly, as noted in 1., with the choices as in Definition~\ref{definition:windowed-recurrence-functions}, we have $y_i=(f_i \circ \ldots \circ f_{i-n+1})(x_{i-n})$ so the windowed recurrence of length $n$, i.e.\, $y_i$, corresponds to a sliding window $\circ$-product of length $n$ applied to the shifted sequence $x_{i-n}$. This is the primary reason for the choice, as it makes the conventions for algorithms for computing windowed recurrences simpler. A second reason is the construction noted in 3., which applies in the frequently occurring case where $*$ has a right identity.

\end{enumerate}
\end{remarks}

\subsection{Set Action Windowed Recurrences}

 We now reformulate the definition of windowed recurrences into the language of set actions. At first sight, this might look like a trivial change of notation, but it provides the notation with a place to denote the {\em data} describing a function, and thus provides us with a language in which we can describe the algebraic properties that such data should satisfy. 

\begin{definition}[Set Action]
An action of a set $A$ on a set $X$ is another name for a function $\bullet\colon A \times X \rightarrow X$, with the convention that the function application is written in infix notation rather than prefix notation. Thus for set actions we write $a \bullet x$ instead of $\bullet(a, x)$.    
\end{definition}

\noindent It also helps us to have a name for the space of functions on a set $X$.

 \begin{definition}[Endomorphisms of a Set]
 Let $X$ be a set. Then we denote the set of functions from $X$ to $X$ by $\Endop(X)$.
 \end{definition}

Now suppose $X$ is a set and we have a sequence of functions $f_1, f_2, \ldots \in \Endop(X)$. If we want to notate the data describing the functions $f_i$, we may call the sequence of data $a_1, a_2, \ldots$, where the $a_i$ come from a second set $A$, and instead of $f_1, f_2, \ldots$, write $f_{a_1}, f_{a_2}, \ldots$, where $f_i = f_{a_i}$. What we really have now is a function $f\colon A \rightarrow \Endop(X)$ and a sequence of elements $a_1, a_2, \ldots$. Another common name for a function $f\colon A\rightarrow \Endop(X)$ is a {\em collection of functions indexed by $A$}, and in places we denote such an indexed collection $\theset{f_a: a\in A}$.
To get from the indexed sequence of function $f_{a_1}, f_{a_2}, \ldots$ to a description of windowed recurrences in terms of set actions we use the well known correspondence between set actions $\bullet\colon A \times X \rightarrow X $ and functions $f\colon A \rightarrow \Endop(X)$ given by
\begin{equation}
a\bullet x = f_a(x)
\end{equation}
for $a \in A$, $x \in X$.%
\footnote{
This is {\em currying}, introduced by Gottlob Frege, and Moses Sch\"onfinkel, popularised by Haskell Curry, and implicit in the work of Cayley.
}
It follows that all results about set actions also hold for functions $f\colon A \rightarrow \Endop(X)$, and in places where it clarifies an explanation or a proof we will switch freely between the two notations. Translating the definition of windowed recurrence to the new setting gives the following definition.

\begin{definition}[Set Action Windowed Recurrence]
\label{definition:windowed-recurrence-set-action}
Let $A$, $X$ be sets, and let $\bullet\colon A \times X \rightarrow X$ be a set action. Let $x_0, x_1, \ldots$ be a sequence of elements of $X$, and $a_1, a_2, \ldots$ be a sequence of elements of $A$. Let $n$ be a strictly positive integer. Then the windowed recurrence of length $n$ corresponding to the action and the sequences $\theset{x_i}, \theset{a_i}$, is the sequence
\begin{equation*}
y_i = \ourcases{
    a_i \bullet(\ldots \bullet(a_1 \bullet x_0) \ldots)          & \text{for $1 \leq i < n$} \\
    a_i \bullet(\ldots \bullet(a_{i-n+1} \bullet x_{i-n})\ldots) & \text {for $i \geq n$} \\
}    
\end{equation*}
\end{definition}
\begin{remarks}
\label{remarks:windowed-recurrence-set-action}
\begin{samepage} \
\begin{enumerate}
\item 
If $x_0, x_1, \ldots$ is a sequence of elements of $X$, and $f_1, f_2, \ldots$ is a sequence of functions $f_i \in \Endop(X)$, then we may define a set action $\bullet\colon \mathbb{Z}_{>0} \times X \rightarrow X\colon (i, x) \mapsto f_i(x)$. I.e., $i \bullet x=f_i(x)$. Then the windowed recurrence for the set action $\bullet$, and the sequence $x_0, x_1,\ldots$, and $1,2,3, \ldots$, is equal to the windowed recurrence for $x_0, x_1, \ldots$ and $f_1, f_2, \ldots$.
\end{enumerate}
\end{samepage}
\begin{enumerate}
\setcounter{enumi}{1}
\item 
Assume $f\colon A \rightarrow \Endop(X)\colon a \mapsto f_a$ is a map from $A$ to functions on $X$, and $x_0, x_1, \ldots \in X$, $a_1, a_2, \ldots \in A$, and $n$ is a strictly positive integer. Then we may define the set action $\bullet\colon A \times X \rightarrow X\colon (a, x) \mapsto f_a(x)$, and the corresponding windowed recurrence of length $n$. We also call this windowed recurrence the windowed recurrence of length $n$ corresponding to $f$, $\theset{x_i}$, and $\theset{a_i}$, or alternatively the windowed recurrence of length $n$ corresponding to $f_a$, $\theset{x_i}$, and $\theset{a_i}$. This is, of course the same as the windowed recurrence for the sequence $x_0, x_1, \ldots$ and the functions $f_{a_1}, f_{a_2}, \ldots$.
  
\item 
As with sliding window $*$-products, and windowed recurrences for sequences of functions, we may use different conventions for defining $y_i$ for $i<n$, for windowed recurrence associated with set actions. This involves making choices for how to define $x_i$, for $i<0$, and $a_i$, for $i<1$, or alternatively to define $y_i$ directly for $i<n$ (e.g., to use an undefined value, or even to choose not to define).

\end{enumerate}    
\end{remarks}

\section{Relating Set Actions to Associative Operations}

The basic technique, which goes back to Cayley, is to relate applications of the set action operation $\bullet$ to functions, and hence to relate successive applications to function composition.

\begin{definition}
Assume $\bullet\colon A \times X \rightarrow X$ is an action of the set $A$ on the set $X$. Then define the corresponding left action operator $\Leftop^\bullet\colon A \rightarrow \Endop(X)$, by
\begin{equation*}
\Leftop_a^\bullet(x) = a \bullet x
\end{equation*}
\end{definition}
\begin{remarks} \
\begin{enumerate}
\item 
We sometimes encounter set actions in prefix notation, i.e., functions $g\colon A \times X \rightarrow X$, and in this case we write the definition of $\Leftop_a^g$ as
\begin{equation*}
\Leftop_a^g(x) = g(a, x) .    
\end{equation*}

\item
We will also write $\Leftop_a$ for $\Leftop_a^\bullet$ or $\Leftop_a^g$ when it is clear which action or function is intended. The notation $\Leftop^\bullet$ refers to the function from $A$ to $\Endop(X)$ defined by $a \mapsto \Leftop_a^\bullet$.

\item 
The windowed recurrence of length $n$ for the set action $\bullet$, initial values $x_0, \ldots$, and elements $a_1, \ldots$, is equal to the windowed recurrence for $x_0, x_1, \ldots$ and the sequence of functions $f_i = \Leftop_{a_i}^\bullet$.
\end{enumerate}
\end{remarks}

\noindent The following result is trivial and immediate.

\begin{lemma}
\label{lemma:set-action-left-action}
Assume $\bullet\colon A \times X \rightarrow X$ is a set action. Then
\begin{enumerate}
\item 
For $a, b, c \in A$
\begin{equation*}
    \Leftop_a^\bullet \circ (\Leftop_b^\bullet \circ \Leftop_c^\bullet) = (\Leftop_a^\bullet \circ \Leftop_b^\bullet) \circ \Leftop_c^\bullet
\end{equation*}
\item
For $a_1, \ldots, a_n \in A, x_0 \in X$

\begin{equation*}
a_n \bullet (a_{n-1} \bullet (\ldots \bullet (a_1 \bullet x_0)\ldots))
    =  (\Leftop_{a_n}^\bullet \circ \ldots \circ \Leftop_{a_1}^\bullet)(x_0)
\end{equation*}
\end{enumerate}
\end{lemma}
\begin{proof}
1.\ follows because function composition is associative. 2.\ follows from the definition of $\Leftop$, as
$a_n \bullet (a_{n-1} \bullet (\ldots \bullet (a_1 \bullet x_0)\ldots))
    = \Leftop_{a_n}^\bullet(\Leftop_{a_{n-1}}^\bullet(\ldots \Leftop_{a_1}^\bullet(x_0)\ldots))
    = (\Leftop_{a_n}^\bullet \circ \ldots \Leftop_{a_1}^\bullet)(x_0)$
\end{proof}

\noindent Lemma~\ref{lemma:set-action-left-action} is trivial, but it tells us that by `lifting' $a_1, \ldots, a_n$ to the functions $\Leftop_{a_1}^\bullet, \ldots, \Leftop_{a_n}^\bullet$, we can replace the set action $\bullet$ by the associative binary operation $\circ$. Thus we get the following algorithm idea.

\begin{algorithm-idea}[Windowed Recurrence]
\label{algorithm-idea:window-recurrence}

Let $\bullet\colon A \times X \rightarrow X$ be a set action. Suppose $a_1, a_2, \ldots \in A$ is a sequence of elements of $A$, and $x_0, x_1, \ldots \in X$ is a sequence of elements of $X$. Define
\begin{equation*}
y_i = \ourcases{
    a_i \bullet(a_{i-1} \bullet \ldots (a_1 \bullet x_0) \ldots), & \text{for $i < n$}\\
    a_i \bullet(a_{i-1} \bullet \ldots (a_{i-n+1} \bullet x_{i-n}) \ldots), & \text{for $i \geq n$}
}
\end{equation*}
Then we can compute $y_1, \ldots, y_N$ as follows.

\begin{description}
\item[Step 1] 
Construct the sequence of functions $\Leftop_{a_1}^\bullet, \ldots \Leftop_{a_N}^\bullet$.

\item[Step 2]
Use Two Stacks, or DEW, or DABA Lite, to compute the length $n$ windowed $\circ$-product functions
\begin{equation*}
Y_i = \ourcases{
    \Leftop_{a_i}^\bullet \circ \ldots \circ \Leftop_{a_1}^\bullet & i \leq n \\
    \Leftop_{a_i}^\bullet \circ \ldots \circ \Leftop_{a_{i-n+1}}^\bullet & i > n
}
\end{equation*}
for $i = 1, \ldots, N$.

\item[Step 3]
Then compute the $y_i$ as
\begin{equation*}
y_i = \ourcases{
    Y_i(x_0), & i \leq n \\
    Y_i(x_{i-n}), & n < i \leq N
    }
\end{equation*}
\end{description}
\end{algorithm-idea}

Algorithm Idea~\ref{algorithm-idea:window-recurrence} looks promising, and function composition is associative, so mathematically this algorithm is correct. But the algorithm idea is is incomplete because it presupposes a method for computing the function composition. A straight-forward approach to finding formulae for computing function compositions is to consider the functions $\Leftop_{a_i}^\bullet$, and start composing them to see what that results in. In practice this means observing how the dimensions of the resulting function spaces grow, and which quantities should be recorded in order to retain the ability to apply the composed functions. Algebraically this corresponds to determining the structure of the subsemigroup of $\Endop(X)$ generated by $\theset{\Leftop_{a_i}^* \colon i=1,2,3,\ldots}$.

We will explore the properties required to compute function compositions in detail in Chapter~\ref{chapter:semi-associativity-and-function-composition}. For the rest of this chapter we consider analogues and examples of Algorithm Idea~\ref{algorithm-idea:window-recurrence}. For a start we note that the idea of lifting to functions also applies to recurrences and reductions, which we also reformulate using set actions.

\begin{definition}[Set Action Recurrence, Set Action Reduction]
\label{definition:set-action-recurrence-reduction}
Let $A$, $X$ be sets, and let $\bullet\colon A \times X \rightarrow X$ be a set action. Let $x_0\in X$ be an element of $X$, and  $a_1, a_2, \ldots$ be a sequence of elements of $A$.
\begin{enumerate}
\item 
The {\em recurrence} corresponding to the element $x_0$ and sequence $\theset{a_i}$ is the sequence defined by 
\begin{equation*}
z_i = \ourcases{
    x_0 & \text{for $i=0$} \\
    a_i \bullet z_{i-1} & \text{for $i \geq 1$}}
\end{equation*}
In other words, $z_0 = x_0, z_1 = a_1 \bullet x_0, z_2 = a_2 \bullet (a_1 \bullet x_0), z_3 = a_3 \bullet (a_2 \bullet (a_1 \bullet x_0)), \ldots$.

\item 
Let $N$ be a strictly positive integer. Then the {\em reduction} corresponding to the element $x_0\in X$, the sequence $\theset{a_i}$, and the integer $N$, is the element of $X$ given by
\begin{equation*}
    a_N \bullet (a_{N-1} \bullet (\ldots \bullet (a_1 \bullet x_0) \ldots ))
\end{equation*}
\end{enumerate}
\end{definition}

\begin{algorithm-idea}(Recurrence) \label{algorithm-idea:recurrence}
Let $\bullet\colon A \times X \rightarrow X$ be a set action. Suppose $a_1, a_2, \ldots \in A$ is a sequence of elements of $A$, and $x_0 \in X$. Define
\begin{equation*}
z_i = a_i \bullet(a_{i-1} \bullet (\ldots \bullet (a_1 \bullet x_0) \ldots))
\end{equation*}
Then we can compute $z_1, \ldots, z_N$ as follows.
\begin{description}
\item[Step 1] 
Construct the sequence of functions $\Leftop_{a_1}^\bullet, \ldots \Leftop_{a_N}^\bullet$.

\item[Step 2]
Use an algorithm (e.g., a parallel prefix sum algorithm) to compute the prefix $\circ$-product functions
\begin{equation*}
Z_i = \Leftop_{a_i}^\bullet \circ \ldots \circ \Leftop_{a_1}^\bullet,
\qquad \text{for $i = 1, \ldots, N$.}
\end{equation*}

\item[Step 3]
Then compute the $z_i$ as
\begin{equation*}
z_i =  Z_i(x_0),
\qquad \text{for $i = 1, \ldots, N$.}
\end{equation*}
\end{description}
\end{algorithm-idea}

\begin{algorithm-idea}(Reduction) \label{algorithm-idea:reduction}
Let $\bullet\colon A \times X \rightarrow X$ be a set action. Suppose $a_1, a_2, \ldots \in A$ is a sequence of elements of $A$, and $x_0 \in X$. Assume $N$ is a strictly positive integer. Then we may compute the reduction 
\begin{equation*}
z_N = a_N \bullet(a_{N-1} \bullet (\ldots \bullet (a_1 \bullet x_0) \ldots))
\end{equation*}
as follows.
\begin{description}
\item[Step 1] 
Construct the sequence of functions $\Leftop_{a_1}^\bullet, \ldots \Leftop_{a_N}^\bullet$.

\item[Step 2]
Use an algorithm (e.g., a parallel product algorithm) to compute the composite function
\begin{equation*}
Z_N = \Leftop_{a_N}^\bullet \circ \ldots \circ \Leftop_{a_1}^\bullet
\end{equation*}

\item[Step 3]
Then compute $z_N$ as
\begin{equation*}
z_N =  Z_N(x_0)
\end{equation*}
\end{description}
\end{algorithm-idea}

\noindent As with the algorithm idea for set action windowed recurrences, the algorithm ideas for recurrences and reductions are only helpful if we have an effective (and efficient) way of computing the function compositions, and representing the composed functions.

\section{Nonassociative Sliding Window $*$-Products}
\label{sec:nonassociative-operations}

We now return to the case of sliding window $*$-products, but consider the situation where the $*$ operation is nonassociative. We saw one example of this in Example~\ref{example:sliding-window-continued-fractions}. When $*$ is nonassociative the algorithms of Chapters \ref{chapter:moving-sums}--\ref{chapter:other-sequential-algorithms}, other than SlickDeque, do not apply directly, and SlickDeque only applies to the case of a selection operator with a corresponding transitive relation.

The technique for nonassociative operations $*\colon A \times A \rightarrow A$ is to observe that these are special cases of set actions where $X = A$, and so Algorithm Idea~\ref{algorithm-idea:window-recurrence} applies, using the left action operations $\Leftop^*$. As we saw in Remarks~\ref{remarks:windowed-recurrences-and-sliding-window-*-products} there are three approaches available to relate a sliding window $*$-product to a windowed recurrence, with two of these depending on the existence of a right identity for $*$ in $A$. The approach which does not assume a right identity is slightly more more general, so we follow this approach below.

\begin{algorithm-idea}[Nonassociative Sliding Window $*$-Product]
\label{algorithm-idea:window-*-product}
Suppose $a_0, a_1, a_2, \ldots$ is a sequence of data and $*$ is a binary operation applying to the $a_i$. Then we can compute the sliding window $*$-product $y_i$, of length $n+1$, i.e.\

\begin{equation*}
y_i = \ourcases{
    a_i*(\ldots (a_1 * a_0)\ldots )            & i < n\\
    a_i*(\ldots (a_{i-n+1} * a_{i-n})\ldots )  & i \geq n}
\end{equation*}
for $i=0, \ldots, N$ as follows.%
\footnote{We have started from index $i=0$ for notational convenience.}
\begin{description}
\item[Step 1] 
Construct the sequence of functions $\Leftop_{a_1}^*, \ldots, \Leftop_{a_N}^*$.

\item[Step 2]
Use either Two Stacks, or DEW, or DABA Lite, or DDABA Lite to compute the length $n$ sliding window $\circ$-product functions
\begin{equation*}
Y_i = \ourcases{
    \Leftop_{a_i}^* \circ \ldots \circ \Leftop_{a_1}^* & i \leq n \\
    \Leftop_{a_i}^* \circ \ldots \circ \Leftop_{a_{i-n+1}}^* & i > n
}
\end{equation*}
for $i = 1, \ldots, N$.

\item[Step 3]
Then compute the length $n+1$ sliding window $*$-products corresponding to $\theset{a_0, a_1, \ldots}$ as
\begin{equation*}
y_i = \ourcases{
    a_0           & i = 0 \\
    Y_i(a_0),     & 1 \leq i \leq n \\
    Y_i(a_{i-n}), & n < i \leq N
    }    
\end{equation*}

\end{description}    
\end{algorithm-idea}
\begin{remarks} \ 
\begin{enumerate}
\item
Note that we started the input sequence at $i=0$, item $a_0$, and worked with the sliding window $*$-products of length $n+1$ rather than length $n$. This differs from the conventions we have used in Chapters \ref{chapter:moving-sums}--\ref{chapter:other-sequential-algorithms} and elsewhere, but it allows us to apply the algorithms of those chapters directly to the functions $\Leftop_{a_i}^*$ with no change of indexing.    

\item
As with the algorithm idea for windowed recurrences, this algorithm idea is only helpful if we have an effective (and efficient) way of computing the function compositions, and representing the composed functions.

\item Algorithm Idea~\ref{algorithm-idea:window-*-product} corresponds to part 2 of Remarks~\ref{remarks:windowed-recurrences-and-sliding-window-*-products}. There are, of course, corresponding algorithm ideas for parts 3 and 4 of Remarks~\ref{remarks:windowed-recurrences-and-sliding-window-*-products}.

\item
It is interesting observe that in Algorithm Idea~\ref{algorithm-idea:window-*-product} we started with a sliding window $*$-product for a possibly nonassociative operation $*$. We then related this to a windowed recurrence for a set action, and then related that windowed recurrence to a sliding window $*$-product for an associative operation $\circ$. In Chapter~\ref{chapter:semi-associativity-and-function-composition} we shall shall see that the associativity property used can be weakened further and shall explore situations where the composition operation $\circ$ is related to another binary operation which may be nonassociative, but for which algorithms that assume associativity still apply. Thus we come full circle from a nonassociative operation through set actions and function composition to another nonassociative operation. However the final nonassociative operation may be used to complete the calculation.

\end{enumerate}

\end{remarks}

\section{Examples}
\label{sec:examples-nonassociative}

We now give some examples of computing function compositions of the left action operators for nonassociative binary operations. In the next chapter we will systematize these examples, and also provide examples of function composition for left action operators of more general set actions.

\begin{example} \label{example:nonassociative-1}
Suppose $*$ has the multiplication table

\begin{center}
\begin{tabular}{l|lll}
$*$ & $a$ & $b$ & $c$ \\
\hline
$a$ & $a$ & $b$ & $a$ \\
$b$ & $a$ & $b$ & $b$ \\
$c$ & $c$ & $c$ & $c$ \\
\end{tabular}
\end{center}

\noindent This is the multiplication table for Example \ref{example:selection-operators-6}, and is nonassociative. We can represent any function in $\Endop(\theset{a, b, c})$ in single row form using the ordering $a, b, c$, so that $x y z$ refers to the function such that $a \mapsto x, b \mapsto y, c \longmapsto z$, where $x, y, z \in \theset{a, b, c}$. For convenience we use the notation $a = aaa, b = bbb, c = ccc$, to represent the constant functions. Using this notation, the mapping $x \mapsto \Leftop_x$, and function composition for the subsemigroup of $\Endop(\theset{a, b, c})$ generated by $\Leftop_a, \Leftop_b, \Leftop_c$, is easily found to be

\begin{center}
\begin{tabular}{cc}
\begin{tabular}[t]{c|c}
$x$ & $\Leftop_x^*$ \\
\hline
$a$ & $aba$\\
$b$ & $abb$\\
$c$ & $ccc$\\
\end{tabular} 
\qquad \qquad \qquad
&
\begin{tabular}[t]{c|ccccc}
$\circ$ & $aba$ & $abb$ & $c$ & $a$ & $b$ \\
\hline
$aba$  & $aba$ & $abb$ & $a$ & $a$ & $b$ \\
$abb$  & $aba$ & $abb$ & $b$ & $a$ & $b$ \\
$c$    & $c$   & $c$   & $c$ & $c$ & $c$ \\
$a$    & $a$   & $a$   & $a$ & $a$ & $a$ \\
$b$    & $b$   & $b$   & $b$ & $b$ & $b$ \\
\end{tabular}
\end{tabular}
\end{center}
\end{example}

\begin{example} \label{example:nonassociative-2}
Let $*$ be the operation of Example \ref{example:selection-operators-7}. Then $*$ is nonassociative, and has multiplication table

\begin{center}
\begin{tabular}{c|ccc}
$*$ & $a$ & $b$ & $c$ \\
\hline
$a$ & $a$ & $b$ & $a$ \\
$b$ & $b$ & $b$ & $b$ \\
$c$ & $c$ & $c$ & $c$
\end{tabular}
\end{center}

\noindent Using single row notation the mapping to left action functions and the function composition table for the subsemigroup of $\Endop(\theset{a, b, c})$ generated by $\Leftop_a, \Leftop_b, \Leftop_c$, are as follows.

\begin{center}
\begin{tabular}[t]{cc}
\begin{tabular}{c|c}
$x$ & $\Leftop_x^*$ \\
\hline
$a$ & $aba$ \\
$b$ & $b$   \\
$c$ & $c$   \\
\end{tabular}
\qquad \qquad \qquad
\begin{tabular}{c|cccc}
$\circ$ & $aba$ & $b$ & $c$ & $a$ \\
\hline
$aba$ & $aba$ & $b$ & $a$ & $a$ \\
$b$   & $b$   & $b$ & $b$ & $b$ \\
$c$   & $c$   & $c$ & $c$ & $c$ \\
$a$   & $a$   & $a$ & $a$ & $a$ \\
\end{tabular}
\end{tabular}
\end{center}
\end{example}

\begin{example} \label{example:nonassociative-3}
Let $*$ be the operation of Example~\ref{example:selection-operators-8}. Then $*$ is nonassociative, and has multiplication table

\begin{center}
\begin{tabular}{c|ccc}
$*$ & $a$ & $b$ & $c$ \\
\hline
$a$ & $a$ & $b$ & $a$ \\
$b$ & $b$ & $b$ & $c$ \\
$c$ & $a$ & $c$ & $c$ 
\end{tabular}    
\end{center}

\noindent In this case, the subsemigroup of $\Endop(\theset{a, b, c})$ generated by $\Leftop_a, \Leftop_b, \Leftop_c$ in this case consists of the set of non-invertible functions in $\Endop(\theset{a, b, c})$ and has $21$ elements.
\end{example}

\begin{example} \label{example:nonassociative-4}
Consider the binary operation $*$ with the following multiplication table.

\begin{center}
\begin{tabular}{c|ccc}
$*$ & $a$ & $b$ & $c$ \\
\hline
$a$ & $b$ & $c$ & $a$ \\
$b$ & $c$ & $b$ & $a$ \\
$c$ & $a$ & $c$ & $c$ 
\end{tabular}    
\end{center}

\noindent Note that $a*(b*c) = b$ and $(a*b)*c = c$. The operation $*$ is nonassociative and is also not a selection operator. The subsemigroup of $\Endop(\theset{a, b, c})$ generated by $\Leftop_a, \Leftop_b, \Leftop_c$ in this case consists of all 27 functions in $\Endop(\theset{a, b, c})$.
\end{example}

\noindent In each of these examples the subsemigroup of $\Endop(\theset{a, b, c})$ generated by the left action operators under composition is strictly larger than the image of $\Leftop^*$, which is $\theset{\Leftop_a, \Leftop_b, \Leftop_c}$. In Examples \ref{example:nonassociative-1} and \ref{example:nonassociative-2} the operators are selection operators whose corresponding relations are transitive, and so the SlickDeque algorithm does apply. For Examples \ref{example:nonassociative-3} and \ref{example:nonassociative-4} the SlickDeque algorithm does not apply. In the case of Example~\ref{example:nonassociative-3} this is because the corresponding relation is intransitive, and in the case of Example~\ref{example:nonassociative-4} this is because the operator is not a selection operator. In these two examples, lifting the calculation to the $\circ$ operator gives an associative operation, but one which is no longer a selection operator. Thus, for these two operations SlickDeque may not be used to compute sliding window $*$-products. However, in all four examples the computation of sliding window $*$-products may be achieved by applying Two Stacks, DEW, DABA or DABA Lite to $\circ$ in conjunction with Algorithm Idea \ref{algorithm-idea:window-*-product} and either the given multiplication tables for function composition or function composition for functions represented in single row form.

%% file: htcams-arxiv-ch07-semi-associativity.tex
\chapter{Semi-Associativity and Function Composition}
\label{chapter:semi-associativity-and-function-composition}

We now formalize the algebraic properties that must be satisfied by any data that represents functions and their composites, and hence the properties required to compute function compositions. The main property is called {\em semi-associativity}, which we now briefly motivate before giving a definition.

When we say (informally) that some data represents a function $X \rightarrow X$, we mean that the data is an element $a\in A$ of some set of possible function data, and the element $a$ determines the function. In other words there is a mapping $f\colon A\rightarrow \Endop(X)$ mapping the function description $a\in A$ to the corresponding function in $\Endop(X)$. This function $f$ corresponds to a set action $\bullet\colon A\times X \rightarrow X$ via the usual definition $a \bullet x = f_a(x)$, and with this definition the set action $\bullet$ corresponds to function application. In order to compute the function composition $f_{a_1} \circ f_{a_2}$ we should have a binary operation corresponding to composition which acts on the set of possible function data, i.e., a binary operation $*\colon A\times A \rightarrow A$ such that $a_1 * a_2$ represents the function composition $f_{a_1} \circ f_{a_2}$. In other words we ask that $f_{a_1} \circ f_{a_2} = f_{a_1 * a_2}$.
Translating this requirement to the notation of set actions, this condition becomes $a_1 \bullet (a_2 \bullet x) = (a_1 * a_2) \bullet x$, for $a_1, a_2 \in A$, $x \in X$. This is the defining condition of semi-associativity, and it characterizes the algebraic properties that must be satisfied by any representation of function application and function composition using data. More formally, we have the following definition.

\begin{definition}[Semi-Associativity]
\label{definition:semi-associativity}
Assume $\bullet\colon A \times X \rightarrow X$ is a set action of $A$ on $X$. Then $\bullet$ is semi-associative if and only if there exists a binary operation $*\colon A\times A \rightarrow A$ such that for all $a_1, a_2 \in A, x \in X$
\begin{equation}
\label{equation:semi-associativity}
    a_1 \bullet (a_2 \bullet x) = (a_1 * a_2) \bullet x    
\end{equation}
If $\bullet$ is semi-associative then any binary operation $*$ on $A$ which satisfies Equation~\ref{equation:semi-associativity} for all $a_1, a_2 \in A, x \in X$ is called a {\em companion operation} of $\bullet$.
\end{definition}

The following lemma provides alternative characterizations of semi-associativity.

\begin{lemma}[Characterizations of Semi-Associativity]
\label{lemma:semi-associativity}
Assume  $\bullet\colon A \times X \rightarrow X$ is a set action of $A$ on X. Then the following are equivalent.

\begin{enumerate}[label=\arabic*.]

\item 
$\bullet$ is semi-associative.

\item 
There exists a binary operation $*$ on $A$ such that for all $a_1, a_2 \in A$, $\Leftop_{a_1}^\bullet \circ \Leftop_{a_2}^\bullet = \Leftop_{a_1 * a_2}^\bullet$.

\item 
There exists a binary operation $*$ on $A$ such that $\Leftop^\bullet\colon A \rightarrow \Endop(X)\colon a \mapsto \Leftop_a^\bullet$ is a morphism of magmas%
\footnote{
Recall that a {\em magma} is a set together with a binary operation on the set. A morphism from a magma $(X_1, *_1)$ to a magma $(X_2, *_2)$ is a function $g\colon X_1 \rightarrow X_2$ such that $g(x *_1 y) = g(x) *_2 g(y)$ for all $x, y \in X_1$.
}
from $(A, *)$ to $(\Endop(X), \circ)$.

\item 
For all $a_1, a_2 \in A$ there exists $b \in A$ such that for all $x \in X$, $a_1 \bullet (a_2 \bullet x) = b \bullet x$.

\item 
For all $a_1, a_2 \in A$ there exists $b \in A$ such that $\Leftop_{a_1}^\bullet \circ \Leftop_{a_2}^\bullet = \Leftop_b^\bullet$.

\item
The image of $\Leftop^\bullet$ in $\Endop(X)$, i.e.\ $\Leftop_{(A)}^\bullet = \theset{\Leftop_a^\bullet \colon a \in A} \subseteq \Endop(X)$, is closed under function composition $\circ$.

\item 
$\Leftop_{(A)}^\bullet = \left\langle \Leftop_{(A)}^\bullet \right\rangle$. I.e., The closure $\left\langle \Leftop_{(A)}^\bullet \right\rangle$ of $\Leftop_{(A)}^\bullet$ in $\Endop(X)$ under function composition is equal to the image $\Leftop_{(A)}^\bullet$ of $\Leftop^\bullet$ in $\Endop(X)$.

\end{enumerate}
\end{lemma}
\begin{proof}
1.\ is clearly equivalent to 2.\ as the statement that for all $a_1, a_2 \in A$, $\Leftop_{a_1}^\bullet \circ \Leftop_{a_2}^\bullet = \Leftop_{a_1 * a_2}^\bullet$ is equivalent to the statement that $*$ is a companion operation of $\bullet$. 3.\ is a restatement of 2.\ in the language of morphisms and magmas. 4.\ is equivalent to 1.\ by the Axiom of Choice, as if we can choose an element $b\in A$ for each $a_1, a_2$ then we can construct a function that maps the pair $(a_1, a_2)$ to our choices of $b$, and this function will be a companion operation. 5.\ is a restatement of 4.\ in terms of the left action operators. 6.\ is equivalent to 5.\ by the definition of what it means for a set to be closed under an operation. 7.\ is equivalent to 6.\ because a subset of a semigroup is closed if and only if it is equal to its closure under the semigroup operation.
\end{proof}

\begin{remarks}  \
\begin{enumerate}[label={\arabic*.}]
\item Informally, parts 2 and 3 of Lemma~\ref{lemma:semi-associativity} tell us that a semi-associative set action $\bullet$ with companion operation $*$ provides a means to represent function application and function composition. More specifically, if $\theset{f_a \colon a \in A}$ is any collection of functions on $X$ then we may define a set action $\bullet$ by $a \bullet x = f_a(x)$. If $\bullet$ is semi-associative with companion operation $*$ then $f_{a_1} \circ f_{a_2} = f_{a_1 * a_2}$ for any $a_1, a_2 \in A$. So $\bullet$ corresponds to function application and $*$ corresponds to function composition. We shall explore this in more detail in Section~\ref{sec:function-composition}

\item 
It is important to note that in Definition~\ref{definition:semi-associativity} there is no requirement that the companion operation be associative.

\item
Any associative binary operation $*$ is also semi-associative as we may take $*$ to be its own companion operation. Furthermore, a binary operation is associative if and only if it is both semi-associative and is a companion operation of itself.

\item 
Trout \cite{Trout1972} describes a related concept which he calls {\em weak associativity}. He defines $\bullet$ to be {\em weakly associative} if for any $a_1, \ldots a_i, b_1, \dots b_j \in A$, we have 
\begin{equation*}
b_j\bullet(\ldots \bullet(b_1 \bullet (a_i \bullet (\ldots (a_1 \bullet x) \ldots )) \ldots ) 
= (b_j \bullet (\ldots (b_1 \bullet x)\ldots )) * (a_i \bullet (\ldots (a_1 \bullet x)\ldots )) 
\end{equation*}
where $x$ is some element of $X$.

\item 
The definition of semi-associativity appears as `Postulate A' in the 1927 paper of Suschkewitsch \cite{Suschkewitsch1927}.
\end{enumerate}
\end{remarks}

Our definition of semi-associativity differs from Blelloch \cite{Blelloch1993} and other authors in that we do not require the companion operation to be associative. This is an unnecessary assumption in the definition of semi-associativity, as we discuss here and in the following section.

There are two possibilities to consider when considering the assumption of associativity of companion operations. One possibility is that by assuming associativity of a companion operation we may be able to derive algorithms that work under that assumption but do not work under the weaker assumption of a nonassociative companion operation. This is a practical concern. A second possibility, which is of more theoretical concern, is whether assuming associativity of companion operations may simplify algorithms or proofs. We show that neither of these two possibilities is the case.

The main reason for the omission of associativity of the companion operator in the definition of semi-associativity is the following lemma, which shows that for the common uses of semi-associativity in algorithms for reductions, recurrences, and windowed recurrences, the weaker assumption of a (possibly) nonassociative companion operation suffices.

\begin{lemma} \label{lemma:semi-associativity-bracketing}
Assume $\bullet\colon A \times X \rightarrow X$ is a semi-associative set action with companion operation $*\colon A \times A \rightarrow A$. ($*$ is not assumed to be associative.) Then 

\begin{enumerate}[label={\arabic*.}]
\item 
For any $a_1, a_2, a_3 \in A$, $x \in X$

\begin{equation*}
((a_1 * a_2) * a_3) \bullet x= (a_1 * (a_2 * a_3)) \bullet x    
\end{equation*}

\item
If $a_1, \ldots, a_n \in A$, then $(a_1 * \ldots * a_n) \bullet x$ does not depend on the order of bracketing of the product $a_1 * \ldots * a_n$.
\end{enumerate}
\end{lemma}
\begin{proof}
For part 1.\
\begin{align*}
((a_1 * a_2) * a_3) \bullet x 
    & = (a_1 * a_2) \bullet (a_3 \bullet x) 
      = a_1 \bullet (a_2 \bullet (a_3 \bullet x)) 
      = a_1 \bullet ((a_2 * a_3) \bullet x) \\
    & = (a_1 * (a_2 * a_3)) \bullet x
\end{align*}
where each step follows from semi-associativity. Part 2.\ is an easy induction on the length of the expressions.
\end{proof}

\begin{remark}
Lemma~\ref{lemma:semi-associativity-bracketing} tells us that we may treat nonassociative companion operations of semi-associative set actions as if they were associative, provided we are using these in a setting where they are eventually applied to an element $x$ via the set action. 
\end{remark}

\section{Companion Operations}

The results of this section are not used elsewhere in the monograph, and may be skipped if the reader's interests lie elsewhere.

\subsection{The Existence of Associative Companions}

We now consider the associativity of companion operations in more detail. First we show that in a strictly logical sense the assumption of associativity is unnecessary as any semi-associative set action has at least one companion operation which is associative. To prove this we start with a well known lemma%
\footnote{See also Gibbon \cite{Gibbons1996}, and also Morita et al.\ \cite{Morita2007}.}
from Set Theory, that allows us to choose a unique representative $a\in A$ for any left action function $f = \Leftop^\bullet_a$. Given a fixed choice of these representatives one may then transfer the associative property from the composition operator $\circ$ on the left action functions of a semi-associative set action to a companion operation constructed from the representatives.

\begin{lemma}[Existence of Sections of Functions/Axiom of Choice%
\footnote{Indeed Lemma~\ref{lemma:existence-of-sections} is easily seen to be equivalent to the Axiom of Choice.}%
] \label{lemma:existence-of-sections}
Let $f\colon X \rightarrow Y$ be any function, and let $f(X) = \theset{f(x)\colon x \in X}$ denote the image of $f$ in $Y$. Then there exists a function $h\colon f(X) \rightarrow X$ such that $f \circ \, s \circ f = f$. Such a function is called a section of $f$.
\end{lemma}
\begin{proof}
To prove the lemma we must show that for any $y$ in the image of $f$ we may choose a value $x\in X$ such that $f(x) = y$. This follows from the Axiom of Choice applied to the collection of sets $\theset{f^{-1}(\theset{y})\colon y \in f(X)}$. Note that for recursively enumerable $X$ we could instead use the enumeration algorithm to search for the value $x$, and therefore not resort to the non-constructive Axiom of Choice.
\end{proof}

\begin{lemma}[Associative Companion Operations] \label{lemma:associative-companion}
Assume $\bullet\colon A \times X \rightarrow X$ is a semi-associative set action of $A$ on $X$, with (possibly nonassociative) companion operator $*$, and assume $s\colon \Endop(X) \rightarrow A$ is any section of $\Leftop^\bullet\colon A \rightarrow \Endop(X)$, i.e., $\Leftop^\bullet \circ \, s \circ \Leftop^\bullet = \Leftop^\bullet$. Then the binary operation $\otimes\colon A \times A \rightarrow A$ defined by $a_1 \otimes a_2 = s(\Leftop_{a_1 * a_2}^\bullet)$ is an associative companion operation of $\bullet$.
\end{lemma}
\begin{proof}
If $a_1, a_2 \in A$ then 
$a_1 \otimes a_2 = s(\Leftop_{a_1 * a_2}^\bullet) = s(\Leftop_{a_1}^\bullet \circ \Leftop_{a_2}^\bullet)$, and also $\Leftop_{a_1 \otimes a_2}^\bullet = (\Leftop^\bullet \circ \, s \circ \Leftop^\bullet)(a_1 * a_2) = \Leftop_{a_1 * a_2}^\bullet = \Leftop_{a_1}^\bullet \circ \Leftop_{a_2}^\bullet$. Now assume that $a_1, a_2, a_3 \in A$. Then
\begin{align*}
(a_1 \otimes a_2) \otimes a_3 
    & = s(\Leftop_{a_1 \otimes a_2}^\bullet \circ \Leftop_{a_3}^\bullet)
      = s((\Leftop_{a_1}^\bullet \circ \Leftop_{a_2}^\bullet) \circ \Leftop_{a_3}^\bullet)\\
    & = s(\Leftop_{a_1}^\bullet \circ (\Leftop_{a_2}^\bullet \circ \Leftop_{a_3}^\bullet))
      = s(\Leftop_{a_1}^\bullet \circ \Leftop_{a_2 \otimes a_3}^\bullet)) \\
    & = a_1 \otimes (a_2 \otimes a_3)
\end{align*}
which is what we wished to prove.
\end{proof}

\begin{corollary} \label{corollary:associative-companion}
Assume  $\bullet\colon A \times X \rightarrow X$ is a set action of $A$ on X. Then $\bullet$ is semi-associative if and only if $\bullet$ has an associative companion operation.
\end{corollary}

We now return to the second possibility when considering associative companion operations, which is whether assuming associativity may simplify algorithms or proofs, and whether we should therefore use Lemmas~\ref{lemma:existence-of-sections}, \ref{lemma:associative-companion} and Corollary~\ref{corollary:associative-companion} to simplify algorithms or proofs. This however presents some subtle difficulties, and it turns out that Lemmas \ref{lemma:existence-of-sections}, \ref{lemma:associative-companion} and Corollary~\ref{corollary:associative-companion} are of more theoretical interest than practical import. The difficulty lies with the use of the Axiom of Choice in Lemma~\ref{lemma:existence-of-sections} which is non-constructive. Furthermore even if a computable section function $s$ can be found there is no guarantee that it will be efficient to compute. In the case where $X$ is recursively enumerable, the obvious algorithm for computing a section function%
\footnote{
The obvious algorithm being to use a fixed computable enumeration to search for the first value $x$ such that $f(x) = y$.
}
may have arbitrary long running times. Even in the finite case functions for which sections are difficult to compute except by exhaustive search are well known.%
\footnote{
For example cryptographic hash functions are constructed to be difficult to invert.
}
A further difficulty of these constructions of a section is that for practical use they rely on a computable equality relation, whereas in our case the function we are sectioning is $\Leftop^\bullet$ whose values lie in a function space and for which equality means equality of functions---which is also notoriously difficult to compute, even when possible.

Thus, we have the situation that while associative companion operations always exist, the relationship between a nonassociative companion operation we can compute with, and the associative companion operation we can construct, or assume, is such that the computing with the nonassociative operation is the more practical or achievable. Fortunately Lemma~\ref{lemma:semi-associativity-bracketing} tells us that for purposes where the companion operation is used to compute with the set action, the associativity assumption is unnecessary.

\subsection{Nonassociativity}
Let's now consider a nonassociative binary operation $*$ on a set $X$, and consider the ways in which it can fail to be associative. The first way it could fail is a failure of semi-associativity. In this case the collection of left action functions $\Leftop_{(X)}^* = \theset{\Leftop_x^* \colon x \in X}$ is not closed under composition, so in a sense the lack of associativity is caused by the set of left actions being `too small', i.e. $X$ itself is in some sense `too small'. There is always, however, a larger collection of functions which is closed under composition, which is the subsemigroup $\left\langle \Leftop_{(X)}^* \right\rangle$ of $\Endop(X)$ generated by $\Leftop_{(X)}^*$. This subsemigroup of $\Endop(X)$ is therefore the object we must understand in order to `compute associatively' with $*$. If  $\left\langle \Leftop_{(X)}^* \right\rangle$ is `too large' (e.g. infinite dimensional), then it may be impractical to find a companion operation to compute compositions of left action operators of $*$. If on the other hand it is `small enough' (e.g. finite dimensional or with slowly growing dimension as operators $\Leftop_x^*$ are added to the set), then computation using companion operations of the left action may be feasible.%
\footnote{
It is interesting to note that for finite sets the ratio of cardinalities $|\langle \Leftop_{(X)}^* \rangle| / |\Leftop_{(X)}^* |$ may be used as a numerical measure of the nonassociativity of $*$ or in the case of finite set actions $|\langle \Leftop_{(X)}^\bullet \rangle| / |\Leftop_{(X)}^\bullet |$ as a numerical measure of non-semi-associativity. For the case of a finite set acting on an infinite set the growth rate of the set of all products $\theset{\Leftop_{a_1}^\bullet \circ \cdots \circ \Leftop_{a_j}^\bullet \colon 1 \leq j \leq k}$ as a function of $k$ is another such measure.
}
Of course, the same considerations apply to any non-semi-associative set action $\bullet$.

The other way that associativity can fail is if companion operations exist, but none of them are equal to the original operation itself. This is a less problematic situation for the purpose of computation, as the properties of semi-associativity, and the associativity of function composition suffice for many algorithms to still be valid.

\subsection{The Relationship between Associativity and Semi-Associativity}

The relation between associativity and semi-associativity in the case of a semi-associative binary operation $*$ involves the collection of companion operations on $X$, which is equal to the set of all binary operations $*'$ on $X$ such that $\Leftop^*$ is a magma morphism from $(X, *')$ to $(\Endop(X), \circ)$. If an example of a companion operation is known, then this collection can be characterized as the collection of operations obtained by all possible replacements for the values $a *' b$ that have the same image under $\Leftop^*$. I.e., you may replace any operation value $a *' b$ with any value $c$ such that $\Leftop_c^* = \Leftop_{a *' b}^*$. Thus the set of companion operators is characterized by the partition of $A$ induced by the inverse image of $\Leftop^*$.%
\footnote{
Strictly speaking it is only the partition of $(\Leftop^*)^{-1}(\theset{\Leftop_a^* \circ \Leftop_b^*\colon a, b \in A}$ that matters here.
}
The original operation $*$ is associative if and only if it is contained in this set of companion operations. Note that the characterization of the set of companion operations using replacement values also holds for set actions $\bullet$. A couple of commonly occurring situations relate to this characterization.

\begin{theorem}%
\footnote{C.f.\ the note after Lemma~4.3 in \cite{Gibbons1996}}
\label{theorem:faithful-action-companions}
Assume $\bullet\colon A \times X \rightarrow X$ is a semi-associative set action. Then
\begin{enumerate}[label={\arabic*.}]
\item If $\Leftop^\bullet$ is 1:1 then collection of companion operations of $\bullet$ is a singleton.%
\footnote{
A common terminology is to say that a set action $\bullet$ is {\em faithful} if and only if $\Leftop^\bullet$ is 1:1. Hence Theorem~\ref{theorem:faithful-action-companions} says that if $\bullet$ is a faithful semi-associative set action then $\bullet$ has a unique companion operation and this companion operation is associative.
}
\item Let 
$(\Leftop^\bullet)^{-1}{\Leftop_{(A)}^\bullet}{\!}^{\circ 2} = \theset{a \in A\colon \Leftop_a^\bullet = \Leftop_b^\bullet \circ \Leftop_c^\bullet \text{ for some $b, c \in A$}}$. Then
$\Leftop^\bullet$ restricted to $(\Leftop^\bullet)^{-1}{\Leftop_{(A)}^\bullet}{\!}^{\circ 2}$ is 1:1 if and only if the collection of companion operations of $\bullet$ is a singleton. 
\item If $\Leftop^\bullet$ is 1:1 then all companion operations of $\bullet$ are associative.
\end{enumerate}
\end{theorem}
\begin{proof}
For 1.\ Suppose $\Leftop^\bullet$ is 1:1. If $*_1$ and $*_2$ are two companion operators of $\bullet$ then for any $a, b \in A$ we have $\Leftop_{a *_1 b}^\bullet = \Leftop_a^\bullet \circ \Leftop_b^\bullet = \Leftop_{a *_2 b}^\bullet$, and hence $a *_1 b = a*_2 b$. For 2.\ the forward implication follows the same argument as 1. For the converse suppose that $\Leftop^\bullet$ is not 1:1 on $(\Leftop^\bullet)^{-1}{\Leftop_{(A)}^\bullet}{\!}^{\circ 2}$, and suppose $*_1$ is a companion operation of $\bullet$. By assumption there is $a_1, b_1, c \in A$ such that $\Leftop_c^\bullet = \Leftop_{a_1 *_1 b_1}^\bullet$ and $c \neq a_1 *_1 b_1$. If we now define $*_2$ by $a *_2 b = a *_1 b$ for $(a, b) \neq (a_1, b_1)$ and $a_1 *_2 b_1 = c$, then $*_2$  is a companion operation for $\bullet$ and $*_2 \neq *_1$. Part 3.\ follows easily from 1.\ and Corollary~\ref{corollary:associative-companion}.
\end{proof}

\begin{theorem} \label{theorem:semi-associativity-right-identity}
Assume $*\colon A \times A \rightarrow A$ is a semi-associative binary operation, and assume that $*$ has a right identity. Then $*$ is associative, $\Leftop^*$ is 1:1, and $*$ is its own unique companion operation.
\end{theorem}
\begin{proof}
Let $1$ be the right identity of $*$. Then setting $x = 1$ in the definition of semi-associativity (Equation~\ref{equation:semi-associativity}) shows that any companion operation of $*$ is equal to $*$, and hence that $*$ is associative. Also $\Leftop^*$ is 1:1 because if $\Leftop_a^* = \Leftop_b^*$ then $a = \Leftop_a^*(1) = \Leftop_b^*(1) = b$.
\end{proof}

\noindent Theorem~\ref{theorem:faithful-action-companions} has partial converses. Here is an example of such a result, which shows that if $\Leftop^\bullet$ is not 1:1 on the set of triple $\circ$ composites then the existence of nonassociative companion operations is the norm.

\begin{theorem} \label{theorem:faithful-action-companions-converse}
Assume $\bullet\colon A \times X \rightarrow X$ is a semi-associative set action, and there are $a, a^\prime, b, c, d \in A$ such that $\Leftop_a^\bullet = \Leftop_b^\bullet \circ \Leftop_c^\bullet \circ \Leftop_d^\bullet$, and $\Leftop_{a^\prime}^\bullet = \Leftop_a^\bullet$ with $a \neq a^\prime$. Then $\bullet$ has a nonassociative companion operation if any one of the following conditions is satisfied.
\begin{enumerate}
\item 
$\Leftop_a^\bullet = \Leftop_a^\bullet \circ \Leftop_d^\bullet$ 
or 
$\Leftop_a^\bullet = \Leftop_b^\bullet \circ \Leftop_a^\bullet$

\item
$\Leftop_b^\bullet \neq \Leftop_b^\bullet \circ \Leftop_c^\bullet$
or
$\Leftop_d^\bullet \neq \Leftop_c^\bullet \circ \Leftop_d^\bullet$

\end{enumerate}
\end{theorem}
\begin{proof}
First note that $\bullet$ has an associative companion operation, which we can call $*$. If we let $a^{\prime\prime} = b * (c * d) = (b * c) * d$ then $\Leftop_{a^{\prime\prime}}^\bullet = \Leftop_a^\bullet$, and so the conditions of the theorem hold with $a = a^{\prime\prime}$. Therefore we may replace $a$ with $a^{\prime\prime}$ and assume we have an element $a = b * (c * d) = (b * c) * d$ with all other conditions of the theorem holding.

We first consider the case $\Leftop_a^\bullet = \Leftop_a^\bullet \circ \Leftop_a^\bullet$. In this case we may define a new binary operation $*^\prime \colon A \times A \rightarrow A$ by
\begin{equation*}
a_1 *^\prime a_2 = \ourcases{
a^\prime & \text{if $a_1 = a$ and $a_2 = a$, or $a_1 = a^\prime$ and $a_2 = a$}  \\
a        & \text{if $a_1 = a$ and $a_2 = a^\prime$}  \\
a_1 * a_2 & \text{otherwise}
}
\end{equation*}
for $a_1, a_2 \in A$. The operation $*^\prime$ is clearly a companion operation of $\bullet$. But $*^\prime$ is nonassociative as $a *^\prime (a *^\prime a) = a *^\prime a^\prime = a \neq a^\prime = a^\prime *^\prime a = (a *^\prime a) *^\prime a$. So for the rest of the proof we now assume $\Leftop_a^\bullet \neq \Leftop_a^\bullet \circ \Leftop_a^\bullet$.

Now consider the case $\Leftop_a^\bullet = \Leftop_a^\bullet \circ \Leftop_d^\bullet$. Since we have assumed $\Leftop_a^\bullet \neq \Leftop_a^\bullet \circ \Leftop_a^\bullet$ we know that therefore $\Leftop_a^\bullet \neq \Leftop_d^\bullet$ and hence also $a \neq d$. We shall split this case into two subcases corresponding to $d \neq d * d$, and $d = d * d$. For the case $d \neq d * d$ we can define 
\begin{equation*}
a_1 *^\prime a_2 = \ourcases{
a        & \text{if $a_1 = a$ and $a_2 = d$}  \\
a^\prime & \text{if $a_1 = a$ and $a_2 = d * d$}  \\
a_1 * a_2 & \text{otherwise}
}
\end{equation*}
for $a_1, a_2 \in A$. Then, again the operation $*^\prime$ is a companion operation of $\bullet$. But $*^\prime$ is nonassociative as $a *^\prime (d *^\prime d) = a *^\prime (d * d) = a^\prime \neq a =  a *^\prime d = (a *^\prime d) *^\prime d$. For the second subcase we assume $d = d * d$, and recall that $\Leftop_d^\bullet \neq \Leftop_a^\bullet$. Then $d \neq a^\prime$ and $\Leftop_{a^\prime}^\bullet \circ \Leftop_d^\bullet = \Leftop_a^\bullet \circ \Leftop_d^\bullet = \Leftop_a^\bullet$, so we may define a companion operation of $\bullet$ by 
\begin{equation*}
a_1 *^\prime a_2 = \ourcases{
a^\prime & \text{if $a_1 = a$ and $a_2 = d$}  \\
a & \text{if $a_1 = a^\prime$ and $a_2 = d$}  \\
a_1 * a_2 & \text{otherwise}
}
\end{equation*}
for $a_1, a_2 \in A$. Then $*^\prime$ is nonassociative as $a *^\prime (d *^\prime d) = a *^\prime (d * d) = a *^\prime d = a^\prime \neq a = a^\prime *^\prime d = (a *^\prime d) *^\prime d$.

The case $\Leftop_a^\bullet = \Leftop_b^\bullet \circ \Leftop_a^\bullet$ is analogous to the case  $\Leftop_a^\bullet = \Leftop_a^\bullet \circ \Leftop_d^\bullet$ with order of operations reversed.

We now consider the remaining cases, $\Leftop_b^\bullet \neq \Leftop_b^\bullet \circ \Leftop_c^\bullet$, and the case $\Leftop_d^\bullet \neq \Leftop_c^\bullet \circ \Leftop_d^\bullet$, and assume that at least one of these two cases is true. Suppose that $(c, d) = (b, c * d)$. Then $c = b$, $d = c * d$, so $a = b * c * d = b * d = c * d = d = b * a$, and so $\Leftop_a^\bullet = \Leftop_b^\bullet \circ \Leftop_a^\bullet$, which is an already handled case. Similarly, if we suppose that $(b, c) = (b, c * d)$, then we have $c = c * d$ and so $a = b * c * d = b * c = a * d$ and hence $\Leftop_a^\bullet = \Leftop_a^\bullet \circ \Leftop_d^\bullet$ which is also an already handled case. Finally if we suppose that $(b * c, d) = (b, c * d)$, then we have $b = b * c$ and also $d = c * d$ and so we have both $\Leftop_b^\bullet = \Leftop_b^\bullet \circ \Leftop_c^\bullet$ and $\Leftop_d^\bullet = \Leftop_c^\bullet \circ \Leftop_d^\bullet$, which we assumed was not the case.

The remaining situation we must consider is therefore the case where $(c, d) \neq (b, c * d)$ and $(b, c) \neq (b, c * d)$ and $(b * c, d) \neq (b, c * d)$. Define a new binary operation $*^\prime \colon A \times A \rightarrow A$ by
\begin{equation*}
a_1 *^\prime a_2 = \ourcases{
a^\prime & \text{if $a_1 = b$ and $a_2 = c * d$,}  \\
a_1 * a_2 & \text{otherwise}
}
\end{equation*}
for $a_1, a_2 \in A$. The operation $*^\prime$ is clearly a companion operation of $\bullet$. We now show that $*^\prime$ is nonassociative. Consider the product $c * d$. We know that $(c, d) \neq (b, c * d)$, and hence $c *^\prime d = c * d$. Thus $b *^\prime (c *^\prime d) = b *^\prime (c * d) = a^\prime$. To calculate $(b *^\prime c) *^\prime d$ we observe that $(b, c) \neq (b, c * d)$ from which we obtain $b *^\prime c = b * c$. But
$(b * c, d) \neq (b, c * d)$ and hence $(b *^\prime c) *^\prime d = (b * c) *^\prime d = (b * c) * d = a$. Thus we have shown that $b *^\prime (c *^\prime d)  = a^\prime \neq a = (b *^\prime c) *^\prime d$.
\end{proof}
\begin{remark}
The condition $\Leftop_a^\bullet \neq \Leftop_b^\bullet \circ \Leftop_d^\bullet$ implies both of the sub-conditions in condition 2 of Theorem~\ref{theorem:faithful-action-companions-converse}, and therefore may also be used as a condition to imply the existence of a nonassociative companion operation.
\end{remark}

\section{Semi-Associativity Examples and Counter-Examples}

Examples of semi-associative set actions arise whenever we have a collection of functions or operations acting on a set, and those functions or operations are closed under composition. Therefore we give only a few examples here. The reader will find many examples of practical importance in Chapter~\ref{chapter:examples} and will have no difficulty finding their own examples.

\begin{example}
Assume $X$ is a set and let $\; \smallblacksquare\colon \Endop(X) \times X \rightarrow X\colon (f, x) \mapsto f \smallbsq x = f(x)$ denote function application. Then $\smallblacksquare$ is a semi-associative set action of $\Endop(X)$ on $X$ with companion operation which is function composition $\circ$. Furthermore, if $F \subseteq \Endop(X)$ is any subset of $\Endop(X)$ which is closed under function composition then the restriction of $\smallblacksquare$ to a set action $F \times X \rightarrow X$ is also semi-associative.
\end{example}

\begin{example}
Assume $R$ is a ring (e.g., integers, rational numbers, real numbers, matrices, see \cite{Jacobson1985}), and let $\operatorname{Mat}_n(R)$ denote the $n \times n$ matrices with entries in $R$. Then the action of $\operatorname{Mat}_n(R)$ on $R^n$ via matrix-vector multiplication is semi-associative, with matrix multiplication as a companion operation. Note that this example does not require that $R$ be commutative, or that $R$ have a multiplicative identity.
\end{example}

\begin{example}
Assume $R$ is a ring and $M$ is any left $R$-module then the action of $R$ on $M$ is semi-associative, and the companion operation is the ring multiplication. (See \cite{Jacobson1985} for definitions.)
\end{example}

\begin{example}
Any group action of a group on a set is semi-associative. (See \cite{Jacobson1985} for definitions.)
\end{example}

\begin{example}
Consider the binary operation $*$ defined by the following multiplication table.

\begin{center}
\begin{tabular}{c|ccc}
$*$ & $a$ & $b$ & $c$ \\
\hline
$a$ & $a$ & $b$ & $c$ \\
$b$ & $a$ & $b$ & $b$ \\
$c$ & $a$ & $c$ & $c$ 
\end{tabular}    
\end{center}

\noindent This is the multiplication table for Example~\ref{example:selection-operators-9}, and is a nonassociative selection operator whose corresponding relation is intransitive. The action of $*$ on $\theset{a, b, c}$ is semi-associative, as shown by the following composition table for the left action operators.

{ 
\renewcommand{\arraystretch}{1.2}
\begin{center}
\begin{tabular}{c|ccc}
$\circ$       & $\Leftop_a^*$ & $\Leftop_b^*$ & $\Leftop_c^*$ \\
\hline
$\Leftop_a^*$ & $\Leftop_a^*$ & $\Leftop_b^*$ & $\Leftop_c^*$ \\
$\Leftop_b^*$ & $\Leftop_b^*$ & $\Leftop_b^*$ & $\Leftop_b^*$ \\
$\Leftop_c^*$ & $\Leftop_c^*$ & $\Leftop_c^*$ & $\Leftop_c^*$ 
\end{tabular}    
\end{center}
} 

\noindent The three left action operators $\Leftop_a^*$, $\Leftop_b^*$, $\Leftop_c^*$ are distinct and therefore the map $\Leftop^*\colon A \rightarrow \Endop(\theset{a, b, c})$ is 1:1. It follows that $*$ has a unique companion operator and this unique companion operator is associative. However $*$ is not equal to the companion operator, and $*$ is nonassociative.
\end{example}

\begin{example}
Consider the binary operations of Examples~\ref{example:nonassociative-1}--\ref{example:nonassociative-4}. These are each non-semi-associative, as in each case the subsemigroup generated by the left action operators was larger than set of left action operators.
\end{example}

\begin{example}
Consider the action $\smat{m\\ a} \bullet x = a + \frac{m}{x}$, where the values are in the integers modulo $2$, extended with $\infty$, and $m, a$ are finite with $m \neq 0$. This action is described by the following table.

{ 
\renewcommand{\arraystretch}{1.5}
\begin{center}
\begin{tabular}{c|ccc}
$\bullet$ & $0$ & $1$ & $\infty$ \\
\hline
\nostretch{$\smat{1\\ 0}$} & $\infty$ & $1$ & $0$ \\
\nostretch{$\smat{1\\ 1}$} & $\infty$ & $0$ & $1$ \\
\end{tabular}
\end{center}
} 
\noindent This action is not semi-associative, because the function compositions of the operators $\Leftop_\tmat{1\\ 0}^\bullet, \Leftop_\tmat{1\\ 1}^\bullet$ form a strictly larger set of functions than $\theset{\Leftop_\tmat{1\\ 0}^\bullet, \Leftop_\tmat{1\\ 1}^\bullet}$. In this case there are many well known representations for these functions, and well known and equivalent ways to compute the function compositions.

{ 
\renewcommand{\arraystretch}{1.5}
\begin{center}
\begin{tabular}{l|ll|llll}
\nostretch{\parbox[t]{1.7in}{Representation of $\Leftop_\tmat{m\\ a}^\bullet$}} &
\nostretch{$\Leftop_\tmat{1\\ 0}^\bullet$} &
\nostretch{$\Leftop_\tmat{1\\ 1}^\bullet$} &
\multicolumn{4}{l}{Additional function compositions}\\
\hline\\[-3.5ex]
\parbox[t]{1.7in}{As matrices over the integers modulo $2$} & 
\nostretch{$\smat{0 & 1\\ 1 & 0}$} & 
\nostretch{$\smat{1 & 1\\ 1 & 0}$} & 
\nostretch{$\smat{1 & 0\\ 0 & 1}$} & 
\nostretch{$\smat{1 & 1\\ 0 & 1}$} & 
\nostretch{$\smat{1 & 0\\ 1 & 1}$} & 
\nostretch{$\smat{0 & 1\\ 1 & 1}$} \\
\parbox[t]{1.7in}{As permutations in disjoint cycle notation} & 
$(0\ \infty)$ & $(0\ \infty\ 1)$ & $(\ )$ & $(0\ 1)$ & $(1\ \infty)$ & $(0\ 1\ \infty)$\\
\parbox[t]{1.7in}{As rational functions} & 
$\frac{1}{x}$ & $1 + \frac{1}{x}$ & $x$ & $x + 1$ & $\frac{x}{x + 1}$ & $\frac{1}{x + 1}$\\
\parbox[t]{1.7in}{As permutations in single row notation for $0, 1, \infty$} & 
$(\infty\ 1\ 0)$ & $(\infty\ 0\ 1)$ & $(0\ 1\ \infty)$ & 
$(1\ 0\ \infty)$ & $(0\ \infty\ 1)$ & $(1\ \infty\ 0)$\\

\end{tabular}
\end{center}
} 

\noindent As the table shows, we started with two functions generating the action, but the subsemigroup of $\Endop(\theset{0, 1, \infty})$ generated by these two functions has $6$ elements.
\end{example}

%

\section{Representations of Function Composition}
\label{sec:function-composition}
We now formalize the approach to the representation of functions and function composition. The key is to embed a set action into another set action which is semi-associative.

\begin{definition}[Representation of Function Composition]
\label{definition:representation-of-function-composition}
Assume that $\bullet\colon A \times X \rightarrow X$ is a set action. Then a representation of function composition for $\bullet$ consists of the following objects.
\begin{enumerate}
\item A set $\Lambda$.
\item A function $\lambda\colon A \rightarrow \Lambda$
\item A binary operation $*\colon \Lambda \times \Lambda \rightarrow \Lambda$
\item A set action $\bullet\colon \Lambda \times X \rightarrow X$
\end{enumerate}
\noindent satisfying the following two properties.
\begin{enumerate}[label=(\alph*)]
\item For all $a \in A$, $x \in X$, $\lambda(a) \bullet x = a \bullet x$.
\item The set action $\bullet\colon \Lambda \times X \rightarrow X$ is semi-associative with companion operation $*$. 
I.e.,\\
$\lambda_1 \bullet (\lambda_2 \bullet x) = (\lambda_1 * \lambda_2) \bullet x$,
for all $\lambda_1, \lambda_2 \in \Lambda$, $x \in X$.
\end{enumerate}
\noindent When necessary we denote this representation of function composition $(\Lambda, \lambda, *, \bullet)$, or $(\lambda, *, \bullet)$.
\end{definition}

\subsubsection*{Terminology}

\begin{itemize}
\renewcommand\labelitemi{--}
  \item We call $\lambda$ the lifting function, or {\tt lift}.
  \item We call $*$ the composition operation or {\tt compose}.
  \item We call $\bullet$ the application operation or {\tt apply}.
\end{itemize}

\noindent When we come to use these in software interfaces, the corresponding procedures to be passed in to our algorithms will be called {\tt lift}, {\tt compose}, {\tt apply}. These names correspond closely to the software interface procedures {\tt lift}, {\tt combine}, {\tt lower}, proposed by Tangwongsan et al.\ \cite{Tangwongsan2015a}, and provide a theoretical underpinning for their design.

\section{Equivalent Formulations}

Definition~\ref{definition:representation-of-function-composition} can be reformulated in several different manners. First we extend the terminology to the equivalent case of a mapping $f\colon A \rightarrow \Endop(X)$.

\begin{definition}[Representation of Function Composition for an Indexed Collection of Functions]
Assume $A$, $X$ are sets and $\theset{f_a\colon a \in A}$ is a collection of functions on $X$, i.e., $f\colon A \rightarrow \Endop(X)$. Define the set action $\bullet\colon A \times X \rightarrow X$ by $a \bullet x = f_a(x)$. Then we also refer to any representation of function composition for the set action action $\bullet$ as a {\em representation of function composition for the functions $\theset{f_a\colon a\in A}$}. 
\end{definition}

The properties (a) and (b) of Definition~\ref{definition:representation-of-function-composition} can be rewritten in many equivalent forms, including those using the function $f = \Leftop^\bullet\colon A \times X \rightarrow X$, and using $\Leftop^\bullet\colon \Lambda \times X \rightarrow X$.

\begin{enumerate}
\item
The properties (a), (b) of Definition~\ref{definition:representation-of-function-composition} are equivalent to the following properties, (a$^\prime$), (b$^\prime$) respectively, with $\text{(a)} \Leftrightarrow \text{(a$^\prime$)}$, $\text{(b)} \Leftrightarrow \text{(b$^\prime$)}$.
\begin{enumerate}
\item[(a$^\prime$)] For all $a \in A$, $\Leftop_{\lambda(a)}^{\bullet} = \Leftop_a^\bullet$
\item[(b$^\prime$)] For all $\lambda_1, \lambda_2 \in \Lambda$, $\Leftop_{\lambda_1}^\bullet \circ \Leftop_{\lambda_2}^\bullet = \Leftop_{\lambda_1 * \lambda_2}^\bullet$.
\end{enumerate}

\item In terms of the functions $f_a$ the condition (a) is equivalent to $\Leftop_{\lambda(a)}^{\bullet} = f_a$, for all $a \in A$. 

\item 
If $f\colon A \rightarrow \Endop(X)$, then an equivalent way of defining a representation of function composition for the functions $\theset{f_a, a \in A}$, is to specify $\lambda\colon A \rightarrow \Lambda$, $*\colon \Lambda \rightarrow \Lambda \times \Lambda$, $F\colon \Lambda \rightarrow \Endop(X)$
such that 
\begin{enumerate}
\item[(a$^{\prime\prime}$)] $F \circ \lambda = f$
\item[(b$^{\prime\prime}$)] For all $\lambda_1, \lambda_2 \in \Lambda$ \quad $F_{\lambda_1} \circ F_{\lambda_2} = F_{\lambda_1 * \lambda_2}$
\end{enumerate}

\noindent In other words $f$ factors as $f = F \circ \lambda$ where $F$ is a {\em magma morphism} from $(\Lambda, *)$ to $(\Endop(X), \circ)$.

\end{enumerate}

\section{Examples}

\begin{example}  \label{example:rofc-examples-1}
Let $\bullet\colon \Lambda \times X \rightarrow X$ be a semi-associative set action, and assume $A \subseteq \Lambda$ is any subset of $\Lambda$. 
Let $\iota\colon A \hookrightarrow \Lambda$ denote the inclusion map from $A$ into $\Lambda$ . Then $(\Lambda, \iota, *, \bullet)$ is a representation of function composition for the restricted set action $\bullet\colon A \times X \rightarrow X$. This follows immediately from semi-associativity, as
\begin{align*}
& \iota(a) \bullet x = a \bullet x = \Leftop_a^\bullet(x) \quad \text{for $a \in A$, $x \in X$} \\
& \lambda_1 \bullet (\lambda_2 \bullet x) = (\lambda_1 * \lambda_2) \bullet x \quad \text{for $\lambda_1, \lambda_2 \in \Lambda$, $x \in X$}
\end{align*}
which are simply the conditions (a), (b) of Definition~\ref{definition:representation-of-function-composition}.
\end{example}

\begin{example}  \label{example:rofc-examples-2}
Let $A$, $X$ be sets, and $f_a \in \Endop(A)$ for $a \in A$. Define 
\begin{equation*}
\Lambda = \theset{\text{The set of finite sequences of length $\geq 1$ of elements of $A$}}
\end{equation*}
Define $*\colon \Lambda \times \Lambda \rightarrow \Lambda$, by
$(a_1, \ldots, a_p) * (b_1, \ldots, b_q) = (a_1, \ldots, a_q, b_1, \ldots, b_q)$.
I.e., $(\Lambda, *)$ is the free semigroup on $A$. Also define $\lambda\colon A \rightarrow \Lambda$ and $\bullet\colon \Lambda \times X \rightarrow X$ by
\begin{align*}
\lambda(a) & = (a) \quad \text {i.e., the sequence of length $1$} \\
(a_1, \ldots, a_p) \bullet x & = f_{a_1}(f_{a_2}(\ldots f_{a_p}(x) \ldots))
\end{align*}
Then $(\Lambda, \lambda, *, \bullet)$ is easily verified to be a representation of function composition for the functions $\theset{f_a, a \in A}$. This shows that any indexed set of functions from a set to itself (equivalently any set action) has a representation of function composition. From an algorithmic efficiency point of view this representation of function composition is not helpful. Even though the composition operation $*$ looks to be cheap---just concatenation of finite sequences---when you come to apply the composed function nothing has been gained, as you must apply all of the functions entering the composition in turn.%
\footnote{
Equivalently we could have started this example with a set action $\bullet\colon A \times X \rightarrow X$ and defined 
\begin{equation*}
(a_1, \ldots, a_p) \bullet x = a_1 \bullet (a_2 \bullet (\ldots \bullet (a_p \bullet x) \ldots))
\end{equation*}
thus showing that any set action has a representation of function composition.
} 
\end{example}

\begin{example}  \label{example:rofc-examples-3}
Assume now that $X$ is a finite set with a total order $\leq$, that $A$ is a set, and that $\theset{f_a\colon a \in A}$ is a collection of functions with $f_a \in \Endop(X)$. Then one can represent function composition of the functions $f_a$ as follows. Let $m=|X|$, and list the elements of $x$ in ascending order $x_1 < \ldots < x_m$. Let

\begin{equation*}
\Lambda =\theset{\text{The set of sequences $(z_1, \ldots, z_m)$ with $z_i \in X$}}\\
\end{equation*}
The interpretation of $\Lambda$ we will use is that elements $\zeta \in \Lambda$, $\zeta=(z_1, \ldots, z_m)$ correspond to functions that map $x_i \mapsto z_i$. Define $\zeta[x_i] = z_i$, and note that $\zeta[x]$ may be computed from $\zeta$ and $x = x_i$ by using a binary search algorithm to locate the position, $i$, of $x$ in the sequence $(x_1, \ldots, x_m)$, and then $z_i$ can be found as the $i^\text{th}$ component of $\zeta$.
Now define $\lambda\colon A \rightarrow \Lambda$, $*\colon \Lambda \times \Lambda \rightarrow \Lambda$, $\bullet\colon \Lambda \times X \rightarrow X$ as follows.
\begin{align*}
\lambda(a) & = (f_a(x_1), \ldots, f_a(x_m)) \\
\zeta * \nu & = (\zeta[\nu[x_1]], \ldots, \zeta[\nu[x_m]]) \\
\zeta \bullet x & =\zeta[x]
\end{align*}
Then $(\Lambda, \lambda, *, \bullet)$ is a representation of function composition for the functions $\theset{f_a\colon a \in A}$.
\end{example}

\begin{example}  \label{example:rofc-examples-4}
Example~\ref{example:rofc-examples-3} can be extended to finite sets $X$ with or without a total order, by setting $\Lambda$ to be the set of dictionary data structures (or associative arrays) whose keys are the entire set $X$. This assumes, of course, that the elements of $X$ have associated operations defined on them allowing dictionary construction and lookup to be defined. E.g.,\ these can be implemented using a hash table, or one of the many tree data structures used to implement dictionaries. Supposing this to be the case, we can then define

\begin{align*}
\lambda(a)  & =\operatorname{Dictionary}[\text{$(x_i, f_a(x_i))$, for $i=1 \ldots, m$}] \\
\zeta * \nu & =\operatorname{Dictionary}[\text{$(x_i, \zeta[\nu[x_i]])$, for $i=1, \ldots, m$}] \\
\zeta \bullet x & =\zeta[x]
\end{align*}
where $\zeta[x]$ denotes dictionary lookup and $\operatorname{Dictionary}[\text{$(x_i, z_i)$, for $i=1, \ldots, m$}]$ denotes construction of a dictionary that maps $x_i$ to $z_i$ for $i=1, \ldots, m$. With these definitions $(\Lambda, \lambda, *, \bullet)$ is a representation of function composition for the functions $\theset{f_a\colon a \in A}$.

In both this example, and the preceding Example~\ref{example:rofc-examples-3}, the size of the dictionaries representing a long function composition $f_{a_1} \circ \ldots \circ f_{a_N}$ does not grow as $N$ increases, but stays constant at $m=|X|$. Also the cost to apply the function composition does not grow beyond a fixed limit. It may vary, but will be bounded by a function of $m$, not $N$.
\end{example}

\begin{example}  \label{example:rofc-examples-5}
Consider the functions
\begin{equation*}
f_a(x)=a+\frac{1}{x}
\end{equation*}
%
where $a$, $x$ come from a field (e.g., real, rational, or complex numbers). Define $\lambda(a) = \smatl{a & 1\\ 1 & 0}$, and $* = 2 \times 2 \text{ matrix multiplication}$, and if $A = \smatl{a & b\\ c & d}$ with $ad - bc \neq 0$, define

\begin{equation*}
A \bullet x = \frac{a x + b}{c x + d}
\end{equation*}
To avoid domain issues, should they arise, extend $f_a$, $\bullet$, to $x = \infty$, $x = 0$ by
\begin{align*}
f_a(\infty)      & = a,           & f_a(0)                                & =\infty\\
A \bullet \infty & = \frac{a}{c}, & A \bullet \left( \frac{-d}{c} \right) & = \infty
\end{align*}
Then $f_a(x)=\lambda(a) \bullet x$, and $A \bullet(B \bullet x) = (A * B) \bullet x$, where $A$, $B$ are $2 \times 2$ matrices with non-zero determinant. Thus $(\lambda, *, \bullet)$ is a representation of function composition for the functions $f_a$. 

In principle we can use $(\lambda, *, \bullet)$ to compute function compositions $f_{a_1}, f_{a_2} \circ f_{a_1}, f_{a_3} \circ f_{a_2} \circ f_{a_1}$ etc., and hence compute sliding window $*$-products for the operation $a * b = a + \frac{1}{b}$. In practice, however, there is a problem. Suppose the $a_i$ are above $1$ and bounded away from $1$, i.e., $a_i > 1 + \tilde{a}$ for some $\tilde{a} > 0$. Then for a long matrix product $A_i * \ldots * A_{i-n+1}$, with $A_j = \smatl{a_j & 1 \\ 1 & 0}$, the coefficients of the matrix will get large, and for finite precision arithmetic this can cause an overflow. So using $(\lambda, *, \bullet)$ as above will not always work in practice, and another $*$ operator must be found.

The solution is to scale the $*$ operation so that the matrix entries remain bounded. For example
\begin{align*}
A *_1 B & = \frac{A B}{\|A B\|_\text{Frob.}}, & 
\left\|\smatl{a & b\\ c & d}\right\|_\text{Frob.}
        & =\sqrt{|a|^2 + |b|^2 + |c|^2 + |d|^2}\\
A *_2 B & = \frac{A B}{\|A B\|_1} & 
\left\|\smatl{a & b\\ c & d}\right\|_1
        & = \max(|a|+|c|,|b|+|d|)
\end{align*}
Both of these operations are associative, and both give representations of function composition for the $f_a$.
As an example to show a nonassociative operation that gives a representation of function composition for the $f_a$, consider
\begin{equation*}
A *_3 B = \frac{A B}{\|A\|_1}
\end{equation*}
Then $(\lambda, *_3, \bullet)$ is a representation of function composition for the $f_a$, but

\begin{equation*}
A *_3 (B *_3 C) = \frac{ABC}{\|A\|_1\|B\|_1}
, \qquad
(A *_3 B) *_3 C = \frac{ABC}{\|AB\|_1}
\end{equation*}
So $*_3$ is not associative, but it may still be used to compute function compositions, and hence sliding window products for the operation $(a, x) \mapsto a + 1/x$, and it also mitigates the matrix multiplication overflow problem.%
\footnote{
Note that the operation $A *_4 B = \frac{A B}{\|A\|_1 \|B\|_1}$ is another nonassociative operation which may be used to mitigate the overflow problem, and $A *_4 (B *_4 C) = \frac{ABC}{\|A\|_1 \|BC\|_1}$ whereas
$(A *_4 B) *_4 C = \frac{ABC}{\|AB\|_1 \|C\|_1}$.
}
\end{example}

We shall describe many more examples and constructions of representations of function composition in Chapter~\ref{chapter:examples}.

\section{Semidirect Products}
\label{sec:semidirect-products}

We have spent the chapter, thus far, investigating the composition of functions, and approaches to computing with set actions. We now turn {\em semidirect products}, which are a method for {\em combining} set actions. There are two reasons for our interest in semidirect products in this work.
\begin{itemize}
\item In Chapters~\ref{chapter:vector-sliding-window-*-products} and \ref{chapter:vector-windowed-recurrences} we shall show that sliding window $*$-products (and hence also prefix sums) are equivalent to computing powers of elements in particular semidirect products, and that windowed recurrences are equivalent to iterated application of particular semidirect product actions. This relates windowed recurrences to powers in semidirect products.
\item Semidirect products provide a basic technique for combining semi-associative set actions to produce new semi-associative set actions, and are thus a source for many of the examples in Chapter~\ref{chapter:examples}.
\end{itemize}
Both of these uses of semidirect products are intimately involved with the relationship between semi-associativity and semidirect products. Interestingly, the same basic results on semidirect products are used both in construction of parallel algorithms for windowed recurrences, and also in the construction of operations to which these algorithms apply, and thus they link two seemingly separate parts of this work. We shall now collect definitions and results on semidirect products and their relation to semi-associativity. 

\subsubsection*{Notations and Conventions}
Recall that a set action $\bullet\colon A \times X \rightarrow X$ is equivalent to a function $L\colon A \rightarrow \Endop(X)$. When we start combining set actions, we will have several set actions in play at the same time, and because of this correspondence between set actions and functions, the results we state have a large number of variants differing only in notation. E.g. for a result in which 3 set actions appear we may have $8 = 2^3$ variants of this result. To keep the number of variations manageable, we state and prove the results below in their pure `set action only' form, and only give a brief indication of other notational variants.

Because of the many set actions and operations interacting in the following, we will follow the common algebraist's convention of using the same symbols for set actions and the same symbols for binary operations despite differing domains of operation, and let the set action or operation that is meant be implied by the objects they are acting on. We find this more helpful than choosing different operator names for each different operator involved, which can make it hard to remember the meaning of the many different symbols. Thus in the following
\begin{equation*}
\begin{array}{ll}
\bullet, \times  & \text{denote set actions}\\
* & \text{denotes a binary operation}
\end{array}
\end{equation*}

The results that follow also involve nonassociative operations, and thus, when stated in full, require a large number of parentheses. To cut down on the notational clutter, we therefore follow the purely notational convention that all set actions and binary operations are treated {\em notationally}%
\footnote{
What this means is that we are leaving brackets out of the notation, and these brackets are assumed to be present with expressions bracketed from right to left. We are {\em not} assuming associativity of the operators, except in places where it is explicitly stated.
}
as {\em right associative}. Thus
\begin{align*}
a_1 \bullet \ldots \bullet a_n \bullet x = a_1 \bullet (a_2 \bullet (\ldots (a_n \bullet x)\ldots))\\
a_1 * \ldots * a_n = a_1 * (a_2 * (\ldots (a_{n-1} * a_n)\ldots))\\
\end{align*}
unless otherwise indicated. We will also need a notation for powers, which we now give.

\begin{definition}[Right-Folded Power]  
Assume $*\colon A \times A \rightarrow A$ is a binary operation, and $n$ is a strictly positive integer, then define the {\em right-folded $n^\text{th}$ power of $a \in A$} to be

\begin{equation*}
a^{* n} = \underbrace{a *(a *(\ldots * (a * a) \ldots))}_\text{$n$ copies of $a$}
\end{equation*}
When $*$ is associative and it is clear which operation is meant, we will also use the standard notation $a^n$. Furthermore, we will also use the notation $a^n$ in nonassociative situations where an equation is valid with any choice of the bracketing.
\end{definition}

We start with the basic operation of combining two set actions, and also define a semidirect product of magmas (sets with binary operations). In each case we give a pure set action based definition as well as a definition that uses a function into a set of endomorphisms.

\begin{definition}[Semidirect Product Action]
\label{definition:semidirect-product-action}
Assume $A$, $B$, $X$ are sets, and assume $\bullet\colon A \times X \rightarrow X$ and $\bullet\colon B \times X \rightarrow X$ are set actions. Then define the {\em semidirect product set action} $\bullet\colon (A \times B) \times X \rightarrow X$ by
\begin{equation*}
\mat{a\\ b} \bullet x = b \bullet (a \bullet x)
\end{equation*}
\end{definition}

\begin{definition}[Semidirect Product Action using a function to Endomorphisms]
\label{definition:semidirect-product-action-using-function}
Assume $A$, $B$, $X$ are sets, and assume $\bullet\colon B \times X \rightarrow X$ is a set action and $L\colon A \rightarrow \Endop(X)$ is a function from $A$ to the set of functions from $X$ to itself. Then define the {\em semidirect product set action} $\bullet\colon (A \times B) \times X \rightarrow X$ by
\begin{equation*}
\mat{a\\ b} \bullet x = b \bullet  L(a)(x)
\end{equation*}
\end{definition}

\begin{definition}[Semidirect Product]
\label{definition:semidirect-product-using-set-action}
Assume $A$, $B$ are sets, and assume $*\colon A \times A \rightarrow A$, $*\colon B \times B \rightarrow B$
are binary operations. Assume $\times\colon A \times B \rightarrow B$ is a set action of $A$ on $B$. Then define the {\em semidirect product} magma $A \ltimes_\times B$ to be the set of ordered pairs $A \times B$ together with the binary operation $*\colon (A \ltimes_\times B) \times (A \ltimes_\times B) \rightarrow (A \ltimes_\times B)$  defined by
\begin{equation*}
\mat{a_1\\ b_1} * \mat{a_2\\ b_2} 
    = \mat{a_1 * a_2\\ b_1 * (a_1\times b_2)}
\end{equation*}
\end{definition}

\begin{definition}[Semidirect Product using a function to Endomorphisms]
\label{definition:semidirect-product-using-function}
Assume $A$, $B$ are sets, and assume $*\colon A \times A \rightarrow A$, $*\colon B \times B \rightarrow B$
are binary operations. Assume $L\colon A \rightarrow \Endop(B)$ is a function from $A$ to the set of functions from $B$ to itself. Then define the {\em semidirect product} magma $A \ltimes_L B$ to be the set of ordered pairs $A \times B$ together with the binary operation $*\colon (A \ltimes_L B) \times (A \ltimes_L B) \rightarrow (A \ltimes_L B)$  defined by
\begin{equation*}
\mat{a_1\\ b_1} * \mat{a_2\\ b_2} 
    = \mat{a_1 * a_2\\ b_1 * L(a_1)(b_2)}
\end{equation*}
\end{definition}
\begin{remarks} \
\begin{enumerate}
\item 
It should be clear that Definitions~\ref{definition:semidirect-product-action} and \ref{definition:semidirect-product-action-using-function} are equivalent, and Definitions~\ref{definition:semidirect-product-using-set-action} and \ref{definition:semidirect-product-using-function} are equivalent.

\item 
We do not assume that any of the operations appearing in Definitions~\ref{definition:semidirect-product-action}--\ref{definition:semidirect-product-using-function} are associative, and we do not assume that any of the set actions that occur are semi-associative, or have other algebraic properties.
  
\item 
In the Definition~\ref{definition:semidirect-product-using-function} we do not assume that $L$ is a magma morphism from $A$ to $\Endop(B)$. I.e., we are not assuming that $L(a_1 * a_2) = L(a_1) \circ L(a_2)$ for all $a_1, a_2 \in A$.
\item 
In the Definition~\ref{definition:semidirect-product-using-function} 
we are also not assuming that $L(a)$ is a magma endomorphism of $B$. I.e., we are not assuming that $L(a)(b_1 * b_2)=L(a)(b_1) * L(a)(b_2)$. Instead we are only assuming that $L(a)$ is a set endomorphism of $B$, i.e.\ a function from $B$ to itself.
  
\end{enumerate}
\end{remarks}

\begin{example}
Assume $A$ is a set and $L_1, L_2, \ldots \in \Endop(A)$ are functions on $A$. Then $L\colon \mathbb{Z}_{>0} \rightarrow \Endop(A)$ is a function from $\mathbb{Z}_{>0}$ to $\Endop(A)$. Furthermore we have the binary operations, $+$ on $\mathbb{Z}_{>0}$, and $\circ$ on $\Endop(A)$, and thus we may form the semidirect product $\mathbb{Z}_{>0} \ltimes_L A$ with respect to these binary operations and the function $L$. The function $L$ corresponds to the set action $\bullet\colon \mathbb{Z}_{>0} \times A \rightarrow A$ defined by $i \bullet a = L_i(a)$, and this set action is semi-associative with companion operation $+$ if and only if $L_i \circ L_j = L_{i+j}$ for $i, j \geq 1$. This is a situation that we will encounter in the following chapters, in the discussion of parallel algorithms for windowed recurrences.
\end{example}

\begin{theorem}
\label{theorem:semidirect-product-set-action}
Assume that $A$, $B$, $X$, are sets, that $\bullet\colon A \times X \rightarrow X$ and $\bullet\colon B \times X \rightarrow X$ are set actions, and that $\smat{a\\b} \bullet x = b \bullet a \bullet x$ is the semidirect product set action. Assume $n$ is a strictly positive integer. Then
\begin{enumerate}
\item For any $a\in A$,  $b\in B$, $x\in X$,
\begin{equation*}
\underbrace{\mat{a\\ b} \bullet \ldots \bullet \mat{a\\ b}}_{\text{$n$ times}} \bullet \, x
= \underbrace{b \bullet a \bullet \ldots  \bullet b \bullet a}_{\text{$n$ times}} \bullet \, x
\end{equation*}

\item 
Assume $\bullet\colon A \times X \rightarrow X$ is semi-associative with companion operation $*\colon A \times A \rightarrow A$, and $\times\colon A \times B \rightarrow B$ is a set action satisfying
$a \bullet b \bullet x = (a \times b) \bullet (a \bullet x)$ for all $a\in A$,  $b\in B$, $x\in X$. Then for any $a\in A$,  $b\in B$, $x\in X$,
\begin{equation*}
 \underbrace{b \bullet a \bullet \ldots  \bullet b \bullet a}_{\text{$n$ times}} \bullet \,x
 = b \bullet (a \times b) \bullet (a^2 \times b) \bullet \ldots \bullet (a^{n-1} \times b) \bullet a^n \bullet x
\end{equation*}
where each $a^i = a * \ldots * a$ is a power of $a$ computed using $*$ which may be bracketed in any order independently of the other powers $a^i$.%
\footnote{
Thus, for example, we may use either $a * (a * a)$ or $(a * a) * a$ for $a^3$ and may use any of $a * (a * (a * a))$, $(a * a) * (a * a)$, $a * ((a * a) * a)$, $(a * (a * a)) * a$, $((a * a) * a) * a$ for $a^4$ and these choices may be made independently. In the case that $*$ is nonassociative, different bracketings of the powers $a^i$ give different elements of $A$, and the statement is true for any of these choices.
}
\item
Assume both $\bullet\colon A \times X \rightarrow X$ and $\bullet\colon B \times X \rightarrow X$ are semi-associative with companion operations $*\colon A \times A \rightarrow A$, $*\colon B \times B \rightarrow B$, and assume $\times\colon A \times B \rightarrow B$ is a set action satisfying
$a  \bullet b \bullet x = (a \times b) \bullet (a \bullet\,  x)$ for all $a\in A$,  $b\in B$, $x\in X$. Then the semidirect product set action $\smat{a\\b} \bullet x = b \bullet a \bullet x$ is an action of the semidirect product $A \ltimes_\times B$ on $X$ and this action is semi-associative with companion operation which is the semidirect product operation of $A \ltimes_\times B$. I.e., the companion operation is 
\begin{equation*}
\mat{a_1\\ b_1} * \mat{a_2\\ b_2} 
    = \mat{a_1 * a_2\\ b_1 * (a_1\times b_2)}
\end{equation*}

\item Under the same assumptions as 3., for any $a \in A$, $b\in B$, $x\in X$, and any strictly positive integer $n$, we have 
\begin{align*}
b \bullet (a \times b) \bullet (a^2 \times b) \bullet \ldots \bullet (a^{n-1} \times b) \bullet a^n \bullet x
& = \underbrace{b \bullet a \bullet \ldots  \bullet b \bullet a}_{\text{$n$ times}} \bullet \, x
& = \mat{a\\ b}^{* n} \bullet x = \mat{a\\ b}^n \bullet x
\end{align*}
where the exponentiation in $A \ltimes_\times B$ may be bracketed in any order, and, independently, the powers $a^i$ may be bracketed in any order independently of the other powers $a^i$.%
\footnote{
If $*$ is not associative, then different bracketings of $\smat{a\\ b}^n$ give different elements of $A \ltimes_\times B$, and the statement is true for each of these elements.
}
\end{enumerate}
\end{theorem}
\begin{remarks} \ 
\begin{enumerate}
\item In part 1 of Theorem~\ref{theorem:semidirect-product-set-action}, neither $\bullet\colon A \times X \rightarrow X$ nor $\bullet\colon B \times X \rightarrow X$ is assumed to be semi-associative. In part 2, $\bullet\colon A \times X \rightarrow X$ is assumed to be semi-associative, but $\bullet\colon B \times X \rightarrow X$ is not assumed to be semi-associative. In parts 3 and 4, both $\bullet\colon A \times X \rightarrow X$ and $\bullet\colon B \times X \rightarrow X$ are assumed to be semi-associative. The set action $\times\colon A \times B \rightarrow B$ occurring in parts 2, 3, 4, of Theorem~\ref{theorem:semidirect-product-set-action} is not assumed to be semi-associative.

\item
The condition $a  \bullet b \bullet x = (a \times b) \bullet (a \bullet x)$ appearing in parts 3 and 4 of Theorem~\ref{theorem:semidirect-product-set-action} is a form of distributivity. This can easily be seen by changing the notation of $\bullet\colon A \times X \rightarrow X$ to $\times\colon A \times X \rightarrow X$. In this new notation the condition becomes $a  \times (b \bullet x) = (a \times b) \bullet (a \times x)$, where we have added the implicit parenthesis back in for clarity. 
\end{enumerate}
\end{remarks}
\begin{proof}[Proof of Theorem~\ref{theorem:semidirect-product-set-action}]\
1.\ is an easy induction using the definition of semidirect product action. For 2.\ we work from the inner part of the expression outwards. To start, observe that
\begin{align*}
(a^{n-1} \times b)  \bullet a^n \bullet x 
  & = (a^{n-1} \times b) \bullet (a^{n-1} * a) \bullet x \\
  & = (a^{n-1} \times b) \bullet (a^{n-1} \bullet (a \bullet x)) \\
  & = a^{n-1}\bullet b \bullet a \bullet x
\end{align*}
where the bracketing on all the occurrences of $a^{n-1}$ in these equations are chosen to match. The bracketing on the $a^n$ may be rearranged in any order because of the semi-associativity of $\bullet\colon A \times X \rightarrow X$.
Now assume we have proved that
\begin{equation*}
(a^i \times b) \bullet (a^{i+1} \times b) \bullet \ldots \bullet (a^{n-1} \times b) \bullet a^n \bullet x
= a^i \bullet \underbrace{b \bullet a \bullet \ldots \bullet b \bullet a}_\text{$n - i$ times} \bullet x
\end{equation*}
and this holds true for any choices of bracketing for the $a^j$, $j = i, \ldots, n$. Then
\begin{align*}
& (a^{i-1} \times b) \bullet (a^i \times b) \bullet (a^{i+1} \times b) \bullet \ldots \bullet (a^{n-1} \times b) \bullet a^n \bullet x\\
& \hspace{1.5in} = 
(a^{i-1} \times b) \bullet a^i \bullet \underbrace{b \bullet a \bullet \ldots \bullet b \bullet a}_\text{$n - i$ times} \bullet x\\
& \hspace{1.5in} =
(a^{i-1} \times b) \bullet (a^{i-1} * a) \bullet \underbrace{b \bullet a \bullet \ldots \bullet b \bullet a}_\text{$n - i$ times} \bullet x\\
& \hspace{1.5in} =
(a^{i-1} \times b) \bullet (a^{i-1} \bullet (a \bullet \underbrace{b \bullet a \bullet \ldots \bullet b \bullet a}_\text{$n - i$ times} \bullet x)\\
& \hspace{1.5in} =
a^{i-1} \bullet \underbrace{b \bullet a \bullet \ldots \bullet b \bullet a}_\text{$n - i + 1$ times} \bullet x
\end{align*}
where as before we may use semi-associativity of $\bullet\colon A \times X \rightarrow X$ to ensure the bracketing on the occurrences of $a^{i-1}$ match. The result now follows by induction. For 3.\ note that
\begin{align*}
\mat{a_1\\ b_1} \bullet \left(\mat{a_2\\ b_2} \bullet x\right) 
& = \mat{a_1\\ b_1} \bullet b_2 \bullet a_2 \bullet x \\
& = b_1 \bullet a_1 \bullet b_2 \bullet a_2 \bullet x \\
& = b_1 \bullet (a_1 \times b_2) \bullet a_1 \bullet a_2 \bullet x\\
& = (b_1 * (a_1 \times b_2)) \bullet (a_1 * a_2) \bullet x\\
& = \mat{a_1 * a_2\\ b_1 * (a_1 \times b_2)} \bullet x
\end{align*}
4.\ is a direct consequence of 1., 2., and 3.
\end{proof}

Theorem~\ref{theorem:semidirect-product-set-action} can be used to prove results about semidirect products of magmas by specializing to the case $X = B$. The following result for semigroups is a well known.

\begin{lemma}
\label{lemma:semidirect-product-semigroup-action-form}
Assume $(A, *),(B, *)$ are semigroups (i.e., $*\colon A \times A \rightarrow A$, $*\colon B \times B \rightarrow B$ are associative), and $\times\colon A \times B\rightarrow B$ is a semi-associative set action with companion operation equal to $*\colon A \times A \rightarrow A$, and which distributes over $\times$. I.e., assume that $a_1 \times (a_2 \times b) = (a_1 * a_2) \times b$, for all $a_1, a_2 \in A$, $b \in B$ and also $a \times (b_1 * b_2) = (a \times b_1) * (a \times b_2)$ for any $a \in A$, $b_1, b_2 \in B$. Then $A \ltimes_\times B$ is also a semigroup. I.e., $*\colon \left(A \ltimes_\times B\right) \times\left(A \ltimes_\times B\right) \rightarrow A \ltimes_\times B$ is associative.
\end{lemma}
\begin{proof}
A direct proof is elementary, but we choose to highlight the relationship to Theorem~\ref{theorem:semidirect-product-set-action}. Using the definition of associativity and of semidirect product, we see that what needs to be shown is that $a_1 * (a_2 * a_3) = (a_1 * a_2) * a_3$ and also that $b_1 * a_1 \times b_2 * a_2 \times b_3 = (b_1 * a_1 \times b_2) * (a_1 * a_2) \times b_3$. The first equation is true by the associativity of $*\colon A \times A \rightarrow A$. The second is a special case of Theorem~\ref{theorem:semidirect-product-set-action} part 3, using $X = B$, and setting $\bullet\colon B \times X \rightarrow X$ to $*\colon B \times B \rightarrow B$ and setting $\bullet\colon A \times X \rightarrow X$ to $\times\colon A \times B \rightarrow B$.
\end{proof}

Here is a more familiar restatement of the same result.

\begin{lemma} 
\label{lemma:semidirect-product-semigroup-function-form}
If $(A, *), (B, *)$ are semigroups (i.e., $*\colon A \times A \rightarrow A$, $*\colon B \times B \rightarrow B$ are associative), and $L\colon A \rightarrow \Endop(B)$ is such that $L(a_1 * a_2) = L(a_1) \circ L(a_2)$, for all $a_1, a_2 \in A$, and also $L(a)(b_1 * b_2)=L(a)(b_1) * L(a)(b_2)$ for any $a \in A$, $b_1, b_2 \in B$. (I.e., $L$ is a semigroup morphism from $A$ to the semigroup of semigroup endomorphisms of $B$.) Then $A \ltimes_L B$ is also a semigroup. I.e., $*\colon \left(A \ltimes_L B\right) \times\left(A \ltimes_L B\right) \rightarrow A \ltimes_L B$ is associative.
\end{lemma}
\begin{proof} 
This is equivalent to Lemma~\ref{lemma:semidirect-product-semigroup-action-form}. It is also a standard result in the theory of semigroups.
\end{proof}

We now give an analog of Theorem~\ref{theorem:semidirect-product-set-action} for semidirect products of binary operations.

\begin{theorem}
\label{theorem:semidirect-product-semi-associativity}
Assume that $A$, $B$ are sets, that $*\colon A \times A \rightarrow A$ and $*\colon B \times B \rightarrow B$ are binary operations, and that $\times\colon A \times B \rightarrow B$ is a set action. Assume $n$ is a strictly positive integer. Then
\begin{enumerate}
\item For any $a\in A$,  $b\in B$, the right-folded $n^\text{th}$ power of $\smat{a\\ b}$ in $A \ltimes_\times B$ is 
\begin{equation*}
\mat{a\\ b}^{*n} 
= \left( \smash[b]{\underbrace{
\begin{matrix} 
a^{*n}\\
b * a \times \ldots  \times b * a \times b
\end{matrix}
}_{\text{$n$ $b$'s and $n-1$ $a$'s}}}
\right)
\vphantom{\underbrace{\ }_{\text{$n$ $b$'s and $n-1$ $a$'s}}}
\end{equation*}

\item 
Assume $\times\colon A \times B \rightarrow B$ is semi-associative, that $*\colon A \times A \rightarrow A$, is a companion operation of $\times$, and assume that $\times$ distributes over $*\colon B \times B \rightarrow B$. I.e., assume that $a_1 \times (a_2 \times b) = (a_1 * a_2) \times b$, for all $a_1, a_2 \in A$, $b \in B$ and also $a \times (b_1 * b_2) = (a \times b_1) * (a \times b_2)$ for any $a \in A$, $b_1, b_2 \in B$.
Then for any $a\in A$,  $b\in B$,
\begin{equation*}
 \underbrace{b * a \times \ldots  \times b * a \times b}_{\text{$n$ $b$'s and $n-1$ $a$'s}}
 = b * (a \times b) * (a^2 \times b) * \ldots * (a^{n-1} \times b)
\end{equation*}
where each $a^i = a * \ldots * a$ is a power of $a$ computed using $*$ which may be bracketed in any order independently of the other powers $a^i$.

\item
With the assumptions as in 2., the right-folded $n^\text{th}$ power of $\smat{a\\ b}$ in $A \ltimes_\times B$ is 
\begin{equation*}
\mat{a\\ b}^{*n} 
= \mat{
a^{*n}\\
b * (a \times b) * (a^2 \times b) * \ldots * (a^{n-1} \times b)
}
\end{equation*}
where each $a^i = a * \ldots * a$ is a power of $a$ computed using $*$ which may be bracketed in any order independently of the other powers $a^i$.
\end{enumerate}
\end{theorem}
\begin{proof}
1.\ follows from the definition of semidirect product, and right-folded power, and an easy induction. 2.\ follows directly from Theorem~\ref{theorem:semidirect-product-set-action} part 2 by setting $X = B$, $x = b$, and setting $\bullet\colon B \times X \rightarrow X$ to $*\colon B \times B \rightarrow B$, and $\bullet\colon A \times X \rightarrow X$ to $\times\colon A \times B \rightarrow B$. 3.\ is a direct consequence of 1.\ and 2.
\end{proof}
\begin{remark}
At no place in the statement or proof of Theorem~\ref{theorem:semidirect-product-semi-associativity} do we assume the associativity of $*\colon A \times A \rightarrow A$ or of $*\colon B \times B \rightarrow B$.
\end{remark}

We complete our discussion of semidirect products and semi-associativity with a result that relates a set action of $A \times A$ on a product of sets $X \times Y$ to the opposite operation of a semidirect product. We will refer to this theorem in the examples of Chapter~\ref{chapter:examples}.

\begin{theorem}
\label{theorem:semidirect-product-xy-action}
Assume $A$, $X$, $Y$ are sets, and $\bullet\colon A \times X \rightarrow X$, $\bullet\colon X \times Y \rightarrow Y$ are semi-associative set actions with companion operations $*\colon A \times A \rightarrow A$, $*\colon X \times X \rightarrow X$. Assume further that there exists a second binary operation $*_2 \colon A \times A \rightarrow A$ such that $(a \bullet x) * (b \bullet x) = (a *_2 b) \bullet x$ for all $a, b \in A$, $x \in X$.%
\footnote{
This condition is equivalent to assuming that the set of functions $\theset{(x \mapsto a \bullet x)\colon a \in A}$ is closed under $*$, where for $f_1, f_2 \in \Endop(X)$ we define $f_1 * f_2 \in \Endop(X)$ by $(f_1 * f_2)(x) = f_1(x) * f_2(x)$. Such a condition could be called {\em right semi-distributivity}.
}
Define $\bullet\colon (A \times A) \times (X \times Y) \rightarrow X \times Y$, and $*\colon (A \times A) \times (A \times A) \rightarrow A \times A$ by
\begin{equation*}
\mat{a\\ b} \bullet \mat{x\\ y} = \mat{a \bullet x\\ (b \bullet x) \bullet y}
, \quad
\mat{a_1\\ b_1} * \mat{a_2\\ b_2} = \mat{a_1 * a_2\\ (b_1 * a_2) *_2 b_2}
\end{equation*}
Then $\bullet\colon (A \times A) \times (X \times Y) \rightarrow X \times Y$ is semi-associative with companion operation $*$.
\end{theorem}
\begin{proof}
\begin{align*}
\mat{a_1\\ b_1} \bullet \mat{a_2\\ b_2} \bullet \mat{x\\ y}
& = \mat{
a_1 \bullet a_2 \bullet x\\
(b_1 \bullet a_2 \bullet x) \bullet (b_2 \bullet x) \bullet y
} 
  = \mat{
(a_1 * a_2) \bullet x\\
(((b_1 * a_2) \bullet x) * (b_2 \bullet x)) \bullet y
} \\
& = \mat{
(a_1 * a_2) \bullet x\\
(((b_1 * a_2) *_2 b_2) \bullet x) \bullet y
}
= \mat{
(a_1 * a_2)\\
(b_1 * a_2) *_2 b_2
}
\bullet \mat{x\\ y}
\end{align*}    
\end{proof}
\begin{remark}
The operation in Theorem~\ref{theorem:semidirect-product-xy-action} is the opposite binary operation $*_\text{op}$ of the semidirect product of $(A, *_\text{op})$ with $(A, (*_2)_\text{op})$ using the set action $*_\text{op}\colon A \times A \rightarrow A$.
\end{remark}

\section{Related Work and References}

The techniques described in Chapter~\ref{chapter:semi-associativity-and-function-composition} relate closely to the work done on the prefix sum problem. This dates back to the work of Trout \cite{Trout1972}, and Blelloch \cite{Blelloch1993}. Fisher and Ghuloum \cite{FisherGhuloum1994} describe techniques for function composition, as do Chin et al.\ \cite{ChinTakano1998}, Chin et al.\ \cite{ChinKhoo2004}, and 
Morita et al.\ \cite{Morita2007}. This is also discussed in Steele \cite{Steele2005}, \cite{Steele2009}.

%% file: htcams-arxiv-ch08-algorithms-for-windowed-recurrences.tex
\chapter{Algorithms for Windowed Recurrences}
\label{chapter:algorithms-for-windowed-recurrences}

\section{The Meta-Algorithm for Computing Windowed Recurrences}

We now put together the definitions and results from the previous chapters to get the following meta-algorithm for computing windowed recurrences. We describe this in set action form using an indexed collection of functions to describe the action.

\begin{algorithm}[Meta-Algorithm for Windowed Recurrences]
\label{algorithm:meta-windowed-recurrence}

Let $A$, $X$ be sets, and let $\theset{f_a \colon a \in A}$ be a collection of functions with $f_a \in \Endop(X)$, indexed by $A$. (I.e., we are given a function $A \rightarrow \Endop(X)$, or equivalently a set action $\bullet\colon A \times X \rightarrow X$ with $a \bullet x = f_a(x)$.) Let $x_0, x_1, \ldots$ be a sequence of elements of $X$, and let $a_1, a_2, \ldots$ be a sequence of elements of $A$. Also let $n$ and $N$ be strictly positive integers. Then the following is an algorithm for computing the items $y_1, \ldots y_N$ of the corresponding windowed recurrence.

\begin{description}
\item[Step 1] 
Choose a representation of function composition for the functions $\theset{f_a\colon a \in A}$. (Or if we are given a set action, then choose a representation of function composition for this set action.) Let this be denoted $(\Lambda, \lambda, *, \bullet)$ where $\lambda\colon A \rightarrow \Lambda$, $*\colon \Lambda \times \Lambda \rightarrow \Lambda$, $\bullet\colon \Lambda \times X \rightarrow X$. Note that $*$ is not required to be associative.

\item[Step 2] 
Choose an algorithm for computing sliding window products. This algorithm must
    \begin{enumerate}
        \item Not require properties other than associativity to work.
        \item Not depend on operations on or functions of the elements of $\Lambda$, other than the product $*\colon \Lambda \times \Lambda \rightarrow \Lambda$.%
\footnote{
In particular it should not depend on equality or comparison relations, or on inverses, and this rules out SlickDeque and Subtract-on-Evict in general.
}
\end{enumerate}
    
Examples of suitable algorithms are Two Stacks, DEW, DABA Lite, though many others exist---see e.g.\ the references in Section~\ref{sec:sliding-window-sum-algorithms}, or the vectorized algorithms of Chapters \ref{chapter:vector-sliding-window-*-products}--\ref{chapter:vector-sliding-window-*-product-algorithms}.

\item[Step 3]
Compute the items $\lambda_1 = \lambda(a_1), \lambda_2 = \lambda(a_2), \ldots, \lambda_i = \lambda(a_i), \ldots$. If desired, this may be done on demand as required during Step 4, and need not be done before those values are needed.

\item[Step 4]
Compute the sliding window $*$-products of $\lambda_1, \lambda_2, \ldots$ using the algorithm chosen in Step 2. 
During the computation, compute the $*$-products as if $*$ was associative, even though is possibly not associative. {\em I.e., pretend $A$ is associative, even if it is not.}
Call these sliding window products $\tilde{Y}_i$, so that
\begin{equation*}
\tilde{Y}_i = \ourcases{
\lambda_i * \ldots * \lambda_1       & \text{with some bracketing, for $i \leq n$}\\
\lambda_i * \ldots * \lambda_{i-n+1} & \text{with some bracketing, for $i > n$}
}    
\end{equation*}
The bracketing of the products for $\tilde{Y}_i$ will depend on the sliding window $*$-product algorithm used.

\item[Step 5] Compute $y_1, \ldots, y_N$ as
\begin{equation*}
y_i = \ourcases{
\tilde{Y}_i \bullet x_0     & \text{for $i \leq n$}\\
\tilde{Y}_i \bullet x_{i-n} & \text{for $i > n$}
}    
\end{equation*}
\end{description}
\end{algorithm}

\begin{proof}[Proof of Algorithm \ref{algorithm:meta-windowed-recurrence} correctness]
By the definitions of windowed recurrence and representation of function composition, we have
%
\begin{align*}
y_i & = f_{a_i}(\ldots f_{a_{i-n+1}}(x_{i-n})) \\
    & = \lambda_i \bullet (\ldots \bullet(\lambda_{i-n+1} \bullet x_{i-n}) \ldots )\\
    & = (\lambda_i * (\ldots * (\lambda_{i-n+2} * \lambda_{i-n+1}) \ldots)) \bullet x_{i-n}
\end{align*}
So the result will follow if $(\lambda_i * \ldots * \lambda_{i-n+1}) \bullet x_{i-n}$ is independent of the bracketing used to evaluate the $*$-product $\lambda_i * \ldots * \lambda_{i-n+1}$. This follows from the semi-associativity of $\bullet$ and Lemma~\ref{lemma:semi-associativity-bracketing}.
\end{proof}

\begin{remark}
Algorithm \ref{algorithm:meta-windowed-recurrence} gives us a method to compute the following:

\begin{enumerate}
\item 
Windowed recurrences, using a collection of functions $f$. I.e. $A$ itself is a set of functions, $A \subseteq \Endop(X)$, and the map $a \mapsto f_a$ is the identity map.
  
\item 
Sliding window $*$-products where $*\colon X \times X \rightarrow X$ is a binary operation, which may be nonassociative.
%
%
%
 This uses the functions $f_x = \Leftop_x^*$, $x \in X$.

\item 
Windowed recurrences for a set action $\bullet\colon A \times X \rightarrow X$, where $\bullet$ may be non-semi-associative.
%
%
%
This uses the functions $\Leftop_a^\bullet$, $a \in A$.
\end{enumerate}
\end{remark}

Similar meta-algorithms exist for computing non-windowed recurrences and reductions, and these stand in the same relationhip to the Algorithm Ideas~\ref{algorithm-idea:recurrence} and \ref{algorithm-idea:reduction} as Algorithm~\ref{algorithm:meta-windowed-recurrence} does to Algorithm Idea~\ref{algorithm-idea:window-recurrence}. The technique for recurrences is to choose a representation of function composition for the set action or recurrence functions, and then apply a prefix $*$-product algorithm to the items $\lambda(a_1), \lambda(a_2), \ldots$ to obtain $\tilde{Z}_i = \lambda(a_i) * \ldots * \lambda(a_1)$, for $i = 1, \ldots, N$, and where the bracketing used will be determined by the algorithm chosen. The recurrence values are then computed as $\tilde{Z}_i \bullet x_0$ where the result is independent of the bracketing used to compute the $\tilde{Z}_i$. For a reduction the procedure is similar, except that only a single value $\tilde{Z}_N$ need be computed and that should be achieved using an algorithm for computing a $*$-product. The reduction is then computed as $\tilde{Z}_N \bullet x_0$ and is independent of the bracketing used to compute the $*$-product.

\section{Examples of the Meta-Algorithm}
\label{sec:examples-meta-algorithm}

\begin{example}
\label{example:moving-sums-with-scale-changes-algorithm}
We now revisit Example~\ref{example:moving-sums-with-scale-changes}, which is a moving sum with scale changes, or equivalently a windowed linear recurrence. We assume there is input data $a_1, a_2, \ldots$ for which we wish compute sliding window sums, and there are multipliers $m_1, m_2, \ldots$, which change the scale of the data. The definition of the sliding window calculation is
\begin{equation*}
y_i = \ourcases{
a_i + m_i(a_{i-1} + m_{i-1}(\ldots + m_3(a_2 + m_2 a_1) \ldots)), & i < n \\
a_i + m_i(a_{i-1} + m_{i-1}(\ldots + m_{i-n+3}(a_{i-n+2} + m_{i-n+2} a_{i-n+1}) \ldots)), & i \geq n
}
\end{equation*}

\noindent In order to describe this in terms of set actions we define

\begin{equation*}
\mat{m\\ a} \bullet x = a + m x
\end{equation*}
so that

\begin{equation*}
y_i = \mat{m_i\\ a_i} \bullet \left(\mat{m_{i+1}\\ a_{i+1}} \bullet \left(\ldots \bullet \left(\mat{m_{i-n+2}\\ a_{i-n+2}} \bullet a_{i-n+1}\right) \ldots\right)\right)
\end{equation*}
We can now proceed with the program above, using this expression, but it is slightly more convenient convention-wise to use the alternative expression below---either approach works.

\begin{equation*}
y_i = \mat{m_i\\ a_i} \bullet \left(\mat{m_{i-1}\\ a_{i-1}} \bullet \left(\ldots \bullet \left(\mat{m_{i-n+1}\\ a_{i-n+1}} \bullet 0\right) \ldots\right)\right)
\end{equation*}
The left action operators have the form
\begin{math}
\Leftop_\tmat{m\\ a}^\bullet(x) = a + m x    
\end{math}
and these are easily composed, as
\begin{align*}
\Leftop_\tmat{m_2\\ a_2}^\bullet \circ \Leftop_\tmat{m_1\\ a_1}^\bullet(x) 
    = \mat{m_2\\ a_2} \bullet (m_1 x + a_1) 
  & = m_2 m_1 x + m_2 a_1 + a_2 \\
  & = \Leftop_\tmat{m_2 m_1\\ m_2 a_1 + a_2}(x)
\end{align*}
So we can compute the function composition of the left action operators using the binary operation%
\footnote{
$*$ is in fact a form of matrix multiplication, as $\Leftop_\tmat{m\\ a}$ is the fractional linear transformation associated with the matrix $\smat{m & a\\ 0 & 1}$. It is also an example of a semidirect product. 
}
\begin{equation*}
\mat{m_2\\ a_2} * \mat{m_1\\ a_1} = \mat{m_2 m_1\\ m_2 a_1 + a_2}    
\end{equation*}
Thus $\bullet$ is semi-associative and has companion operator $*$. This gives the following algorithm for computing the windowed recurrence $y_i$.

\begin{description}

\item[Step 1]
Form the pairs $\smat{m_1\\ a_1}, \smat{m_2\\ a_2}, \ldots$

\item[Step 2]
Compute the sliding window $*$-products of the $\smat{m_i\\ a_i}$ using Two Stacks, DEW, DABA Lite, or another algorithm requiring only associativity. Call these $\tilde{Y}_i$ where

\begin{equation*}
\tilde{Y}_i= \ourcases{
\fmat{m_i\\ a_i} * \ldots * \fmat{m_1\\ a_1} & i \leq n \\[1.5ex] 
\fmat{m_i\\ a_i} * \ldots * \fmat{m_{i-n+1}\\ a_{i-n+1}} & i>n
}
\end{equation*}
These satisfy $Y_i = \Leftop_{\tilde{Y}_i}^\bullet$, where the $Y_i$ are as in Algorithm Idea~\ref{algorithm-idea:window-recurrence}.

\item[Step 3]
Compute $y_i$ as
\begin{align*}
y_i  & = \tilde{Y}_i \bullet 0  = \Leftop_{\tilde{Y}_i}^\bullet(0) = Y_i(0)\\
     & = \operatorname{proj}_2(\tilde{Y}_i) 
\end{align*}
where $\operatorname{proj}_2(\smat{m\\ a}) = a$ denotes the second component of a two-element vector.

\end{description}





\end{example}

\begin{example}
\label{example:continued-fraction-algorithm}

Suppose $\bullet$ is a set action of pairs acting on the real numbers extended by infinity, defined by

\begin{equation*}
\mat{m\\ a} \bullet x = \ourcases{
    a + \frac{m}{x}, & \text{if $x \neq 0$} \\
    \infty,          & \text{if $x = 0$}    \\
    a,               & \text{if $x = \infty$}
}
\end{equation*}
where $m \neq 0$, and $a$, $m$  are finite. Then is not semi-associative. I.e, there does not exist an operator $*$ such that $a \bullet (b \bullet c) = (a * b) \bullet c$.
The reason is that

\begin{equation*}
\mat{m_2\\ a_2} \bullet \left(\mat{m_1\\ a_1} \bullet x\right)
    =\frac{(a_2 a_1 + m_2) x + a_2 m_1}{a_1 x+m_1}
\end{equation*}
which is not of the form $a + \frac{m}{x}$ for any $a, m$, unless $m_1 = 0$. However, we may still find representatives for the compositions of the left action operator, as we now show.

The left action operator $\Leftop_\tmat{m\\ a}^\bullet$ is a special case of a fractional linear transformation. For any $2 \times 2$ matrix $A = \smatl{a_{11} & a_{12} \\ a_{21} & a_{22}}$, define the corresponding fractional linear transformation $T_A$, by $T_A(x)=\frac{a_{11} x + a_{12}}{a_{21} x + a_{22}}$.%
\footnote{
As discussed in Example~\ref{example:rofc-examples-5} we should extend the definition of $T_A$ so that $T_A(\infty) = \frac{a_{11}}{a_{21}}$, and $T_A(\frac{-a_{22}}{a_{21}}) = \infty$. See also Chapter~\ref{chapter:examples} Examples~49 and 50.
}
Then
\begin{equation*}
\Leftop_\tmat{m\\ a}^\bullet (x) = a + \frac{m}{x} = \frac{a x + m}{1 \cdot x + 0} = T_\tmat{a & m\\ 1 & 0}(x)
\end{equation*}
so
\begin{equation*}
\Leftop_\tmat{m\\ a}^\bullet = T_\tmat{a & m\\ 1 & 0}
\end{equation*}
The rule for composing fractional linear transformations is well known to be simply matrix multiplication, 
and therefore we obtain the following representation of function composition for $\bullet$.
\begin{equation*}
\lambda(\mat{m\\ a}) = \mat{a & m\\ 1 & 0}
, \quad
A \bullet x = T_A(x)
, \quad
* = \text{matrix multiplication}
\end{equation*}
Now consider the windowed recurrence
\begin{displaymath}
y_i = \ourcases{
\fmat{m_i\\ a_i} \bullet \left(\fmat{m_{i-1}\\ a_{i-1}} \bullet \left( \ldots
        \bullet \left( \fmat{m_1\\ a_1} \bullet x_0 \right)\right)\right)
& \text{if $1 \leq i < n$} \\[1.5ex]
\fmat{m_i\\ a_i} \bullet \left(\fmat{m_{i-1}\\ a_{i-1}} \bullet \left( \ldots
        \bullet \left( \fmat{m_{i-n+1}\\ a_{i-n+1}} \bullet x_{i-n} \right)\right)\right)
& \text{if $i \geq n$} 
}
\end{displaymath}
where $x_0, x_1, \ldots$ are in the real numbers extended by $\infty$. According to Meta-Algorithm~\ref{algorithm:meta-windowed-recurrence} we can compute the $y_i$ using the following algorithm.
\begin{description}
\item[Step 1] 
Form the matrices $\matl{a_1 & m_1\\ 1 & 0}, \matl{a_2 & m_2\\ 1 & 0}, \ldots$.\\

\item[Step 2]
Compute the length $n$ sliding window matrix products $\tilde{Y}_i$. These satisfy $Y_i=T_{\tilde{Y_i}}$, where $Y_i$ are as in Algorithm Idea~\ref{algorithm-idea:window-recurrence}.

\item[Step 3]
Compute
\begin{math}
y_i = \ourcases{
T_{\tilde{Y}_i}(x_0)
& \text{if $1 \leq i < n$}\\ 
T_{\tilde{Y}_i}(x_{i-n})
& \text{if $i \geq n$}\\ 
}
\end{math}
\end{description}

This example illustrates again that for a set action $\bullet\colon A \times X \rightarrow X$, the space of functions generated under composition by $\Leftop_{(A)}^\bullet$ will in general be larger than $\Leftop_{(A)}^\bullet$. However in this example we were able to embed $A$ in a larger space with set action which was semi-associative and use that larger set action to compute the windowed recurrence according to Algorithm Idea \ref{algorithm-idea:window-recurrence} and Meta-Algorithm~\ref{algorithm:meta-windowed-recurrence}.

\end{example}

\section{Final Notes on Chapters \ref{chapter:windowed-recurrences}--\ref{chapter:algorithms-for-windowed-recurrences} }

The theory of Chapters \ref{chapter:windowed-recurrences}--\ref{chapter:algorithms-for-windowed-recurrences} itself also applies to prefix sums as these are simply sliding window $*$-products where the window length at least as great as the data length. This shows how to compute those in nonassociative and non-semi-associative settings. In the next chapter we extend these ideas slightly further to cover situations where the functions being applied to compute the windowed recurrence have different domains and codomains, and also to extend to category-like algebraic systems where the set of allowed operations in the recurrence is allowed to vary with the index $i$. Then in subsequent chapters, we return to binary operations on sets, to set actions, and sequences of functions $f_i \in \Endop(X)$, and develop vectorized algorithms for the corresponding sliding window $*$-products and windowed recurrences.

%% file: htcams-arxiv-ch09-categories-and-magmoids.tex
\chapter{Categories and Magmoids}
\label{chapter:categories}

This chapter is independent of the following chapters, and may be safely skipped if the reader's interests lie elsewhere. In it we generalize the setting of windowed recurrences to category-theoretic settings. In keeping with the spirit of category theory, this chapter contains many definitions and equivalences.

\section{Windowed Recurrences with Multiple Domains}

In the preceding sections and chapters we have considered three closely related classes of problems.

\begin{enumerate}

\item 
Windowed recurrences for a sequence of functions on a set $X$
\begin{equation*}
y_i = f_i(f_{i-1}(\ldots f_{i-n+1}(x_{i-n}) \ldots))    
\end{equation*}

\item Windowed recurrences for a set action $\bullet\colon A \times X \rightarrow X$.
\begin{equation*}
    y_i = a_i \bullet (a_{i-1} \bullet (\ldots \bullet (a_{i-n+1} \bullet x_{i-n}) \ldots))    
\end{equation*}

\item Sliding window $*$-products for a binary operation $*\colon X \times X \rightarrow X$
\begin{equation*}
y_i = a_i * (a_{i-1} * (\ldots * (a_{i-n+2} * a_{i-n+1}) \ldots))    
\end{equation*}
\end{enumerate}

We now consider the situation where the domain $X$ of the functions varies with $i$, or in the case of set actions $\bullet$, or binary operations $*$, where the allowed operations vary with $i$. For windowed recurrences this means we have a chain of composable functions

\begin{equation*}
X_0 \xrightarrow{f_1} X_1 \xrightarrow{f_2} X_2 \xrightarrow{f_3} \cdots
\end{equation*}
and the definition of the $y_i$ generalizes in the obvious manner.

\begin{definition}[Windowed Recurrence, Multi-domain Version]

Let $X_0, X_1, \ldots$ be a sequence of sets, $x_0, x_1, \ldots$ be a sequence of set elements with $x_i \in X_i$, and let $f_1, f_2, \ldots$ be a sequence of composable functions $f_i\colon X_{i-1} \rightarrow X_i$. Let $n$ be a strictly positive integer. Then the windowed recurrence of length $n$ corresponding to the sequences $\theset{x_i}$, $\theset{f_i}$ is the sequence

\begin{equation*}
y_i = \ourcases{
f_i(f_{i-1}(\ldots f_1(x_0) \ldots)) & \text{for $1 \leq i < n$} \\
f_i(f_{i-1}(\ldots f_{i-n+1}(x_{i-n}) \ldots)) & \text{for $i \geq n$}
}
\end{equation*}
\end{definition}

\bigskip

\noindent One obvious way to extend Meta-Algorithm \ref{algorithm:meta-windowed-recurrence} to this new definition is to define 
\begin{displaymath}
X=\bigcup_{i \geq 0} X_i \cup \theset{\undefined}    
\end{displaymath}
and extend $f_i\colon X_{i-1} \rightarrow X_i$ to $\tilde{f}_i\colon X \rightarrow X$ by $\tilde{f}_i(\undefined) = \undefined$ and  $\tilde{f}_i(x)= \undefined$ if $x \notin X_{i-1}$. Another, ultimately equivalent, approach is to translate the constructions we have developed for composition of functions on a single set to compositions of functions on multiple sets. In translating Meta-Algorithm \ref{algorithm:meta-windowed-recurrence} we immediately run into questions to resolve: What are the analogs of a representation of function composition, semi-associativity, set actions, magmas, semigroups, etc.? What are the new algebraic structures involved, and do the algorithms of Chapters \ref{chapter:moving-sums}--\ref{chapter:algorithms-for-windowed-recurrences} apply to these? Fortunately these have easy solutions and the answers take us on a quick detour through elementary category theory.

\section{Quivers, Categories, and Windowed Recurrences}

We start with some definitions, adapted from Jacobsen \cite{Jacobson1989}.

\begin{definition}[Quiver]
A quiver, $Q$, consists of

\begin{enumerate}
\item 
A class%
\footnote{
Here we mean `class' in the sense used in Set Theory.
},
$\Ob(Q)$, of objects,

\item 
For each pair of objects $X, Y$, a set $\Hom_Q(X, Y)$, whose elements are called {\em morphisms} with domain $X$ and codomain $Y$,
\end{enumerate}
\noindent such that if $X, Y, U, V \in \Ob(Q)$ and $(X, Y) \neq (U, V)$ then $\Hom_Q(X, Y)$ and $\Hom_Q(U, V)$ are disjoint. When the quiver $Q$ being referred to is clear we also write $\Hom(X, Y)$ for $\Hom_Q(X, Y)$.
\end{definition}

\begin{definition}[$\Hom$, Composability, Quiver Maps]
\
\begin{enumerate}  
\item For any quiver, $Q$, define
\begin{align*}
\Hom(Q) & = \bigcup_{X, Y \in \Ob(Q)} \Hom_Q(X, Y)\\
        & = \text{ The collection of all morphisms of $Q$}
\end{align*}

\item 
For any quiver $Q$, and object $X \in \Ob(Q)$, define
\begin{align*}
& \Endop(X)        = \Hom(X, X) \\
& \UHom(X, -) = \bigcup_{Y \in \Ob(Q)} \Hom(X, Y) \\
& \UHom(-, X) = \bigcup_{W \in \Ob(Q)} \Hom(W, X)
\end{align*}

\item 
If $Q$ is a quiver and $a \in \Hom(Q)$ is a morphism in $Q$, then define
\begin{align*}
\dom(a) & = \text{domain of $a$} \\
\cod(a) & = \text{codomain of $a$}
\end{align*}

\item A sequence of morphisms $a_1, a_2, \ldots$ in a quiver $Q$ is said to be composable if $\cod(a_i) = \dom(a_{i+1})$ for $i=1,2, \ldots$.

\item 
Assume $Q$, $R$ are quivers. Then a quiver map from $Q$ to $R$, also called a morphism of quivers from $Q$ to $R$, is a pair of functions $\cF\colon \Ob(Q) \rightarrow \Ob(R)$, $\cF\colon \Hom(Q) \rightarrow \Hom(R)$, such that for any $f \in \Hom(Q)$ we have
\begin{equation*}
\dom(\cF(f)) = \cF(\dom(f)), \qquad \cod(\cF(f)) = \cF(\cod(f))
\end{equation*}
\end{enumerate}
\end{definition}

\begin{definition}[Magmoid, Semigroupoid, Category]
\label{definition:magmoid} \begin{samepage} \
\begin{enumerate}
\item A {\em magmoid} $\cM = (Q, *)$ consists of a quiver $Q$, together with binary operations $*\colon \Hom_Q(Y, Z) \times \Hom_Q(X, Y) \rightarrow \Hom_Q(X, Z)$ defined for any $X, Y, Z \in \Ob(Q)$.
For a magmoid $\cM$ we denote $\Ob(\cM) = \Ob(Q)$, $\Hom(\cM)=\Hom(Q)$ etc., regarding magmoids as a special case of quivers with additional structure.
\end{enumerate}
\end{samepage}
\begin{enumerate}
\setcounter{enumi}{1}
\item 
A {\em semigroupoid} $\cS$ is a magmoid whose binary operations are associative. I.e., if $f, g, h \in \Hom(\cS)$ are composable, then $(f * g) * h = f *(g * h)$.

\item 
A {\em category} $\cC$, is a semigroupoid such that for every object $X \in \Ob(\cC)$ there is an element $1_X \in \Endop(X)$ such that for any $Y \in \Ob(\cC)$ we have $f * 1_X = f$ for all $f \in \Hom(X, Y)$, and $1_X * g = g$ for $g \in \Hom(Y, X)$.
\end{enumerate}
    
\end{definition}

\begin{remarks} \
\begin{enumerate}
\item 
Quivers are simply directed graphs where loops and multiple edges are allowed. With this interpretation
\begin{align*}
\Ob(Q)       & = \text{The set of vertices of the graph $Q$.} \\
\Hom(Q)      & = \text{The set of (directed) edges of $Q$.} \\
\Hom_Q(X, Y) & = \text{The set of edges from the vertex $X$ to the vertex $Y$.} \\
\cod(f)      & = \text{The head of the edge $f$. I.e., the vertex that $f$ points towards.}\\
\dom(f)      & = \text{The tail of the edge $f$. I.e., the vertex that $f$ points away from.}
\end{align*}
Other names for `codomain' are `target', `head', `tip'. Other names for `domain' are `source', `tail'.

\item 
Another name for `morphism' in a quiver is `arrow'. We will use `arrow' for `morphism' interchangeably.

\item 
A quiver map (morphism of quivers) is a map from one directed graph to another which preserves the incidence relations of the graph. Quiver maps map sequences of composable arrows in a quiver to sequences of composable arrows in a quiver.

\item 
We denote the category of sets, whose objects are sets and whose morphisms are functions between sets, as $\Set$.

\item 
Let $Q$ be a quiver, then the free semigroupoid, on $Q$, denoted $\Free_\text{Semi}(Q)$ is the semigroupoid, whose objects are the objects of $Q$, and whose arrows are finite sequences of length $\geq 1$ of composable arrows in $Q$. We list these sequences in reverse order of composition in order to match the composability convention for magmoids in Definition~\ref{definition:magmoid} (i.e., to match the usual conventions for function composition). If $a_1, \ldots, a_n$ is a composable sequence of arrows in $Q$, and $a = (a_n, \ldots a_1) \in \Free_\text{Semi}(Q)$ then define $\dom(a) = \dom(a_1), \cod(a)=\cod(a_n)$. If $b=(b_m, \ldots, b_1)$, where $b_1, \ldots, b_m$ are composable, and with $\dom(b)=\cod(a)$, i.e., $\dom(b_1) = \cod(a_n)$, then the semigroupoid operation on $\Free_\text{Semi}$ is defined to be
\begin{equation*}
b * a = (b_m, \ldots b_1, a_n, \ldots, a_1)
\end{equation*}
\end{enumerate}
\end{remarks}

\begin{definition}[Magmoid Morphism, Semi-Functor, Functor]
\
\begin{enumerate}

\item 
Let $\cM$, $\cN$ be magmoids. Then a {\em magmoid morphism} from $\cM$ to $\cN$ is a quiver morphism $\cF$ from $\cM$ to $\cN$, such that for any two composable arrows $a, b$ of $\cM$, we have
\begin{equation*}
\cF(a * b) = \cF(a) * \cF(b)
\end{equation*}

\item 
A magmoid morphism between semigroupoids is also called a {\em semi-functor}.

\item Let $\cC$, $\cD$ be categories. Then a {\em functor} from $\cC$ to $\cD$ is a magmoid morphism from $\cC$ to $\cD$ that also satisfies $f(1_X) = 1_{\cF(X)}$ for all $X \in \Ob(\cC)$.
\end{enumerate}
\end{definition}

\begin{definition}[Representation]
\
\begin{enumerate}
  \item A representation of a quiver $Q$ in a magmoid $\cM$, is a quiver map from $Q$ to $\cM$.
  \item A representation of a magmoid $\cM$ in a magmoid $\cN$ is a magmoid morphism from $\cM$ to $\cN$.
  \item A representation of a category $\cC$ in a category $\cD$, is a functor from $\cC$ to $\cD$.
\end{enumerate}
\end{definition}

\begin{definition}[Quiver Action, Semi-Associativity]
\label{definition:quiver-action} \begin{samepage} \
\begin{enumerate}
\item 
Let $Q$ be a quiver. Then a quiver action $(\cF, \bullet)$ of $Q$ on the category $\Set$ is a function $\cF\colon \Ob(Q) \rightarrow \Set$, together with, for each $X, Y \in \Ob(Q)$ with nonempty $\Hom(X, Y)$, an operation
\begin{equation*}
\Hom_Q(X, Y) \times \cF(X) \longrightarrow \cF(Y) \colon (a, x) \longmapsto a \bullet x
\end{equation*}
\end{enumerate}
\end{samepage}
\begin{enumerate}
\setcounter{enumi}{1}
\item 
A quiver action $(\cF, \bullet)$ of $Q$ on $\Set$ is semi-associative if there is a magmoid operation $*$ such that $(Q, *)$ is a magmoid, and for any composable arrows $a_1, a_2 \in \Hom(Q)$, and $x \in \cF(\dom(a_1))$, we have
\begin{equation*}
a_2 \bullet (a_1 \bullet x) = (a_2 * a_1) \bullet x    
\end{equation*}
In such case $*$ is said to be a companion operation of $\bullet$.

\item 
Let $(\cF, \bullet)$ be a quiver action of $Q$ on $\Set$. Define the left action morphism $\Leftop^\bullet$ corresponding to $(\cF, \bullet)$ to be the quiver morphism from $Q$ to $\Set$ defined by
\begin{align*}
\Leftop^\bullet(X) & = \cF(X) & \text{\parbox[t]{2.0in}{
for $X \in \Ob(Q)$
}} \\
(\Leftop^\bullet(a))(x) & = a \bullet x & 
\text{\parbox[t]{2.0in}{
    for $a \in \Hom_Q(X, Y)$,\\
    $x \in \cF(X)$, $X, Y \in \Ob(Q)$
}}
\end{align*}
We also denote $\Leftop^\bullet(a)$ by $\Leftop_a^\bullet$.
\end{enumerate} 
\end{definition}

\begin{remarks} \
\begin{enumerate}
\item 
Part 3 of Definition~\ref{definition:quiver-action} contains the claim that $\Leftop^\bullet$ is indeed a quiver morphism, but this follows immediately from the defining equations of $\Leftop^\bullet$.

\item
The correspondence between quiver actions on $\Set$ and their left action morphisms shows that quiver actions are equivalent to quiver morphisms from a quiver to $\Set$, or equivalently to quiver representations in $\Set$.
\end{enumerate}
\end{remarks}

The following lemma is analogous to Lemma~\ref{lemma:semi-associativity}.

\begin{lemma}
Assume $(\cF, \bullet)$ is a quiver action of $Q$ on $\Set$. Then the following are equivalent.
\begin{enumerate}
\item 
$(\cF, \bullet)$ is semi-associative.

\item 
There exists a magmoid operation on $Q$ such that $\Leftop^\bullet$ is a magmoid morphism from $(Q, *)$ to $\Set$.

\item
For any composable arrows $a_1, a_2 \in \Hom(Q)$ there is an arrow $b \in \Hom(Q)$ with $\dom(b) = \dom(a_1)$, and $\cod(b) = \cod(a_2)$ such that for any $x \in \cF(\dom(a_1))$, we have
\begin{equation*}
a_2 \bullet (a_1 \bullet x) = b \bullet x
\end{equation*}

\item
The image of $\Leftop^\bullet$ in $\Set$ is closed under function composition.
\end{enumerate}
\end{lemma}
\begin{proof}
The proof is analogous to the proof of Lemma~\ref{lemma:semi-associativity}
\end{proof}

\begin{definition}[Sliding Window $*$-Products for Magmoids]
    
Let $\cM$ be a magmoid with operation $*$. Let $a_1, a_2, \ldots$ be a sequence of composable arrows in $\cM$, and let $n$ be a strictly positive integer. Then the sliding window $*$-product of length $n$ is the sequence
\begin{equation*}
y_i = \ourcases{
a_i * (a_{i-1} * (\ldots * (a_2 * a_1) \ldots)) & \text{for $1 \leq i < n$} \\
a_i * (a_{i-1} *(\ldots * (a_{i-n+2} * a_{i-n+1}) \ldots)) & \text{for $i \geq n$}
}
\end{equation*}
\end{definition}

\begin{definition}[Windowed Recurrences for Quiver Actions]
 Let $X_0, X_1, \ldots$ be a sequence of sets, let $x_0, x_1, \ldots$ be a sequence of set elements with $x_i \in X_i$. Let $(\cF, \bullet)$ be a quiver action of a quiver $Q$ on $\Set$, and let $a_1, a_2, \ldots$ be a composable sequence of arrows in $Q$, with $q_0 \xrightarrow{a_1} q_1 \xrightarrow{a_2} q_2 \xrightarrow{a_3} \cdots$, such that $\cF(q_i) = X_i$. Let $n$ be a strictly positive integer. Then the windowed recurrence of length $n$ corresponding to the sequences $\theset{x_i}$, $\theset{a_i}$ and the quiver action $(\cF, \bullet)$ is the sequence
\begin{equation*}
y_i = \ourcases{
a_i \bullet (a_{i-1} \bullet (\ldots \bullet (a_1 \bullet x_0) \ldots)) & \text{for $1 \leq i < n$} \\
a_i \bullet (a_{i-1} \bullet (\ldots \bullet (a_{i-n+1} \bullet x_{i-n}) \ldots)) & \text{for $i \geq n$}
}
\end{equation*}
\end{definition}

\noindent We will now consider algorithms for sliding window $*$-products in the associative cases. After that we consider algorithms for windowed recurrences for quiver actions, and functions with multiple domains. Finally, we round up with the nonassociative magmoid sliding window $*$-product case. With the exception of the nonassociative magmoid case these follow the same approach as we saw for binary operations, set actions and functions.

\begin{theorem}
The Two Stacks, DEW, and DABA-Lite algorithms may be used to compute sliding window $*$-products for semigroupoids and categories.    
\end{theorem}
\begin{proof}
First note that for semigroupoids (and categories) $*$ is associative when applied to composable arrows. So for the theorem to hold true, the algorithms must only apply the $*$ operation to composable arrows in the semigroupoid. This is indeed the case, as these algorithms only ever compute products of the form $b * c$, where $b = a_l * \ldots * a_{k+1}$, $c = a_k * \ldots * a_{j+1}$ for some $j, k, l$ with $j < k < l$.
\end{proof}

The same result holds for other sliding window $*$-product algorithms which only rely on associativity.
To handle windowed recurrences with multi-domain function sequences, and to handle quiver actions, we need to generalize the definition of a representation of function composition.

\begin{definition}[Representation of Function Composition for Quivers] 
Let $f\colon Q \rightarrow \Set$ be a quiver map (also called a morphism of quivers, or a representation of $Q$) from the quiver $Q$ to the category of sets. Denote its action on objects of $Q$ by $q \mapsto X_q$ for $q \in \Ob(Q)$, and on arrows by $a \mapsto f_a$ for $a \in \Hom(Q)$. I.e., $\theset{f_a}$ is a collection of functions indexed by the arrows of the quiver $Q$ in a way that preserves composability. Then a representation of function composition for the functions $\theset{f_a}$ consists of the following.
\begin{enumerate}
\item 
A magmoid $\cM$ with binary operations $*$.

\item 
A quiver map $\lambda\colon Q \rightarrow \cM$. (I.e.\  a representation of $Q$ in $\cM$.)

\item 
A quiver action $(\cF, \bullet)$ of $\cM$ on $\Set$.
\end{enumerate}
\noindent satisfying the following properties
\begin{enumerate}[label=(\alph*)]
\item 
For $a \in \Hom(Q)$, $x \in \dom(f_a)$, $\lambda(a) \bullet x = f_a(x)$. For $q \in \Ob(Q)$, $X_q = \cF(\lambda(q))$.
\item
$\bullet$ is semi-associative with companion operation $*$. I.e., if 
$M_0 \xrightarrow{\mu_1} M_1 \xrightarrow{\mu_2} M_2$ in $\cM$ and $x_0 \in \cF(M_0)$, then
$\mu_2 \bullet (\mu_1 \bullet x_0) = (\mu_2 * \mu_1) \bullet x_0$.
\end{enumerate}    
\end{definition}

\begin{remark}
A representation of function composition for $f\colon Q \rightarrow \Set$ is equivalent to a factoring of $f$ through a magmoid morphism.

\begin{center}
\begin{tikzcd}[row sep=large, column sep=large]
    Q \arrow[r, "f"] \arrow[d, "\lambda"'] &  \Set \\
    \cM \arrow[ru,"\Leftop^\bullet" below right, rightarrow,]\\
\end{tikzcd} 
\vspace{-0.35in}
\end{center}

\noindent Property (a) states that $f = \Leftop^\bullet \circ \lambda$. Property (b) states that $\Leftop^\bullet$ is a magmoid morphism.
\end{remark}

\begin{theorem}
Meta-Algorithm \ref{algorithm:meta-windowed-recurrence} may be used to compute windowed recurrences for quiver actions on $\Set$, and for function sequences with multiple domains, with the indexed set of functions $\theset{f_a\colon a\in A}$ replaced by a quiver map $Q \rightarrow \Set$, and the representation of function composition for $\theset{f_a\colon a \in A}$ replaced by a representation of function composition for the quiver map $f\colon Q \rightarrow \Set$.
\end{theorem}
\begin{proof}
The arguments involving semi-associativity are the same, and only composable arrows or functions with the correct domains are applied. For the function case the quiver may be taken to be that formed from the sets $X_0, X_1, \ldots$ and the functions $f_1, f_2, \ldots$ themselves. For the action case use the quiver map $\Leftop^\bullet$ where $\bullet$ is the action operator.
\end{proof}
\begin{remark}
 It should be clear that analog of (non-windowed) recurrences and reductions may be defined for a quiver action $(\cF, \bullet)$ on $\Set$, and a sequence of composable arrows $q_0 \xrightarrow{a_1} q_1 \xrightarrow{a_2} q_2 \xrightarrow{a_3} \ldots$, and an element $x_0$ in the domain of $\cF(a_1)$. These may be computed using a representation of function composition for the quiver action, by first computing either a prefix product (for recurrences) or a product (for reductions) of the arrows, and then applying these to $x_0$.
\end{remark}

\noindent This leaves the case of sliding window magmoid $*$-products in the nonassociative case. The answer here is less satisfying. We would like to define left action functions `$\Leftop_a^*$' that embed the magmoid in the category of sets as functions on the $\Hom$-sets of $\cM$. But this doesn't quite work, as $\Hom(-, X)$ depends on an object in the magmoid, so there are many functors and not just one.
This can be remedied by considering the action of $*$ on the union of the $\Hom$-sets.

\begin{definition}
Let $\cM$ be a magmoid with operation $*$. Then define the left action quiver morphism $\Leftop^* \colon \cM \rightarrow \Set$ as follows. For $X \in \Ob(\cM)$
\begin{equation*}
\Leftop^*(X) = \UHom(-, X) =\bigcup_{W \in \Ob(\cM)} \Hom(W, X)
\end{equation*}
For $a \in \Hom(X, Y)$, define $\Leftop^*(a)\colon \UHom(-, X) \rightarrow \UHom(-, Y)$, by 
%
$\left(\Leftop^*(a)\right)(b) = a * b$
for $b \in \UHom(-, X)$. The function $\Leftop^*(a)$ is also written as $\Leftop_a^*$.
\end{definition}

\begin{remark}
$\Leftop^*$ is not a magmoid morphism unless $\cM$ is a semigroupoid. It is, however, a quiver morphism 
(i.e., a quiver map, or representation).
\end{remark}

We can now describe a procedure for computing sliding window $*$-products for nonassociative magmoids. Suppose
\begin{equation*}
X_0 \xrightarrow{a_1} X_1 \xrightarrow{a_2} X_2 \xrightarrow{a_3} \cdots
\end{equation*}
is a sequence of composable arrows in $\cM$. Now consider the sequence of functions
\begin{equation*}
\UHom(-, X_1) \xrightarrow{\Leftop_{a_2}^*} \UHom(-, X_2) \xrightarrow{\Leftop^*_{a_3}} \cdots    
\end{equation*}
This is a sequence of functions in $\Set$, and the left action morphism $\Leftop^*$ is a quiver map $\cM \rightarrow \Set$. Given a representation of function composition for $\Leftop^*\colon \cM \rightarrow \Set$ we may compute the windowed recurrence of length $n-1$ for $\Leftop_{a_2}^*, \Leftop_{a_3}^*, \ldots$, and the sequence $a_1, a_2, a_3, \ldots$. This yields the sliding window $*$-products $y_i$ as
\begin{align*}
y_i 
& = \ourcases{
    a_i * (\ldots * (a_2 * a_1)) \ldots), & 1 \leq i \leq n\\
    a_i * (\ldots * (a_{i-n+2} * a_{i-n+1})) \ldots), & i > n\\
}\\
& = \ourcases{
    a_1, & i = 1\\
    \Leftop_{a_i^*}(\ldots \Leftop_{a_2}^*(a_1) \ldots), & 2 \leq i \leq n\\
    \Leftop_{a_i^*}(\ldots \Leftop_{a_{i-n+2}}^*(a_{i-n+1}) \ldots), & i > n\\
}
\end{align*}

\noindent Here is a table of correspondences of concepts relating the Category Theory based theory to the one for binary operations and set actions.

\begin{center}
{\bf{Category Theory Correspondences}} 
\nopagebreak
\medskip

{ 
\renewcommand{\arraystretch}{2.0}
\begin{tabular}{|m{3.0in}|m{3.0in}|}
\hline binary operation, magma
                & magmoid\\
\hline semigroup
                & semigroupoid\\
\hline monoid
                & category\\
\hline index set $A$
                & quiver $Q$\\
\hline indexed set of functions $f_a$
                & quiver map $Q \rightarrow \Set$\\
\hline set action $\bullet\colon A\times X \rightarrow X$
                & \parbox[t]{3.0in}{
                    quiver action, \\ quiver map/morphism/representation 
                    $Q \rightarrow \Set$
                    }\\[0.17in]
\hline left action operators for set action
                & quiver morphism corresponding to quiver action\\
\hline representation of function composition
                & \parbox[t]{3.0in}{
                    factorization of quiver map $f\colon Q \rightarrow \Set$ as
                    $f = \cF \circ \lambda$ where $\lambda\colon Q \rightarrow \cM$ is
                    a quiver map to a magmoid $\cM$ and $\cF$ is a magmoid morphism
                    $\cM \rightarrow \Set$, and $\cF = \Leftop^\bullet$ for 
                    a semi-associative quiver action of $\cM$ acting on $\Set$
                    }\\[0.67in]
\hline left action functions for magma
                & \parbox[t]{3.0in}{
                    left action quiver morphism $\Leftop^* \colon \cM \rightarrow \Set$ for 
                    magmoid $\cM$
                    }\\[0.17in]
\hline
\end{tabular}
} 
\end{center}

%% file: htcams-arxiv-ch10-introduction-to-vector-algorithms.tex

\chapter{Introduction to Vector Algorithms for Windowed Recurrences}
\label{chapter:vector-algorithms-guide}
%
%
%

A vector algorithm for a {\em windowed recurrence}, or a {\em sliding window $*$-product} 
is an algorithm that computes the windowed recurrence or sliding window $*$-product using only operations that operate on collections of objects. As discussed in Section~\ref{sec:what-is-next}, such algorithms are important not only because they are parallel algorithms described in a manner that abstracted from the details of how the vector operations themselves are computed, but also because they present a user interface where the recurrence function or $*$-operation may itself be defined in terms of vector operations. The papers and monograph of Blelloch \cite{Blelloch1996} \cite{Blelloch1990} contain extensive discussions of vector models of computation. For our algorithms, we require a limited model of vector computation which allows element-wise operations, and also `shift' or `lag' operations. Because the useful models for vector computation are varied, we proceed by defining the mathematical properties we require for our vector algorithms to work, and indicate by way of examples how these relate to the windowed recurrences and sliding window $*$-products defined in Chapters \ref{chapter:moving-sums}--\ref{chapter:categories}.

Our plan is as follows:

\begin{enumerate}

  \item Define vector sliding window $*$-products.
  
  \item Describe how to relate (non-vector) sliding window $*$-products to the vector versions of these. There are multiple ways to do this, corresponding to different models of vector computation, different conventions, and different computational settings---in short, corresponding to different applications and use cases.
  
  \item We relate vector sliding window $*$-products to powers of an element in a semidirect product semigroup or magma%
  %
  \footnote{Recall that a {\em magma} is a set with a binary operation.}.
  The main results here are Theorem~\ref{theorem:semidirect-product-vector-sliding-window}, and Algorithms \ref{algorithm:vector-sliding-window-*-product} and \ref{algorithm:general-vector-sliding-window-*-product}.

  \item In the associative case we may compute the power of this semidirect product element using any of the known methods for fast exponentiation in semigroups, e.g., sequential or parallel algorithms for binary exponentiation (see \cite{Knuth1998}), or Brauer's method \cite{Brauer1939}, Thurber's method \cite{Thurber1973b}, Yao's method \cite{Yao1976}, or optimal addition chain exponentiation. In particular using parallel binary exponentiation gives an algorithm of depth $\left\lceil\log_2 n\right\rceil$ where $n$ is the window length and $n$ is not required to be a power of $2$.
  
  \item We next turn to vector windowed recurrences in both a function recurrence and set action setting. We start with definitions of vector windowed recurrences, and examples and constructions relating these to the non-vector cases. We then relate vector windowed recurrences to sliding window vector $*$-products, and this requires a brief further study of semi-associativity (i.e., models of function application and composition), semidirect products, and shift operations. From these results we then obtain vector algorithms for windowed recurrences, as well as for vector sliding window $*$-products in the nonassociative case. The main results here are Theorems \ref{theorem:vector-set-actions-and-semi-associativity}, \ref{theorem:vector-windowed-recurrence-algorithm}, and Algorithm \ref{algorithm:vector-windowed-recurrence}.
  
\end{enumerate}

\medskip

\noindent The results of Chapters \ref{chapter:vector-sliding-window-*-products}--\ref{chapter:vector-sliding-window-*-product-algorithms} yield the following:

\begin{enumerate}
  \item Vector and parallel algorithms for sliding window $*$-products, with complexity given in a number of vector $*$ operations depending on the exponentiation method used. Here $n$ is the window length.
  
{ 
\renewcommand{\arraystretch}{1.5}
\begin{tabular}{ll}
Exponentiation method & \hspace{-1.0pt}Number of operations\\
\hline
Binary exponentiation & \hspace{-1.0pt}$2\left\lfloor\log_2 n\right\rfloor$ vector $*$ operations\\
Brauer's method & \hspace{-1.0pt}$(\log_2 n)
                   \left(1 + \frac{1}{\log_2\log_2(n+2)} 
                           + \littleoh{\frac{1}{\log_2\log_2(n+2)}}
                   \right)$ \hspace{-1.0pt}vector $*$ operations\\
Thurber's method & \hspace{-1.0pt}$(\log_2 n)
                   \left(1 + \frac{1}{\log_2\log_2(n+2)} 
                           + \littleoh{\frac{1}{\log_2\log_2(n+2)}}
                   \right)$ \hspace{-1.0pt}vector $*$ operations\\
Parallel binary exponentiation & \hspace{-1.0pt}$\left\lceil\log_2 n \right\rceil$ parallel steps (depth)
\end{tabular}
} 

\item Algorithms for computing windowed recurrences in a number of vector or parallel operations corresponding to the sliding window $*$-product operation counts in the table above. These algorithms perform their operations in a vector representation of function composition and require an additional vector function application at the end. They also apply to nonassociative sliding window $*$-products.

\item New algorithms for parallel prefix sums, i.e., parallel prefix $*$-products, parallel prefix recurrences.

\item Algorithms for simultaneous vector or parallel computation of windowed recurrences at multiple window lengths---this is the {\em multi-query} problem. This includes the simultaneous computation of parallel prefix sums and windowed recurrences.
\end{enumerate}

\noindent The definitions and the proofs in this chapter are abstract, but they lead to compact and simple code for computing sliding window $*$-products and windowed recurrences. The abstraction is a symptom of the general purpose nature of the code. For practitioners more interested in the code than the proofs, here is a complete implementation of the algorithms in pseudo-code, and examples of its use.

\begin{alltt}
window_compose(compose, shift, a, n, exponentiate):
    define semidirect_product(u, v):
        return (u[1] + v[1], compose(u[2], shift(u[1], v[2])))
    return exponentiate(semidirect_product, (1, a), n)[2]

window_apply(compose, apply, lift, shift, shiftx, n, a, x, exponentiate):
    function_data = window_compose(compose, shift, lift(a), n, exponentiate)
    return apply(function_data, shiftx(n, x))

binary_exponentiate(op, x, n, flip):
    q=x, z=x, first=true
    repeat indefinitely
        if n is odd
            if first is true
                q=z, first=false
            else
                q = op(q, z) if flip is true else op(z, q)
        n = \(\lfloor\)n/2\(\rfloor\)                                    \(\,\)This is a logical right shift n >> 1
        if n = 0
            return q
        z = op(z, z)
        
\end{alltt}

We now demonstrate how to use this code to compute a sliding window sum (a moving sum). First we need to decide how to vectorize the problem, so for this example, we can choose to let the input to {\tt window\_compose}, {\tt a}, be an array of numbers of length $N$. Then define functions as follows 

%
%
\begin{samepage}
\begin{alltt}
compose(a, b):
    return a + b                                        This is vector addition of arrays
\end{alltt}
\end{samepage}

\begin{alltt}
shift(i, a):
    j = min(i, N)
    return \(\underbrace{\text{\tt{0,...,0}}}\sb{\tt{j}}\),a[1],a[2],...,a[N-j]

binary_exponentiate_no_flip(op, x, n):
    return binary_exponentiate(op, x, n, flip=false)

window_sum(a, n):
    return window_compose(compose, shift, a, n, binary_exponentiate_no_flip)

\end{alltt}

\noindent With these definitions the {\tt window\_sum} procedure computes a sliding window sum given the input sequence in the array {\tt a}. We shall describe the properties required of the inputs to the {\tt window\_compose}, and {\tt window\_apply} procedures in Chapters~\ref{chapter:vector-sliding-window-*-products}, \ref{chapter:vector-windowed-recurrences} and \ref{chapter:vector-pseudo-code}. But even with the one example above, we can already begin to describe how to modify the inputs to solve other problems or produce algorithms with different properties.

\begin{enumerate}

\item By replacing the $+$ in {\tt compose} with another associative operation, we can compute sliding window $*$-products. (Though in this specific example we used an identity element $0$.)
  
\item By varying the definition of {\tt compose} and {\tt shift} we can support other vectorization schemes. E.g., if we define {\tt compose}, {\tt shift} as follows

\begin{alltt}
compose(a, b):
    M = length(a), N = length(b)
    return a[1],...,a[M-N],a[M-N+1]+b[1],...,a[M]+b[N]

shift(i, a):
    N = length(a)
    return a[1],a[2],...,a[N-i]
\end{alltt}

\noindent then we obtain an algorithm for sliding window sums that uses slightly fewer operations and generalizes to semigroups rather than monoids (i.e., it does not require an identity element). 
Note that this version of {\tt compose} requires ${\tt{length(a)}} \geq {\tt{length(b)}}$, but this is never an obstacle as during the course of the computation {\tt compose} will only ever be passed arrays satisfying this condition; this is because the definition of the semidirect product applies the shift operator to the array in the pair on the right, and this holds true regardless of whether we set {\tt flip=true} or {\tt flip=false} in the call to {\tt binary\_exponentiate}.

\item By varying the exponentiate procedure we can improve the complexity (and parallel depth) of the algorithm, as well as varying requirements and access patterns. Additional exponentiation procedures are described in Chapter~\ref{chapter:exponentiation-in-semigroups}.
  
\end{enumerate}
  
The use of the {\tt window\_apply} function is similar to {\tt window\_compose}, though a point of warning about the required properties is in order. When using {\tt window\_compose} to compute sliding window $*$-products directly, we will assume {\tt compose} is associative, as well as some properties of the shift operators. When using {\tt window\_apply} we do not make the same assumptions, and the call from {\tt window\_apply} to {\tt window\_compose} may pass a nonassociative operator. Also the input {\tt a} to {\tt window\_apply} is frequently of a different type to the input {\tt a} to {\tt window\_compose}.

Let's consider the computation of a windowed linear recurrence, or equivalently, a `moving sum with scale changes'. This was considered in Examples~\ref{example:moving-sums-with-scale-changes} and \ref{example:moving-sums-with-scale-changes-algorithm}, and is the computation of
\begin{equation*}
y_i = \ourcases{
v_i + u_i\left(v_{i-1} + u_{i-1}\left(\ldots + u_3\left(v_2+u_2 v_1\right) \ldots\right)\right) & \text{if $i < n$} \\
v_i + u_i\left(v_{i-1} + u_{i-1}\left(\ldots + u_{i-n+3}\left(v_{i-n+2}+u_{i-n+2} v_{i-n+1}\right)\ldots \right)\right) & \text{if $i \geq n$}
}
\end{equation*}

\noindent To compute this we define

\begin{alltt}
compose(a, b):
    u = a[1], v = a[2], w = b[1], z = b[2]
    return (u * w, v + u * z)                vector addition and multiplication of arrays
\end{alltt}

\noindent In {\tt compose}, {\tt a=(u, v), b=(w, z)} are pairs of arrays, and {\tt *, +} are component-wise multiplication and addition respectively.

%
%
\begin{alltt}
apply(a, x):
    u = a[1], v = a[2]
    return v + u * x                         vector addition and multiplication of arrays

shift(i, a):
    u = a[1], v = a[2], N = length(v), j = min(i, N)
    return ([\(\underbrace{\text{\tt{1,...,1}}}\sb{\tt{j}}\),u[1],\dots,u[N-j]], [\(\underbrace{\text{\tt{0,...,0}}}\sb{\tt{j}}\),v[1],\ldots,v[N-j])

shiftx(i, x):
    N = length(x), j = min(i, N)
    return [\(\underbrace{\text{\tt{0,...,0}}}\sb{\tt{j}}\),x[1],...,x[N-j]]

lift = identity function

window_sum_with_scale_changes(u, v, n): Here u, v are arrays of length N.
    if n = 1
        return v 
    else 
        return window_apply(compose, apply, identity, shift, shiftx, n - 1, (u, v), v,
                            binary_exponentiate_no_flip)
\end{alltt}

\noindent Note the $n-1$ in the call to {\tt window\_apply}. An alternative approach is to use the same compose, apply, and shift operators to define

\begin{alltt}
window_sum_with_scale_changes(u, v, n): 
    N = length(v)
    return window_apply(compose, apply, identity, shift, shiftx, n, (u, v), 
                        [\(\underbrace{\text{\tt{0,...,0}}}\sb{\tt{N}}\)], binary_exponentiate_no_flip)
\end{alltt}

\noindent See Examples~\ref{example:windowed-linear-recurrence-pseudo-code-fixed-length} and \ref{example:windowed-linear-recurrence-pseudo-code-variable-length} for further approaches to this calculation.

%% file: htcams-arxiv-ch11-vector-sliding-window-star-products.tex
\chapter{Vector Sliding Window $*$-Products} 
\label{chapter:vector-sliding-window-*-products}

\section{Definitions}

\begin{definition}[Vector Product]
Let $A$ be a set, and $*\colon A \times A \rightarrow A$ be a binary operation on $A$, and assume $L_1, L_2, \ldots \in \Endop(A)$ be functions on $A$ such that
\begin{align*}
L_i \circ L_j & = L_{i+j}      & & \text{for $i, j \geq 1$, and}\\
L_i(a * b) & = L_i(a) * L_i(b) & & \text{for $a, b \in A, i \geq 1$}
\end{align*}

\noindent Then $*$ is called a {\em vector product} on $A$ with {\em shift operators} $L_i$, $i \geq 1$.\\
\end{definition}

\begin{definition}[Vector Sliding Window $*$-Product]
Let $*$ be a vector product on $A$ with shift operators $L_i$, $i \geq 1$. Let $a \in A$, and let $n \geq 1$ be a strictly positive integer. Then the {\em vector sliding window $*$-product} of length $n$ corresponding to the element $a \in A$ is the element $y \in A$ computed as

\begin{equation*}
y = a * (L_1(a) * (L_2(a) * ( \ldots * (L_{n-2}(a) * L_{n-1}(a)) \ldots )))
\end{equation*}
\end{definition}

\begin{remarks} \
\begin{enumerate}
  \item When we want to emphasize the $*$ notation we also say {\em $*$ is a vector $*$-product on $A$ with shift operators $L_i$}.
  \item We also call the operators $L_i$ {\em lag operators}.
  \item If $L_i \circ L_j = L_{i+j}$ for $i, j \geq 1$, then necessarily $L_i=\underbrace{L_1 \circ \ldots \circ L_1}_{i \text{ times}}=L_1^i$, and hence if $L_1(a * b)=L_1(a) * L_1(b)$, then we must also have $L_i(a * b) = L_i(a) * L_i(b)$ for all $i \geq 1$.
  \end{enumerate}
  \item The condition $L_i(a * b) = L_i(a) * L_i(b)$ says that $L_i$ is a {\em magma endomorphism} of the magma $(A, *)$.
\end{remarks}
  
\section{Examples and Constructions}
\label{sec:vector-product-examples}

\begin{example}
Assume $*\colon A \times A \rightarrow A$ and $L\colon A \rightarrow A$ satisfy $L(a * b)=L(a) * L(b)$ for all $a, b \in A$, i.e.\ $L$ is a magma endomorphism, then we may set $L_i=\underbrace{L_1 \circ \ldots \circ L_1}_{i \text{ times}}=L^i$ and with this choice of $L_i$, $*$ is a vector $*$-product with shift operator $L_i$.
\end{example}

\begin{example}
Let $\mathcal{C}$ be a category (or magmoid), and let $L\colon \mathcal{C} \rightarrow \mathcal{C}$ be a functor from $\mathcal{C}$ to $\mathcal{C}$ (i.e., $L$ is an endofunctor). Let $X$ be an object in $\mathcal{C}$. Then $\circ\colon \Endop_\mathcal{C}(X) \times \Endop_\mathcal{C}(X) \rightarrow \Endop_\mathcal{C}(X)$ is a vector $\circ$-product with shift operators $L_i=\underbrace{L_1 \circ \ldots \circ L_1}_{i \text{ times}}=L^i$. This example relates to the intuition that communication operations are functors.
\end{example}

\begin{example}
\label{example:vector-*-product-fixed-length-sequences}

Let $A$ be a set and $A^N=\overbrace{A \times \ldots \times A}^N$ be the $N$-fold cartesian product of $A$. Assume $*$ is a binary operation on $A$. A left identity in $A$ is an element $1_A$ such that $1 * a = a$ for all $a \in A$. Let
%
\begin{equation*}
A_1 = \ourcases{
A & \text{if $A$ has a left identity with respect to $*$}\\ 
A\cup \theset{1} & \text{otherwise}
}
\end{equation*}
where in the latter case $*$ is extended to $1$ by $1 * a = a * 1 = 1$ for $a\in A$, and $1\not\in A$. In this case we let $1_{A_1} = 1$. Define $*^N$ on $A_1^N$ by applying $*$ componentwise. I.e.,
\begin{equation*}
\component{i}{a *^N b} = a_i * b_i \quad \text{for $a, b \in A_1^N$, \ $i=1, \ldots, N$}
\end{equation*}

\footnote{
We use the notation $\component{i}{\hspace{1ex}}$ to indicate extraction of the $i^\text{th}$ component of a vector, array, or list, so e.g., $\component{i}{x}$ indicates the $i^\text{th}$ component of the vector $x$.
}%
\ Define shift operators $L_i$, $i \geq 1$ on $A_1^N$ by

\begin{equation*}
\component{j}{L_i(a)} = \ourcases{
    a_{j-i} & \text {if $j - i \geq 1$}\\
    1_{A_1} & \text {otherwise}
}
\end{equation*}
Then $*^N$ is a vector product on $A_1^N$ with shift operators $L_i$. Furthermore, the $i^\text{th}$ component of the vector sliding window $*$-product of length $n$ for $a = a_1, \ldots, a_N \in A^N$, is
\begin{equation*}
y_i = \ourcases{
a_i * \left(\ldots * \left(a_2 * a_1 \right) \ldots \right) & \text{if $i < n$} \\
a_i * \left(\ldots * \left(a_{i-n+2} * a_{i-n+1}\right) \ldots\right) & \text{if $i \geq n$}
}
\end{equation*}
and hence the vector sliding window $*$-product is equal to the (non-vector) sliding window $*$-product sequence.
\end{example}

\begin{example}
\label{example:vector-*-product-variable-length-sequences}

Let $A$ be a set, and $*\colon A \times A \rightarrow A$ be a binary operation on $A$. Let
\begin{equation*}
V_N(A) = \bigcup_{i=0}^N A^i 
       = \theset{\text{sequences of elements in $A$ of length $\leq N$}}
\end{equation*}
For $u \in A^p, v \in A^{q}$ with $p, q \leq N$, define
\begin{equation*}
u * v = \ourcases{
    u_1 * v_1, \ldots u_p * v_p & \text{if $p = q$} \\
    u_1, u_2, \ldots u_{p-q}, u_{p-q+1} * v_1, \ldots, u_p * v_q & \text{if $q < p$} \\
    v_1, v_2, \ldots v_{q-p}, u_1 * v_{q-p+1}, \ldots, u_p * v_q & \text{if $q > p$}
}
\end{equation*}
If $u \in A^p, p \leq N$, and $i \geq 1$, then define
\begin{equation*}
L_i(u) = \ourcases{
    u_1, \dots , u_{p-i}          & \text{if $i < p$}\\
    \text{the empty sequence ( )} & \text{if $i \geq p$}
}
\end{equation*}
Then $*$ is a vector $*$-product on $V_N(A)$, and if $a \in V_N(A)$ then the vector sliding window $*$-product corresponding to $a$ is exactly the (non-vector) sliding window $*$-product. I.e.\
\begin{equation*}
y_i = \ourcases{
    a_i * (a_{i-1} * (\ldots * (a_2 * a_1) \ldots)) 
        & \text{if $i < n$}\\
    a_i * (a_{i-1} * (\ldots *(a_{i-n+2} * a_{i-n+1)}) \ldots)) 
        & \text{if $i \geq n$}
}
\end{equation*}
where $1 \leq i \leq p$, and $p = \operatorname{length}(a)$.
\end{example}

\begin{example}
\label{example:vector-*-product-unbounded-length-sequences}

Example \ref{example:vector-*-product-variable-length-sequences} extends to the infinite union 
\begin{equation*}
V_{\infty}(A) = \bigcup_{i=0}^\infty A^i = \theset{\text{all finite sequences of elements of $A$}}
\end{equation*}
with the same definitions and results. In particular, the components of the vector sliding window $*$-product for any $a \in V_{\infty}(A)$ are precisely the components of the (non-vector) sliding window $*$-product sequence.

\end{example}

\section{Vector Sliding Window $*$-Products and Semidirect Products}
\label{sec:vector-semidirect-products}

Assume that $*\colon A \times A \rightarrow A$ is a vector $*$-product with shift operators $L_i$, $i \geq 1$. Then the function $L\colon i \mapsto L_i$ is a mapping of $\mathbb{Z}_{>0}$ into $\Endop(A)$, and hence we may form the semidirect product $\mathbb{Z}_{>0} \ltimes_L A$ whose semidirect product operation is
\begin{equation*}
\mat{i\\ a} * \mat{j\\ b} = \mat{i + j\\ a * L_i(b)}
\end{equation*}
The condition that $L_i \circ L_j = L_{i+j}$ for $i, j \geq 1$ says that $L\colon \mathbb{Z}_{>0} \rightarrow \Endop(A)$ is a magma morphism from $(\mathbb{Z}_{>0}, +)$ to $(\Endop(A), \circ)$, or equivalently that the set action $i \bullet a = L_i(a)$ is semi-associative. The condition that $L_i(a * b) = L_i(a) * L_i(b)$ says that $L$ maps $(\mathbb{Z}_{>0}, +)$ into the semigroup of magma endomorphisms of $A$, and also translates into the equation $i \bullet (a * b) = (i \bullet a) * (i \bullet b)$ which is a form of distributivity. These properties allow us to apply the results of Section~\ref{sec:semidirect-products}, on semidirect products, to vector $*$-products, and hence to sliding window $*$-products.

The following theorem describes the basic algebraic facts about vector sliding window $*$-products, and shows that they are equivalent to computing powers of elements in the semidirect product $\mathbb{Z}_{>0} \ltimes_L A$.

\begin{theorem}
\label{theorem:semidirect-product-vector-sliding-window}
\
\begin{enumerate}
\item
Assume $*\colon A \times A \rightarrow A$ is a binary operation and $L_1, L_2, \ldots \in \Endop(A)$ are functions on $A$. Then the right-folded $n^\text{th}$ power of $\smat{1\\ a}$ in the semidirect product $\mathbb{Z}_{>0} \ltimes_L A$ is
\begin{align*}
\mat{1\\ a}^{*n} & =\mat{1\\ a} *\left(\mat{1\\ a} *\left(\ldots *\left(\mat{1\\ a} *\mat{1\\ a}\right) \ldots\right)\right)\\
& = \mat{n \\ a * L_1(a * L_1(\ldots * L_1(a * L_1 a) \ldots))}
\end{align*}

\item
Assume that $*\colon A \times A \rightarrow A$ is a vector $*$-product on $A$ with shift operators $L_i$, $i \geq 1$. Then, for any $a \in A$, $n \geq 1$,
\begin{equation*}
\underbrace{a * L_1(a * L_1(\ldots * L_1(a * L_1 a) \ldots))}_\text{$n$ copies of $a$}
   = a * (L_1(a) * (L_2(a) * (\ldots * (L_{n-2}(a) * L_{n-1}(a) ) \ldots )))
\end{equation*}

\item
Assume again that $*$ is a vector $*$-product on $A$ with shift operators $L_i$, $i \geq 1$. Then the 
vector sliding window $*$-product of length $n$ corresponding to the element $a \in A$ is the second component of the $n^\text{th}$ right-folded power of $\smat{1 \\ a}$ in $\mathbb{Z}_{>0} \ltimes_L A$. I.e.,\ the right-folded $n^\text{th}$ power of $\smat{1\\ a}$ in the semidirect product $\mathbb{Z}_{>0} \ltimes_L A$ is
\begin{equation*}
\mat{1\\ a}^{*n} 
  = \mat{n \\ 
  a *\left(L_1(a) *\left(L_2(a) *\left(\ldots *\left(L_{n-2}(a) * L_{n-1}(a)\right) \ldots\right)\right)\right)
}
\end{equation*}
\end{enumerate}
\end{theorem}
\begin{proof} 
This follows directly from Theorem~\ref{theorem:semidirect-product-semi-associativity} applied to the semidirect product $\mathbb{Z}_{>0} \ltimes_L A$ where $L\colon \mathbb{Z}_{>0} \rightarrow \Endop(A)\colon i \mapsto L_i$.
\end{proof}
%




%

\begin{remark} 
At no place in the statement or proof of Theorem~\ref{theorem:semidirect-product-vector-sliding-window} do we assume associativity of $*$. However, if $*$ is an associative vector $*$ product on $A$ with shift operators $L_i$, $i \geq 1$, then by by Lemma~\ref{lemma:semidirect-product-semigroup-function-form} the semidirect product $\mathbb{Z}_{>0} \ltimes_L A$ is also associative.
\end{remark}

\section{Algorithms for Vector Sliding Window $*$-Products}
\label{sec:algorithms-for-vector-sliding-window-*-products-section}

It is well known (see e.g., \cite{Knuth1998} Section 4.6.3), that an $n^\text{th}$ power in a semigroup can be computed in at most $2\left\lfloor\log_2 n\right\rfloor$  $*$-operations, using binary exponentiation. Thus, Lemma~\ref{lemma:semidirect-product-semigroup-function-form} and Theorem~\ref{theorem:semidirect-product-vector-sliding-window} together give us an algorithm for computing vector sliding window-products using at most $2\left\lfloor\log_2 n\right\rfloor$ vector $*$ operations. This therefore gives us a parallel algorithm for computing sliding window  $*$-products, with depth $\leq 2\left\lfloor\log_2 n\right\rfloor$, under the assumption that $*$ is associative. Note that we will describe more efficient algorithms in later sections.

\begin{algorithm}
\label{algorithm:vector-sliding-window-*-product}

Assume $*\colon A \times A \rightarrow A$ is a vector $*$-product with shift operators $L_i$, $i \geq 1$, and assume $*$ is associative. Then the vector sliding window $*$-product of length $n \geq 1$ for $a \in A$ can be computed as follows.
\begin{description}
\item[Step 1] 
Form $z = \smat{1\\ a} \in \mathbb{Z}_{>0} \ltimes_L A$

\item[Step 2] Calculate the binary expansion of $n$
\begin{equation*}
    n = p_1 + \ldots + p_l 
\end{equation*}
where $p_1 < \ldots < p_l$, and $p_1, \ldots, p_l$ are distinct powers of $2$. 

\item[Step 3] Successively square $z$ until $z^{p_l}$ is reached
\begin{align*}
z_1 & =z \\
z_2 & =z_1 * z_1 \\
z_{4} & =z_2 * z_2 \\
\vdots & \\
z_{p_l} & = z_{p_l / 2} * z_{p_l / 2}
\end{align*}

\item[Step 4] Compute
\begin{align*}
& q_1=z_{p_1} \\
& q_2=q_1 * z_{p_2} \\
& \vdots \\
& q_l=q_{l-1} * z_{p_l - 1}
\end{align*}

\item[Step 5] Extract the second component of the $n^\text{th}$ power $\smat{1\\ a}^{*n}$ just computed.
\end{description}
\end{algorithm}

\begin{theorem}
\label{theorem:vector-sliding-window-*-product}
Algorithm~\ref{algorithm:vector-sliding-window-*-product} computes the vector sliding window $*$-product in at most $2\left\lfloor\log_2 n\right\rfloor$ (vector) $*$ operations.
\end{theorem}
\begin{proof}
First note that $\mathbb{Z}_{>0} \ltimes_L A$ is a semigroup by Lemma~\ref{lemma:semidirect-product-semigroup-function-form}, and that steps 2{--}4 compute the power $\smat{1\\ a}^n$ in this semigroup using binary exponentiation.
Step 3 uses $\left\lfloor\log_2 n\right\rfloor$ (vector) $*$ operations, and 
Step 4 uses $(\text{\# of nonzero binary digits in $n$})-1$ (vector) $*$ operations.
\end{proof}
\begin{remarks} \
\begin{enumerate}
  \item Steps 3 and 4 can be combined, as seen in the implementation of {\tt binary\_exponentiate} in Chapter~\ref{chapter:vector-algorithms-guide}. This saves memory.
  \item There are several obvious variants of Algorithm \ref{algorithm:vector-sliding-window-*-product} that are useful in practice and which correspond to different ways of ordering the product computed in Step 4. To illustrate consider the case $n=7$

{ 
\renewcommand{\arraystretch}{1.5}
\begin{center}
\begin{tabular}{l|l|l|l}
   digit extractions & fold  & $z^7$ & Step 4\\
    \hline 
up    &  left & $(z * z^2)*z^4 $   & $q_1 = z_{p_1}$, $q_i = q_{i-1} * z_{p_i}$\\

up    & right & $z^4 * (z^2 * z)$ & $q_1 = z_{p_1}$, $q_i = z_{p_i} * q_{i-1}$ \\
 
down  &  left & $(z^4 * z^2)*z$    & $q_1 = z_{p_l}$, $q_i = q_{i-1} * z_{p_{l-i+1}}$ \\
 
down  & right & $z * (z^2 * z^4)$  & $q_1 = z_{p_l}$, $q_i = z_{p_{l-i+1}} * q_{i-1}$ \\
\end{tabular}
\end{center}
} 

\noindent These approaches to the computation of the $n^\text{th}$ power are all variants of Algorithm A in \cite{Knuth1998} Section 4.6.3. The four algorithms corresponding to up/down, left/right are useful in different situations depending on the $L$ and $*$ operators used. Clearly the `up' algorithms require less working space, as Step 4 can be combined with Step 3 so as to consume the $z_{p_i}$ as soon as they are produced. These `up' variants only require enough memory to store two elements of $A$ plus bookkeeping data. On the other hand, if we write out the expressions computed in the second component of $z^n=\smat{1\\ a}^n$, in terms of $a$, $*$, and $L_1$, then we obtain the following for $n=7$.

{ 
\renewcommand{\arraystretch}{1.5}
\begin{center}
\begin{tabular}{l|l|l}      
digit extractions & fold & second component of $z^7$\\
    \hline
    up   & left  & $(a * L_1(a * L_1 a)) * L_1((a * L_1 a) * L_1(a * L_1 a))$  \\
    up   & right & $((a * L_1 a) * L_1(a * L_1 a)) * L_1((a * L_1 a) * L_1 a)$ \\
    down & left  & $(((a * L_1 a) * L_1(a * L_1 a)) * L_1(a * L_1 a)) * L_1 a$ \\
    down & right & $a * L_1((a * L_1 a) * L_1((a * L_1 a) * L_1(a * L_1 a)))$  \\
\end{tabular}
\end{center}
} 

In these expressions $L_i$ often represents movement of data, and for a database system with data sharing, more deeply nested $L_i$ operators can cause additional retrievals. In other words, the depth of the $L_i$ operators in these expressions matters for some applications, whereas for other applications it does not. The maximum depth of the expressions depends on whether $*$ operations are counted, $L_i$ operations, or both.

{ 
\renewcommand{\arraystretch}{1.5}
\begin{center}
\begin{tabular}{ll|c|c|c}
\multicolumn{2}{l|}{variant} & $*$ depth & L depth &  combined $*$, $L$ depth \\
\hline
up   & left & 3 & 3 & 6 \\
up   & right & 3 & 2 & 5 \\
down & left  & 4 & 2 & 6 \\
down & right & 4 & 4 & 8 \\
\end{tabular}
\end{center}
} 

\item 
An equivalent approach is to work with operators $*_i$ defined by $a *_i b=a * L_i b$. These are nonassociative but satisfy $a *_i(b *_j c)=(a *_i b) *_{i+j} c$. One can define $a^{* n}=a *_1\left(a *_1\left(\ldots *_1(a *_1 a) \ldots\right)\right)$ which also equals $(\ldots ((a *_1 a) *_2 a) *_3 \ldots) *_{n-1} a$, and it follows that $a^{* n}=a^{*i} *_i a^{* j}$ when $i+j=n$. However, if we now define $\smat{i\\ a} * \smat {j\\ b} = \smat{i+j\\  a *_i b}$ we get back the semidirect product again, and the first component of the semidirect product keeps track of which operator $*_i$ to use.
  
\item 
It is interesting to note that we can `flip' the order of the semidirect product operation to get a different algorithm, but that flipping this operation by replacing it by its opposite does not change the final result computed. This results from the 
observation that for any associative operation $*$, we have $z^{* n}=z^{*_\text{op} n}$, where $z_1 *_\text{op} z_2 = z_2 * z_1$ is the opposite operation.

This invariance to the use of the opposite operation applies to the semidirect product operator used in the algorithm, but not to the original $*$ operation on $A$, 
as if $*$ is noncommutative 
then in general $a * L_1 a * L_2 a * \ldots * L_{n-1} a$ will not equal $L_{n-1} a * L_{n-2} a * \ldots * L_1 a * a$. 
If we use Example \ref{example:vector-*-product-fixed-length-sequences} or Example \ref{example:vector-*-product-variable-length-sequences}, then computing in $A_1^N$ or $V_N(A)$, with the opposite $*$ operation will give $((\ldots(a_{i-n+1} * a_{i-n+2}) * \ldots) * a_{i-1}) * a_i$ instead of $a_i *(a_{i-1} *(\ldots *(a_{i-n+2} * a_{i-n+1}) \ldots))$. 
In contrast, using the opposite semidirect product operation does not change the final result computed, but only changes the method of computation and the intermediate results. It is equivalent to choosing a different algorithm to compute the power $\smat{1\\ a}^{*n}$.
\end{enumerate}
\end{remarks}

\noindent Algorithm \ref{algorithm:vector-sliding-window-*-product} is not the most efficient vector algorithm possible in the number of vector operations it uses or the parallel depth. This is because there are more efficient ways to exponentiate an element of a semigroup than (sequential) binary exponentiation. To prepare for this, we first state the obvious generalization of Algorithm \ref{algorithm:vector-sliding-window-*-product}.

\begin{algorithm}
\label{algorithm:general-vector-sliding-window-*-product}
Assume $*\colon A \times A \rightarrow A$ is a vector $*$-product with shift operators $L_i$, $i \geq 1$, and assume $*$ is associative.
Assume further, that $\exponentiate(z, n)$ is a procedure for computing strictly positive powers of an element of a given semigroup, that returns $z^n$ in the semigroup. 
Then the vector sliding window $*$-product of length $n \geq 1$ for $a \in A$ can be computed as follows.

\begin{description}
\item[Step 1] 
Form $z= \smat{1\\ a} \in \mathbb{Z}_{>0} \ltimes_L A$

\item[Step 2] Call $\exponentiate(z, n)$ to compute $z^n=z^{* n}$ in the semigroup $\mathbb{Z}_{>0} \ltimes_L A$.

\item[Step 3] Extract the second component of the $n^\text{th}$ power $\smat{1\\ a}^{*n}$ just computed.
\end{description}
\end{algorithm} 

\noindent We now take a quick detour into the theory of exponentiation in semigroups.

%% file: htcams-arxiv-ch12-exponentiation-in-semigroups.tex
\chapter{Exponentiation in Semigroups}
\label{chapter:exponentiation-in-semigroups}

In 1894 H.\ Dellac (\cite{Dellac1894} p.\ 20, question 49] asked the question `What is the minimum number of multiplications to perform to raise the number $A$ to the power $m$?'.%
\footnote{`Quel est le nombre minimum de multiplications à effectuer pour élever le nombre $A$ à la puissance $m$?'} 
To see that this is not a trivial problem, consider computing $a^{15}$. We have already seen 4 methods for this (all were variants of binary exponentiation), each of which take 6 multiplications, e.g.,
\begin{equation*}
a^{15}=a^8 *\left(a^4 *\left(a^2 * a\right)\right)
\end{equation*}
where $a^2, a^4, a^8$ are computed by successive squaring. Essentially we are computing a sequence of powers
\begin{equation*}
a, a^2, a^3, a^4, a^7, a^8, a^{15}
\end{equation*}
where we have arranged the powers computed in ascending order.%
\footnote{This ordering allows the computation to proceed with working space of just two variables to compute the successive powers.}
This is not the least number of multiplications required to compute $a^{15}$, however, as the following sequence uses only 5 multiplications.
\begin{equation*}
a, a^2, a^3, a^6, a^{12}, a^{15}
\end{equation*}
where these are computed as $a^2 = a * a$, $a^3=a^2 * a, a^6=a^3 * a^3, a^{12}=a^6 * a^6$, and $a^{15}=a^{12} * a^3$.

Dellac's question was partially answered in 1894 by E.\ de Jonquiéres (\cite{deJonquieres1894} pp.\ 162-164, response 49), and research has continued up to the present day. Notable advances are the introduction of {\em addition chains} by Scholz \cite{Scholz1937}, an asymptotically optimal solution by Brauer \cite{Brauer1939}, a proof of asymptotic optimality for almost all $n$ by Erd\"{o}s \cite{Erdos1960}, a detailed survey and exposition with new results by Knuth (first published in 1968, a later edition is \cite{Knuth1998}), improvements to Brauer's method by Thurber \cite{Thurber1973b}, algorithms for computing with more than one desired exponent by Yao \cite{Yao1976}, and optimal solutions for $n$ up to $2^{32}$ by Clift \cite{Clift2011}.%
\footnote{These calculations have also been extended to $n \leq 2^{39}$ by Clift as recorded in \cite{Flammenkamp2022}.}
There are surveys of the theory in \cite{Knuth1998} (Volume II of {\em The Art of Computer Programming}), \cite{Doche2005} (Chapter 9 of {\em Handbook of Elliptic and Hyperelliptic Cryptography}), \cite{Cohen1996} ({\em A Course in Computational Algebraic Number Theory}), and \cite{Bernstein2002}.

Interest in the problem has come from cryptography (see e.g., \cite{Doche2005}, \cite{Gordon1998}), but we find that the techniques are perhaps even better suited to the use case of sliding window calculations, as the computation of a vector sliding window $*$-product involves operations that may be both expensive at an individual level (e.g., matrix multiplication), but also are vectorized, and the length $N$ of the vector may be long. So the overhead of bookkeeping (i.e., keeping track of which powers are combined 
to form new powers) or searching for the optimal algorithm (e.g., choosing the optimal base in Brauer's method or Thurber's method), may be small compared to the cost of the vector $*$-products themselves.

\section{Addition Chains}

\begin{definition}[Addition Chain]
\label{definition:addition-chain}
Let $n$ be a strictly positive integer. An {\em addition chain} for $n$ is a finite sequence of positive integers\\
\begin{equation*}
e_0 = 1, \ldots, e_l=n
\end{equation*} 
such that for all $i$ with $1 \leq i \leq l$,
\begin{equation*}
e_i = e_j + e_k \quad  \text{for some $j$, $k$ with $0 \leq j$, $k<i$.}
\end{equation*}
The integer $l$ is called the length of the addition chain. 
\end{definition}
Given an addition chain and an element of a semigroup, there is a unique sequence of powers to which it corresponds.

\begin{definition}[Power Sequence]

Assume $A$ is a semigroup, $a \in A$, and $e_0 = 1, \ldots, e_l = n$ is an addition chain for the strictly positive integer $n$. Then the {\em power sequence} of $a$ 
%
%
%
%
corresponding to the addition chain is
\begin{equation*}
a^{e_1}, a^{e_2}, \ldots, a^{e_l}.
\end{equation*}
\end{definition}

\noindent The idea is that an addition chain encodes a method for computing $a^n$. E.g., the addition chain $1$, $2$, $3$, $6$, $12$, $15$ corresponds to $a^2=a$, $a^3=a^2 * a$, $a^6=a^3 * a^3$, $a^{12}=a^6 * a^6$, $a^{15}=a^{12} * a^3$.

Many authors (e.g., \cite{Brauer1939}, \cite{Clift2011}) require that addition chains be increasing or non-decreasing, and Knuth \cite{Knuth1998} requires that $k \leq j$ in Definition~\ref{definition:addition-chain}. We give here several reasons for dropping these conditions in our definition.

\begin{enumerate}
  \item We wish to expand to the noncommutative setting, so order of operations matters in specifying what calculations are actually performed. Even though the end result $a^{e_j} * a^{e_k}$ is the same as $a^{e_k} * a^{e_j}$, the computation itself is different. 

  \item Although the bracketing of an expression for a power will not affect the value of that computation in a semigroup, differently bracketed expressions for a power do correspond to different computations, and have different intermediate results. When we peel back the curtain behind the computation of $*$ we may see different complexities, different patterns of memory access, and different patterns of data communication.
  
  \item When computing with nonassociative operations, the order of bracketing affects the result. It is sometimes useful to use nonassociative operations to represent associative operations, and this leads to exponentiation calculations where the operator used is nonassociative. See, for example, Example~\ref{example:rofc-examples-5}.
\end{enumerate}

\bigskip

An addition chain does not unambiguously determine a procedure for computing an $n^\text{th}$ power. There is firstly the question of order $a^{e_j} * a^{e_k}$ or $a^{e_k} * a^{e_j}$. Beyond this, as noted by Clift \cite{Clift2011}, there may be multiple $j$, $k$, for which $e_j + e_k = e_i$. For example, consider the addition chain $1$, $2$, $3$, $4$, $7$, which is a minimal length chain for 7. Since $4 = 2 + 2 = 1 + 3$, specifying only the chain does not uniquely determine which powers need to be multiplied at each step.  To account for this ambiguity we make the following definition, approximately following Clift \cite{Clift2011}.

\begin{definition}[Formal Addition Chain]
\label{definition:formal-addition-chain}  

An {\em addition chain index sequence} is a finite sequence of integers $(i_1, j_1), \ldots, (i_l, j_l)$ with $0 \leq i_k, j_k < k$ for $1 \leq k \leq l$. 
A {\em formal addition chain} is an addition chain $1=e_0, e_1, \ldots, e_l=n$ together with an addition chain index sequence $(i_1, j_1), \ldots, (i_l, j_l)$, such that
\begin{align*}
& e_0 = 1 \\
& e_k = e_{i_k} + e_{j_k} \quad \text{for $1 \leq k \leq l$}
\end{align*}
\end{definition}
It is not difficult to see that an addition chain index sequence uniquely determines a formal addition chain. The reason for the definition is that they also uniquely determine algorithms for computing powers in a semigroup.

\begin{algorithm}[Formal Addition Chain Algorithm]
\label{algorithm:formal-addition-chain}
Assume $A$ is a semigroup, $a \in A$, and $(i_1, j_1), \ldots$, $(i_l, j_l)$ is an addition chain index sequence with corresponding addition chain $1=e_0, \ldots, e_l=n$. Then $a^n$ may be computed as follows:
\begin{align*}
q_0 & =a \\
q_1 & =q_{i_1} * q_{j_1} \\
q_2 & =q_{i_2} * q_{j_2} \\
\vdots & \\
q_l & =q_{i_l} * q_{j_l} = a^n
\end{align*}
\end{algorithm}

\begin{remarks} \ 
\begin{enumerate} 
\item Although it is stated for a semigroup, Algorithm \ref{algorithm:formal-addition-chain} may be applied to compute a power for a nonassociative operation, i.e., it may used to compute powers in a magma which is not a semigroup. In this case the result will depend not only on $n$, but on the entire addition chain index sequence used. I.e., different addition chain index sequences for the same $n$, and even corresponding to the same addition chain will in general give different results when $*$ is nonassociative.

In Chapter~\ref{chapter:vector-windowed-recurrences} we will apply Algorithm~\ref{algorithm:formal-addition-chain} in the nonassociative case to compute vector windowed recurrences.

\item All the commonly used methods for exponentiation in semigroups correspond to Algorithm \ref{algorithm:formal-addition-chain} for some choice of index sequence depending on the method. This includes Brauer's method, Thurber's method, and binary exponentiation. Collectively we call these {\em addition chain methods for exponentiation}.

\item The memory usage of Algorithm \ref{algorithm:formal-addition-chain} depends on the particular addition chain index sequence, though for a given formal addition chain it is easy to calculate the required memory use and avoid unnecessary storage.

\end{enumerate}
\end{remarks}

\section{Brauer's Algorithm}
\label{sec:brauers-algorithm}

In 1939 Brauer \cite{Brauer1939} demonstrated an addition chain of length $(k + 1) l + 2^k - 2$, where $k \geq 1$ is a strictly positive integer, and $l$ satisfies $(2^k)^l \leq n<2^{k(l+1)}$. By choosing $k=\lfloor(1-\varepsilon) \log_2 \log_2(n+2)\rfloor + 1$, we obtain%
%
\footnote{
Note that the conclusion of this chain of inequalities also holds for $n=1$.
}

\begin{align*}
(k + 1)l + 2^k - 2 
& \leq \left\lfloor \log_2 n\right\rfloor
    \left(1 + \frac{1}{k} \right) + 2^k - 2 \\
& \leq \left( \log_2 n \right)
    \left(1 + \frac{1}{(1-\varepsilon) \log_2 \log_2(n + 2)}
            + \frac{1}{\log_2 n} 2^{(1-\varepsilon) \log_2 \log_2(n + 2) + 1} 
    \right) \\
& \leq \left( \log_2 n \right) 
    \left(1 + \frac{1}{\log_2 \log_2(n+2)}
            + \frac{\varepsilon}{1-\varepsilon} \frac{1}{\log_2 \log_2(n + 2)}
            + 2\frac{\log_2(n+2)}{\log_2 n}
                \left(\log_2(n+2)\right)^{-\varepsilon}\right) \\
& = \left( \log_2 n \right) 
    \left(1 + \frac{1}{\log_2 \log_2(n+2)}
            + \littleoh{\frac{1}{\log_2 \log_2(n+2)}}\right)
\end{align*}
Erd\"{o}s \cite{Erdos1960} later proved this was asymptotically optimal for `almost all $n$'. Together with improvements suggested by Knuth \cite{Knuth1998} and Thurber \cite{Thurber1973b} (also see \cite{Bernstein2002}), Brauer's method results in the algorithm we now describe.

\medskip

In this section and the following section on Thurber's algorithm, we assume that $*$ is a semigroup operation and we adopt the notation
\begin{align*}   
\text{\tt n <{<}  k} \ & = \ 2^k n & \text {left bitwise shift}\\
\text{\tt n >{>} k} \ & = \ \left\lfloor \frac{n}{2^k} \right\rfloor & \text{right bitwise shift}\\
      \text{\tt \&} \ & = \ \text{bitwise and}
\end{align*}

Our implementation of Brauer's algorithm will require three helper procedures (subroutines), which are {\tt repeated\_square}, {\tt extract\_powers\_of\_two}, and {\tt digits\_base\_2\_k}. As before we use Landin's off-side rule \cite{Landin1966} to indicate the end of code blocks.

%
%
\begin{alltt}
repeated_square(op, z, k): op represents a binary operation which is passed in
    for i = 1 to k
        z = op(z, z)
    return z

extract_powers_of_two(n): Finds \(j,b\) such that \(n=2\sp{j}b\) with \(b\) odd or \(b=0\)
    j = 0
    if n = 0
        return (j, n)
    while n is even                                                   Test with n & 1 = 0         
        j = j + 1
        n = n >> 1
    return (j, n)
    
digits_base_2_k(n, k): Computes the digits of n base \(2\sp{k}\)
    mask = (1 << k) - 1                                                         i.e. \(2\sp{k}-1\)
    digits = an empty array
    while n > 0
        d = n & mask
        append d to digits
        n = n >> k
    return digits

\end{alltt}

\noindent We can now give an algorithm for Brauer's method of exponentiation, 

%
%
\begin{alltt}
brauer_exponentiate(op, x, n, k, flip):
    digits = digits_base_2_k(n, k)                          Least significant digit first
    split_digits = [extract_powers_of_two(d) for d in digits]
    
    define eop(y, z):
        return (y[1] + z[1], (op(z[2], y[2]) if flip else op(y[2], z[2])))
        
    max_b = max(b for (j, b) in split_digits)
    n_precompute = (max_b + 1) >> 1
    precomputed = array of length n_precompute
    precomputed[1] = (1, x)
    if n_precompute > 1
        x_squared = (2, op(x, x))
        for i in 2,...,n_precompute
            precomputed[i] = eop(precomputed[i - 1], x_squared)
            
    define repeated_square_no_duplicates(z, j):
        if n_precompute > 1 and z[1] = 1 and j > 0
            return repeated_square(eop, x_squared, j - 1)
        return repeated_square(eop, z, j)

    i = length(digits)
    j, b = split_digits[i]
    z = repeated_square_no_duplicates(precomputed[(b >> 1) + 1], j)
    i = i - 1
    while i > 0
        j, b = split_digits[i]
        if b = 0
            z = repeated_square_no_duplicates(z, k)
        else
            exponent = b + z[1] \(\times\) (1 << (k - j))              \(\,\times\) = integer multiplication
            if exponent \(\leq\) max_b
                z = precomputed[(exponent >> 1) + 1]
            else
                z = repeated_square_no_duplicates(z, k - j)
                z = eop(z, precomputed[(b >> 1) + 1])
            z = repeated_square(eop, z, j)
        i = i - 1
        
    return z[2]
\end{alltt}

\section{Thurber's Algorithm}
\label{sec:thurbers-algorithm}

In 1973 Thurber \cite{Thurber1973b} published a related algorithm, commonly known as the {\em sliding window algorithm}.%
\footnote{
This terminology has no relation to the sliding window algorithms discussed in this monograph.
}
In addition to the {\tt repeated\_square} procedure, Thurber's algorithm also depends on a procedure we call {\tt thurber\_windows}.

%
%
\begin{alltt}
thurber_windows(n, k):
    windows = empty array of triples        
    i = (number of binary digits in n) - 1                                    I.e. \(\lfloor\log\sb2{n}\rfloor\)
    bit = 1 << i
    while i \(\geq\) 0
        start = max(i - k + 1, 0)
        start_bit = 1 << start
        while start_bit & n = 0
            start = start + 1
            start_bit = start_bit << 1
        width = i - start + 1
        value = ((1 << width) - 1) & (n >> start)
        i = i - width
        bit = bit >> width
        gap = 0
        while (n & bit = 0) and i \(\geq\) 0
            i = i - 1
            gap = gap + 1
            bit = bit >> 1
        append the triple (width, value, gap) to windows
    return windows
\end{alltt}

The {\tt thurber\_windows} procedure dices the binary digits of $n$ into strings of digits of length $\leq k$ which correspond to odd numbers $\leq 2^k-1$, and are separated by strings of zeros. These are called the {\em windows}. For example, with $n = 2630 = 101001000110_2$, $k=3$, the {\tt thurber\_windows} procedure splits the binary digits as 
$\ourunderbracket{101} \mid 00 \mid \ourunderbracket{1} \mid 000\mid \ourunderbracket{11}\mid0$. The {\tt thurber\_windows} procedure then returns for each window the triple ({\em width, value, gap}) where {\em width} is the number of digits in the window, {\em value} is the value of the digits as a number, and {\em gap} is the number of consecutive zeros following the window. For $n=2630, k=3$, this is $(3,5,2)$, $(1,1,3)$, $(2,3,1)$ as per the windows
$\ourunderbracket{\ourunderbracket{\overbrace{101}^3}_{=5} \mid \overbrace{00}^2} \left| \ourunderbracket{\ourunderbracket{\overbrace{1}^1}_{=1} \mid \overbrace{000}^3}\right| \ourunderbracket{\ourunderbracket{\overbrace{11}^2}_{=3}\mid\overbrace{0}^1}$.

\medskip
%
\begin{alltt}
thurber_exponentiate(op, x, n, k, flip):
    windows = thurber_windows(n, k)
    max_value = max(value for (width, value, gap) in windows)
    n_precompute = (max_value + 1) >> 1
    precomputed = array of length n_precompute
    precomputed[1] = x
    if n_precompute > 1
        x_squared = op(x, x)
        for i in 2,...,n_precompute
            precomputed[i] = (op(x_squared, precomputed[i-1]) if flip else 
                              op(precomputed[i-1], x_squared))
    window = windows[1]
    width = window[1], value = window[2], gap = window[3]
    if value = 1 and gap > 0 and n_precompute > 1
        z = repeated_square(op, x_squared, gap - 1)
    else
        z = repeated_square(op, precomputed[(value >> 1) + 1], gap)
    for i in 2,...,length(windows)
        window = windows[i]
        width = window[1], value = window[2], gap = window[3]
        z = repeated_square(op, z, width)
        z = (op(precomputed[(value >> 1) + 1], z) if flip else
             op(z, precomputed[(value >> 1) + 1]))
        z = repeated_square(op, z, gap)
    return z
\end{alltt}

\subsection{Notes on Brauer's algorithm and Thurber's algorithm}

\begin{enumerate}
\item 
  To compute a power $x^n$, choose an integer $k \geq 1$, and choose {\tt flip=true} or {\tt flip=false}, and set $\operatorname{op} = * = \text{the semigroup operation}$. Then
\begin{align*}
x^n & = \text{\tt brauer\_exponentiate(op, x, n, k, flip)}\\
    & = \text{\tt thurber\_exponentiate(op, x, n, k, flip)}
\end{align*}

\item 
For many computing languages and systems, the procedure {\tt digits\_base\_2\_k}, which computes the digits of $n$ base $2^k$ from least significant first to most significant last, is a built-in function or a standard library procedure. The implementation given above is only included to unambiguously demonstrate the desired behavior.
  
\item 
The pseudo-code given for both Brauer's algorithm and Thurber's algorithm is adapted from \cite{Doche2005} and \cite{Cohen1996}, where we have added logic to avoid unnecessary or duplicate $*$ operations. For exponentiation of numbers, as in cryptography applications, the cost of this extra logic would need to be balanced against the savings of avoiding the extra $*$ operations. In our use case of vector sliding window $*$-products, the $*$ operations ({\tt{op}} in the code) represent potentially expensive vector operations, so it is more useful to include the de-duplication logic.
  
\item 
The case $k=1$ for both Brauer's and Thurber's algorithm is equivalent to the method for exponentiation which Knuth \cite{Knuth1998} calls the `S-and-X binary method', also known as the `left-to-right binary method' or the `square and multiply' method \cite{Doche2005}.
  
\item 
We have included a {\tt flip} argument to allow computing powers using the opposite operator, without having to define and pass in the opposite operator explicitly.

\end{enumerate}

\section{Choosing $k$ in the Algorithms of Brauer and Thurber}

Both Brauer's algorithm and Thurber's algorithm have a parameter, $k$, which must be chosen. The choice $k=1$ corresponds to the square and multiply (S-and-X) method of binary exponentiation. The standard advice \cite{Erdos1960} \cite{Cohen1996} \cite{Doche2005} is to choose $k$ to be close to $\log_2 \log n$, and this works well for large $n$, on average. For example Doche \cite{Doche2005} gives the following table for Brauer's method.

{ 
\renewcommand{\arraystretch}{1.5}
\begin{center}
\begin{tabular}{c|c|c|c|c|c}
$k$ & 1 & 2 & 3 & 4 & 5 \\
\hline
\#binary digits of $n$ & $1$--$9$ & $10$--$25$ & $26$--$70$ & $70$--$197$ & $197$--$539$ \\
\end{tabular}
\end{center}
} 

\noindent In our situation, where the operation $*$ is expensive, it is helpful to choose $k$ with more care. It is easy to write routines to count the $*$ operations performed by both the Brauer and the Thurber algorithms, and to search for the optimal $k$ for each $n$. When we compute the optimal $k$ for $n$ up to $n=10^{10}$, we find the following.

{ 
\renewcommand{\arraystretch}{1.25}
\begin{samepage}
\begin{center}
Brauer's method $n$ for first (smallest) occurrence of optimal $k$, for $n$ up to $10^{10}$    
\end{center}
{
\noindent \indent 
\begin{tabular}{c|c|c|c|c|c|c|c|c|c|c|c|c}
$k$ & 1 & 2 & 3 & 4 & 5 & 7 & 8 & 9 & 11 & 13 & 17 & 19 \\
\hline
first $n$ & 1 & 15 & 30 & 83 & 120 & 480 & 4832 & 5984 & 7680 & 30720 & 491520 & 1966080
\end{tabular}

\nobreak \medskip
\indent
\begin{tabular}{l|l|l|l}
$k$ & 23 & 29 & 31 \\
\hline
first $n$ & 31457280 & 2013265920 & 8053063680
\end{tabular}
} 
\end{samepage}
} 

\medskip
{ 
\renewcommand{\arraystretch}{1.25}
\begin{samepage}
\begin{center}
Thurber's method $n$ for first (smallest) occurrence of optimal $k$, for $n$ up to $10^{10}$
\end{center}
{
\noindent \indent 
\begin{tabular}{c|c|c|c|c|c}
$k$ & 1 & 2 & 3 & 4 & 5 \\
\hline
first $n$ & 1 & 15 & 23 & 151 & 9413609 \\
\end{tabular}
} 
\end{samepage}
} 

\bigskip

\noindent So we see that for Brauer's method larger optimal $k$ occur much earlier than the rule of thumb would suggest, and also the optimal $k$ for Thurber's method is $\leq 5$ for $n \leq 10^{10}$.

The results of Brauer and Erd\"{o}s show that for large $n$ the optimal choice of $k$ for Brauer's method is approximately $\log_2 \log n$, and that this yields a chain close to optimal over all addition chains, except for $n$ in a set of density (asymptotically) $0$. If we consider $n$ from $10^9+1$ through $10^9+10^6$, then the best $k$ for Brauer's method is $k=3$ in 988750 cases ($98.9\%$) and $k=2$ in 11250 cases ($1.1\%$). For smaller $n$, however, the optimal choice of $k$ is more evenly distributed between different values of $k$. In the following tables 0 means `identically 0'. 

\medskip


{ 
\renewcommand{\arraystretch}{1.25}
\begin{samepage}
\begin{center}
Brauer's algorithm \% of $n$ for which $k$ is the smallest best $k$
\end{center}
{\footnotesize \
\begin{tabular}{l|cccccccccccccc}
$n$ range \textbackslash\ $k$ & $1$ & $2$ & $3$ & $4$ & $5$ & $6$ & $7$ & $8$ & $9$ & $10$ & $11$ & $13$ & $17$ & $19$ \\
\hline
$1$--$10^3$ & $42.5$ & $31.2$ & $19.3$ & $4.1$ & $2.3$ & $0.3$ & $0.3$ & $0$ & $0$ & $0$ & $0$ & $0$ & $0$ & $0$ \\
$1$--$10^4$& $28.6$ & $36.1$ & $22.4$ & $8.0$ & $4.6$ & $0.14$ & $0.14$ & $<{\!}0.1$ & $<{\!}0.1$ & $0$ & $<{\!}0.1$ & $0$ & $0$ & $0$ \\
$1$--$10^5$ & $15.9$ & $39.3$ & $28.1$ & $13.1$ & $2.1$ & $1.2$ & $0.3$ & $<{\!}0.1$ & $<{\!}0.1$ & $0$ & $<{\!}0.1$ & $<0.1 $ & $0$ & $0$ \\
$1$--$10^6$ & $7.0$ & $38.1$ & $32.3$ & $17.6$ & $4.5$ & $0.3$ & $0.2$ & $ <0.1$ & $<{\!}0.1$ & $0$ & $ <0.1$ & $<{\!}0.1$ & $<{\!}0.1$ & $0$  \\
$1$--$10^7$ & $4.6$ & $36.5$ & $37.9$ & $16.3$ & $3.4$ & $1.1$ & $<{\!}0.1$ & $<{\!}0.1$ & $<{\!}0.1$ & $<{\!}0.1$ & $<{\!}0.1$ & $<0.1 $ & $<{\!}0.1$ & $<{\!}0.1$\\
\end{tabular}
} 
\end{samepage}
} 

\bigskip

{ 
\renewcommand{\arraystretch}{1.25}
\begin{samepage}
\begin{center}
Thurber's algorithm \% of $n$ for which $k$ is the smallest best $k$
\end{center}
{\footnotesize \ 
\begin{tabular}{l|cccccc}
$n$ range \textbackslash\ $k$ & $1$ & $2$ & $3$ & $4$ & $5$ & $6$ \\
\hline
$1$--$10^3$ & $37.9$ & $41.6$ & $19.8$ & $0.7$ & $0$         & $0$ \\
$1$--$10^4$ & $24.1$ & $49.0$ & $23.2$ & $3.7$ & $0$         & $0$ \\
$1$--$10^5$ & $11.0$ & $47.7$ & $36.8$ & $4.6$ & $0$         & $0$ \\
$1$--$10^6$ & $4.8$  & $49.2$ & $41.5$ & $4.5$ & $0$         & $0$ \\
$1$--$10^7$ & $2.8$  & $46.5$ & $44.2$ & $6.6$ & $<{\!}0.1$ & $0$ \\
\end{tabular}
} 
\end{samepage}
} 
\bigskip

\noindent Thus for Brauer's and Thurber's method, choosing optimal rather than `rule of thumb' $k$ leads to fewer operations in more than half of cases for $1 \leq n \leq 10^7$. This also holds over the range $1 \leq n \leq 2^{25}$, as shown by the following tables. 

{ 
\renewcommand{\arraystretch}{1.25}
\begin{samepage}
\begin{center}
Brauer's algorithm \% of $n$ for which $k$ is the smallest best $k$
\end{center}
{
\noindent \indent
\begin{tabular}{l|cccccc}
$n$ range \textbackslash\ $k$ & $1$ & $2$ & $3$ & $4$ & $5$ & $>5$ \\
\hline
$1$--$2^9$ & $46.5$ & $24.8$ & $23.0$ & $3.5$ & $2.0$ & $0.2$ \\
$(2^9+1)$--$2^{25}$ & $2.6$ & $31.9$ & $45.2$ & $15.4$ & $4.2$ & $0.9$ \\
\end{tabular}
} 
\end{samepage}
} 

\medskip

{ 
\renewcommand{\arraystretch}{1.25}
\begin{samepage}
\begin{center}
Thurber's algorithm \% of $n$ for which $k$ is the smallest best $k$
\end{center}
\noindent \indent
{ 
\begin{tabular}{l|cccccc}
$n$ range  \textbackslash\ $k$ & $1$ & $2$ & $3$ & $4$ & $5$ & $>5$ \\
\hline
$1$--$2^9$ & $44.3$ & $38.5$ & $16.6$ & $0.6$ & $0$ & $0$ \\
$(2^9+1)$--$2^{25}$ & $1.4$ & $39.7$ & $52.7$ & $6.2$ & $<{\!}0.1$ & $0$ \\
\end{tabular}
} 
\end{samepage}
} 

\bigskip

\noindent Neither Brauer's method nor Thurber's method dominates the other. The first $n$ for which Brauer's method beats Thurber's (at optimal $k$) is $n = 349$, and the first $n$ for which Thurber's method beats Brauer's method is $n = 23$. For $n \leq 10^7$ Thurber's method beats Brauer's method in $47.4\%$ of cases whereas Brauer's method beats Thurber's method in $2.3\%$ of cases. Thus, Thurber's method is more frequently better for $n \leq 10^7$.
For $n \leq 10^7$ both Brauer's and Thurber's methods give only modest improvements in performance over binary exponentiation, on average, with an average improvement of $13.2\%$ for Thurber's method and $11.5\%$ for Brauer's method. For individual $n$ the improvement may be much greater, as the following examples show.

\begin{example} \
\begin{enumerate}
\item For $n = 63$, Brauer's method and Thurber's method, both with $k=2$, give the addition chain $1$, $2$, $3$, $6$, $12$, $15$, $30$, $60$, $63$, of length $8$, whereas binary exponentiation gives $1$, $2$, $3$, $4$, $7$, $8$, $15$, $16$, $31$, $32$, $63$ (using {\tt binary\_exponentiate} as in Chapter~\ref{chapter:vector-algorithms-guide}), which has length $10$. 

\item For $n=2^{20}-1=1048575$, Brauer's method and Thurber's method, both with $k=3$, have addition chains of length $28$ and $27$ respectively, whereas binary exponentiation has length $38$. In this case Thurber's method is a $29\%$ improvement over binary exponentiation.
\end{enumerate}
\end{example}

\noindent Here are procedures to find the best $k$ for Thurber's method.

\begin{alltt}
thurber_count(n, k):
    windows = thurber_windows(n, k)
    max_value = max(value for (width, value, gap) in windows)
    n_precompute = (max_value + 1) >> 1
    count = n_precompute - 1 + (n_precompute > 1)
    window = windows[1]
    width = window[1], value = window[2], gap = window[3]
    count = count + gap - (value = 1 and gap > 0 and n_precompute > 1)
    for i in 2,...,length(windows)
        window = windows[i]
        width = window[1], value = window[2], gap = window[3]
        count = count + width + 1 + gap
    return count

thurber_best_k(n): Uses results in the tables above to speed the search
    k_max = (1 if n < 15 else 2 if n < 23 else 3 if n < 151 else 4 if n < 9413609 else
             5 if n < 10000000000 else
             number of binary digits in n)
    k_best = 1
    count_best = thurber_count(n, 1)
    for k in 2,...,k_max
        count = thurber_count(n, k)
        if count < count_best
            count_best = count
            k_best = k
    return k_best
\end{alltt}

\noindent Brauer's method and Thurber's method do not always yield an optimal addition chain even with their best choices of $k$. The first $n$ for which they are not optimal are $n=23$ for Brauer's method (shortest length $7$ at $k=1$ and $k=2$ vs optimal length $6$) and $n=39$ for Thurber's method (shortest length $8$ at $k=1$, $2$, $4$ vs optimal length $7$). Note that Brauer's method is also not optimal at $n=39$. In both cases ($n=23$ and $n=39$) an optimal addition chain can be read from the tree figures in \cite{Knuth1998}. Optimal addition chains for $n$ have been computed for all $n \leq 2^{32}$ by Clift \cite{Clift2011}.%
\footnote{
Also for $n \leq 2^{39}$ by Clift as recorded in \cite{Flammenkamp2022}.
}

\section{Parallel Algorithms for Exponentiation in Semigroups}

The length of the minimal addition chain for $n$ is the smallest number of $*$ product steps in a sequential program that computes $a^n$ in a semigroup in general, though for specific semigroups faster approaches may be possible.%
\footnote{
For example, exponentiation in the trivial $1$ element semigroup requires no computation, and exponentiation in the cyclic group $\mathbb{Z}/2\mathbb{Z}$ can be computed by looking at the last bit of the exponent $n$.
}
If we allow parallel computation, however, then a smaller number of steps is possible. In other words, the minimal depth of a parallel algorithm for computing $a^n$ may have shorter depth than the length of a minimal addition chain for $n$. To see this we can simply use a parallel version of binary exponentiation. That is to say, the {\tt binary\_exponentiate} procedure from Chapter~\ref{chapter:vector-algorithms-guide} can be parallelized.%
\footnote{This is the algorithm Knuth \cite{Knuth1998} calls Algorithm A.}

\begin{algorithm}[Parallel Binary Exponentiation] \
\label{algorithm:parallel-binary-exponentiation} 
\begin{alltt}
parallel_binary_exponentiate(op, x, n, flip):
    q = x
    z = x
    first = true
    repeat indefinitely
        n_next = n >> 1
        if (not first) and (n odd) and n_next \(\neq\) 0
            compute in parallel:
                (1) q = op(q, z) if flip else op(z, q)
                (2) z = op(z, z)
        else if (not first) and (n odd)
            q = op(q, z) if flip else op(z, q)
        else
            if n odd
                q = z
                first = false
            if n_next = 0
                return q
            z = op(z, z)
        n = n_next
\end{alltt}
\end{algorithm}

\begin{lemma}
Algorithm \ref{algorithm:parallel-binary-exponentiation} computes $x^n$ in $\left\lceil\log_2 n\right\rceil$ parallel steps when $*$ is associative. I.e., the depth of the parallel algorithm is $\left\lceil\log_2 n\right\rceil$.
\end{lemma}

\section{Multiple Exponents}
\label{sec:multiple-exponents}

Suppose that instead of a single power $a^n$, that we need to compute multiple powers $a^{n_1}, a^{n_2}, \ldots, a^{n_p}$. In order to do this, we should find an (efficient) addition chain containing all the numbers $n_1, \ldots, n_p$. This problem was solved by Andrew Yao and published in \cite{Yao1976}.

\begin{theorem}(Yao \cite{Yao1976} Theorem~3 and its Corollary)
For any set of strictly positive integers, $\theset{n_1, \ldots, n_p}$, the collection of powers $\theset{x^{n_1}, \ldots, x^{n_p}}$ is computable from input $\theset{x}$ in less than

\begin{equation}
 \log_2 m+c \sum_{i=1}^p \frac{\log_2 n_i}{\log_2 \log_2\left(n_i+2\right)} \leq \log_2 m+c p \frac{\log_2 m}{\log_2 \log_2(m+2)}
\end{equation}
\noindent multiplications, for some constant $c$  where $m=\max\theset{n_1, \ldots, n_p}$.
\end{theorem}
\begin{proof}
For the proof we refer to \cite{Yao1976}, where the corresponding algorithm and addition chains are constructed.
\end{proof}

The other cases of multiple exponents that interest us, for applications to vector sliding window $*$-products, are those where we have a collection of exponents $n_1, \ldots, n_p$ and an additional exponent $N$, and we wish to compute $a^{n_1}, \ldots, a^{n_p}$ and also to compute $a^M$ for some $M \geq N$, but where we do not care which $M \geq N$ is computed as long as $a^M$ is computed for at least one $M \geq N$. For this, the following simple result suffices.

\begin{lemma}
\label{lemma:multiple-exponent-and-higher-power}
Suppose $\theset{n_1, \ldots, n_p}$ is a set of strictly positive integers,  and there is an addition chain of length $l$ containing $n_1, \ldots, n_p$. Suppose $N \geq n_1, \ldots, n_p$, and $m=\max \theset{n_1, \ldots, n_p}$. Then there is an addition chain of length $l+\left\lceil\log_2 \frac{N}{m}\right\rceil$ which contains $n_1, \ldots, n_p, M$ for some $M \geq N$. 
\end{lemma}
\begin{proof}
Suppose $1, e_1, \ldots, e_l$ is an addition chain containing $n_1, \ldots, n_p$. Let $j=\left\lceil\log_2 \frac{N}{m}\right\rceil$, and $M = 2^j m$. Then $1, e_1, \ldots, e_l, 2 m, 2^2 m, \ldots, 2^j m = M$, is an addition chain of length $l+\left\lceil\log_2 \frac{N}{m}\right\rceil$ containing $n_1, \ldots, n_p, M$, and $M \geq N$.
\end{proof}

%% file: htcams-arxiv-ch13-vector-sliding-window-star-products-algorithms.tex
\chapter{Vector Sliding Window $*$-Products -- Algorithms and Multi-Query Algorithms}
\label{chapter:vector-sliding-window-*-product-algorithms}

\section{Vector Sliding Window $*$-Product Algorithms}
\label{sec:vector-sliding-window-*-product-algorithms-section}

The results of Sections~\ref{sec:vector-semidirect-products} and \ref{sec:algorithms-for-vector-sliding-window-*-products-section}, and Chapter~\ref{chapter:exponentiation-in-semigroups} combine to give algorithms for sliding window $*$-products in several different ways.

\begin{enumerate}
  \item Given an algorithm for computing exponentiation in a semigroup, such as Brauer's method, Thurber's method, or an optimal formal addition chain, we may use this algorithm together with Algorithm~\ref{algorithm:general-vector-sliding-window-*-product} to obtain an algorithm for vector sliding window $*$-products.
  \item Using the parallel binary exponentiation algorithm, Algorithm~\ref{algorithm:parallel-binary-exponentiation}, together with Algorithm~\ref{algorithm:general-vector-sliding-window-*-product}, we obtain a parallel algorithm for sliding window $*$-products, of minimal depth.
  \item Given a list of window lengths $n_1, \ldots, n_p$, we may use the addition chain of Yao \cite{Yao1976} together with Algorithm~\ref{algorithm:general-vector-sliding-window-*-product} to obtain an algorithm for jointly computing the sliding window $*$-products of lengths $n_1, \ldots, n_p$. This is an example of a vector algorithm to solve the multi-query problem for vector sliding window $*$-products. (See \cite{Shein2019} for terminology.)
\end{enumerate}
  
\noindent These give the following results.


\begin{theorem}
Assume $*$ is associative. A vector sliding window $*$-product of length $n$ may be computed using at most $\left(\log_2 n\right)\left(1+\frac{1}{\log_2 \log_2(n+2)}+\littleoh{\frac{1}{\log_2 \log_2(n+2)}}\right)$ vector $*$-products.
\end{theorem}
\begin{proof}
Use Brauer's algorithm and Algorithm~\ref{algorithm:general-vector-sliding-window-*-product}.
\end{proof}

\begin{theorem}
Assume $*$ is associative. 
A sliding window vector $*$-product of length $n$ may be computed in parallel in no more than $\left\lceil\log_2 n \right\rceil$ parallel steps involving vector $*$-products. I.e., there is a parallel algorithm of depth at most $\left\lceil\log_2 n\right\rceil$.
\end{theorem}
\begin{proof}
Use parallel binary exponentiation, Algorithm~\ref{algorithm:parallel-binary-exponentiation}, and Algorithm~\ref{algorithm:general-vector-sliding-window-*-product}.    
\end{proof}

\begin{corollary}
Assume $*$ is associative. A sliding window $*$-product of length $n$ may be computed in parallel in no more than $\left\lceil \log_2 n \right\rceil$ parallel steps.
\end{corollary}

\section{Multi-Query Algorithms}

\begin{theorem}
Assume $*$ is associative. Assume $n_1, \ldots, n_p$ are strictly positive integers. Then the vector sliding window $*$-products of length $n_1,\dots, n_p$ may be jointly computed using no more than

\begin{equation*}
\log_2 m+c \sum_{i=1}^p \frac{\log_2 n_i}{\log_2 \log_2\left(n_i+2\right)} \leq \log_2 m+c p \frac{\log_2 m}{\log_2 \log_2(m+2)}
\end{equation*}
vector $*$-products, where $m=\max\theset{n_1, \ldots, n_p}$, and $c$ is a constant.
\end{theorem}
\begin{proof}
Use Yao's algorithm and Algorithm~\ref{algorithm:general-vector-sliding-window-*-product}.
\end{proof}

\section{Parallel Prefix Sum Algorithms}
\label{sec:parallel-prefix-sum-algorithms}

\subsection{Prefix Sums}


\begin{definition}[Prefix Sum, Prefix $*$-Product] 
Assume $*\colon A \times A \rightarrow A$ is a binary operation and $a_1, \ldots, a_N \in A$. 
Then the {\em prefix sum}, also called the {\em prefix $*$-product}, of the sequence $a_1, \ldots, a_N$ is the sequence $z_1, \ldots, z_N$ defined by

\begin{align*}
z_1 & =a_1 \\
z_2 & =a_2 * a_1 \\
z_3 & =a_3 *\left(a_2 * a_1\right) \\
\vdots & \;\;\;\;\;\;\;\;\vdots \\
z_N & =a_N *\left(a_{N-1} *\left(\ldots *\left(a_2 * a_1\right) \ldots\right)\right)
\end{align*}    
\end{definition}

\noindent Parallel algorithms for computing prefix sums in the associative case have been extensively studied by Kogge and Stone \cite{KoggeStone1973}, Ladner and Fischer \cite{LadnerFischer1980}, Hillis and Steele \cite{HillisSteele1986}, and Blelloch \cite{Blelloch1993}, among other authors, with precursor work by Ofman \cite{Ofman1962}.
Prefix sums/prefix $*$-products are also known as {\em cumulative sums}, {\em cumulative products}, {\em partial sums}, {\em running sums}, {\em running totals}, or {\em a scan}.
It should be clear from the definition that prefix sums are identical to sliding window $*$-products where the window length $n$ is greater than the data length $N$.

\subsection{Parallel Algorithms for Prefix Sums}

The vector algorithms for sliding window $*$-products of Sections \ref{sec:algorithms-for-vector-sliding-window-*-products-section} and \ref{sec:vector-sliding-window-*-product-algorithms-section} give new algorithms for computing prefix sums (prefix $*$-products) in parallel in the associative case.%
\footnote{For the nonassociative case see Chapter~\ref{chapter:vector-windowed-recurrences}.}
To see this, define vector $*$-products and shift operators using Example \ref{example:vector-*-product-fixed-length-sequences} or Example \ref{example:vector-*-product-variable-length-sequences}, using a data length $N$ which is less than or equal to the window length $n$. If we then use $n=2^{\left\lceil\log_2 n \right\rceil}$, and use binary exponentiation (i.e., successive squaring) to perform the exponentiation in the semidirect product semigroup in Algorithm~\ref{algorithm:general-vector-sliding-window-*-product}, we obtain the algorithms of Kogge and Stone \cite{KoggeStone1973}, and Hillis and Steele \cite{HillisSteele1986}, and this yields a new proof of correctness for these algorithms. If, however, we use a different exponentiation algorithm, such as those of Brauer or Thurber, or an optimal formal addition chain, or if we choose a window length $n$ which is not a power of $2$ (and also $n \geq N$), then we obtain new algorithms for computing prefix sums.

To see that these algorithms are indeed different and new, we now demonstrate algorithms for the joint computation of a prefix sum (prefix $*$-product) with a sliding window $*$-product.

\begin{theorem}
\label{theorem:joint-vector-window-and-prefix-sum}
Assume $*$ is an associative binary operation, and $n$, $N$ are strictly positive integers with $n \leq N$. Then a sliding window $*$-product of length $n$ on $N$ data points may be computed together with a prefix $*$-product (prefix sum) on the same $N$ data points in a total of no more than
\begin{equation*}
\log_2 N + \frac{\log_2 n}{\log_2 \log_2(n+2)} + \littleoh{\frac{\log_2 n}{\log_2 \log_2(n+2)}}
\end{equation*}
vector $*$ operations of length $\leq N$, and hence in no more than an equal number of parallel steps.\\
\end{theorem}
\begin{proof}
Use either the construction of Example \ref{example:vector-*-product-fixed-length-sequences} or Example \ref{example:vector-*-product-variable-length-sequences} to relate the sliding window $*$-product to a sliding window vector $*$-product on a semigroup $V$ with shift operators $L_i$, $i \geq 1$. Compute the vector sliding window $*$-product using Algorithm~\ref{algorithm:general-vector-sliding-window-*-product}, using Brauer's algorithm (or Thurber's algorithm, or an optimal addition chain) on $\mathbb{Z}_{>0} \ltimes_L V$. If $a \in V$ is the input data, then we have computed $z=\smat{1\\ a}^{*n}$. Now raise $z$ to the power of $2^{\ceiling{\log_2 \frac{N}{n}}}$ by successively squaring $\ceiling{\log_2 \frac{N}{n}}$ times. The total number of vector operations used is
\begin{equation*}
\log_2 n + \frac{\log_2 n}{\log_2 \log_2(n+2)}
        +\littleoh{\frac{\log_2 n}{\log_2 \log_2(n+2)}}+\ceiling{\log_2 \frac{N}{n}}
\end{equation*}
The result now follows trivially.
\end{proof}
\begin{remark}
The algorithm described in Theorem~\ref{theorem:joint-vector-window-and-prefix-sum} only uses vector $*$ operations and shift operations, other than bookkeeping only depending on $n$ and $N$. Therefore, this algorithm also applies to non-parallel vectorized settings.     
\end{remark}

Theorem~\ref{theorem:joint-vector-window-and-prefix-sum} can be improved using parallel binary exponentiation.

\begin{theorem}
\label{theorem:joint-parallel-window-and-prefix-sum}
Assume $*$ is an associative binary operation and $n$, $N$ are strictly positive integers with $n \leq N$. Then a sliding window $*$-product of length $n$ on $N$ data points may be computed together with a prefix $*$-product (prefix sum) on the same $N$ data points in a total of precisely
$\ceiling{\log_2 n}+\ceiling{\log_2 \frac{N}{n}}$
parallel steps, and 
\begin{equation*}
\ceiling{\log_2 n}+\ceiling{\log_2 \frac{N}{n} } 
\leq \ourcases{
    \ceiling{\log_2 N}     & \text{if $n$ is a power of $2$, else} \\[0.5ex]
    \ceiling{\log_2 N} + 1 & \text{otherwise}    
    }
\end{equation*}
\end{theorem}
\begin{proof}
Similar to the proof of Theorem~\ref{theorem:joint-vector-window-and-prefix-sum} except using parallel binary exponentiation in place of Brauer's algorithm. For the final inequality, let $x=\log_2 n - \floor{\log_2 n}$, and note that if $n$ is not a power if 2 , then
\begin{align*}
\ceiling{\log_2 n} + \ceiling{\log_2 \frac{N}{n}} 
    & = \floor{\log_2 n} + 1 + \ceiling{\log_2 N - \floor{\log_2 n} - x} \\
    & = \ceiling{\log_2 N - x} + 1 \\
    & \leq \ceiling{\log_2 N} + 1
\end{align*}
\end{proof}

\begin{theorem}
\label{theorem:joint-vector-multiquery-window-and-prefix-sum}
Assume $*$ is an associative binary operation and $n_1, \ldots, n_p \leq N$ are strictly positive integers. Then the sliding window $*$-products of lengths $n_1, \ldots, n_p$ on $N$ data points may be computed together with a prefix $*$-product (prefix sum) on the same data points in a total of no more than
\begin{align*}
\log_2 m + c \sum_{i=1}^p \frac{\log_2 n_i}{\log_2 \log_2(n_i + 2)}
         + \ceiling{\log_2 \frac{N}{m}} 
    & \leq \log_2 N + c\left(\sum_{i=1}^p \frac{\log_2 n_i}{\log_2 \log_2\left(n_i+2\right)}\right)+1 \\
    & \leq \log_2 N + c p \frac{\log_2 m}{\log_2 \log_2(n + 2)} + 1
\end{align*}
vector $*$ operations of length $\leq N$, where $c$ is a constant, and $m=\max \theset{n_1, \ldots, n_p}$. And hence it requires no more than an equal number of parallel steps.
\end{theorem}
\begin{proof}
Similar to the proof of Theorem~\ref{theorem:joint-vector-window-and-prefix-sum} but use Yao's algorithm and  Lemma~\ref{lemma:multiple-exponent-and-higher-power}. 
\end{proof}

\begin{remark}
The algorithm of Theorem~\ref{theorem:joint-vector-multiquery-window-and-prefix-sum} only uses shift operators and vector $*$-products (other than bookkeeping only depending on $n$ and $N$) and hence also applies to non-parallel vector settings.    
\end{remark}

%% file: htcams-arxiv-ch14-vector-windowed-recurrences.tex
\chapter{Vector Windowed Recurrences} 
\label{chapter:vector-windowed-recurrences}

\section{Definitions}

\begin{definition}[Vector Function Action]
\label{definition:vector-function-action}
Let $X$ be a set, and $F \subseteq \Endop(X)$, i.e., $F$ is a set of functions $X \rightarrow X$,
and let $L_{X, 1}, L_{X, 2}, \ldots \in \Endop(X)$ be functions $L_{X, i}\colon X \rightarrow X$, and let $L_1, L_2, \ldots \in \Endop(F)$ be functions $F \rightarrow F$, such that
\begin{enumerate}
\item
$L_{X, i} \circ L_{X, j}=L_{X, i+j}$ for $i, j \geq 1$ 

\item
$L_{X, i}(f(x))=\left(L_i f\right)\left(L_{X,i} x\right)$ for $x \in X, f \in F, i \geq 1.$\\
\end{enumerate}
Then $F$ together with $X$, $\theset{L_i}$, $\theset{L_{X,i}}$ is called {\em a vector function action of $F \subseteq \Endop(X)$ on $X$ with shift operators $L_i$, $L_{X,i}$}. 
\end{definition}

\begin{definition}[Vector Windowed Recurrence]
\label{definition:vector-windowed-recurrence-function-actions}
Assume $F$, $X$, $\theset{L_i}$, $\theset{L_{X,i}}$ is a vector function action of $F \subseteq \Endop(X)$ on $X$ with shift operators $L_i$, $L_{X,i}$. If $f \in F$ and $x \in X$, and $n \geq 1$ is a strictly positive integer, then the {\em vector windowed recurrence of length $n$} is defined to be
\begin{align*}
y & = f(L_1 f(L_2 f(\ldots L_{n-1} f(L_{X, n} x) \ldots))) \\
  & = (f \circ L_1 f \circ L_2 f \circ \ldots \circ L_{n-1} f)(L_{X, n} x).
\end{align*}
\end{definition}

\begin{definition}[Vector Set Action]
\label{definition:vector-set-action}
Let $A$, $X$ be sets and $\bullet\colon A \times X \rightarrow X$ be a set action of $A$ on $X$, and assume $L_1, L_2, \ldots \in \Endop(A)$ are functions on $A$, and $L_{X, 1}, L_{X, 2}, \ldots \in \Endop(X)$ are functions on $X$ such that
\begin{enumerate}
    \item $L_{X, i} \circ L_{X, j} = L_{X, i + j}$ for $i, j \geq 1$
    \item $L_{X, i}(a \bullet x) = L_i a \bullet L_{X, i} x$ for $a \in A, x \in X, i \geq 1$
\end{enumerate}
Then $\bullet$ together with $\theset{L_i}$, $\theset{L_{X,i}}$ is called a {\em vector set action of $A$ on $X$} with shift operators $L_i$, $L_{X, i}$, $i \geq 1$.   
\end{definition}

\begin{definition}[Vector Windowed Recurrence for Set Actions]
\label{definition:vector-windowed-recurrence-set-actions}
Assume $\bullet$ is a vector set action of $A$ on $X$ with shift operators $L_i$, $L_{X, i}$, $i \geq 1$. Let $a \in A, x \in X$, and let $n \geq 1$ be a strictly  positive integer. Then the {\em vector windowed recurrence of length $n$ corresponding to $a$ and $x$} is
\begin{equation*}
y = a \bullet (L_1(a) \bullet (L_2(a) \bullet (\ldots \bullet(L_{n-1}(a) \bullet L_{X, n}(x)) \ldots)))
\end{equation*}    
\end{definition}

\begin{remarks} \
\begin{enumerate}
\item 
The $L_i$ and $L_{X, i}$ are also called {\em lag operators}.

\item
In Definitions \ref{definition:vector-function-action} and \ref{definition:vector-windowed-recurrence-function-actions} we do not assume that $L_i \circ L_j = L_{i+j}$,  and also do not assume that $L_i(f \circ g) = L_i f \circ L_i g$.

\item In Definitions \ref{definition:vector-set-action} and \ref{definition:vector-windowed-recurrence-set-actions} we do not assume that $L_i \circ L_j = L_{i+j}$.

\item 
If we know that $L_{X, i} \circ L_{X, j} = L_{X, i+j}$ for $i, j \geq 1$, and in addition that $L_i \circ L_j = L_{i+j}$ for $i, j \geq 1$ then in order to show 2.\ of Definition~\ref{definition:vector-function-action} we need only show that $L_{X, 1}(f(x)) = (L_1 f)(L_{X, 1} x)$ for all $f \in F$ and $x \in X$. Similarly, in order to show 2.\ of Definition~\ref{definition:vector-set-action} we would only need to show that $L_{X, 1}(a \bullet x) = L_1 a \bullet L_{X, 1} x$ for all $a \in A$ and $x \in X$.

\item 
Suppose that $L_i \circ L_j = L_{i+j}$ for $i, j \geq 1$, and also that $L_i(f \circ g) = L_i f \circ L_i g$ for $f, g \in F$, and also that $F$ is closed under composition. Then we could in principle compute the vector windowed recurrence by computing $f \circ L_1 f \circ \ldots \circ L_{n-1} f$ as a vector sliding window $\circ$-product, and then apply the result to $L_{X, n}(x)$. In general there are two difficulties to overcome in order to use this approach in practice.
\begin{enumerate}[label=(\alph*)]
    \item $F$ may not be closed under composition.
    \item We need an effective (i.e., efficient) way of representing and computing function compositions.
\end{enumerate}

We will develop these ideas further in this chapter. As it turns out we will not need to assume either $L_i \circ L_j = L_{i+j}$, or $L_i(f \circ g)  = L_i f \circ L_i g$, but instead can work with only the assumptions of Definition~\ref{definition:vector-function-action} or Definition~\ref{definition:vector-set-action}.
\end{enumerate}
\end{remarks}



\section{Examples and Constructions}

\subsection{Vector Function Actions -- Examples and Constructions}

\begin{example}
Assume $X$ is a set, $L_X \in \Endop(X)$, $F \subseteq \Endop(X)$, $L\in \Endop(F)$, and that $L_X(f(x)) = L(f)(L_X(x))$ for $f \in F$, $x \in X$. Then if we define $L_i=L^i$, $L_{X, i}=(L_X)^i$, this defines a vector function action of $F$ on $X$ with shift operators $L_i$, $L_{X,i}$.
\end{example}

\begin{example}
Let $\bullet$ be a vector set action of $A$ on $X$ with shift operators $L_i$, $L_{X, i}$. Assume further that for all $a, b \in A$, $\Leftop_a^\bullet = \Leftop_b^\bullet \Rightarrow \Leftop_{L_i(a)}^\bullet = \Leftop_{L_i(b)}^\bullet$. Then we can lift the shift operators $L_i$ to operators $\tilde{L}_i$ on $F = \theset{\Leftop_a^\bullet \colon a \in A}$ by defining $\tilde{L}_i(\Leftop^\bullet_a) =  \Leftop_{L_i(a)}$. Thus $\tilde{L}_i\colon F\rightarrow F$. With these definitions $F$, $X$, $\theset{\tilde{L}_i}$, $\theset{L_{X, i}}$ is a vector function action of $F$ on $X$ with shift operators $\tilde{L}_i$, $L_{X, i}$, $i \geq 1$.
\end{example}

\begin{example}
\label{example:vector-function-action-from-functions}
Let $X$ be a set, and $f_1, f_2, \ldots, f_N \in F \subseteq \Endop(X)$ and assume $\id \in F$.  E.g., we could choose $F = \Endop(X)$.  We can define a function $f\colon \Endop(X^N) \rightarrow \Endop(X^N)$ by 
$\component{i}{f(x)} = f_i(x_i)$ for $i=1, \ldots, N$, where $x = (x_1, \ldots, x_N) \in X^N$. This defines an embedding $F^N \hookrightarrow \Endop(X)^N  \hookrightarrow \Endop(X^N)\colon (f_1, \ldots, f_N) \mapsto f_1 \times \ldots \times f_N = f$. Now define shift operators $L_j$ on $F^N \subseteq \Endop(X)^N \subseteq \Endop(X^N)$ by
\begin{equation*}
L_j(f_1 \times \ldots \times f_N) = \ourunderbracket{\id \times \ldots \times \id}_{\min(j, N) \text{ times}} \times f_1 \times \ldots \times f_{N-j}
\end{equation*}
Let $x_0 \in X$, and define
\begin{equation*}
\component{i}{L_j^{x_0} x} = \ourcases{
x_{i-j} & i-j \geq 1 \\
x_0 & i-j<1
}    
\end{equation*}
for $i=1, \ldots, N$, $x = (x_1,\dots,x_N)$. Then $F^N$ is a vector function action on $X^N$ with shift operators $L_i, L_i^{x_0}$. If $x = (x_1, \ldots x_N)$, and $f = f_1 \times \ldots \times f_N \in F^N$, and $n \geq 1$ then the corresponding vector windowed recurrence is
\begin{equation*}
y = (f_1(x_0), f_2f_1(x_0), \ldots, f_n \cdots f_1(x_0), 
     f_{n+1} f_n \cdots f_2(x_1), \ldots, f_N \cdots f_{N-n+1}(x_{N-n})) 
\end{equation*}
So the vector windowed recurrence for $f_1 \times \ldots \times f_N,\left(x_1, \ldots, x_N\right)$ is equal to the non-vector windowed recurrence for $x_0, x_1, \ldots, x_N$, $f_1, f_2, \ldots, f_N$ of Definition~\ref{definition:windowed-recurrence-functions}. 
\end{example}


\begin{example} \label{example:vector-function-action-from-set-automorphism}
Assume $X$ is a set, and $F \subseteq \Endop(X)$ is a collection of functions from $X$ to itself. Assume that $L_X \in \Endop(X)$ is an {\em invertible} element of $\Endop(X)$. Let $\overline{F}$ denote the closure of $F$ under the operation $f \mapsto L_X \circ f \circ L_X^{-1}$. I.e., $\overline{F} = \theset{(L_X)^i \circ f \circ (L_X^{-1})^i\colon f\in F, i\geq 0}$. Let $L_{X,i} = (L_X)^i$, and define $L_i(f) = (L_X)^i \circ f \circ (L_X^{-1})^i$ for $f\in \overline{F}$, $i \geq 1$. Then $\overline{F}$, $\theset{L_i}$, $\theset{L_{X,i}}$ is a vector function action of $\overline{F}$ on $X$ with shift operators $L_i$, $L_{X, i}$. In particular, this example shows that the shift operators for vector function actions (and vector set actions) may be periodic.
\end{example}

\subsection{Vector Set Actions -- Examples and Constructions}

\begin{example}
Assume $A$, $X$ are sets, and $\bullet\colon A \times X \rightarrow X$ is a set action of $A$ on $X$, and $L\colon A \rightarrow A$, $L_X\colon X \rightarrow X$ satisfy
\begin{equation*}
L_X(a \bullet x)=L(a) \bullet L_X(x).
\end{equation*}
Then if we define $L_i = L^i, L_{X, i} = L_X^i$, then $\bullet$ is a vector set action of $A$ on $X$ with shift operators $L_i$, $L_{X, i}$, $i \geq 1$. 
\end{example}

\begin{example}
\label{example:vector-set-action-from-vector-*-product}
Assume $A$ is a set and $*$ is a vector $*$-product on $A$ with shift operators $L_i$. Then define $X = A$, $L_{X, i} = L_i$. Then
\begin{enumerate}
\item
The action $*\colon A \times A \rightarrow A$ is a vector set action of $A$ on itself with shift operators $L_i$, $L_{X, i}$, $i \geq 1$.

\item
The vector sliding window $*$-product of length $n \geq 2$ corresponding to $a \in A$ is equal to the vector windowed recurrence of length $n-1$, for $a$, $x=a$. I.e.,
\begin{equation*}
a * (L_1 a * (\ldots * (L_{n-2} a * L_{n-1} a) \ldots))
    = a * (L_1 a * (\ldots * (L_{n-2} a * L_{X, n-1} a)\ldots )    
\end{equation*}

\item
If $*$ has a right identity $1_A$, and $L_i(1_A)=1_A$ for $i \geq 1$, then the vector $*$-product of length $n$ for $a$ is equal to the vector windowed recurrence for $a$, $x=1_A$ of length $n$. I.e.,
\begin{equation*}
a * (L_1 a * (\ldots * (L_{n-2}(a) * L_{n-1}(a))\ldots ))
    = a * (L_1 a * (\ldots * (L_{n-2}(a) * (L_{n-1}(a) * L_{X, n}(1_A))) \ldots))
\end{equation*}

\end{enumerate}
\noindent Note that for this example we do not assume that $*$ is associative, and this gives an approach to computing vector $*$-products for nonassociative operations by relating them to vector windowed recurrences for vector set actions.    
\end{example}

\begin{example}
\label{example:vector-set-action-from-vector-function-action}
Assume $F$ is a vector function action on $X$ with shift operators $L_i$, $L_{X, i}$, $i \geq 1$. Define $\bullet\colon F \times X \rightarrow X\colon (f, x) \mapsto f \bullet x=f(x)$. Then $\bullet$ is a vector set action of $F$ on $X$ with shift operators $L_i$, $L_{X, i}$, $i \geq 1$, and the windowed recurrence for $f \in F, x \in X$ for the vector function action of $F$ on $X$ is equal to the windowed recurrence for $f \in X, x \in X$ for the vector set action $\bullet$ of $F$ on $X$.
\end{example}

\begin{example}
\label{example:vector-set-action-fixed-length-sequences}

Let $\bullet\colon A \times X \rightarrow X$ be a set action, and $N \geq 1$. Define
\begin{equation*}
A_1 = \ourcases{
    A & \text{if $A$ has a left identity $1_A$ with respect to $\bullet$, and thus $\Leftop_{1_A}^\bullet = \id_X$.}\\
    A \cup \theset{1} & \text{if $A$ does not have a left identity with respect to $\bullet$.}
}
\end{equation*}
where in the latter case $\bullet$ is extended to $1$ by $1 \bullet x = x$, for $x \in A$, and $1 \not\in A$, and we set $1_{A_1} = 1$. Let 
\begin{equation*}
A_1^N=\underbrace{A_1 \times \ldots \times A_1}_{\text{$n$ times}}
, \qquad X^N=\underbrace{X \times \ldots \times X}_{\text{$n$ times}}
\end{equation*}
Define $\bullet\colon A_1^N \times X^N \rightarrow X^N$, $L_j\colon A_1^N \rightarrow A_1^N$, 
$L_{X, j}\colon X^N \rightarrow X^N$ by
$\component{i}{a \bullet x} = a_i \bullet x_i$ for $i=1, \ldots N$, where 
$a=(a_1,\ldots a_N)$, $x= (x_1, \ldots, x_N)$ with $a \in A_1^N, x \in X^N$, and
\begin{equation*}
\component{i}{L_j a} = \ourcases{
    a_{i-j} & \text{if $i - j \geq 1$} \\
    1_{A_1} & \text{if $i - j < 1$}
}
\end{equation*}
for $j \geq 1$, $i=1, \ldots, N$, and
\begin{equation*}
\component{i}{L_j^{x_0} x} = \ourcases{
    x_{i-j} & \text{if $i - j \geq 1$} \\
    x_0     & \text{if $i - j < 1$}
}
\end{equation*}
for $x_0 \in X$,  $j \geq 1$, $x = (x_1, \dots, x_N) \in X^N$. Then
\begin{equation*}
L_i^{x_0} \circ L_j^{x_0}=L_{i+j}^{x_0}    
\end{equation*}
for $i, j \geq 1$, and
\begin{equation*}
L_i^{x_0}(a \bullet x) = L_i a  \bullet L_i^{x_0}(x)
\end{equation*}
for $i \geq 1$, $a \in A_1^N$, $x \in X^N$. So $\bullet$ is a vector set action of $A_1^N$ on $X^N$. Furthermore, the vector windowed recurrence for $a \in A^N$, $x \in X^N$ of length $n \geq 1$ is
\begin{align*}
& \left(
   a_1 \bullet x_0, a_2 \bullet(a_1 \bullet x_0), \ldots, 
   a_n \bullet \ldots \bullet(a_1 \bullet x_0), a_{n+1} \bullet(a_n \bullet \ldots(a_2 \bullet x_1) \ldots),
   \right.\\
& \left.\qquad \ldots, a_N \bullet(a_{N-1} \bullet(\ldots \bullet(a_{N-n+1} \bullet x_{N-n}) \ldots ))
  \right)
\end{align*}
and therefore the vector windowed recurrence for $a = (a_1,\dots,a_N) \in A^N$, $x = (x_1,\dots,x_N)  \in X^N$ is equal to the non-vector windowed recurrence for $x_0, x_1, \ldots, x_N$, and $a_1, \ldots a_N$, of Definition~\ref{definition:windowed-recurrence-set-action}. Note that in this example we also have $L_i \circ L_j = L_{i+j}$.
\end{example}

\begin{example}
\label{example:vector-set-action-variable-length-sequences}
Let $\bullet\colon A \times X \rightarrow X$ be a set action, and $N \geq 1$. Define 
\begin{equation*}
V_N(A)= \bigcup_{i=0}^N A^i
      = \theset{\text{sequences of elements in $A$ of length $\leq N$}}
\end{equation*}
Define $\bullet\colon V_N(A) \times X^N \rightarrow X^N$ by
$
u \bullet x = (x_1, \ldots x_{N-p}, u_1 \bullet x_{N-p+1}, \ldots, u_p \bullet x_N)
$
for $u = (u_1, \ldots, u_p) \in A^p$, $x = (x_1, \ldots, x_N) \in X^N$, $p \leq N$. Also define $L_i\colon V_N(A) \rightarrow V_N(A)$, and $L_i^{x_0}\colon X^N \rightarrow X^N$, for $i \geq 1$ by
\begin{equation*}
L_i u = \ourcases{
(u_1, \ldots, u_{p-i})       & \text{if $i < p$}\\
\text{the empty sequence ( )} & \text{if $i \geq p$}
}   
\end{equation*}
for $u=(u_1,\dots,u_p) \in A^p$, $p \leq N$ and
$
L_i^{x_0}(x) = \!\!\overbrace{x_0, \ldots, x_0}^{\text{$\min(i,N)$ times}}\!\! ,x_1, \ldots, x_{N-i}     
$
where $x_0 \in X$, $x = (x_1, \ldots, x_N) \in X^N$, and $i \geq 1$.
Then $\bullet\colon V_N(A) \times X^N \rightarrow X^N$ is a vector set action of $V_N(A)$ on $X^N$ with shift operators $L_i$, $L_i^{x_0}$, $i \geq 1$, and the vector windowed recurrence for $a=(a_1,\ldots,a_N) \in A^N \subseteq V_N(A)$, $x = (x_1,\dots , x_N) \in X^N$ is equal to the non-vector windowed recurrence for $x_0, x_1, \ldots, x_N$, and $a_1, \ldots, a_N$ of Definition~\ref{definition:windowed-recurrence-set-action}.
\end{example}

\begin{example}
\label{example:vector-set-action-unbounded-length-sequences}
We now consider the infinite union
\begin{equation*}
V_\infty(A) = \bigcup_{i=0}^{\infty} A^i= \theset{\text{all finite sequences of elements of $A$}}
\end{equation*}
acting on 
\begin{equation*}
V_\infty(X) = \bigcup_{i=0}^{\infty} X^i=\theset{\text{all finite sequences of elements of $X$}}    
\end{equation*}
Choose an $x_0\in X$ and define the action of $V_\infty(A)$ on $V_\infty(X)$ via
\begin{equation*}
u \bullet x = \ourcases{
x_1, \ldots, x_{q-p}, u_1 \bullet x_{q-p+1}, \ldots, u_p \bullet x_q 
    & \text{if $p < q$}\\
u_1 \bullet x_1, \ldots, u_p \bullet x_p
    & \text{if $p = q$} \\
u_1 \bullet x_0, \ldots, u_{p-q} \bullet x_0, u_{p-q+1} \bullet x_1, \ldots, u_p \bullet x_q 
    & \text{if $p > q$}
}    
\end{equation*}
where $u \in V_\infty(A), x \in V_\infty(X)$, $p=\operatorname{length}(u)$, and $q=\operatorname{length}(x)$. Note that then $u \bullet x \in X^{\max\theset{p, q}} \subseteq V_\infty(X)$.
We define $L_i$ as in Example \ref{example:vector-set-action-variable-length-sequences} noting that this extends to $V_\infty(A)$. However, for $L_{V_\infty(X), i}$ we define 
\begin{equation*}
L_{V_\infty(X), i} x = \ourcases{
(x_1, \ldots, x_{q-i})       & \text{if $i < q$}\\
\text{the empty sequence ( )} & \text{if $i \geq q$}
}   
\end{equation*}
where $x\in V_\infty(X)$ and $q=\operatorname{length}(x)$. Then $\bullet\colon V_\infty(A) \times V_\infty(X) \rightarrow V_\infty(X)$ is a vector set action of $V_\infty(A)$ on $V_\infty(X)$, with shift operators $L_i$, $L_{V_\infty(X), i}$, and the vector window recurrence for $a \in A^N$, $x \in X^N$, for $N \geq 1$, of length $n \geq 1$, is equal to the non-vector windowed recurrence of length $n$ for $x_0, x_1, \ldots, x_N$, and $a_1, \ldots, a_N$, where $x=(x_1, \ldots, x_N)$ and $a=(a_1, \ldots a_N)$, according to Definition~\ref{definition:windowed-recurrence-set-action}.
\end{example}

\section{Vector Set Actions, Semi-Associativity and Semidirect Products}
We start by proving an analog of part 2 of Theorem~\ref{theorem:semidirect-product-vector-sliding-window}.

\begin{lemma}
\label{lemma:vector-set-actions-and-windowed-recurrences}
Assume $\bullet\colon A \times X \rightarrow X$ is a vector set action of $A$ on $X$ with shift operators $L_i \in \Endop(A)$, $L_{X, i} \in \Endop(X)$, $i \geq 1$. Then for $a \in A$, $x \in X$, $n \geq 1$ we may express the vector windowed recurrence of length $n$ for $a$, $x$, as
\begin{align*}
a \bullet (L_1 a \bullet (L_2 a \bullet (\ldots \bullet (L_{n-1} a \bullet  L_{X, n} x) \ldots ))) 
& = \underbrace{a \bullet L_{X, 1}(a \bullet  L_{X, 1}(\ldots L_{X, 1}(a \bullet  L_{X, 1}}_{\text{$n$ times}} x) \ldots)) \\
& = \Bigl(\underbrace{(\Leftop_a\circ  L_{X, 1}) \circ \ldots \circ (\Leftop_a  \circ L_{X, 1})}_{\text{$n$ times}}\Bigr)(x) \\
& = (\Leftop_a^\bullet \circ L_{X, 1})^n(x) 
\end{align*}
\end{lemma}
\begin{proof}
This is a special case of Theorem~\ref{theorem:semidirect-product-set-action} Part 2, obtained by substituting $\mathbb{Z}_{>0}$ for $A$ in that theorem and $A$ for $B$. The set actions used are $\bullet\colon \mathbb{Z}_{>0} \times X \rightarrow X\colon i \bullet x = L_{X, i} x$, and $\bullet\colon A \times X \rightarrow X$, and $\times\colon \mathbb{Z}_{>0} \times A \rightarrow A\colon i \times a = L_i(a)$. The set action $\bullet\colon \mathbb{Z}_{>0} \times X \rightarrow X$ is semi-associative with companion operation $+$, and $i \bullet (a \bullet x) = L_{X, i}(a \bullet x) = L_i(a) \bullet L_{X, i}(x) = (i \times a) \bullet (i \bullet x)$. So the conditions of Theorem~\ref{theorem:semidirect-product-set-action} Part 2 are satisfied, and the result follows from applying this part to $1 \in \mathbb{Z}_{>0}$, and $a \in A$.
\end{proof}

%
\begin{corollary}
Assume $F \subseteq\Endop(X)$ is a vector function action on $X$ with shift operators $L_i$, $L_{X, i}$, $i \geq 1$. Assume $f \in F$, and $x \in X$, and $n \geq 1$. Then the windowed recurrence of length $n$ for $f, x$ is
\begin{equation*}
f(L_1 f (L_2 f(\ldots L_{n-1}f(x) \ldots)))  
=  \underbrace{f(L_{X, 1}(f(L_{X, 1}(\ldots f(L_{X, 1}(x)) \ldots)))}_{\text{$n$ $f$'s}} \\
= (f \circ L_{X, 1})^n(x)
\end{equation*}
\end{corollary}

\noindent The main result on vector set actions and semi-associativity is the following.

\begin{theorem}[]
\label{theorem:vector-set-actions-and-semi-associativity}
Assume $\bullet \colon A \times X \rightarrow X$ is a vector set action of $A$ on $X$ with shift operators $L_i$, $L_{X, i}$, $i \geq 1$, and assume $n$ is a strictly positive integer. Then
\begin{enumerate}
\item 
Define an action of $\mathbb{Z}_{>0} \times A$, and hence $\mathbb{Z}_{>0} \ltimes_L A$, on $X$, by
\begin{equation*}    
\mat{i\\ a} \bullet x = a \bullet L_{X,i}(x)
\end{equation*}
for $a \in A$, $i \geq 1$, $x \in X$. Then
\begin{align*}
\smat{1\\ a} \bullet \left(\smat{1\\ a} \bullet \left(\ldots \bullet \left(\smat{1\\ a} \bullet x \right) \ldots \right) \right) 
& = \underbrace{a \bullet L_{X, 1}(a \bullet L_{X, 1}(\ldots(a \bullet L_{X, 1} x) \ldots))}_{\text{$n$  $a$'s}}\\
& = \text{the windowed recurrence of length $n$ for $a$, $x$}
\end{align*}

\item
Assume $\bullet$ is semi-associative with companion operator $*$. I.e., $*\colon A \times A \rightarrow A$ is a binary operation on $A$ and $a \bullet(b \bullet x) = (a * b) \bullet x$ for all $a, b \in A$, $x \in X$.
Then the operation $\smat{i\\ a} \bullet x = a \bullet L_{X, i} x$ is a set action of the semidirect product $\mathbb{Z}_{>0} \ltimes_L A$ on $X$ and this action is semi-associative with companion operator 
$*\colon \left(\mathbb{Z}_{>0} \ltimes_L A\right) \times \left(\mathbb{Z}_{> 0} \ltimes_L A\right) \rightarrow \left(\mathbb{Z}_{>0} \ltimes_L A\right)$ given by
\begin{equation*}
\mat{i\\ a} * \mat{j\\ b} = \mat{i + j\\ a * L_i b}
\end{equation*}

\item
Assume $\bullet$ is semi-associative as in 2., then for $n \geq 1$, the vector windowed recurrence for $a \in A$, $x \in X$, of length $n$ is
\begin{align*}
a \bullet  \left(L_1 a \bullet\left(L_2 a\bullet\left(\ldots \bullet\left(L_{n-1} a \bullet L_{X, n} a\right) \ldots\right)\right)\right) 
& = a \bullet L_{X, 1}(a \bullet L_{X, 1}(\ldots L_{X, 1}(a \bullet L_{X, 1} x) \ldots )) \\
& = \smat{1\\ a}^{* n} \bullet x = \smat{1\\ a}^n \bullet x
\end{align*}
where the exponentiation in $\mathbb{Z}_{>0} \ltimes_L A$ can be bracketed in any order.%
\footnote{
If $*$ is not associative, then different bracketing of $\smat{1\\ a}^n$ give different elements of $\mathbb{Z}_{>0} \ltimes_L A$, and the statement is true for each of these elements.
}

\end{enumerate}
\end{theorem}
\begin{proof}
This follows from Lemma~\ref{lemma:vector-set-actions-and-windowed-recurrences}, and from Theorem~\ref{theorem:semidirect-product-set-action} using the same substitutions as for the proof of Lemma~\ref{lemma:vector-set-actions-and-windowed-recurrences}.
\end{proof}

%
%

\noindent Lemma~\ref{lemma:vector-set-actions-and-windowed-recurrences} and Theorem~\ref{theorem:vector-set-actions-and-semi-associativity} suggest approaches to computing vector windowed recurrences.

\begin{enumerate}

\item 
Suppose $F \subseteq \Endop(X)$ is a vector function action on $X$ with shift operators $L_i$, $L_{X, i}$, $i \geq 1$, and suppose $F$ is closed under function composition. Then letting $\bullet\colon F \times X  \rightarrow X$ with $(f, x) \mapsto f \bullet x=f(x)$, the set action $\bullet$ is semi-associative with companion operator $\circ$. Hence we can define
\begin{equation*}
\mat{i\\ f} \bullet X= f(L_{X, i} x)
, \qquad
\mat{i\\ f} *\mat{j\\ g} = \mat{i+j\\ f \circ L_i g}, 
\end{equation*}
for $i, j \geq 1$, $f, g \in F$, $x \in X$, and the vector windowed recurrence for $f \in F, x \in X$, of window length $n \geq 1$ is then $\smat{1\\ f}^n \bullet x$, where the exponentiation in $\mathbb{Z}_{>0} \ltimes_L F$ may be bracketed in any order.

In order for this to be useful for computation, however, we must have an effective (i.e., efficient) way to represent and compute composite functions, and $F$ must be closed under function composition. 

\item  
Assume $\bullet\colon A \times X \rightarrow X$ is a vector set action of $A$ on $X$ with shift operators $L_i$, $L_{X, i}$, $i \geq 1$. Assume further that $\Leftop_a^\bullet= \Leftop_b^\bullet \Rightarrow \Leftop_{L_i a}^\bullet= \Leftop_{L_i b }^\bullet$ for $a, b \in A$, $i \geq 1$. Then we may define 
\begin{equation*}
\mat{i\\ \Leftop_a^\bullet} \bullet x =  a \bullet L_{X, i}(x)
, \quad
\mat{i\\ \Leftop_a^\bullet} *\mat{j\\ \Leftop_b^\bullet} 
    = \mat{i+j\\ \Leftop_a^\bullet \circ \Leftop_{L_i b}^\bullet}
\end{equation*} 
and then the vector windowed recurrence for $a \in A, x \in X$, of length $n \geq 1$, should be
%
\begin{equation*}
y = \smat{1\\ \Leftop_a^\bullet}^n \bullet x
\end{equation*}
However, this exponentiation can only be computed if we can extend $\bullet$ and $*$ to compositions of the left action operators and shifts of those compositions, and so on. 

\item  
More generally, we would like $\smat{i\\ a}\bullet x = a \bullet L_{X, i} x$  to be semi-associative with companion operator $\smat{i\\ a} * \smat{j\\ b} = \smat{i+j\\ a * L_i b}$. But there are several problems with this. Such a binary operation $*$ on $A$ may not exist, and even if replace $a$ with $\Leftop_a^\bullet$ and form $\smat{i\\ \Leftop_a^\bullet}$ there may be no $c$ such that $\Leftop_{c}= \Leftop_a \circ \Leftop_b$. We need a semi-associative action for this to work. Function composition would seem to provide the necessary companion operator, but as in the non-vector case (Chapter~\ref{chapter:semi-associativity-and-function-composition}) $A$ may not be big enough to describe the necessary function compositions. 
\end{enumerate}
\noindent As in the non-vector case, what we need is a representation of function composition. We now define vector representations of function composition.

\section{Vector Representations of Function Composition}
\label{sec:vector-representations-of-function-composition}

\begin{definition}[Vector Representation of Function Composition for a Set Action]
\label{definition:vector-representation-of-function-composition}
Assume $\bullet\colon A \times X \rightarrow X$ is a vector set action of $A$ on $X$ with shift operators $L_i$, $L_{X, i}$, $i \geq 1$. Then a vector representation of function composition for the vector set action $\bullet$ consists of the following objects.
\begin{enumerate}
\item A representation of function composition $(\Lambda, \lambda, *, \bullet)$ for the set action $\bullet\colon A \times X \rightarrow X$.
\item
Shift operators $L_i\colon \Lambda \rightarrow \Lambda$ for $i \geq 1$, such that $\bullet\colon \Lambda \times X \rightarrow X$ is a vector set action with shift operators $L_i$, $L_{X, i}$, $i \geq 1$.
\end{enumerate}

\end{definition}

\begin{remarks} \
\begin{enumerate}
\item 
A vector representation of function composition is a representation of function composition for a vector set action such that the lifted set action $\bullet\colon \Lambda \times X \rightarrow X$ is itself a vector set action with the same shift operators on $X$.

\item 
By definition, if $(\Lambda, \lambda, *, \bullet)$ is a vector representation of function composition with shift operators $L_i$, then 
$\lambda\colon A \rightarrow \Lambda$,
$*\colon \Lambda \times \Lambda \longrightarrow \Lambda$, 
$\bullet\colon \Lambda \times X \longrightarrow X$,
$L_i\colon \Lambda \rightarrow \Lambda$, 
and
\begin{enumerate}[label=(\alph*)]
    \item For all $a \in A$, $x \in X$,  $\lambda(a) \bullet x=a \bullet x$
    \item For all $\lambda_1, \lambda_2 \in \Lambda$, $x \in X$, $\lambda_1 \bullet (\lambda_2 \bullet x) =(\lambda_1 * \lambda_2) \bullet x$. I.e., $\bullet$ is semi-associative with companion operator $*$.
    \item For all $\lambda_1 \in \Lambda$, $x \in X$, $i \geq 1$, $L_{X, i}(\lambda_1 \bullet x) = L_i(\lambda_1) \bullet L_{X, i}(x)$
\end{enumerate}

\item As before, $\lambda$ is called {\tt lift}, $*\colon \Lambda \times \Lambda \rightarrow \Lambda$ is called {\tt compose}, and $\bullet\colon \Lambda \times X \rightarrow X$ is called {\tt apply}.

\end{enumerate}

\end{remarks}

\begin{definition}[Vector Representation of Function Composition for an Indexed Collection of Functions]
Assume $X$ is a set and $\theset{f_a \colon a \in A}$ is an indexed collection of functions on $X$. Define the set action $\bullet\colon A \times  X \rightarrow X$ by $a \bullet x = f_a(x)$. Assume there are shift operators $L_i$, $L_{X, i}$ such that $\bullet$ is a vector set action with shift operators $L_i$, $L_{X, i}$. Then we also refer to a vector representation of function composition for the vector set action $\bullet$ as a {\em vector representation of function composition for the functions $\theset{f_a \colon a \in A}$}.%
\footnote{Note that in this case $f_a = \Leftop_a^\bullet$.}
\end{definition}

\begin{theorem}
\label{theorem:vector-windowed-recurrence-algorithm}

Assume  $\bullet\colon A \times X \rightarrow X$ is a vector set action with shift operators $L_i$, $L_{X, i}$, $i \geq 1$, and $(\Lambda, \lambda, *, \bullet)$ is a vector representation of function composition for $\bullet\colon A \times  X \rightarrow X$ with shift operators $L_i\colon \Lambda \rightarrow \Lambda$. Define the semi-associative action of $\mathbb{Z}_{>0} \ltimes_L \Lambda$ on $X$ by
\begin{equation*}
\mat{i\\ \lambda_1} \bullet x = \lambda_1 \bullet L_{X, i}(x)
, \quad
\mat{i\\ \lambda_1} *\mat{j\\ \lambda_2} = \mat{i + j\\ \lambda_1 * L_i(\lambda_2)}
\end{equation*}
Then the vector windowed recurrence of length $n \geq 1$ for $a \in A, x \in X$ is
\begin{align*}
y & = a \bullet (L_1(a) \bullet (\ldots \bullet (L_{n-1}(a) \bullet L_{X, n}(x)) \ldots )) \\
& = \left(\smat{1\\ \lambda(a)} * \ldots *\smat{1\\ \lambda(a)}\right) \bullet x \\
& = \smat{1\\ \lambda(a)}^n \bullet x
\end{align*}
where the exponentiation in $\mathbb{Z}_{>0} \ltimes_L \Lambda$ can be bracketed in any order.
\end{theorem}
\begin{proof}
By Definition~\ref{definition:vector-representation-of-function-composition} the {\tt apply} action $\bullet\colon \Lambda \times X \rightarrow X$ is a vector set action of $\Lambda$ on $X$ with shift operators $L_i$, $L_{X, i}$, and hence Theorem~\ref{theorem:vector-set-actions-and-semi-associativity} applies to this set action.
\begin{align*}
y & = a \bullet (L_1 a \bullet (\ldots \bullet (L_{n-1} a \bullet L_{X, n} x) \ldots)) \\
& = \ourunderbracket{a \bullet L_{X, 1}(a \bullet L_{X, 1}(\ldots \bullet L_{X, 1}(a \bullet L_{X, 1}}_n x) \ldots)) 
& \text{ by Lemma~\ref{lemma:vector-set-actions-and-windowed-recurrences}}\\
%
& = \ourunderbracket{\lambda(a) \bullet L_{X, 1}(\lambda(a) \bullet L_{X, 1}(\ldots \bullet L_{X, 1}(\lambda(a) \bullet L_{X, 1}}_n x) \ldots)) 
& \text{ by Definition~\ref{definition:representation-of-function-composition} (a)}\\
& = \ourunderbracket{\smat{1\\ \lambda(a)} \bullet \left(\smat{1\\ \lambda(a)} \bullet \left(\ldots \bullet \left(\smat{1\\ \lambda(a)} \, \bullet \right. \right. \right.}_n x 
\left. \left. \left. \vphantom{\smat{1\\ \lambda(a)}} \right) \ldots \right) \right)
& \text { by Theorem~\ref{theorem:vector-set-actions-and-semi-associativity} Part 1 } \\
& =\mat{1\\ \lambda(a)}^{* n} \bullet x=\mat{1\\ \lambda(a)}^n \bullet x 
& \text{ by Theorem~\ref{theorem:vector-set-actions-and-semi-associativity} Part 3}
\end{align*}
Theorem~\ref{theorem:vector-set-actions-and-semi-associativity} shows that the set action $\bullet$ of  $\mathbb{Z}_{>0} \ltimes_L \Lambda$ on $X$ is semi-associative with companion operator $*$ on $\mathbb{Z}_{>0} \ltimes_L \Lambda$, and hence the exponentiation in $\mathbb{Z}_{>0} \ltimes_L \Lambda$ may be bracketed in any order. 
\end{proof}

\begin{remark}
The proof of Theorem~\ref{theorem:vector-windowed-recurrence-algorithm} be made less 
delicate by assuming that $*$ is associative, that $L_i(\lambda(a))=\lambda(L_i(a))$, that $L_i \circ L_j = L_{i+j}$, and that $L_i(\lambda_1 * \lambda_2) = L_i \lambda_1 * L_i \lambda_2$. For in that case $\mathbb{Z}_{>0} \ltimes_L \Lambda$ is associative by Lemma~\ref{lemma:semidirect-product-semigroup-function-form}. Furthermore,
\begin{align*}
y & = a \bullet(L_1 a \bullet(\ldots \bullet(L_{n-1} a \bullet L_{X, n} x) \ldots )) \\
  & = \left(\lambda(a) * L_1(\lambda(a)) * \ldots * L_{n-1}(\lambda(a))\right) \bullet L_{X, n} x
\end{align*}
and Algorithms~\ref{algorithm:vector-sliding-window-*-product} and \ref{algorithm:general-vector-sliding-window-*-product} therefore apply directly to the vector sliding window $*$-product $\lambda(a) * L_1(\lambda(a)) * \ldots * L_{n-1}(\lambda(a))$. However the proof of Theorem~\ref{theorem:vector-windowed-recurrence-algorithm} shows that these additional assumptions are unnecessary.
\end{remark}

Theorem~\ref{theorem:vector-windowed-recurrence-algorithm} immediately gives us an algorithm for computing windowed recurrences by computing $\smat{1\\ \lambda(a)}^n$ in the semidirect product magma $\mathbb{Z}_{>0} \ltimes_L \Lambda$, which may be nonassociative, using a semigroup algorithm for exponentiation, which assumes associativity, and then applying the result to $L_{X, n} x$ using $\bullet$. Any addition chain method may be used to compute the exponentiation, including binary exponentiation, Brauer's method, Thurber's method, or an optimal addition chain. The element of $\mathbb{Z}_{>0} \ltimes_L \Lambda$ computed by the exponentiation will depend on the method used in the nonassociative case, but the vector windowed recurrence will still be correctly computed due to semi-associativity. We write this algorithm out more explicitly in Section~\ref{sec:algorithms-for-vector-windowed-recurrences} and Chapter~\ref{chapter:vector-pseudo-code}. First we provide examples and constructions of vector representations of function composition.

\section{Constructions of Vector Representations of Function Composition}

The first four constructions we give are of theoretical interest, because they show that vector representations of function composition always exist, but they are of less use computationally. Examples~\ref{example:handwritten-notes-iii-7-5-5}--\ref{example:handwritten-notes-iii-7-5-8}, on the other hand, show that (non-vector) representations of function composition can be vectorized, and are of direct use constructing vector algorithms for windowed recurrences.

\begin{lemma}
Assume $F \subseteq\Endop(X)$ is a vector function action on $X$ with shift operators $L_i$, $L_{X, i}$, $i \geq 1$. Let
\begin{align*}
\Lambda & = \text{The free semigroup on $F$}\\
        & = \text{The set of finite sequences of length $\geq 1$ of elements of $F$}
\end{align*}
with semigroup product which is concatenation of sequences. I.e.,
\begin{equation*}
(f_1, \ldots, f_p) * (g_1, \ldots, g_q)=(f_1, \ldots, f_p, g_1, \ldots, g_q)
\end{equation*}
for $f_i, g_j \in F$. Let $\lambda(f) = (f)$ be the sequence of length $1$ consisting of $f$, 
and also define 
\begin{equation*}
(f_1, \ldots, f_p) \bullet x  = f_1(\ldots f_{p-1}(f_p(x))\ldots)
, \quad 
L_i\left((f_1, \ldots, f_p)\right) = (L_i f_1, \ldots, L_i f_p)
\end{equation*} 
where $f, f_1, \ldots, f_p \in F$, $x \in X$, $i \geq 1$. Then $(\Lambda, \lambda, *, \bullet)$ is a vector representation of function composition with shift operators $L_i \colon \Lambda \rightarrow \Lambda$ for the functions $F$ acting on $X$.
\end{lemma}
\begin{proof} 
$(\Lambda, \lambda, *, \bullet)$ is a representation of function composition by Example~\ref{example:rofc-examples-2}. To show that $(\Lambda, \lambda, *, \bullet)$ is a vector representation of function composition, observe that
\begin{align*}
L_{X, i}\left((f_1, \ldots, f_p) \bullet x\right) 
& = L_{X, i}\left(f_1(\ldots f_p(x) \ldots)\right) 
  = (L_i f_1)\left((L_i f_2)\left(\ldots (L_i f_p)\left(L_{X, i} x\right) \ldots \right)\right) \\
& = \left(L_i f_1, \ldots, L_i f_p\right) \bullet \left(L_{X, i} x\right)
  = L_i\left((f_1, \ldots, f_p)\right) \bullet L_{X, i} x
\end{align*}
\end{proof}
\begin{lemma}
Assume $\bullet\colon A \times X \rightarrow X$ is a vector set action of $A$ on $X$ with shift operators $L_i$, $L_{X, i}$, $i \geq 1$. Let 
\begin{align*}
\Lambda & = \text{The free semigroup on $A$}\\
        & = \text{The set of finite sequences of length $\geq 1$ of elements of $A$}
\end{align*}
with semigroup product which is concatenation of sequences. I.e.,
\begin{equation*}
(a_1, \ldots, a_p) * (b_1, \ldots, b_q) = (a_1, \ldots, a_p, b_1, \ldots, b_q)
\end{equation*}
for $a_j, b_j \in A$. Define $\lambda(a)=(a)$, i.e., the sequence of length $1$, and define
\begin{equation*}
(a_1, \ldots, a_p) \bullet x = a_1 \bullet (a_2 \bullet \ldots (a_p \bullet x) \ldots)
, \quad
L_i\left((a_1, \ldots, a_p)\right) = (L_i a_1, \ldots, L_i a_p)
\end{equation*}
Then $(\Lambda, \lambda, *, \bullet)$ is a vector representation of function composition with shift operators $L_i\colon \Lambda \rightarrow \Lambda$ for the vector set action $\bullet\colon A \times X \rightarrow X$ of $A$ on $X$.
\end{lemma}
\begin{proof}
$(\Lambda, \lambda, *, \bullet)$ is a representation of function composition by Example~\ref{example:rofc-examples-2}. To show that $(\Lambda, \lambda, *, \bullet)$ is a vector representation of function composition, observe that
\begin{align*}
L_{X, i}\left((a_1, \ldots, a_p) \bullet x\right) 
& = L_{X, i}\left(a_1 \bullet (a_2 \bullet \ldots (a_p \bullet x) \ldots )\right) 
  = L_i a_1 \bullet (L_i a_2 \bullet \ldots (L_i a_p \bullet L_{X, i} x) \ldots) \\
& = \left(L_i a_1, \ldots, L_i a_p\right) \bullet L_{X, i} x 
  = L_i\left((a_1, \ldots, a_p)\right) \bullet L_{X, i} x
\end{align*}
\end{proof}
\begin{lemma}
Assume $F \subseteq \Endop(X)$ is a vector function action on $X$ with shift operators $L_i$, $L_{X, i}$, $i \geq 1$. Assume further that $F$ is closed under function composition. Let $\; \smallblacksquare\colon F \times X \rightarrow X\colon (f, x) \mapsto f \smallbsq x = f(x)$ denote function application, and $\circ$ denote function composition. Then $(F, \id, \circ, \smallbsq)$, is a vector representation of function composition for the functions $F$, with shift operators $L_i$.
\end{lemma}
\begin{proof}
Function application is semi-associative with companion operator which is function composition. The result now follows directly from the definition of vector representation of function composition.
\end{proof}
\begin{lemma}
Assume  $\bullet\colon A \times X \rightarrow X$ is a vector set action of $A$ on $X$ with shift operators $L_i$, $L_{X, i}$, $i \geq 1$. Let $F=\left\langle \theset{\Leftop^\bullet_a \colon a \in A}\right\rangle$ be the subsemigroup of $\Endop(X)$ generated by the left action operators of $\bullet$. 
Assume further that there are operators $L_i\colon F \rightarrow F$, such that $L_i(\Leftop_a^\bullet) = \Leftop_{L_i a}^\bullet$ for $a \in A$, $i \geq 1$, and $L_i(f \circ g) = L_i f \circ L_i g$ for $f, g \in F$, $i \geq 1$. Let $\smallbsq$ denote function application, i.e., $f \smallbsq x = f(x)$. Then $(F, \Leftop^\bullet, \circ, \smallbsq)$ is a vector representation of function composition with shift operators $L_i$ for the vector set action $\bullet$ of $A$ on $X$, where $\Leftop^\bullet$ denotes the function $(a \longmapsto \Leftop^\bullet_a)\colon A \longrightarrow F$.
\end{lemma}
\begin{proof}
Let $\lambda=\Leftop^\bullet$. Then
\begin{align*}
& \lambda(a) \smallbsq x = \Leftop_a^\bullet \smallbsq x = \Leftop_a^\bullet(x) = a \bullet x \\
& f \smallbsq (g \smallbsq x) = f(g(x)) = (f \circ g)(x) = (f \circ g) \smallbsq  x \\
%
%
& L_{X, i}\left((\Leftop_{a_1}^\bullet \circ \ldots  \circ \Leftop_{a_p}^\bullet) \smallbsq x \right)
 = L_{X, i}\left(a_1 \bullet\left(\ldots \bullet\left(a_p \bullet x\right) \ldots\right)\right)\\
& \qquad \qquad 
  = \left(L_i a_1\right) \bullet \left(\ldots \bullet \left((L_i a_p) \bullet  L_{X, i} x\right) \ldots\right) 
 = \left(\Leftop_{L_i a_1} \circ  \ldots \circ \Leftop_{L_i a_p}\right) \smallbsq L_{X, i} x \\
 & \qquad \qquad 
 = \left(L_i(\Leftop^\bullet_{a_1}) \circ \ldots \circ L_i(\Leftop^\bullet_{a_1})\right) \smallbsq  L_{X, i} x 
 = L_i\left(\Leftop_{a_1}^\bullet \circ \ldots \circ \Leftop_{a_p}^\bullet\right) \smallbsq  L_{X, i} x
\end{align*}    
\end{proof}

\begin{example}
\label{example:handwritten-notes-iii-7-5-5}
Consider the vector function action of $\Endop(X)^N \subseteq \Endop(X^N)$ on $X^N$ of Example \ref{example:vector-function-action-from-functions}. Assume $F \subseteq \Endop(X)$, and $\id  \in F$. Then we have a vector function action of $F^N \subseteq \Endop\left(X^N\right)$ on $X^N$. 
Assume $(\Lambda, \lambda, *, \bullet)$ is a (non-vector) representation of function composition for $F \subseteq \Endop(X)$ acting on $X$. Define
\begin{align*}
& \lambda^N=\ourunderbracket{\lambda \times \ldots \times \lambda}_N \colon F^N \rightarrow \Lambda^N 
\text{ via }
(f_1, \ldots, f_N) \mapsto(\lambda(f_1), \ldots, \lambda(f_N)) \\
& L_i\colon \Lambda^N \rightarrow \Lambda^N 
\text{ via } 
(\lambda_1, \ldots, \lambda_N)   \mapsto (\ourunderbracket{\lambda(\id), \ldots, \lambda(\id)}_i, \lambda_1, \ldots, \lambda_{N-i})  \\
& *^N \colon \Lambda^N \times \Lambda^N \rightarrow \Lambda^N
\text{ via }
(\lambda_1, \ldots, \lambda_N) *(\mu_1, \ldots, \mu_N) = (\lambda_1*\mu_1, \ldots, \lambda_N * \mu_N) \\
& \bullet^N \colon \Lambda^N \times X^N \rightarrow X^N 
\text{ via } 
(\lambda_1, \ldots, \lambda_N) \bullet (x_1, \ldots, x_N) = (\lambda_1 \bullet x_1, \ldots, \lambda_N \bullet x_N)
\end{align*}
Then $(\Lambda^N, \lambda^N, *^N, \bullet^N)$ is a vector representation of function composition for the functions $F^N \subseteq\Endop(X)^N \subseteq \Endop(X^N)$ acting on $X^N$, with shift operators $L_i\colon \Lambda^N \rightarrow \Lambda^N$.    
\end{example}

\begin{example}
\label{example:handwritten-notes-iii-7-5-6}
Consider the vector set action of $A_1^N$ on $X^N$ of Example \ref{example:vector-set-action-fixed-length-sequences}. Assume $(\Lambda, \lambda, *, \bullet)$ is a (non-vector) representation of function composition for the set action $\bullet\colon A_1 \times X \rightarrow X$.\\
Define

\begin{align*}
& \lambda^N = \ourunderbracket{\lambda \times \ldots \times \lambda}_N \colon A_1^N \rightarrow \Lambda^N 
\text{ via } 
(a_1, \ldots, a_N) \mapsto (\lambda(a_1), \ldots, \lambda (a_N)) \\
& L_i\colon \Lambda^N \rightarrow \Lambda^N 
\text{ via }
(\lambda_1, \ldots, \lambda_N) \longmapsto (\ourunderbracket{\lambda(1_{A_1}), \ldots, \lambda(1_{A_1})}_i, \lambda_1, \ldots, \lambda_{N-i}) \\
& *^N \colon \Lambda^N \times \Lambda^N \rightarrow \Lambda^N 
\text{ via } 
(\lambda_1, \ldots \lambda_N) * (\mu_1, \ldots, \mu_N)=(\lambda_1 * \mu_1, \ldots, \lambda_N*\mu_N) \\
& \bullet^N \colon \Lambda^N \times X^N \rightarrow X^N 
\text{ via } 
(\lambda_1, \ldots, \lambda_N) \bullet(x_1, \ldots, x_N)=(\lambda_1 \bullet x_1, \ldots, \lambda_N \bullet x_N)
\end{align*}
Then $(\Lambda^N, \lambda^N, *^N, \bullet^N)$ is a vector representation of function composition for the set action of $A_1^N$ on $X^N$, with shift operators $L_i\colon \Lambda^N \rightarrow \Lambda^N$.
    
\end{example}

\begin{example}
\label{example:handwritten-notes-iii-7-5-7}
Consider the vector set action of $V_N(A)$ on $X^N$ of Example~\ref{example:vector-set-action-variable-length-sequences}. Assume $(\Lambda, \lambda, *, \bullet)$ is a (non-vector) representation of function composition for the set action $\bullet\colon A \times X \rightarrow X$. Let 
$
V_N(\Lambda) = \bigcup_{i=0}^N \Lambda^i = \theset{\text{sequences of elements of $\Lambda$ of length $\leq N$}}
$
and define, $\lambda\colon V_N(A) \longrightarrow V_N(\Lambda)$, $L_i\colon V_N(\Lambda) \rightarrow V_N(\Lambda)$,
$*\colon   V_N(\Lambda) \times V_N(\Lambda) \rightarrow V_N(\Lambda)$, and $\bullet\colon  V_N(\Lambda) \times X^N \rightarrow X^N$, by
\begin{align*}
& \lambda((a_1, \ldots, a_p)) = (\lambda(a_1), \ldots, \lambda(a_p)) \\
& L_i((\lambda_1, \ldots, \lambda_p)) 
    = \ourcases{
        (\lambda_1, \ldots, \lambda_{p-i}) & \text{if $i < p$}\\
        \text{the empty sequence ( )} & \text{if $i \geq p$}
} \\
& (\lambda_1, \ldots, \lambda_p) * (\mu_1, \ldots, \mu_q)
    = \ourcases{
        (\mu_1, \ldots, \mu_{q-p}, \lambda_1 * \mu_{q - p + 1}, \ldots, \lambda_p * \mu_q) 
        & \text{if $p < q$} \\
        (\lambda_1 * \mu_1, \ldots, \lambda_p * \mu_p)
        & \text{if $p = q$} \\
        (\lambda_1, \ldots, \lambda_{p - q}, \lambda_{p - q + 1} * \mu_1, \ldots, \lambda_p * \mu_q) 
        & \text{if $p > q$}
} \\
& (\lambda_1, \ldots, \lambda_p) \bullet(x_1, \ldots, x_N) 
    = (x_1, \ldots, x_{N - p}, \lambda_1 \bullet x_{N - p + 1}, \ldots, \lambda_p \bullet x_N)
\end{align*}
Then $(V_N(\Lambda), \lambda, *, \bullet)$ is a vector representation of function composition for the set action of $V_N(A)$ on $X^N$, with shift operators $L_i\colon V_N(\Lambda) \rightarrow V_N(\Lambda)$.
\end{example}

\begin{example}
\label{example:handwritten-notes-iii-7-5-8}
Consider the vector set action of $V_\infty(A)$ on $V_\infty(X)$ of Example~\ref{example:vector-set-action-unbounded-length-sequences}, and recall that this action depends on a choice of $x_0\in X$. Assume $(\Lambda, \lambda, *, \bullet)$ is a (non-vector) representation of function composition for the set action $\bullet\colon A \times X \rightarrow X$. The definitions of $\lambda$, $L_i$, $*$ given in Example~\ref{example:handwritten-notes-iii-7-5-7} clearly extend to functions $\lambda\colon V_\infty(A) \rightarrow V_\infty(\Lambda)$, $L_i\colon V_\infty(\Lambda) \rightarrow V_\infty(\Lambda)$, $*\colon V_\infty(\Lambda) \times V_\infty(\Lambda) \rightarrow V_\infty(\Lambda)$. Also define $\bullet\colon V_\infty(\Lambda) \times V_\infty(X) \rightarrow V_\infty(X)$ by
\begin{equation*}
(\lambda_1, \ldots, \lambda_p) \bullet (x_1, \ldots, x_q) = \ourcases{
(x_1, \ldots, x_{q-p}, \lambda_1 \bullet x_{q - p + 1}, \ldots, \lambda_p \bullet x_q)
    & \text{if $p < q$} \\
(\lambda_1 \bullet x_1, \ldots, \lambda_p \bullet x_p) 
    & \text{if $p = q$} \\
(\lambda_1 \bullet x_0, \ldots, \lambda_{p - q} \bullet x_0, \lambda_{p - q + 1} \bullet x_1, \ldots, \lambda_p \bullet x_q) 
    & \text{if $p > q$}
}
\end{equation*}
Then $(V_\infty(\Lambda), \lambda, *, \bullet)$ is a vector representation of function composition for the set action of $V_\infty(A)$ on $V_\infty(X)$, with shift operators $L_i\colon V_\infty(\Lambda) \rightarrow V_\infty(\Lambda)$.%
\footnote{Note again that the shift operators $L_i^{x_0}$ for Examples~\ref{example:vector-set-action-variable-length-sequences} and \ref{example:handwritten-notes-iii-7-5-7} differ from the shift operators $L_{V_{\infty}(X), i}$ used in Examples~\ref{example:vector-set-action-unbounded-length-sequences} and \ref{example:handwritten-notes-iii-7-5-8}.
}
\end{example}

\section{Algorithms for Vector Windowed Recurrences}
\label{sec:algorithms-for-vector-windowed-recurrences}

The following algorithm is an immediate consequence of Theorem~\ref{theorem:vector-windowed-recurrence-algorithm}.

\begin{algorithm}
\label{algorithm:vector-windowed-recurrence}
Assume $\bullet\colon A \times X \rightarrow X$ is a  vector set action of $A$ on $X$ with shift operators $L_i$, $L_{X, i}$, $i \geq 1$. Then the vector windowed recurrence of length $n \geq 1$ corresponding to $a \in A$, $x \in X$ may be computed as follows.

\begin{description}
\item[Step 1] 
Choose a vector representation of function composition $(\Lambda, \lambda, *, \bullet)$ with shift operator $L_i\colon \Lambda \rightarrow \Lambda$, for the set action $\bullet\colon A \times X \rightarrow X$.

\item[Step 2]
Choose a semigroup exponentiation procedure $\exponentiate(*, x, n)$ that exponentiates solely by computing products in a pattern determined by the strictly positive integer $n$. I.e., an addition chain exponentiation procedure. E.g., Binary exponentiation, or Brauer's method or Thurber's method, or an optimal addition chain exponentiation. The operator $*$ passed to this procedure, however, is not required to be associative.

\item[Step 3]
Form $\smat{1 \\ \lambda(a)} \in \mathbb{Z}_{>0} \ltimes_L \Lambda$.

\item[Step 4]
Call  $\exponentiate(*, \smat{1 \\ \lambda(a)}, n)$, where $*$ is the semidirect product operator on $\mathbb{Z}_{>0} \ltimes_L \Lambda$. During the call to $\exponentiate$, compute any $*$ products as if they were associative, even though $*$ may not be associative. I.e., pretend that
$\left(\mathbb{Z}_{>0} \ltimes_L \Lambda\right) \times
 \left(\mathbb{Z}_{>0} \ltimes_L \Lambda\right)
 \rightarrow \mathbb{Z}_{>0} \ltimes_L \Lambda$
is associative even if it is not. This will compute $\zeta = \smatl{1\\ a}^n$ with some bracketing which depends on the exponentiation algorithm used.

\item[Step 5]
Compute $\zeta \bullet x = z \bullet L_{X, n} x$, where $z$ is given by $\zeta = \smat{1\\ \lambda(a)}^n = \smat{n\\ z}$.
\end{description}
\end{algorithm}

\begin{remarks} \
\begin{enumerate}
\item 
The complexity of this algorithm is determined by the complexity of the exponentiation algorithm used and the complexity of operations in $\Lambda$. E.g., if parallel binary exponentiation is used then the number of parallel steps involving a $*$-product in $\Lambda$ is $\ceiling{\log_2 n}$. If Brauer's method or Thurber's method is used then no more than $(\log_2 n)(1 + \frac{1}{\log_2 \log_2(n+2)} + o(\frac{1}{\log_2 \log_2(n + 2)}))$ $*$-operations in $\Lambda$ are required. The method of Yao \cite{Yao1976} may also be used to compute windowed recurrences with multiple window lengths simultaneously, and binary exponentiation or Yao's method may also be combined with successive squaring in $\mathbb{Z}_{>0} \ltimes_L \Lambda$ to compute non-windowed (i.e., prefix) recurrences simultaneously with windowed recurrences.

\item
This algorithm also gives new algorithms for computing parallel prefix sums (parallel prefix $*$-products) in the nonassociative case and for computing parallel non-windowed recurrences. I.e., for computing
$a_i \bullet(a_{i-1} \bullet(\ldots \bullet(a_2 \bullet a_1) \ldots))$, or 
$f_i(f_{i-1}(\ldots f_2(f_1(x_0)) \ldots))$.

\item
Algorithm \ref{algorithm:vector-windowed-recurrence} gives a family of vectorized and parallel algorithms for computing windowed recurrences, and set action windowed recurrences, when combined with the constructions in Examples \ref{example:handwritten-notes-iii-7-5-5}, \ref{example:handwritten-notes-iii-7-5-6}, \ref{example:handwritten-notes-iii-7-5-7}, \ref{example:handwritten-notes-iii-7-5-8}.

\end{enumerate}    
\end{remarks}

%% file: htcams-arxiv-ch15-pseudo-code.tex
\chapter{Pseudo-Code for the Vector Algorithms} 
\label{chapter:vector-pseudo-code}

We now return to the pseudo-code for Algorithms~\ref{algorithm:general-vector-sliding-window-*-product} and \ref{algorithm:vector-windowed-recurrence} that we gave in Sections~\ref{sec:algorithms-for-vector-sliding-window-*-products-section} and \ref{sec:algorithms-for-vector-windowed-recurrences}.

\section{Pseudo-Code}



\begin{alltt}
window_compose(compose, shift, a, n, exponentiate):
    define semidirect_product(u, v):
        return (u[1] + v[1], compose(u[2], shift(u[1], v[2]))) 
    return exponentiate(semidirect_product, (1, a), n)[2] 

window_apply(compose, apply, lift, shift, shiftx, n, a, x, exponentiate):
    function_data = window_compose(compose, shift, lift(a), n, exponentiate) 
    return apply(function_data, shiftx(n, x))
\end{alltt}

\subsubsection*{How to use {\tt window\_compose}}

The {\tt window\_compose} procedure has two uses. One is to compute vector sliding window $*$-products in the case where $*$ is associative, and the other use is to be called by {\tt window\_apply} as part of the computation of a windowed recurrence, and which may involve nonassociative operations. We describe the sliding window $*$-product case here, and leave the windowed recurrence case to the description of {\tt window\_apply} usage. 

To use {\tt window\_compose} to compute a vector sliding window $*$-product one must define procedures {\tt compose}, {\tt shift}, and {\tt exponentiate}, and these will be passed in to the {\tt window\_compose} procedure. To describe the conditions these procedures shall satisfy, it is also helpful to think of objects as having types.

\bigskip
\begin{minipage}[t]{\dimexpr.3\textwidth-1.0\columnsep}
{\tt{compose(a$_1$, a$_2$)}}
\end{minipage}
\begin{minipage}[t]{\dimexpr.7\textwidth-1.0\columnsep}
The {\tt compose} procedure takes two objects of type $A$ and returns an object of the same type $A$.
\end{minipage} 

\smallskip

\begin{minipage}[t]{\dimexpr.3\textwidth-1.0\columnsep}
{\tt{shift(i, a)}}
\end{minipage}
\begin{minipage}[t]{\dimexpr.7\textwidth-1.0\columnsep}
Takes a strictly positive integer {\tt i}, and an object of type $A$ and returns an object of type $A$. 
\end{minipage} 

\smallskip

\begin{minipage}[t]{\dimexpr.3\textwidth-1.0\columnsep}
{\tt{a}}
\end{minipage}%
\begin{minipage}[t]{\dimexpr.7\textwidth-1.0\columnsep}
Is an object of type $A$. This is the data from which the vector sliding window $*$-product will be computed. 
\end{minipage} 

\smallskip

\begin{minipage}[t]{\dimexpr.3\textwidth-1.0\columnsep}
{\tt{n}}
\end{minipage}%
\begin{minipage}[t]{\dimexpr.7\textwidth-1.0\columnsep}
Is a strictly positive integer. {\tt n} is the window length. 
\end{minipage} 

\smallskip

\begin{minipage}[t]{\dimexpr.3\textwidth-1.0\columnsep}
{\tt{exponentiate(*, u, n)}}
\end{minipage}
\begin{minipage}[t]{\dimexpr.7\textwidth-1.0\columnsep}
The {\tt exponentiate} procedure is a semigroup exponentiation procedure that takes a binary operation $*$, and computes ${\tt{u}} * \ldots * {\tt{u}} = {\tt{u}}^{*{\tt{n}}}$ according to some bracketing scheme. E.g., binary exponentiation (sequential or parallel), Brauer's method, Thurber's method, or optimal addition chain exponentiation. {\tt{u}} will be a pair {\tt{(i, a)}}  of a strictly positive integer and an object of type $A$. 
\end{minipage} 

\bigskip

\noindent In order for {\tt window\_compose} to correctly compute a vector sliding window $*$-product with {\tt compose} as the $*$ operation, {\tt compose} and {\tt shift} should satisfy the following properties.

\begin{alltt}
      \(\hspace*{0.075em}\) compose(a\(\sb{1}\),compose(a\(\sb{2}\),a\(\sb{3}\))) = compose(compose(a\(\sb{1}\),a\(\sb{2}\)),a\(\sb{3}\))
           \;\(\hspace*{0.08em}\) shift(i, shift(j,a)) = shift(i+j,a)
          shift(i,compose(a\(\sb{1}\),a\(\sb{2}\))) = compose(shift(i,a\(\sb{1}\)),shift(i,a\(\sb{2}\)))
\end{alltt}

\noindent This follows from Theorem~\ref{theorem:semidirect-product-vector-sliding-window} and Lemma~\ref{lemma:semidirect-product-semigroup-function-form}.

\subsubsection*{How to use {\tt window\_apply}}

To use {\tt window\_apply} to compute a vector windowed recurrence (for a vector function action or vector set action) one must define procedures {\tt compose}, {\tt apply}, {\tt lift}, {\tt shift}, {\tt shiftx}, and {\tt exponentiate}, and these will be passed to the {\tt window\_apply} procedure. There are three types of objects involved in the algorithm, in addition to integer, Boolean and tuple types, and we denote these types $A$, $\Lambda$, and $X$.

\bigskip

\begin{minipage}[t]{\dimexpr.3\textwidth-1.0\columnsep}
{\tt{lift(a)}}
\end{minipage}
\begin{minipage}[t]{\dimexpr.7\textwidth-1.0\columnsep}
Takes an object  of type $A$ and returns an object of type $\Lambda$. This is the lift function of a vector representation of function composition.
\end{minipage} 

\smallskip

\begin{minipage}[t]{\dimexpr.3\textwidth-1.0\columnsep}
{\tt{compose(z$_1$, z$_2$)}}
\end{minipage}
\begin{minipage}[t]{\dimexpr.7\textwidth-1.0\columnsep}
Takes two objects of type $\Lambda$ and returns an object of type $\Lambda$. This is the compose operation of a vector representation of function composition.
\end{minipage}

\smallskip

\begin{minipage}[t]{\dimexpr.3\textwidth-1.0\columnsep}
{\tt{apply(z, x)}}
\end{minipage}
\begin{minipage}[t]{\dimexpr.7\textwidth-1.0\columnsep}
Takes an object of type $\Lambda$ and an object of type $X$ and returns an object of type $X$. This is the apply operation of a vector representation of function composition.
\end{minipage}

\smallskip

\begin{minipage}[t]{\dimexpr.3\textwidth-1.0\columnsep}
{\tt{shift(i, z)}}
\end{minipage}
\begin{minipage}[t]{\dimexpr.7\textwidth-1.0\columnsep}
Takes a strictly positive integer {\tt i}, and an object {\tt z} of type $\Lambda$ and returns an object of type $\Lambda$. 
\end{minipage} 

\smallskip

\begin{minipage}[t]{\dimexpr.3\textwidth-1.0\columnsep}
{\tt{shiftx(i, x)}}
\end{minipage}
\begin{minipage}[t]{\dimexpr.7\textwidth-1.0\columnsep}
Takes a strictly positive integer {\tt i}, and an object {\tt x} of type $X$ and returns an object of type $X$. 
\end{minipage} 

\smallskip

\begin{minipage}[t]{\dimexpr.3\textwidth-1.0\columnsep}
{\tt{a}}
\end{minipage}%
\begin{minipage}[t]{\dimexpr.7\textwidth-1.0\columnsep}
Is an object of type $A$. This is the function data from which the windowed recurrence will be computed. The element {\tt a} represents the left action functions of the windowed recurrence and is lifted by the {\tt lift} procedure to a potentially alternative representation for which function composition can be computed. 
\end{minipage} 

\smallskip

\begin{minipage}[t]{\dimexpr.3\textwidth-1.0\columnsep}
{\tt{n}}
\end{minipage}%
\begin{minipage}[t]{\dimexpr.7\textwidth-1.0\columnsep}
Is a strictly positive integer. {\tt n} is the window length. 
\end{minipage} 

\smallskip

\begin{minipage}[t]{\dimexpr.3\textwidth-1.0\columnsep}
{\tt{x}}
\end{minipage}%
\begin{minipage}[t]{\dimexpr.7\textwidth-1.0\columnsep}
Is an object of type $X$. It is the initial value data from which the windowed recurrence will be computed. 
\end{minipage} 

\smallskip

\begin{minipage}[t]{\dimexpr.3\textwidth-1.0\columnsep}
{\tt{exponentiate(*, u, n)}}
\end{minipage}
\begin{minipage}[t]{\dimexpr.7\textwidth-1.0\columnsep}
The {\tt exponentiate} procedure that takes a binary operation $*$  and computes ${\tt{u}} * \ldots * {\tt{u}} = {\tt{u}}^{*{\tt{n}}}$ according to some bracketing scheme. The bracketing does not affect the result of the result produced by {\tt window\_apply} provided the properties listed below hold. {\tt exponentiate} may be any addition chain method for exponentiation. E.g., binary exponentiation (sequential or parallel), Brauer's method, Thurber's method, or optimal addition chain exponentiation. {\tt u} will be a pair {\tt (i, z)}  of a strictly positive integer {\tt i} and an object {\tt z} of type $\Lambda$. 
\end{minipage} 

\bigskip

\noindent In order for {\tt window\_apply} to correctly compute a vector windowed recurrence, the procedures {\tt compose}, {\tt apply}, {\tt shift}, and {\tt shiftx} should satisfy the following properties.


\begin{alltt}
          apply(w,apply(z,x)) = apply(compose(w,z),x)
         shiftx(i,apply(z,x)) = apply(shift(i,z),shiftx(i,x))
        shiftx(i,shiftx(j,x)) = shiftx(i+j,x)
\end{alltt}

\noindent This follows from Theorem~\ref{theorem:vector-windowed-recurrence-algorithm}, and the set action for which the windowed recurrence is computed is $\text{\tt{(a,x)}} \mapsto \text{\tt{apply(lift(a),x)}}$. Note that the procedure {\tt apply} which is passed in to {\tt window\_apply} is only ever applied to the {\tt{x}} which is passed in, and therefore for many applications a full implementation of apply need not be provided, but instead only an implementation that works for the {\tt{x}} value (or values) in which we are interested. For example, for the evaluation of a nonassociative windowed $*$-product where the $*$ operation has a right unit $1$, {\tt apply} can be set to the function ${\tt{apply(z, x)}} = {\tt{z}} * 1$ (see Example~\ref{example:vector-set-action-from-vector-*-product}), as {\tt apply} will only be called with ${\tt{x}} = 1$ (or ${\tt{x}} = 0$ if the $*$ operation is interpreted as being in `additive notation').

\section{Examples}

\begin{example}
Assume $*$ is an associative binary operation with right unit $1$, and let $N$ be a strictly positive integer. We define {\tt compose} to operate on pairs of arrays of length $N$, and {\tt shift} to operate on an integer and an array of length $N$.
 



%
%

\begin{alltt}
compose(a, b):
    return (a[1]*b[1],..., a[N]*b[N])                                  a vector *-product     

shift(i, a):  
    j = min(i, N)
    return (\(\underbrace{\text{\tt{1,...,1}}}\sb{\tt{j}}\),a[1],a[1],...,a[N-j]) 
\end{alltt}
Let {\tt exponentiate} be any addition chain exponentiation procedure, e.g., we could choose the procedure {\tt binary\_exponentiate\_no\_flip}, or {\tt parallel\_binary\_exponentiate} with {\tt flip = false}, etc. Then the following is a procedure for computing sliding window $*$-products.

\begin{alltt}
window_product(a, n):
    return window_compose(compose, shift, a, n, exponentiate)
\end{alltt}
\end{example}

\begin{example}
\label{example:windowed-linear-recurrence-pseudo-code-fixed-length}
Assume that $*$ is an associative binary operation which may or may not have a right unit (i.e.\ a right unit is not required). Define {\tt compose} and {\tt shift} to operate on arrays of finite length as follows

\begin{alltt}
compose(a, b): 
    M = length(a), N = length(b)
    if M >= N
        return (a[1],...,a[M-N],a[M-N+1]*b[1],...,a[M]*b[N])  
    else
        return (b[1],...,b[N-M],a[1]*b[N-M+1],..., a[M]*b[N])

shift(i, a):
    N = length(a) 
    return (a[1], a[2],..., a[N-i])
\end{alltt}

\noindent Then the following procedure computes sliding window $*$-products 

\begin{alltt}
window_product(a, n):  
    return window_compose(compose, shift, a, n, exponentiate) 
\end{alltt}
\end{example}

\noindent During the computation of {\tt window\_product(a, n)} the procedure {\tt compose} is only ever called with $\text{\tt{length(a)}} \geq \text{\tt{length(b)}}$, and therefore the definition of {\tt compose} may be simplified to the following partial definition.

\begin{alltt}
compose(a, b): 
    M = length(a), N = length(b)
    return (a[1],...,a[M-N],a[M-N+1]*b[1],...,a[M]*b[N])  
\end{alltt}

\begin{example}
To compute a windowed linear recurrence
\begin{equation*}
v_i + u_i (v_{i-1} + u_{i-1} (\ldots + u_{i-n+2}(v_{i-n+1} + u_{i-n+1} x_{i-n}) \ldots))
\end{equation*}
we define
\begin{alltt}
window_linear_recurrence(u, v, x, n):
    return window_apply(compose, apply, identity, shift, shiftx, n, (u, v), x, 
                        exponentiate)
\end{alltt}
where the inputs {\tt u, v, x} are arrays of length $N$, and {\tt compose}, {\tt apply}, {\tt shift}, {\tt shiftx} are as follows, and {\tt identity(a) = a}. We need a mechanism to pass in the initial value $x_0$ in addition to $x_1,\ldots, x_N$ and a convenient way to do this is to pass the values $x_0, \ldots, x_{N-1}$ in the array {\tt x}. Thus, the array {\tt x} should contain the initial values $x_0, \ldots, x_{N-1}$, and so already has a shift of $1$. This is reflected in the definition of {\tt shiftx}.%
\footnote{
 An alternative would be to pass the initial value $x_0$ separately, and instead pass $x_1, \ldots, x_{N-1}$ in the array {\tt x}. This would require a definition of {\tt shiftx} different from the one given, and which shifts by $i$, as expected, rather than $i-1$. Both approaches are practical.
}
\begin{alltt}
compose(a, b):
    u = a[1], v = a[2], w = b[1], z = b[2]
    return (u * w, v + u * z)                vector addition and multiplication of arrays
\end{alltt}
where {\tt a=(u,v), b=(w,z)} are pairs of arrays and $*$, $+$ are componentwise multiplication and addition respectively.
\begin{alltt}
apply(a, x):
    u = a[1], v = a[2] 
    return v + u * x                         vector addition and multiplication of arrays
\end{alltt}

\begin{alltt}
shift(i, a): 
    u = a[1], v = a[2], N = length(v), j = min(i, N) 
    return ([\(\underbrace{\text{\tt{1,...,1}}}\sb{\tt{j}}\),u[1],...,u[N-j]],[\(\underbrace{\text{\tt{0,...,0}}}\sb{\tt{j}}\),v[1],...,v[N-j]])
\end{alltt}
For the definition of {\tt shiftx}, we shift by $i - 1$ rather than $i$ in order to simplify the handling of the initial value $x_0$. This works because {\tt shiftx} is only applied to ${\tt{x}} = (x_0, \ldots, x_{N-1})$, which is already shifted by $1$.
\begin{alltt}
shiftx(i, x):
    N = length(x), j = max(0, min(i - 1, N)), x0 = x[1] 
    return [\(\underbrace{\text{\tt{x0,...,x0}}}\sb{\tt{j}}\),x[1],...,x[N-j]]
\end{alltt}
These procedures may also be used to compute sliding window sums with scale changes, using either of the following procedures.
\begin{alltt}
window_sum_with_scale_changes(u, v, n):
    N = length(v)
    return window_linear_recurrence(u, v, [\(\underbrace{\text{\tt{0,...,0}}}\sb{\tt{N}}\)], n)
\end{alltt}
\begin{alltt}
window_sum_with_scale_changes(u, v, n):
    if n = 1
        return v
    else
        N = length(v)
        return window_linear_recurrence(u, v, [0,v[1],...,v[N-1]], n - 1)
\end{alltt}
\end{example}

\begin{example}
\label{example:windowed-linear-recurrence-pseudo-code-variable-length}
Another approach to windowed linear recurrences, following Example \ref{example:handwritten-notes-iii-7-5-8} is to define
\begin{alltt}
compose(a, b):
    u = a[1], v = a[2], w = b[1], z = b[2], M = length(v), N = length(z) 
    if M >= N  
        return ([u[1],...,u[M-N],u[M-N+1]*w[1],          ...,u[M]*w[N]     ],
                [v[1],...,v[M-N],v[M-N+1]+u[M-N+1]*z[1], ...,v[M]+u[M]*z[N]])
    else
        return ([w[1],...,w[N-M],u[1]*w[M-N+1],      ...,u[M]*w[N]     ],  
                [z[1],...,z[N-M],v[1]+u[1]*z[N-M+1], ...,v[M]+u[M]*z[N]]) 

apply(a, x): 
    u = a[1], v = a[2], M = length(v), N = length(x), i = abs(M - N), x0 = x[1]
    if  M >= N 
        return (v[1]+u[1]*x0,...,v[i]+u[i]*x0,v[i+1]+u[i+1]*x[1],  ...,v[M]+u[M]*x[N])
    else 
        return (x[1],        ...,x[i],        v[1] + u[1] * x[i+1],...,v[M]+u[M]*x[N])

shift(i, a):
    u = a[1], v = a[2], N = length(v)
    return ([u[1],...,u[N-i]],[v[1],..., v[N-i]])

shiftx(i, x):
    N = length(x), j = max(0, i - 1), k = max(1, N - j) 
    return (x[1],...,x[k])
\end{alltt}
\end{example}
\noindent where {\tt abs} denotes the absolute value function. In this example the array {\tt x} again contains the initial values $x_0, \ldots, x_{N-1}$. 
We have again modified {\tt shiftx} so that it shifts by $i-1$ rather than $i$, to compensate for the presence of $x_0$ in entry $1$ of the array {\tt x}. In the implementation of {\tt apply} we use the assumption that $\text{\tt{x[1]}}= x_0$, which will be true for the array it receives during the computation of the windowed linear recurrence.
Also note that in the definitions of {\tt compose} and {\tt apply} only the cases $\text{\tt{M}} \geq \text{\tt{N}}$ are used during the evaluation of {\tt window\_linear\_recurrence(u, v, x, n)}, and therefore these procedures may be simplified if desired.

\begin{example}
We now consider the parallel (vector) computation of sliding window continued fractions, as in Example~\ref{example:sliding-window-continued-fractions}, and Example~\ref{example:rofc-examples-5}.
Assume, for simplicity, that {\tt a} is an array of strictly positive numbers of length {\tt N} with elements ${\tt{a[1]}}, {\tt{a[2]}}, \ldots > 0$. Define the following procedures.

\begin{alltt}
lift(a):
    return (\(\mat{{\text{\tt{a[1] 1}}}\\{\text{\tt{  1  0}}}}\),\(\mat{{\text{\tt{a[2] 1}}}\\{\text{\tt{  1 
\;\; 0}}}}\),...,\(\mat{{\text{\tt{a[N] 1}}}\\{\tt{  1  0}}}\))
\end{alltt}
For two arrays {\tt{A, B}} of length {\tt{N}}, of $2 \times 2$ matrices, define 
\begin{alltt}
compose(A, B):
    Z = A[1]*B[1], ..., A[N]*B[N]                              * is matrix multiplication
    return Z[1]/\(\|\)Z[1]\(\|\sb1\), ..., Z[N]/\(\|\)Z[N]\(\|\sb1\)           / is division of a matrix by a scalar
\end{alltt}
where {\tt A[i]*B[i]} is a $2 \times 2$ matrix product and 
\begin{equation*}
\left\|\matl{z_{11} & z_{12}\\ z_{21} & z_{22}}\right\|_1
    = \max\theset{|z_{11}| + |z_{21}|, |z_{12}| + |z_{22}|}
\end{equation*}
For {\tt apply} we only implement the case corresponding to ${\tt{x}}=\ourunderbracket{\infty, \ldots, \infty}_N$, as $\infty$ is a right unit for $a * b = a + 1 / b$, but to do this we do not need to define $\infty$ or even represent $\infty$. Instead we need only define {\tt apply} so that it returns what would be returned if ${\tt{x}} = \infty, \ldots, \infty$ were to be defined and were to be passed to the procedure.

\begin{alltt}
apply_infinite_x(A, x):
    return (A[1]\({\sb{\!11}}\)/A[1]\({\sb{\!21}}\),..., A[N]\({\sb{\!11}}\)/A[N]\({\sb{\!21}}\))                       / is division of numbers
\end{alltt}
We also don't need to define {\tt shiftx} for all cases, since we will be ignoring {\tt x} and assuming {\tt shiftx(i, x)} is $(\infty, \ldots, \infty)$.
\begin{alltt}
dummy_shift_x(i, x):
    return x                                                        \(\,\)This is a dummy value
\end{alltt}
Finally we define {\tt shift} which we do by inserting the identity matrix, which is the lift of the action of the identity function.%
\footnote{
Note that in this example none of original functions in the recurrence is the identity, but the lift of the identity function is present in our set of objects supported by {\tt compose}.
}

\begin{alltt}
shift(i, A):
    j = min(i, N)
    return (\(\underbrace{\mat{{\tt{1 0}}\\{\tt{0 1}}}{\text{\tt{,...,}}}\mat{{\tt{1 0}}\\{\tt{0 1}}}}\sb{\tt{j}}\),A[1],..., A[N-j])
\end{alltt}
Then we may compute a sliding window continued fraction using the following procedure.
\begin{alltt}
window_continued_fraction(a, n):
    return window_apply(compose, apply_infinite_x, lift, shift, dummy_shift_x, 
                        n, a, a, exponentiate)
\end{alltt}
Note that the second {\tt{a}} that we pass is ignored, so it could be replaced by any object of the correct type, e.g., an array of zeros of length {\tt{N}}. 
\end{example}

\section{Multi-Query Pseudo-Code}

The code for simultaneously computing vector sliding window $*$-products or vector windowed recurrences with multiple window lengths is as compact and simple as for the single window length case.

\begin{alltt}
multi_window_compose(compose, shift, a, window_lengths, multi_exponentiate): 
    define semidirect_product(u, v):
        return (u[1] + v[1], compose(u[2], shift(u[1], v[2])))
    powers = multi_exponentiate(semidirect_product, (1, a), window_lengths) 
    p = length(window_lengths)
    return (powers[1][2], powers[2][2], ..., powers[p][2])

multi_window_apply(compose, apply, lift, shift, shiftx, window_lengths, a, x,
                   multi_exponentiate): 
    fd = multi_window_compose(compose, shift, lift(a), window_lengths,
                              multi_exponentiate)
    p = length(window_lengths) 
    return (apply(fd[1], shiftx(window_lengths[1], x)), 
            apply(fd[2], shiftx(window_lengths[2], x)),
            ...
            apply(fd[p], shiftx(window_lengths[p], x)))
\end{alltt}

Here is an explanation of the parameters {\tt window\_lengths}, and {\tt multi\_exponentiate}.

\bigskip

\begin{minipage}[t]{\dimexpr.43\textwidth-1.0\columnsep}
{\tt{window\_lengths}}
\end{minipage}
\begin{minipage}[t]{\dimexpr.57\textwidth-1.0\columnsep}
Is a tuple of strictly positive integers $n_1, \ldots, n_p \geq 1$, which are the window lengths for which the sliding window $*$-products or windowed recurrences are to be computed.
\end{minipage}

\smallskip

\begin{minipage}[t]{\dimexpr.43\textwidth-1.0\columnsep}
{\tt{multi\_exponentiate(*, u,\\\phantom{multi\_exponentiate(}window\_lengths)}}
\end{minipage}
\begin{minipage}[t]{\dimexpr.57\textwidth-1.0\columnsep}
Is a procedure which takes a binary operation $*$, a tuple of window lengths $(n_1, \ldots, n_p)$, and computes $\ourunderbracket{{\tt{u}} * \ldots * {\tt{u}}}_{n_i} = u^{*n_i}$ for $i=1, \ldots, p$ for some 
bracketing of each power $u^{*n_i}$. E.g.,\ this could be an implementation of Yao's algorithm \cite{Yao1976}, or for the case $p=2$, $n_2 = 2^k n_1$ (such as used for simultaneous computation of a prefix $*$-product or prefix recurrence with a sliding window $*$-product or windowed recurrence) could be binary exponentiation (sequential or parallel), or Brauer's or Thurber's method, followed by successive squaring.
\end{minipage}

%% file: htcams-arxiv-ch16-examples-and-constructions.tex
\chapter{Representations of Function Composition -- Examples and Constructions}
\label{chapter:examples}

\section{Guide to the Examples}

We now present a range of examples, of practical interest, of representations of function composition, and present constructions for producing new examples from existing examples. These are intended primarily for practitioners interested in applying the algorithms of Chapters \ref{chapter:moving-sums}--\ref{chapter:vector-pseudo-code} to their calculations at hand. The examples fall broadly into the following (overlapping) categories: 

\begin{enumerate}[label=\alph*.]
\item 
Semi-associative set actions with their companion operations.

\item 
Representations of function composition for non-semi-associative set actions and for left actions of nonassociative binary operations.

\item 
Representations of function composition for collections of functions acting on sets.

\item 
Associative binary operations, i.e., semigroups.

\item 
Constructions for combining examples in categories a-d to produce new examples.

\end{enumerate}

\noindent The examples, with few exceptions (notably the first example), follow the following pattern. They start with a recurrence relation
\begin{equation*}
x_i = f_{a_i} (x_{i-1})
\end{equation*}
from which a mapping $a \mapsto f_a$ or an equivalent set action $(a, x) \mapsto a \bullet x = f_a(x)$ is extracted. We then describe the associativity or semi-associativity properties of the mapping or set action, or alternatively describe a representation of function composition. If warranted, we also discuss the interpretation of the corresponding windowed recurrence.

Given these examples, the practitioner will be able to perform the following calculations for the corresponding recurrences.

\begin{enumerate}[label=\alph*.]
\item 
Parallel Reduction. I.e., the computation of $f_{a_N}(\ldots f_1(x_0)\ldots)$ or $a_N \bullet(\ldots \bullet (a_1 \bullet x_0) \ldots)$ or $a_N * (\ldots * (a_2 * a_1) \ldots)$.

\item 
Parallel Scans/Prefix Sums. I.e., the computation of $f_{a_i}(\ldots f_1(x_0)\ldots)$ or $a_i \bullet (\ldots \bullet (a_1 \bullet x_0) \ldots )$ or $a_i * (\ldots * (a_2 * a_1) \ldots )$, for $i=1, \ldots, N$, using parallel algorithms.

\item
Sequential Windowed Recurrences. I.e., the computation of $f_{a_i}(\ldots f_{i-n+1}(x_{i-n})\ldots)$ or $a_i \bullet(\ldots \bullet (a_{i-n+1} \bullet x_{i-n}) \ldots )$ or $a_i * (\ldots * (a_{i-n+2} * a_{i-n+1}) \ldots )$, for $i=1, \ldots, N$, using sequential algorithms such as Two Stacks, DEW, or DABA Lite.

\item
Vector and Parallel Windowed Recurrences. I.e., the computation of $f((L_1 f)(\ldots (L_{n-1} f)(L_{X, n} x) \ldots))$ or $a \bullet (L_1(a) \bullet (\ldots \bullet(L_{n-1}(a) \bullet L_{X, n}(x)) \ldots))$ or $a * (L_1(a) * ( \ldots * (L_{n-2}(a) * L_{n-1}(a)) \ldots ))$, where $a$, and $x$ refer to the entire sequences of data, $f$ operates on the data in a vectorized fashion, and the $L_i$, $L_{X, i}$ are shift operators.
\end{enumerate}

Our guiding principle in producing these examples is to treat them algebraically. Given a set action, or equivalently a collection of functions acting on a set, the direct way to find a representation of function composition is to compose two functions and look for a common parameterization of the functions and their composition. In cases where no efficient parameterization exists it may also be possible to prove that the amount of information required to describe successive iterations of function composition grows in a way that precludes efficient parametrization. In the case where the collection of functions contains a parametrized semigroup of functions, the problem becomes one of finding a larger parameterized semigroup that contains the original semigroup together with the functions not in the semigroup. I.e., we are looking for a semigroup extension. Furthermore, in all these set action cases, we must also describe function application as well as function composition.



\section{Examples and Constructions}
\label{sec:examples-and-constructions}

\subsubsection*{Example 1. Common Associative Operators}

We start with a listing of commonly used associative operators.
\medskip

{ 
\renewcommand{\arraystretch}{1.1}
\begin{center}
\begin{tabular}{l|m{3.75in}}
Operation &  Notes\\
\hline 
addition & Addition of numbers. There are many number systems.\\
multiplication & Multiplication of numbers. \\
matrix multiplication & This subsumes most of our examples. \\
and & Acting on $\theset{T, F}$, or alternatively bitwise on integers. \\
or & Acting on $\theset{T, F}$, or alternatively bitwise on integers. \\
exclusive or & Acting on $\theset{T, F}$, or alternatively bitwise on integers. \\
first & $\operatorname{first}(x, y) = x$ \\
last & $\operatorname{last}(x, y) = y$ \\
coalesce & $\coalesce(x, y) = (\text{$y$ if $x$ is undefined else $x$})$. \\
max & \parbox[t]{3.75in}{$\max(x, y) = (\text{$y$ if $xRy$ else $x$})$, where $R$ is reflexive, connected, and transitive. \vspace{1.0ex}} \\
min & \parbox[t]{3.75in}{$\min(x, y) = (\text{$y$ if $xRy$ else $x$})$, where $R$ is reflexive, connected, and transitive. Same as $\max$ but used for $R_{\operatorname{op}}$. } \\
list concatenation & $(a_1, \ldots, a_m) \cdot (b_1, \ldots, b_n) = (a_1, \ldots, a_m, b_1, \ldots, b_n)$. \\
string concatenation & This is a special case of list concatenation. \\
union & Union of sets. \\
intersection & Intersection of sets. \\
symmetric difference & Symmetric difference of sets. \\
function composition & This presupposes a representation for the composite function. \\
\end{tabular}
\end{center}
} 

\subsubsection*{Example 2. Averages}

To compute an average one must keep track of both a sum and the number of observations in the sum. Thus we have a two-variable recurrence

\begin{equation*}
\mat{x_i\\ w_i}=\mat{x_{i-1} + a_i\\ w_{i-1}+1} = f_{a_i}(\mat{x_{i-1}\\ w_{i-1}}) = a_i \bullet \mat{x_{i-1}\\ w_{i-1}}
\end{equation*}
where each step in the recurrence introduces new information $a_i$, and the average is computed as $\frac{x_i}{w_i}$. The action of $a_i$ on $\smat{x_{i-1}\\ w_{i-1}}$ is $a \bullet \smat{x\\ w} = \smat{x + a\\ w + 1}$, but this action is not semi-associative. A representation of function composition is easily found.
\begin{equation*}
\lambda(a) = \mat{a\\ 1}
  , \quad
\mat{a\\ u} \bullet \mat{x\\ w} = \mat{x+a\\ w+u}
  , \quad
\mat{a_1\\ u_1} *\mat{a_2\\ u_2} = \mat{a_1+a_2\\ u_1+u_2}
\end{equation*}
This is a deliberately trivial example that illustrates the definitions.

\subsubsection*{Example 3. Weighted Averages}

Again we have a two-variable recurrence on $\smat{x_i\\ w_i}$ and the average is $\frac{x_i}{w_i}$. The recurrence is
\begin{equation*}
\mat{x_i\\ w_i}=\mat{x_{i-1} + u_i a_i\\ w_{i-1} + u_i}
\end{equation*}
where $a_i$ are the observations being averaged and $u_i$ are nonnegative weights. At this point it might be tempting to set 
\begin{equation*}
\mat{a\\ u} \bullet \mat{x\\ w}=\mat{x+u a\\ w+u}  
\end{equation*}
which is semi-associative with associative companion operation 
\begin{equation*}
\mat{a_1\\ u_1} * \mat{a_2\\ u_2}
   = \mat{
    \text{$\frac{u_1 a_1 + u_2 a_2}{u_1 + u_2}$ if $u_1+u_2 \neq 0$ else $0$}\\ 
    u_1 + u_2}
\end{equation*}
but this would be unnecessarily complicated. Instead it is simpler and more efficient to define
\begin{equation*}
\lambda(\mat{a\\ u}) =\mat{u a\\ u}
, \quad
\mat{b\\ u} \bullet \mat{x\\ w} = \mat{x + b\\ w + u}
, \quad
\mat{b_1\\ u_1} * \mat{b_2\\ u_2} = \mat{b_1+b_2\\ u_1+u_2}
\end{equation*}
so that 
$\mat{x_i\\ w_i} = \lambda(\mat{a_i\\ u_i}) \bullet \mat{x_{i-1}\\ u_{i-1}}$.

\medskip

This is an example where a representation of function composition is used to simplify the calculation rather than to rectify non-semi-associativity, as the original operation was semi-associative.

\begin{remark}
The weights, $u_i$, in this example, vary with $i$, and such weighted averages occur frequently in practice (e.g., observations that are weighted by observation). In general, however, the corresponding sliding window weighted averages are not convolutions unless the weights are constant. We shall see more examples which are convolutions later in these examples, but the general convolution problem requires other techniques.%
\footnote{I.e., Fourier Analysis and Fast Fourier Transforms.}
\end{remark}

\subsubsection*{Example 4. Lags and the Trivial Semigroup}

It is instructive to observe the case of the trivial semigroup acting trivially on a set. In this case the recurrence is
\begin{equation*}
x_i = x_{i-1} = a_i \bullet x 
\end{equation*}
where $a_i = 1 \in (\text{The Trivial Semigroup})$, and $1 \bullet x = x$, $1 * 1 = 1$.
In this case the corresponding windowed recurrence of length $n$ is
\begin{equation*}    
y_i = \ourcases{
x_0     & \text{if $i \leq n$, else} \\
x_{i-n} &
}
\end{equation*}
This observation becomes more interesting when one considers that 
many operations that represent `no computation' nevertheless represent some kind of action. Examples are copies, data-moves, and even operations that definitely do involve computation are frequently thought of that way, such as re-indexing, format changes, and data joins. For these kinds of operations the windowed recurrence of length $n$ can be thought of as a lag of length $n$.

E.g., if $f_i$ represents a format change operation from the format used at index $i-1$ to the format used at index $i$ (one hopes that such changes are infrequent), then
\begin{equation*}
y_i = \ourcases{
f_i(\ldots       f_1(x_0    ) \ldots) & \text{if $i \leq n$} \\
f_i(\ldots f_{i-n+1}(x_{i-n}) \ldots) & \text{if $i > n$}
}
\end{equation*}
is a `lag with format changes' operation.
Why might one do this? The reason is it allows you to store the data in its original form.

\subsubsection*{Example 5. Multiplication}
The recurrence for a product is
\begin{equation*}
x_i = m_i x_{i-1}
\end{equation*}
which corresponds to the multiplication operation which is associative. So to use the definition of windowed recurrence we may define.
\begin{equation*}
m \bullet x = m x, \quad m_1 * m_2 = m_1 \bullet m_2 = m_1 m_2
\end{equation*}
The windowed recurrence in this case is $y_i =  m_i \cdots m_{i-n+1} \tilde{x}_{i-n}$, where $\tilde{x}_i$ is a sequence of starting values.%
\footnote{
The starting values are denoted $x_i$ in Chapters \ref{chapter:moving-sums}--\ref{chapter:vector-pseudo-code} but here we have used $x_i$ to denote the recurrence variable, so we use $\tilde{x}_i$ for the starting values instead.}
If $\tilde{x}_i=1$, we recover the sliding window product. If, however, we have a different sequence of initial values $\tilde{x}_i$, then $m_i \cdots m_{i-n+1} \tilde{x}_{i-n}$ can be interpreted as applying a sequence of scale changes to the original sequence $\tilde{x}_{i-n}$, bringing it from the `scale at $i-n$' to the `scale at $i$'. Such calculations are common with financial time series where the scale changes result from corporate or government actions. Thus the windowed recurrence corresponding to the multiplication operator can be thought of also as a `lag operator with scale changes'.
More generally, the general windowed recurrence of length $n$ corresponding to a sequence of functions $f_1, f_2, \ldots$ and (initial) data $\tilde{x}_0, \tilde{x}_1, \ldots$ can be thought of as `lag with updating'.

\subsubsection*{Example 6. Fill Forward}

Working with data frequently means also working with missing data. One common technique for handling missing data in a sequence of data is to fill forward the last known (non-missing) value until a non-missing value is encountered again. This is also a useful building block for data operations with hysteresis (e.g., one can easily build a `Schmitt trigger' or `latch' from such an operation). The recurrence for filling forward is
\begin{equation*}
x_i = \coalesce(a_i, x_{i-1})
    = \ourcases{
x_{i-1} & \text{if $a_i$ is undefined, else} \\
a_i     &
}
\end{equation*}
where $a_1, a_2, \ldots$ is the data to be filled forward. As noted in Example 1, $\coalesce$ is associative so we can apply prefix sum or sliding window $*$-product algorithms directly. The sliding window $*$-product of length $n$ in this case corresponds to  filling forward $n - 1$ steps, whereas the windowed recurrence with length $n$ and initial data $\tilde{x}_i = a_i$ corresponds to filling forward $n$ steps. See also Example~\ref{example:fill-forward-I}.

\subsubsection*{Example 7. Fill Forward with Updating}

If we are filling forward data but an updating function must be applied to earlier data in the sequence to bring the data `up to date with' later data, then we are filling forward with updating. The recurrence for filling forward with updating is
\begin{equation*}
x_i = \coalesce(a_i, f_i(x_{i-1}))
\end{equation*}
where $a_i$ is the data being filled forward and $f_i$ is the $i^\text{th}$ updating function. If we let 
\begin{equation*}
\coalesce_{a_i} = \Leftop^{\coalesce}_{a_i}\colon x \longmapsto \coalesce(a_i, x),
\end{equation*} 
then our task is to represent and compute compositions of the functions $\coalesce_{a_i} \circ f_i$.
If the functions $f_i$ preserve missingness in the sense that $f_i(x)$ is undefined if and only if $x$ is undefined then we have the equation
\begin{equation*}
f_i(\coalesce(x, y)) = \coalesce(f_i(x), f_i(y))
\end{equation*}
and this will enable us to find a representation of function composition. But first we characterize functions for which such an equation holds.

\begin{lemma}
\label{lemma:coalesce-with-updating}
Assume $f\colon X \rightarrow Y$ is a function, and there are elements $\undefined_X \in X$, $\undefined_Y \in Y$. Let $\coalesce_X$, and $\coalesce_Y$ be defined by
\begin{align*}
\coalesce_X(x_1, x_2) & \ = \ \text{$x_2$ if $x_1=\undefined_X$ else $x_2$}, & \text{ for $x_1, x_2 \in X$} \\
\coalesce_Y(y_1, y_2) & \ = \ \text{$y_2$ if $y_1=\undefined_Y$ else $y_2$}, & \text{ for $y_1, y_2 \in Y$}
\end{align*}
Then the equation $f(\coalesce_X(x_1, x_2)) = \coalesce_Y(f(x_1), f(x_2))$ holds for all $x_1, x_2 \in X$ if and only if $f$ is constant or $f(x)= \undefined_Y  \Leftrightarrow x=\undefined_X$ for all $x \in X$.
\end{lemma}
\begin{proof}
Clearly if $f$ is constant then $f(\coalesce_X(x_1, x_2)) = \coalesce_Y(f(x_1), f(x_2))$, 
and if for all $x \in X$, we have $x=\undefined_X \Leftrightarrow f(x) = \undefined_Y$, then
\begin{align*}
f(\coalesce_X(x_1, x_2)) 
& = f(\text{$x_2$ if $x_1 =\undefined_X$ else $x_1$}) \\
& = \text{$f(x_2)$ if $x_1 =\undefined_X$ else $f(x_1)$} \\
& = \text{$f(x_2)$ if $f(x_1) =\undefined_X$ else $f(x_1)$} \\
& = \coalesce_Y(f(x_1), f(x_2))
\end{align*}
To prove the other direction we assume that $f(\coalesce_X(x_1, x_2)) = \coalesce(f(x_1), f(x_2))$ and that there is some $x \in X$ such that the assertion $(f(x) = \undefined_Y \Leftrightarrow x = \undefined_X)$ is false, and then prove $f$ is constant. But if $f(x) = \undefined_Y$ and $x \neq \undefined_X$, then for any $y \in X$ we have
\begin{equation*}
f(y) = \coalesce_Y(f(x), f(y)) = f(\coalesce_X(x, y)) = f(x)
\end{equation*}
Similarly if $f(x) \neq \undefined_Y$ and $x = \undefined_X$, then
\begin{equation*}
f(y) = f(\coalesce_X(x, y)) = \coalesce_Y(f(x), f(y)) = f(x) 
\end{equation*}
\end{proof}

We now proceed with the example assuming $f_i(\coalesce(x, y))=\coalesce(f_i(x), f_i(y))$.
Define
\begin{equation*}
\mat{f\\ a} \bullet x = \coalesce(a, f(x))
, \quad
\mat{f\\ a} * \mat{g\\ b} = \mat{f \circ g\\ \coalesce(a, f(b))}
\end{equation*}
Then it follows that
\begin{equation*}
\mat{f\\ a} \bullet\left(\mat{g\\ b} \bullet x\right)=\left(\mat{f\\ a} *\mat{g\\ b}\right) \bullet x
\end{equation*}
where $f, g$ are any compositions of the $f_i$. I.e., $\bullet$ is semi-associative with companion operation $*$. Of course to compute $f \circ g$ we would need a representation of function composition for the functions $f_i$ and their composites. So this gives us a method for constructing a representation of function composition for the functions $x \mapsto \coalesce(a, f(x))$ given a representation of function composition for the functions $f_i$. Specifically, if $f_i=f_{\zeta_i}$, and $f_\zeta \circ f_\nu = f_{\zeta * \nu}$, and $f_\zeta(\coalesce(x, y)) = \coalesce(f_\zeta(x), f_\zeta(y))$, and $f_{\zeta}(x)=\zeta \bullet x$, then we can set
\begin{equation*}
\mat{\zeta\\ a} \bullet x
    = \coalesce(a, \zeta \bullet x)
, \quad
\mat{\zeta\\ a} * \mat{\nu\\ b} 
    = \mat{\zeta * \nu\\ 
           \coalesce(a, \zeta \bullet b)}
\end{equation*}
and then it will follow that
\begin{equation*}
\mat{\zeta\\ a} \bullet\left(\mat{\nu\\ b} \bullet x\right)
    = \left(\mat{\zeta\\ a} * \mat{\nu\\ b}\right) \bullet x
\end{equation*}

We will describe how to generalize this example to cases where $f$ is arbitrary in later examples, but first we generalize the arguments used so far.

\subsubsection*{Example 8. $*$-Products with Updating} 

The technique of Example 7 is easy to generalize to other recurrences. The result that justifies this is a variant of Section 1.4.1 of \cite{Blelloch1993}, or Theorem~2.4 of \cite{Trout1972}.


\smallskip

\begin{lemma}
\label{lemma:*-products-with-updating}
Assume $\bullet\colon \Lambda \times X \rightarrow X$ is a semi-associative action of $\Lambda$ on $X$ with companion operation $*\colon \Lambda \times \Lambda \rightarrow \Lambda$ (\underline{possibly nonassociative}), and $*\colon X \times X \rightarrow X$ is an \underline{associative} binary operation on $X$. Assume further that $\zeta \bullet(x * y)=(\zeta \bullet x) * (\zeta \bullet y)$ for all $\zeta \in \Lambda$ and all $x, y \in X$. Define
\begin{equation*}
\bullet\colon (\Lambda \ltimes_{\Leftop^\bullet} X) \times X \rightarrow X\colon 
          \mat{\zeta\\ x} \bullet y = x * (\zeta \bullet y)
\end{equation*}
Then $\bullet\colon (\Lambda \ltimes_{\Leftop^\bullet} X) \times X \rightarrow X$ is semi-associative with companion operation which is the semidirect product operation 
\begin{equation*}
\mat{\zeta_1\\ x_1} * \mat{\zeta_2\\ x_2} 
    = \mat{\zeta_1 * \zeta_2\\ x_1 * (\zeta_1 \bullet x_2)}
\end{equation*}
Furthermore, if the binary operation $*$ on $\Lambda$ is associative, then the semidirect product operation on $\Lambda\ltimes_{\Leftop^\bullet} X$ is also associative.
\end{lemma}
\begin{proof}
This is a special case of Theorem~\ref{theorem:semidirect-product-set-action} with $A = \Lambda$, $B = X$, $*\colon X \times X \rightarrow X$ as $\bullet\colon B \times X \rightarrow X$, and $\bullet\colon \Lambda \times X \rightarrow X$ as $\times\colon A\times B \rightarrow B$. The final statement on associativity of the semidirect product follows from Lemma~\ref{lemma:semidirect-product-semigroup-action-form}, or equivalently Lemma~\ref{lemma:semidirect-product-semigroup-function-form}.
\end{proof}

%

\begin{remark}
Note that in Lemma~\ref{lemma:*-products-with-updating}, $*\colon \Lambda \times \Lambda \rightarrow \Lambda$ may be nonassociative, but $*\colon X \times X \rightarrow X$ is assumed to be associative.
\end{remark}

Now suppose $\bullet\colon A \times X \rightarrow X$ is a set action and $(\Lambda, \lambda, *, \bullet)$ is a representation of function composition for $\bullet$, and assume $*\colon X \times X \rightarrow X$ is an associative binary operation. We consider the recurrence 
\begin{equation*}
    x_i=z_i *\left(a_i \bullet x_{i-1}\right)
\end{equation*}
where $a_i \in A$ and $z_i \in X$. 
Assume further that $\zeta \bullet (x * y)=(\zeta \bullet x) * (\zeta \bullet y)$ for $\zeta \in \Lambda$, $x, y \in X$. 
Then we may define $\bullet\colon \left(A \times X\right) \times X \rightarrow X$, by $\smat{a\\ z} \bullet x = z * (a \bullet x)$. 
If we now define $\lambda\colon A \times X \rightarrow \Lambda \times X$ by  
$\lambda(\smat{a\\ z}) = \smat{\lambda(a)\\ z}$ for $a \in A$, $z \in X$,
then 
\begin{align*}
& (\Lambda \ltimes_{\Leftop^\bullet} X
, \quad 
    \lambda\colon A \times X \rightarrow \Lambda \times X
,  \quad
    *\colon \left(\Lambda \ltimes_{\Leftop^\bullet} X\right)
         \times
         \left(\Lambda \ltimes_{\Leftop^\bullet} X\right)
         \rightarrow 
         \Lambda \ltimes_{\Leftop^\bullet} X
, \\
& \qquad   \bullet\colon 
        \left(\Lambda \ltimes_{\Leftop^\bullet} X\right)
        \times X \rightarrow X
)
\end{align*}
is a representation of function composition for $\bullet\colon ( A \times X) \times X \rightarrow X$. 
Thus, we may write
\begin{align*}
x_i   = z_i *\left(a_i \bullet x_{i-1}\right)
    & = \mat{\lambda\left(a_i\right)\\ z_i} \bullet x_{i-1}\\
\mat{\lambda\left(a_2\right)\\ z_2} \bullet\left(\mat{\lambda\left(a_1\right)\\ z_1} \bullet x\right) & =
\left(\mat{\lambda\left(a_2\right)\\ z_2} *\mat{\lambda\left(a_1\right)\\ z_1}\right) \bullet x
\end{align*}

\begin{remark}
If we restrict to the submagma $\Lambda_A$ of $\Lambda$ generated by 
$\theset{\lambda(a) \colon a \in A}$  then it is easy to see that the condition 
$\zeta \bullet (x * y)= (\zeta \bullet x) * (\zeta  \bullet y)$ holds on this submagma provided 
$a \bullet (x * y) = (a \bullet x) * (a \bullet y)$ for all $a \in A$. I.e., the left distributivity of $\bullet$ over $*$ need only be shown for $a \in A$.
\end{remark}

\subsubsection*{Example 9. Fill Forward with Scale Changes}

The recurrence for filling forward with scale changes is
\begin{equation*}
x_i = \coalesce (a_i, m_i \cdot x_{i-1})
\end{equation*}
Here we assume $a_i, m_i, x_i$ are numbers and $\cdot$ is multiplication, and the $a_i, m_i, x_i$ may also be $\undefined$. As we have seen, this generalizes to the situation where $\bullet$ is semi-associative, and either the function $x \mapsto m_i \bullet x$ is constant or we have $(m_i \bullet x = \undefined \Leftrightarrow x = \undefined)$. By Lemmas~\ref{lemma:coalesce-with-updating} and \ref{lemma:*-products-with-updating}, a representation of function composition for the functions $x \longmapsto \coalesce(a, m \cdot x) = f_{\tmat{m\\ a}}$, is
\begin{equation*}
\lambda\colon \mat{m\\ a} \longmapsto \mat{m\\ a} 
, \quad 
\mat{m\\ a} \bullet x =\coalesce(a, m \cdot x)
, \quad 
\mat{m_2\\ a_2} * \mat{m_1\\ a_1} = \mat{m_2 \cdot m_1\\ \coalesce(a_2, m_2 \cdot a_1)}
\end{equation*}
Note that $*$ is also associative in this example.

\subsubsection*{Example 10. Linear Recurrences}

These are recurrences of the form
\begin{equation*}
x_i = a_i + m_i x_{i-1}    
\end{equation*}
and these exactly match the form of Example 8. Thus we may write
\begin{equation*}
  x_i = \mat{m_i\\ a_i} \bullet x_{i-1}  
\end{equation*}
with
\begin{equation*}
\mat{m\\ a} \bullet x = a + m x
, \quad
\mat{m_1\\ a_1} * \mat{m_2\\ a_2} = \mat{m_1 m_2\\ a_1 + m_1 a_2}    
\end{equation*}
and then $\bullet$ is semi-associative with associative companion operation $*$.

\subsubsection*{Example 11. Sums with Scale Changes}
These occur frequently with financial time series. They are equivalent to linear recurrences.

\subsubsection*{Example 12. Sums with Missing Data}

\begin{align*}
x_i & = \ourcases{
    x_{i-1}       & \text{if $a_i$ is $\undefined$, else} \\
    a_i + x_{i-1} &  
    }\\
& = \coalesce(a_i, 0) + x_{i-1}
\end{align*}
These are easily handled using $\lambda(a)=\coalesce(a, 0)$ as the lifting function.

\subsubsection*{Example 13. Sums with Scale Changes and Missing Data}

\begin{equation*}
x_i= \ourcases{
    m_i x_{i-1}       & \text{if $a_i$ is $\undefined$, else} \\
    a_i + m_i x_{i-1} & 
    }
\end{equation*}
A representation of function composition is given as
\begin{equation*}
\lambda (\mat{m\\ a}) = \mat{m\\ \coalesce(a, 0)}
, \quad
\mat{m\\ b} \bullet x = b + m x 
, \quad
\mat{m_1\\ b_1} * \mat{m_2\\ b_2} = \mat{m_1 m_2\\ b_1 + m_1 b_2}
\end{equation*}

\subsubsection*{Example 14. Fill Forward with Scale Changes and Additive Adjustments}

This combines a linear recurrence with $\coalesce$.
\begin{equation*}
    x_i = \coalesce(b_i, a_i + m_i x_{i-1})
\end{equation*}
The semi-associative action giving a representation of function composition is

\begin{equation*}
 \mat{m\\ a\\ b} \bullet x = \coalesce(b, a + m x) 
 , \quad
\mat{m_1\\ a_1\\ b_1} * \mat{m_2\\ a_2\\ b_2} 
    = \matl{m_1 m_2\\ a_1 + m_1 a_2\\ \coalesce(b_1, a_1 + m_1 b_2)}
\end{equation*}
which follows by applying the construction of Example 8 twice. In this case $*$ is associative.

\subsubsection*{Example 15. Averages with Missing Data}
\begin{equation*}
\mat{x_i\\ w_i} = \matl{
    \text{$x_{i-1}$ if $a_i$ is $\undefined$ else $x_{i-1} + a_i$} \\
    \text{$w_{i-1}$ if $a_i$ is $\undefined$ else $w_{i-1} + 1$}
}   
\end{equation*}
A representation of function composition is given by
\begin{equation*}
\lambda(a) = \mat{\coalesce(a, 0)\\ \text{$0$ if  $a$ is $\undefined$ else $1$}}
, \quad
\mat{a\\ u} \bullet \mat{x\\ w} = \mat{a + x\\ u + w}
, \quad
\mat{a_1\\ u_1} * \mat{a_2\\ u_2} = \mat{a_1 + a_2\\ u_1 + u_2} 
\end{equation*}
and the average is $\frac{x_i}{w_i}$.

\subsubsection*{Example 16. Weighted Averages with Missing Data}

\begin{equation*}
\mat{x_i\\ w_i} = \matl{
    \text{ $x_{i-1}$ if $a_i$ is $\undefined$ else $x_{i-1} + u_i a_i$} \\
    \text{ $w_{i-1}$ if $a_i$ is $\undefined$ else $w_{i-1} + u_i$}
}
\end{equation*}
Use $\bullet$, $*$ as in Example 15, but for the lifting operation instead use
\begin{equation*}
\lambda(\mat{a\\ u}) = \matl{
    \text{$0$ if $a$ is $\undefined$ else $u a$} \\
    \text{$0$ if $a$ is $\undefined$ else $u$ \ }
}
\end{equation*}
This assumes the weights are not undefined. If we want to handle undefined weights (by dropping them) then we use the following.
\begin{align*}
& \mat{x_i\\ w_i} = \matl{
    \text{$x_{i-1}$ if $a_i$ is $\undefined$ or $u$ is $\undefined$ else $x_{i-1} + u_i a_i$} \\
    \text{$w_{i-1}$ if $a_i$ is $\undefined$ or $u$ is $\undefined$ else $w_{i-1} + u_i$}
}\\
& \lambda(\mat{a\\ u}) = \matl{
    \text{$0$ if $a$ is $\undefined$ or $u$ is $\undefined$ else $ua$} \\
    \text{$0$ if $a$ is $\undefined$ or $u$ is $\undefined$ else $u$}
}
\end{align*}

\subsubsection*{Example 17. Weighted Average with Missing Data and Scale Changes and Additive Adjustments}

\begin{equation*}
\mat{x_i\\ w_i} = \matl{
u_i b_i + a_i + m_i x_{i-1} & \text{if $b_i \neq \undefined$ else $a_i + m_i x_{i-1}$} \\
u_i + w_{i-1}               & \text{if $b_i \neq \undefined$ else $w_{i-1}$}
}
\end{equation*}
A representation of function composition is given as follows. 
\begin{equation*}
\lambda(\fmat{m \\ a \\ b \\ u})
    = \fmatl{
    m\\ 
    \text{$a$ if $b$ is $\undefined$ else $ub+a$}\\ 
    \text{$0$ if $b$ is $\undefined$ else $u$}\\    
    }
, \quad
\fmat{m\\ a\\ u} \bullet \fmatl{x\\ w} = \fmatl{a + m x\\ u + w}
, \quad
\fmat{m_1\\ a_1\\ u_1} * \fmat{m_2\\ a_2\\ u_2} 
    = \smatl{m_1 m_2\\ a_1 + m_1 a_2\\ u_1 + u_2}
\end{equation*}
This assumes the $m_i, a_i$ and $u_i$ are not missing. As usual the average is $x_i / w_i$.

\subsubsection*{Example 18. Exponentially Weighted Moving Averages of Type I}

There are two kinds of exponentially weighted moving averages which differ in how they behave on finite windows. The first kind satisfies the following recurrence.%
\footnote{
This puts a heavy weight on the initial point which can take many steps to decay if $c$ is close to 1. The corresponding weights are therefore not geometric!
}
\begin{equation*}
x_i = (1-c) a_i + c x_{i-1}
\end{equation*}
This recurrence is a special case of a linear recurrence (Example 10), and so we may obtain a representation of function composition as follows.
\begin{equation*}
\lambda(a) = \mat{c\\ (1-c) a}
, \quad
\mat{m\\ b} \bullet x = b + m x
, \quad
\mat{m_1\\ b_1} * \mat{m_2\\ b_2} = \mat{m_1 m_2\\ b_1 + m_1 b_2}
\end{equation*}

\subsubsection*{Example 19. Exponentially Weighted Moving Averages of Type II}

A second type of exponentially weighted moving average actually has geometrically decaying weights. This is defined by the recurrence

\begin{equation*}
\mat{x_i\\ w_i} = \mat{a_i + c x_{i-1}\\ 1 + c w_{i-1}}
\end{equation*}
where the recurrence for $w_i$ is started with $w_0 = 1$, and the average is computed as $x_i/w_i$. For this recurrence we have a representation of function composition given by
\begin{equation*}
\lambda(a) = \mat{c\\ a\\ 1}
, \quad
\mat{m\\ a\\ u} \bullet \mat{x\\ w} = \mat{a + m x\\ u + m w}
, \quad
\mat{m_1\\ a_1\\ u_1} * \mat{m_2\\ a_2\\ u_2} 
    = \mat{m_1 m_2\\ a_1 + m_1 a_2\\ u_1+m_1 u_2}
\end{equation*}
An easy way to see this is to note that
\begin{equation*}
\mat{m & 0 & a\\ & m & u \\ & & 1} \cdot \mat{x\\ w\\ 1}
    = \mat{a + m x\\ u + m w\\ 1}
\end{equation*}
and use matrix multiplication to compute the compositions.

\subsubsection*{Example 20. Exponentially Weighted Moving Averages with Missing Data}

For this example we work with Type II exponentially weighted moving averages. As with other weighted averages missing data can be handled using the lifting function $\lambda$. There are two ways we might handle a missing data point, depending on how the decay is handled.
\medskip

\noindent If the recurrence is

\begin{equation*}
\mat{x_i\\ w_i}= \matl{
\text{$a_i + c x_{i-1}$ if $a_i$ is defined else $c x_{i-1}$} \\
\text{$1 + c w_{i-1}  $ if $a_i$ is defined else $c w_{i-1}$}
}
\end{equation*}
then we use 
\begin{equation*}
\text{$\lambda(a) = \mat{c\\ 0\\ 0}$ if $a$ is $\undefined$ else $\mat{c\\ a\\ 1}$}
\end{equation*}
If the recurrence is
\begin{equation*}
\matl{x_i\\ w_i} = \matl{
\text{$a_i + c x_{i-1}$ if $a_i$ is defined else $x_{i-1}$} \\
\text{$1 + c w_{i-1}$   if $a_i$ is defined else $w_{i-1}$}
}
\end{equation*}
then we use 
\begin{equation*}
\text{$\lambda(a) = \matl{1 \\ 0\\ 0}$ if $a$ is $\undefined$ else $\mat{c\\ a\\ 1}$}
\end{equation*}
The apply and compose operators, $\bullet$ and $*$ are defined as in Example 19.

\subsubsection*{Example 21. Exponentially Weighted Moving Averages with Scale Changes and Additive Adjustments}

The recurrence is
\begin{equation*}
\mat{x_i\\ w_i} = \matl{b_i + c (a_i + m_i x_{i-1})\\ 1 + c w_{i-1}}
\end{equation*}
and a representation of function composition is

\begin{equation*}
\lambda\fmat{m\\ a\\ b} 
    = \fmatl{c m\\ b + c a\\ c\\ 1}
, \quad
\fmat{m\\ a\\ \nu\\ b} \bullet \fmat{x\\ w} 
    = \matl{a + m x\\ b + \nu w}
, \quad
\fmat{m_1\\ a_1\\ \nu_1\\ b_1} * \fmat{m_2\\ a_2\\ \nu_2\\ b_2}
    = \fmatl{m_1 m_2\\ a_1 + m_1 a_2\\ \nu_1 \nu_2\\ b_1 + \nu_1 b_2}
\end{equation*}
As with other examples, missing data can be handled by modifying the lifting function $\lambda$.

\subsubsection*{Example 22. Exponentially Weighted Moving Sums}

The recurrence for exponentially weighted moving sums is

\begin{equation*}
x_i = a_i + c x_{i-1}    
\end{equation*}
which is a special case of a linear recurrence (Example 10) where the multipliers are constant. A representation of function composition is
\begin{equation*}
\lambda(a) = \mat{c\\ a}
, \quad
\mat{m\\ a} \bullet x = a + m x
, \quad
\mat{m_1\\ a_1} * \mat{m_2\\ a_2} = \mat{m_1 m_2\\ a_1 + m_1 a_2}
\end{equation*}
The corresponding sliding window $*$-product of length $n$ computes a convolution of the sequence $a_1, a_2, \ldots$ with $(1, c, c^2, \ldots, c^{n-1})$, as
\begin{equation*}
\mat{c\\ a_i} * \ldots * \mat{c\\ a_{i-n+1}}
    = \mat{c^n\\ a_i + c a_{i-1} + \ldots + c^{n-1} a_{i-n+1}}
\end{equation*}
when $i \geq n$, and 
\begin{equation*}
\mat{c\\ a_i} * \ldots * \mat{c\\ a_1} 
    = \mat{c^i\\ a_i + c a_{i-1} + \ldots + c^{i-1} a_1}
\end{equation*}
when $i<n$. The corresponding windowed recurrence initialized off the same sequence $a_i$ (and with $a_0$ defined to be zero) gives the convolution of $a_1, a_2, \ldots$ with $(1, c_1, c^2, \ldots, c^n)$, as

\begin{equation*}
\left( \mat{c\\ a_i} * \ldots * \mat{c\\ a_{i-n+1}} \right) \bullet a_{i-n} = a_i + c a_{i -1} + \ldots + c^n a_{i-n} .
\end{equation*}

\subsubsection*{Example 23. Convolutions}
 
The previous example provides a limited capability to compute more general convolutions by computing exponentially weighted moving sums with different decay constants and summing. Suppose the sequence we wish to convolve with is $c_0, \ldots, c_n$, where $c_i = \sum_{j=1}^k b_j (z_j)^i$, and $b_1, \ldots, b_k$, $z_1, \ldots , z_k$ are constants.  Then we may compute a windowed recurrence for each $c=z_j$ as in Example 22, and then combine these using the $b_j$ to obtain the convolution. The recurrence we use is
\begin{equation*}
\mat{x_{1 i} \\ \vdots \\ x_{k i}} 
    = \matl{a_i + z_1 x_{1 i-1} \\ \vdots \\ a_i + z_k x_{k i-1}}
\end{equation*}
and a corresponding representation of function composition uses $2 k$ variables.
\begin{equation*}
\lambda(a)= \fmat{z_1\\ \vdots\\ z_k\\ a\\ \vdots\\ a}
, \quad
\fmat{m_1\\ \vdots\\ m_k\\ v_1\\ \vdots\\ v_k} 
    \bullet \fmat{x_1\\ \vdots\\ x_k}
    = \fmat{v_1 + m_1 x_1\\ \vdots\\ v_k + m_k x_k},
, \quad
\fmat{m_1\\ \vdots\\ m_k\\ v_1\\ \vdots\\ v_k} 
    * \smat{m_1'\\ \vdots\\ m_k'\\ v_1'\\ \vdots\\ v_k'}
    = \smat{m_1 m_1'\\ \vdots\\ m_k m_k'\\ v_1 + m_1 v_1'\\
             \vdots\\ v_k + m_k v_k'}
\end{equation*}
To obtain the convolution we compute the windowed recurrences of length $n$ to obtain $y_{1i}, \ldots y_{ki}$ where $y_{ji}$ is the convolution of $a_1, a_2, \ldots$ with $(1, z_j, (z_j)^2, \ldots, (z_j)^n)$. Then the desired convolution is
\begin{equation*}
y_i = \sum_{j=1}^k b_j y_{ji} = c_0 a_i + c_1 a_{i-1} + \ldots + c_n a_{i-n}
\end{equation*}
where we choose $a_i = 0$ for $i \leq 1$. Clearly this is most useful when $k$ is small.

\subsubsection*{Example 24. Max and Min}

The recurrence for a running maximum is
\begin{equation*}
    x_i = \max(a_i, x_{i-1})
\end{equation*}
We recap some results here. It is well known that $\max$ and $\min$ are associative. More generally if $R$ is a binary relation on a set $X$ we may define
\begin{equation*}
x *_R y = (y \text { if } x R y \text { else } x)
\end{equation*}
Then 
{ 
\renewcommand{\arraystretch}{1.25}
\begin{center}
\begin{tabular}{l}
$*_R$ is associative and $R$ is reflexive  $\Rightarrow R$ is transitive\\
$R$ is reflexive, connected and transitive $\Rightarrow R$ is associative
\end{tabular}
\end{center}
} 

\noindent However, there are cases where $R$ is reflexive and transitive and not connected but is still associative. For example, $\coalesce = *_R$ where $x R y \Leftrightarrow x = \undefined \text{ or } y = x$. There are also cases where $R$ is transitive but $*_R$ is not associative. See Theorem~\ref{theorem:selection-operators-and-reflexive-relations} and Examples~\ref{example:selection-operators-6} and \ref{example:selection-operators-7} for details.

\subsubsection*{Example 25. Argmax and Argmin}

In order to obtain an associative argmax or argmin operation it  is necessary to make further assumptions about the relation $R$. We therefore assume that $R$ is a binary relation that is reflexive, connected, antisymmetric and transitive, {\em i.e., $R$ is a  total order}. In this case the associated $\max$ operation, which we call $*_R$, is a selection operator which is idempotent, associative, and commutative. There are three associative operators for argmax corresponding to whether the position of the first maximum found is recorded, or the position of the last maximum found is recorded, or all of the maxima positions are recorded.

\begin{description}

\item[argmax earlist]

\begin{align*}
\mat{m_i\\ k_i} & = \mat{m_{i-1}\\ k_{i-1}} \text{ if } a_i R m_{i-1} \text{ else } \mat{a_i\\ i}\\
& = \matl{
    a_i *_R m_{i-1} \\
    \text{$k_{i-1}$  if $a_i *_R m_{i-1} = m_{i-1}$ else $i$}}\\
& = \mat{a_i\\ i} *\mat{m_{i-1}\\ k_{i-1}}
\end{align*}
where

\begin{equation*}
\mat{m_1\\ k_1} * \mat{m_2\\ k_2} 
    = \matl{
        m_1 *_R m_2\\ 
        \text{$k_2$  if $m_1 *_R m_2 = m_2$ else $k_1$}}
\end{equation*}
and the condition $m_1 *_R m_2 = m_2$ is equivalent to $m_1 R m_2$.

\item[argmax latest]

\begin{equation*}
\mat{m_i\\ k_i} = \mat{a_i\\ i} * \mat{m_{i-1}\\ k_{i-1}}    
\end{equation*}
where 

\begin{equation*}
\mat{m_1\\ k_1} * \mat{m_2\\ k_2}
    = \matl{m_1 *_R m_2\\ 
           \text{$k_1$ if $m_1 *_R m_2 = m_1$ else $k_2$}}
\end{equation*}
and the condition $m_1 *_R m_2 = m_1$ is equivalent to $m_2 R m_1$ by the commutativity of $*_R$. This is the opposite operation of the one used for {\em argmax earliest}.

\item[argmax set]

\begin{equation*}
\mat{m_i\\ K_i} = \mat{a_i\\ \theset{i}} * \mat{m_{i-1}\\ K_{i-1}}    
\end{equation*}
where 
\begin{equation*}
\mat{m_1\\ K_1} * \mat{m_2\\ K_2}
    = \mat{m_1 *_R m_2\\ 
           \text{$K_1 \cup K_2$ if $m_1 = m_2$ else $K_2$ if $m_1 *_R m_2 = m_2$ else $K_1$}}
\end{equation*}
and the condition $m_1 *_R m_2 = m_2$ is equivalent to $m_1 R m_2$.
\end{description}

\noindent The operations for {\em argmax earliest} and {\em argmax latest} keep track of 
the maximum and a single index where it occurs. The operation for {\em argmax set} keeps 
track of the maximum and the set of indices where it occurs. The proofs of associativity 
are straight forward. Note however that commutativity of $*_R$ is required for the proof,
and this is why we require $R$ to be a total order. See Theorem~\ref{theorem:selection-operators-and-reflexive-relations} for the relationship between 
properties of $R$ and $*_R$. A simple counterexample when $*_R$ is not commutative is 
the equality relation on a two element set.

\subsubsection*{Example 26. Max Count and Min Count}

To count maxima or minima in a sequence (or in the windowed recurrence case, to count maxima or minima in a sliding window), we again assume $R$ is a total order. We use a recurrence that keeps track of the maximum and the count.
\begin{align*}
\mat{m_i\\ c_i} & = \matl{
    a_i *_R m_{i-1} \\
    \text{$c_{i-1} + 1$ if $a_i = m_{i-1}$ else $c_{i-1}$ if $a_i R m_{i-1}$ else $1$}}\\
& = a_i \bullet \mat{m_{i-1}\\ c_{i-1}}
\end{align*}
A representation of function composition is given by

\begin{equation*}
\lambda(a) = \mat{a\\ 1}
, \quad
\mat{m_1\\ c_1} * \mat{m_2\\ c_2} 
    = \mat{m_1\\ c_1} \bullet \mat{m_2\\ c_2} 
    = \mat{m_1 *_R m_2\\ 
      \text{$c_1 + c_2$ if $m_1 = m_2$ else $c_2$ if $m_1 R m_2$ else $c_1$}}
\end{equation*}

\subsubsection*{Example 27. Max or Min with Updating}
 
The recurrence for $\max$ with updating can be described in set action notation or function notation as
\begin{equation*}
x_i = \max(z_i, a_i \bullet x_{i-1}) 
\text{ \quad or \quad } 
x_i = \max(z_i, f_i(x_{i-1}))
\end{equation*}
where $\bullet$ is the update operator. Intuitively, if $x \mapsto a \bullet x$ or $f_i$ is non-decreasing then $a_i \bullet \max(x, y) = \max(a_i \bullet x, a_i \bullet y)$, or in the notation with functions $f_i(\max(x, y)) = \max(f_i(x), f_i(y))$. If this is the case then the results of Example 8 hold and we may use a semidirect product to construct a representation of function composition for the {\em max with updating} recurrence from a representation of function composition for the updating action $\bullet$ (or for the functions $f_i$).

To be more precise, we again assume we have a reflexive binary relation $R$ on a set $X$, and consider the `max operator' defined by
\begin{equation*}
x *_R y = \text{$y$ if $xRy$ else $x$}
\end{equation*}
The following lemma describes conditions under which a function $f\colon X \rightarrow X$ (e.g. $x \mapsto a \bullet x$) satisfies $f(x *_R y) = f(x) *_R f(y)$

\begin{lemma}[Selection Operator Distributivity]
\label{lemma:selection-operator-distributivity}
Assume $R$ is a reflexive binary relation on the set $X$ and $f\colon X \rightarrow X$. Then
\begin{enumerate}
\item 
$f(x *_R y) = f(x) *_R f(y)$ for all $x, y \in X$ if and only if for all $x, y \in X$, we have 
$(x R y \Leftrightarrow f(x) R f(y))$ or $(f(x) = f(y))$.

\item
Suppose that $x R y \Leftrightarrow f(x) R f(y)$ for all $x, y \in X$, 
then $f(x *_R y) = f(x) *_R f(y)$ for all $x, y \in X$. 

\item 
Suppose $R$ is connected and antisymmetric, then if $x R y \Rightarrow f(x) R f(y)$ for all $x, y \in X$, then $f(x *_R y) = f(x) *_R f(y)$ for all $x, y \in X$.
\end{enumerate}
\end{lemma}
\begin{proof}
1.\ is a consequence of the definition of $*_R$. 2.\ is a direct consequence of 1. For 3.\ assume $xRy \Rightarrow f(x) R f(y)$ for all $x, y \in X$. Suppose $xRy \neq f(x) R f(y)$. We will show that $f(x) = f(y)$, and then the result will follow from 1. If $x R y$ was true then $f(x) R f(y)$ would be true and so $x R y = \operatorname{true} = f(x) R f(y)$. Therefore $x R y$ must be false. So then $y R x$ by connectedness and hence $f(y) R f(x)$. But $x R y \neq f(x) R f(y)$ so $f(x) R f(y)$ must also be true. The result follows by antisymmetry.
\end{proof}
\begin{remarks} \
\begin{enumerate}

\item
Lemma~\ref{lemma:selection-operator-distributivity} is easily extended to the case $f\colon X \rightarrow Y$ and relations $R$ on $X$ and $S$ on $Y$.    

\item 
Note that we also require $*_R$ to be associative for the algorithms to work, and hence that $R$ is transitive. So when applying Lemma~\ref{lemma:selection-operator-distributivity} Part 3 in practice, one requires that $R$ be a total order.
\end{enumerate}
\end{remarks}

\subsubsection*{Example 28. Max or Min with Scale Changes}
The recurrence is
\begin{equation*}
x_i= \max(a_i, m_i x_{i-1})
\end{equation*}
where $m_i > 0$. As per Examples 8 and 27, a representation of function composition is given by the following.

\begin{equation*}
\lambda(\mat{m\\ a}) = \mat{m\\ a}
, \quad
\mat{m\\ a} \bullet x = \max(a, m x)\\
, \quad
\mat{m_1\\ a_1} * \mat{m_2\\ a_2} = \mat{m_1 m_2\\ \max(a_1, m_1 a_2)}
\end{equation*}

\subsubsection*{Example 29. Max or Min of a Sum}

The recurrence for the maximum of a sum requires we keep track of both the sum and 
its maximum so far

\begin{equation*}
\mat{z_i\\ x_i} 
    = \matl{a_i + z_{i-1}\\ \max(a_i + z_{i-1}, x_{i-1})}
    = a \bullet\mat{z_{i-1}\\ x_{i-1}}
\end{equation*}
where 

\begin{equation*}
a \bullet \mat{z\\ x} = \mat{a + z\\ \max (a + z, x)}
\end{equation*}
Iterating the action quickly yields the following representation of function composition for $\bullet$.

\begin{equation*}
\lambda(a) = \mat{a\\ a}
, \quad
\mat{a\\ b} \bullet \mat{z\\ x} = \mat{a + z\\ \max(b + z, x)}
, \quad
\mat{a_1\\ b_1} * \mat{a_2\\ b_2} = \mat{a_1 + a_2\\ \max(b_1 + a_2, b_2)}
\end{equation*}

\subsubsection*{Example 30. Combining Recurrences}

Suppose we have a recurrence $z_i = f_i(z_{i-1})$, and we wish to accumulate the results of that recurrence using a binary operator $*$ on a set $X$, where the values $z_i$ come from $X$. The combined recurrence for this is
\begin{equation*}
\mat{z_i\\ x_i}
    = \matl{
        f_i(z_{i-1})\\ 
        f_i(z_{i-1}) * x_{i-1}}
    = f_i \bullet \matl{z_{i-1}\\ x_{i-1}} 
\end{equation*}
where
\begin{equation*}
f \bullet \mat{z\\ x} = \matl{f(z)\\ f(z) * x}
\end{equation*}
{\em We now assume $*$ is associative for the rest of this example.} Iterating the application of $\bullet$ shows that to find a representation of function composition we should consider the slightly more general recurrence

\begin{equation*}
\mat{z_i\\ x_i} = \mat{f_i(z_{i-1})\\ g_i(z_{i-1}) * x_{i-1}}
\end{equation*}
So define 
\begin{equation*}
    \lambda(f) = \mat{f\\ f}
, \quad
\mat{f\\ g} \bullet \mat{z\\ x} = \matl{f(z)\\ g(z) * x}
\end{equation*}
Also for any two functions $f, g\in \Endop(X)$   define the function $f * g$ by $(f * g)(z) = f(z) * g(z)$ for any $z \in X$ where $*\colon X \times X \rightarrow X$. Call this operator $*\colon \Endop(X) \times \Endop(X) \rightarrow \Endop(X)$ the associated operator of $*$ acting on $\Endop(X)$. We now use these definitions and look for a companion operation for $\bullet$. We have

\begin{align*}
\mat{f_1\\ g_1} \bullet \left(\mat{f_2\\ g_2} \bullet \mat{z\\ x} \right)
& = \mat{f_1\\ g_1} \bullet \matl{f_2(z)\\ g_2(z) * x}\\
& = \matl{(f_1 \circ f_2)(z)\\
          g_1(f_2(z))) * g_2(z) * x}\\
& = \matl{(f_1 \circ f_2)(z)\\
          ((g_1\circ f_2)(z) * g_2(z)) * x}\\
& = \matl{f_1 \circ f_2\\ (g_1\circ f_2) * g_2} \bullet \mat{z\\ x}
\end{align*}
So if we define 
\begin{equation*}
\mat{f_1\\ g_1} * \mat{f_2\\ g_2} = \mat{f_1 \circ f_2\\ (g_1 \circ f_2) * g_2}
\end{equation*} 
then $(\lambda, *, \bullet)$ is a representation of function composition for $\smat{z\\ x} \mapsto \smat{f(z)\\ f(z) * x}$, provided the $f$ and $g$ functions come from a subset $F \subseteq \Endop(X)$ that is closed under both function composition $\circ$ and the function product $*\colon \Endop(X) \times \Endop(X) \rightarrow \Endop(X)$ associated to $*$.
Of course, what we want in order to compute is a representation of function composition for the functions $f$, $g$, as well as a parametrization of $f * g$. I.e., $f$, $g$ should come from a parameterized family of functions $f_\zeta$ and there should be binary operations $*_1, *_2$ (which need not be associative) such that $f_{\zeta_1} \circ f_{\zeta_2} = f_{\zeta_1 *_1 \zeta_2}$, and
$f_{\zeta_1} * f_{\zeta_2} = f_{\zeta_1 *_2 \zeta_2}$.

The argument above is easily reformulated into the language of set actions and representations of function composition, and yields the following theorem.

\begin{theorem}
\label{theorem:combined-recurrence}
Assume $\bullet\colon A \times X \rightarrow X$ is a set action, and $*\colon X \times X \rightarrow X$ is an associative binary operation on $X$. Define a set action of $A$ on $X \times X$ by
\begin{equation*}
a \bullet \mat{z\\ x} = \mat{a \bullet z\\ (a \bullet z) * x}
\end{equation*}
for $a\in A$, $z, x\in X$. Let $(\lambda, \Lambda, *_1, \bullet)$ be a representation of function composition for $\bullet\colon A \times X \rightarrow X$, and assume that the collection of functions $\theset{(x\mapsto \zeta \bullet x)\colon \zeta \in \Lambda}$ is closed under the associated operator $*$, so that there exists binary operation $*_2$ such that $(\zeta_1 \bullet x) * (\zeta_2 \bullet x) = (\zeta_1 *_2 \zeta_2) \bullet x$ for all $\zeta_1, \zeta_2 \in \Lambda$, $x \in X$. Define $\lambda'\colon A \rightarrow \Lambda \times \Lambda$, $\bullet\colon (\Lambda \times \Lambda) \times (X \times X) \rightarrow X \times X$, and $*\colon (\Lambda \times \Lambda) \times (\Lambda \times \Lambda) \rightarrow \Lambda \times \Lambda$ by
\begin{equation*}
\lambda'(a) = \mat{\lambda(a)\\ \lambda(a)}
, \quad
\mat{\zeta\\ \chi} \bullet \mat{z\\ x} = \matl{\zeta \bullet z\\ (\chi \bullet z) * x}
, \quad
\mat{\zeta_1\\ \chi_1} * \mat{\zeta_2\\ \chi_2}
    = \matl{\zeta_1 *_1 \zeta_2\\ (\chi_1 *_1 \zeta_2) *_2 \chi_2}
\end{equation*}
for $a \in A$, $\zeta, \zeta_i, \chi, \chi_i \in \Lambda$, $z, x \in X$. Then $(\Lambda \times \Lambda, \lambda', *, \bullet)$ is a representation of function composition for the set action $\bullet\colon A \times (X \times X) \rightarrow X \times X$.
\end{theorem}
\begin{proof}
This is a special case of Theorem~\ref{theorem:semidirect-product-xy-action}.
\end{proof}

\begin{remarks} \ 
\begin{enumerate}
\item 
At the start of the example we did not assume any algebraic relationship between the original operation and the functions $f_i$. Similarly, in Theorem~\ref{theorem:combined-recurrence} we did not assume any algebraic relation between the set action $\bullet\colon A \times X \rightarrow X$ and the associative binary operation $*\colon X \times X \rightarrow X$. We did, however, assume that the collection of left action operations of the representation of function composition of $\bullet$ was closed under the associated operation of $*: X \times X \rightarrow X$ on functions in $\Endop(X)$.

\item
Theorem~\ref{theorem:combined-recurrence} is easily reformulated into the language of parameterized function families. In particular, if $\theset{f_\zeta\colon \zeta \in \Lambda} \subseteq \Endop(X)$ is closed under both function composition and the function product given by $(f * g)(x) = f(x) * g(x)$, then we may write the apply operation for the representation of function composition constructed in Theorem~\ref{theorem:combined-recurrence} as
\begin{equation*}
\mat{\zeta\\ \chi} \bullet \mat{z\\ x} = \matl{f_\zeta(z)\\ f_\chi(z) * x}
\end{equation*}
\end{enumerate}

\end{remarks}

\subsubsection*{Example 31. Maximum Contiguous Subsequence Sum}

This is treated in \cite{ChinKhoo2004} and \cite{FisherGhuloum1994}. The original problem comes from \cite{Bentley1984} and \cite{Bentley1986}. As a recurrence the problem to compute is

\begin{equation*}
\mat{z_i\\ x_i}
    =\matl{\max(z_{i-1} + a_i, 0)\\ 
           \max(\max(z_{i-1} + a_i, 0), x_{i-1})}
    = a_i \bullet\matl{z_{i-1}\\ x_{i-1}}
\end{equation*}
where
$
a \bullet \smat{z\\ x} = \smatl{\max(z + a, 0)\\ 
           \max(\max(z + a, 0), x)}
$.
This clearly has the form $\smatl{f_i\left(z_{i-1}\right)\\ f_i\left(z_{i-1}\right) * x_{i-1}}$ of Example 30, where $* = \max$, so to find a representation of function composition we must find the closure of the functions $z \mapsto \max (z + a, 0)$ under composition and maximum. Taking compositions quickly leads to the functions

\begin{equation*}
f_{\tmat{a\\  b}}(z) = \max(z + a, b)
\end{equation*}
and so we let $\smat{a\\ b} \bullet z = f_{\tmat{a\\ b}}(z) = \max(z + a, b)$.

\begin{lemma}
\label{lemma:max-sum-closure}
The collection of functions $f_{\tmat{a\\ b}}$ is closed under function composition and maximum, and

\begin{align*}
f_{\tmat{a_1\\ b_1}} \circ f_{\tmat{a_2\\ b_2}} 
    & = f_{\tmatl{a_1 + a_2\\ \max(b_1, a_1 + b_2)}} \\
\max\left( f_{\tmat{a_1\\ b_1}}, f_{\tmat{a_2\\ b_2}} \right)
    & = f_{\tmatl{\max(a_1, a_2)\\ \max(b_1, b_2)}} 
\end{align*}
\end{lemma}
\begin{proof}
This is an easy and direct calculation.
\end{proof}

\begin{corollary}
\label{corollary:max-sum-closure}
The operation $\smat{a\\ b} \bullet z = \max(z + a, b)$ is semi-associative with companion operation $*$ given by
\begin{equation*}
\mat{a_1\\ b_1} * \mat{a_2\\ b_2} = \matl{a_1+a_2\\ \max(b_1, a_1 + b_2)}
\end{equation*}
\end{corollary}
\begin{proof}
This follows directly from Lemma~\ref{lemma:max-sum-closure} and Theorem~\ref{theorem:combined-recurrence}.
\end{proof}

\noindent Now we can use the construction of Example 30 to find a representation of function composition for the original action

\begin{equation*}
a \bullet\mat{z\\ x} = \matl{\max(z + a, 0)\\ \max(\max(z+a, 0), x)}
\end{equation*}
This is

\begin{align*}
& \lambda(a) = \fmat{a\\ 0\\ a\\ 0}
, \quad
  \fmat{a\\ b\\ c\\ d} \bullet \fmat{z\\ x}
    = \fmatl{\max(z + a, b)\\ \max(z + c, d, x)}
    = \fmatl{\smatl{a\\ b} \bullet z\\ 
          \max(\smatl{c\\ d} \bullet z, x)} \\
& \fmat{a_1\\ b_1\\ c_1\\ d_1} * \fmat{a_2\\ b_2\\ c_2\\ d_2}
    = \fmatl{a_1+a_2 \\
             \max(b_1, a_1 + b_2) \\
             \max(c_1 + a_2, c_2) \\
             \max(d_1, c_1 + b_2, d_2)}
    =\fmatl{\smat{a_1\\ b_1} * \smat{a_2\\ b_2}\\ 
            \max(\smat{c_1\\ d_1} * \smat{a_2\\ b_2}, \smat{c_2\\ d_2})}
\end{align*}
where $\smat{a\\ b} \bullet z = \max(z + a, b)$, and the max of vectors in the last equation is taken componentwise.

\subsubsection*{Example 32. Cusum Test}

Cusum tests are common statistical tests for exchangeability which can be used to detect regime shifts in sequential data. See for example \cite{Basseville1993}. A basic cusum test satisfies the recurrence

\begin{equation*}
x_i=\max( 0, x_{i-1} + z_i - \omega_i)
\end{equation*}
where $z_i$ denotes the data to be tested and $\omega_i$ is estimated from the mean of the data and the change to be detected. This clearly has the general form $\max(x+a, b)$, and so a representation of function composition may be found using Lemma~\ref{lemma:max-sum-closure} and Corollary~\ref{corollary:max-sum-closure}. This gives the following representation of function composition

\begin{equation*}
\lambda\mat{z\\ w} = \mat{z - \omega\\ 0}
, \quad
\mat{a\\ b} \bullet x = \max(x + a, b)\\
, \quad
\mat{a_1\\ b_1} * \mat{a_2\\ b_2} = \mat{a_1 + a_2\\ \max(b_1, a_1 + b_2)}
\end{equation*}

\subsubsection*{Counter-Example 33. Sum of Max}

It is instructive to consider how sum of max differs from max of sum. The recurrence for sum of max is
\begin{equation*}
\mat{z_i\\ x_i} = \matl{\max(a_i, z_{i-1})\\ \max(a_i, z_{i-1}) + x_{i-1}}
\end{equation*}
which has the from $\mat{f(z)\\ g(z) + x}$ of the construction in Example 30. According to that example, a representation of function composition may be found by taking the closure of the collection of functions $z \longmapsto \max (a, z)$ under the operations of function composition and addition of functions. We have seen that if $\max_a(z) = \max (a, z)$ then $\max_a \circ \max_{b} = \max_{\max(a, b)}$ so these functions are closed under function composition ($\max$ is associative). But under addition the dimensionality of the closure can increase without limit.
\begin{align*}
& \max(a, x) + \max(b, x) \
    = \max(a+b, \max(a, b) + x, 2 x)\\
& \max(a, x) + \max(b, x) + \max (c, x) 
    = \max(a + b + c, \max(a + b, a + c, b + c) + x, \max(a, b, c) + 2 x, 3 x)\\
& \ldots \text{ and so on.}
\end{align*}
In the case where we are using real numbers topological considerations show that in general there is no parametrization of the closure with a finite number of real parameters.%
\footnote{Of course we need to specify the properties and meaning of parameterization to be precise here.}
For particular subsets of the allowed values $a$, however, a parametrization is possible. For example if we only allow $a = 0$, then the closure is the set of functions, $x \mapsto \max(0, m x)$ for $m > 0$, $m \in \mathbb{Z}$. Of course in this case the corresponding recurrence is trivial to compute.

A more interesting case where the increasing number of parameters needed to describe the closure is not a problem is the case where the operations are on a finite set, as then the size of the closure is limited by the total number of endomorphisms of the set being operated on. For example if $X = \mathbb{Z} / 3\mathbb{Z} = \theset{\text{the integers modulo } 3}$, and $\max(x, y)$ for $x, y \in \mathbb{Z}/3\mathbb{Z}$ is the maximum with respect to the total order with $0 \leq 1 \leq 2$, then any function $f\colon X \rightarrow X$ can be represented by an array of length $3$ with entries in $\theset{0,1,2}$. In particular, the functions $\operatorname{max}_a\colon z \longmapsto \max(a, x)$ correspond to the following arrays.
%
\begin{align*}
\operatorname{max}_0 & \longleftrightarrow (0,1,2) \\
\operatorname{max}_1 & \longleftrightarrow (1,1,2) \\
\operatorname{max}_2 & \longleftrightarrow (2,2,2)
\end{align*}
and function application, function composition, and function addition correspond to
\begin{align*}
u \bullet i & = u[i+1] \\
      u * v & = (u[v[1]+1], u[v[2]+1], u[v[3]+1]) \\
      u + v & = (u[1]+v[1], u[2]+v[2], u[3]+v[3])
\end{align*}
Here $u, v$ are arrays of length $3$ and entries in $\theset{0,1,2}$, with indexing starting at 1. The set action $\bullet$ corresponds to function application and the binary operation $*$ to function composition.
So in this case the infinite increase in the number of variables required to parameterize the closure is averted as we can parameterize the entire space of functions $\Endop(X)$ directly.

\subsubsection*{Counter-Example 34. Max of a Linear Recurrence}

The recurrence to consider is

\begin{equation*}
\mat{z_i\\ x_i}=\mat{a_i+m_i z_{i-1}\\ \max \left(a_i+m_i z_{i-1}, x_{i-1}\right)}
\end{equation*}
and to find a representation of function composition we must parameterize the closure of the collection of functions $z \mapsto a + m z$ under function composition and function maximum. In the case where $a, m, z$ are real numbers, consider that the function $z \mapsto \max(a_1 + m_1 z, \ldots, a_k + m_k z)$ is piecewise linear and has up to $k$ different slopes, so as $k$ increases there will in general be no parametrization (not locally smooth or locally 1:1 and continuous) with a bounded number of variables. As with Example 33 there are subsets of these functions with boundedly parameterizable closures (e.g. $m_i=1$ ), and when this example is applied to functions on finite sets the functions may be represented directly by arrays or tables.

\subsubsection*{Example 35. Argmax or Argmin with Updating}

We describe the situation for {\em argmax earliest}, as the cases for {\em argmax latest} and {\em argmax set} are similar. As with Example 25 we assume $R$ is a binary relation on $X$ which is reflexive, connected, transitive, and antisymmetric, i.e. $R$ is a total order. Let's also assume that $f_i\colon X \rightarrow X$ are functions such that $f_i(x) R f_i(y) \Leftrightarrow x R y$ for all $x, y \in X$. The recurrence for {\em argmax earliest with updating} is then
%
\begin{equation*}
\matl{m_i\\ k_i} 
    = \matl{a_i *_R f_i(m_{i-1}) \\
            \text{$k_{i-1}$ if $a_i *_R f_i(m_{i-1}) = f_i(m_{i-1})$ else $i$}}
 = \matl{a_i\\ i} * \tilde{f}_i\matl{m_{i-1}\\ k_{i-1}}
\end{equation*}
where

\begin{equation*}
\mat{m_1\\ k_1} * \mat{m_2\\ k_2}
    = \matl{m_1 *_R m_2\\
            \text{$k_2$ if $m_1 R m_2$ else $k_1$}}
, \quad \text{ and } \quad
\tilde{f}(\mat{m\\ k}) = \mat{f(m)\\ k }.
\end{equation*}
We can therefore apply the construction of Example 8 to get a representation of function composition provided we can show that 
$\tilde{f}(\smat{m_1\\ k_1} * \smat{m_2\\ k_2}) 
   =\tilde{f}(\smat{m_1\\ k_1}) * \tilde{f}(\smat{m_2\\ k_2})$ 
whenever $f$ is a function that satisfies $f(x) R f(y) \Leftrightarrow xRy$ for all $x, y \in X$. But this is immediate, as
\begin{align*}
\tilde{f}(\matl{m_1\\ k_1} * \matl{m_2\\ m_2}) 
& =\tilde{f}(\matl{m_1 *_R m_2 \\
            \text{$k_2$ if $m_1 R m_2$ else $k_1$}} \\
& = \matl{f(m_1 *_R m_2)\\ 
         \text{$k_2$ if $m_1 R m_2$ else $k_1$}} \\
& = \matl{f(m_1) *_R f(m_2)\\ 
        \text{$k_2$ if $f(m_1) R f(m_2)$ else $k_1$}} \\
& = \tilde{f}(\matl{m_1\\ k_1}) * \tilde{f}(\matl{m_2\\ k_2})
\end{align*}
Therefore the action
\begin{equation*}
\fmat{f\\ a\\ i} \bullet \fmat{m\\ k} 
    = \fmatl{a *_R f(m)\\
             \text{$k$ if $a R f(m)$ else $i$}}
    = \fmat{a\\ i} * \tilde{f}(\fmat{m\\ k})
\end{equation*}
is semi-associative with companion operation $*$ given by
\begin{equation*}
\fmat{f_1\\ a_1\\ i_1} * \fmat{f_2\\ a_2\\ i_2}
    = \fmatl{f_1 \circ f_2\\
             a_1 *_R f_1(a_2) \\
             \text{$i_2$ if $a_1 R f_1(a_2)$ else $i_1$}}
\end{equation*}
where the functions $f_1, f_2$ are assumed to satisfy $f_i(x) R f_i(y) \Leftrightarrow x R y$. If, in addition, we have a representation of function composition for the closure of the original functions $f_i$ under composition, then we may get a representation of function composition of the form
\begin{equation*}
\fmat{\zeta\\ a\\ i} \bullet \fmat{m\\ k} 
    = \fmatl{a *_R (\zeta \bullet m)\\
             \text{$k$ if $a R (\zeta \bullet m)$ else $i$}}
, \quad 
\fmat{\zeta_1\\ a_1\\ i_1} * \fmat{\zeta_2\\ a_2\\ i_2}
    = \fmatl{\zeta_1 * \zeta_2\\
             a_1 *_R (\zeta_1 \bullet a_2) \\
             \text{$i_2$ if $a_1 R (\zeta_1 \bullet a_2)$ else $i_1$}}
\end{equation*}
and the lift of $\smat{f\\ a\\ i}$ may be written in terms of the lift of $f$ as 
$\lambda(\smat{f\\ a\\ i}) = \smat{\lambda(f)\\ a\\ i}$.

\subsubsection*{Example 36. Argmax or Argmin of Sum}

The recurrence for {\em argmax earliest} of a sum is
\begin{equation*}
\fmat{z_i\\ x_i \\k_i}
    = \fmatl{a_i + z_{i-1}\\
            \max(a_i + z_{i-1}, x_{i-1})\\
            \text{$k_{i-1}$ if $a_i + z_{i-1} \leq x_{i-1}$ else $i$}}
\end{equation*}
Here we assume $\leq$ is a total order (reflexive, connected, antisymmetric and transitive) and $\max =*_{\leq}$. We write
\begin{equation*}
\fmat{z_i\\ x_i\\ k_i}
    = \fmat{a_i\\ i} \bullet \fmat{z_{i-1}\\ x_{i-1}\\ k_{i-1}}
\end{equation*}
where
\begin{equation*}
\fmat{a\\ i} \bullet \fmat{z\\ x\\ k}
    = \fmatl{a + z\\
            \max(a + z, x)\\
            \text{$k$ if $a + z \leq x$ else $i$}}
\end{equation*}
Then a representation of function composition for $\bullet$ is the following
\begin{equation*}
\lambda(\fmat{a\\ i}) = \fmat{a\\ a\\ i}
, \quad
\fmat{a\\ b\\ j} \bullet \fmat{z\\ x\\ k}
    = \fmatl{a + z\\ \max(b + z, x)\\
             \text{$k$ if $b + z \leq x$ else $j$}}
, \quad
\fmat{a_1\\ b_1\\ j_1} * \fmat{a_2\\ b_2\\ j_2}
    = \fmatl{a_1 + a_2\\ 
             \max(b_1 + a_2, b_2)\\
             \text{$j_2$ if $b_1 + a_2 \leq b_2$ else $j_1$}}
\end{equation*}
and in fact $\bullet = *$ is associative.

\subsubsection*{Example 37. Fill Forward with Updating: Updates that Fail}

We now return to Example 7, fill forward with updating, which had the recurrence

\begin{equation*}
x_i = \coalesce(a_i, f_i(x_{i-1}))
\end{equation*}
where $x_i, a_i \in X$, and $X$ is a set containing an undefined element, and $f_i\colon X \rightarrow X$. This time, however, we do not make any further assumptions on the functions $f_i$. Instead we use a variant of a technique developed by Blelloch \cite{Blelloch1993}, in the context of segmented scans. We start by writing

\begin{equation*}
x_i = \coalesce(a_i, f_i(x_{i-1}))\\
    = \ourcases{
            f_i(x_{i-1}) & \text{if $a_i$ is undefined, else}\\
            a_i          & } 
\end{equation*}
and then note this is a special case of
\begin{equation*}
x_i = \ourcases{
         f_i(x_{i-1}) & \text{if $c_i$ else}\\
         a_i          & }
\end{equation*}
where $c_i$ is a truth-valued variable. To rewrite this in a more algebraic notation, define
\begin{equation*}
\case(c, x, y) = (\text{$x$ if $c$ else $y$})
\end{equation*}
and also define

\begin{equation*}
\case_\tmat{c\\ y}^{(1 3)}(x) = \case(c, x, y)
\end{equation*}
The case function has many nice algebraic properties, but the properties we will use in this example are that
\begin{enumerate}
\item 
$f(\case(c, x, y)) = \case(c, f(x), f(y))$ for any function $f$, and

\item 
$\case(c_1, \case(c_2, x, y_2), y_1) = \case(c_1 \wedge c_2, x, \case(c_1, y_2, y_1))$, 
where $\wedge$ is logical `and'. 
\end{enumerate}
These imply the following properties for $\case_{\tmat{c\\ y}}^{(13)}$.
\begin{enumerate}
\item 
$f \circ \case_\tmat{c\\ y}^{(13)} = \case_\tmat{c\\ f(y)}^{13} \circ f$, and

\item 
$\case_\tmat{c_1\\ y_1}^{(13)} \circ \case_\tmat{c_2\\ y_2}^{(13)}
    = \case_\tmat{c_1 \wedge c_2\\ \case(c_1, y_2, y_1)}^{(13)}
$
\end{enumerate}
Applying these to our recurrences we can write
\begin{equation*}
x_i =\case(c_i, f_i(x_{i-1}), a_i) 
    = \fmat{f_i\\ c_i\\ a_i} \bullet x_{i-1}
\end{equation*}
where
\begin{equation*}
\fmat{f\\ c\\ a} \bullet x
    = \case(c, f(x), a)
    = (\case_\tmat{c\\ a}^{(13)} \circ f)(x)
\end{equation*}
To find the companion operation of $\bullet$ we compose
\begin{align*}
\mat{f_1\\ c_1\\ a_1} \bullet \left(\mat{f_2\\ c_2\\ a_2} \bullet x \right)
    & = \left(\case_\tmat{c_1\\ a_1}^{(13)} \circ f_1 \circ  
              \case_\tmat{c_2\\ a_2}^{(13)} \circ f_2 \right)(x)\\
    & = \left(\case_\tmat{c_1\\ a_1}^{(13)} \circ \case_\tmatl{c_2\\ f_1(a_2)}^{(13)} \circ
              f_1 \circ  f_2 \right)(x)\\
    & = \left(\case_\tmatl{c_1 \wedge c_2\\ \case(c_1, f_1(a_2), a_1)}^{(13)} \circ 
              (f_1 \circ  f_2) \right)(x)\\
    & = \matl{f_1 \circ f_2\\ c_1 \wedge c_2\\ \case(c_1, f_1(a_2), a_1)}
            \bullet x
\end{align*}
Hence $\bullet$ is semi-associative with companion operation, $*$ given by
\begin{equation*}
\fmat{f_1\\ c_1\\ a_1} * \fmat{f_2\\ c_2\\ a_2}
    = \fmat{ f_1 \circ f_2\\ 
              c_1 \wedge c_2\\
              \case(c_1, f_1(a_2), a_1)}
\end{equation*}
A short calculation also shows that $*$ is associative. To relate the action to the original recurrence, we use the lifting function
\begin{equation*}
\lambda(\fmat{f\\ a}) = \fmatl{f\\ \text{$a$ is $\undefined$}\\ a}
\end{equation*}
Writing our calculation in set action form we get the following result.

\begin{lemma} 
\label{lemma:updating-with-case}
Assume $\bullet\colon A \times X \rightarrow X$ is a set action and $(\Lambda, \lambda, *, \bullet)$ is a representation of function composition for $\bullet$. Define a set action $\bullet$ by
\begin{equation*}
A' = A \times \theset{T, F} \times X\
, \quad
\bullet\colon A' \times X \rightarrow X \\
, \quad
\fmat{a\\ c\\ z} \bullet x= \case(c, a \bullet x, z)
\end{equation*}
Define 
\begin{align*}
& \Lambda' = \Lambda \times\theset{T, F} \times X\\
& \lambda' \colon A' \rightarrow \Lambda' \colon \fmat{a\\ c\\ z} 
    \longmapsto \fmat{\lambda(a)\\ c\\ z}\\
& \bullet' \colon \Lambda' \times X \rightarrow X\colon
    \fmat{\zeta\\ c\\ z} \bullet x = \case(c, \zeta \bullet x, z)\\
& *' \colon \Lambda' \times \Lambda' \rightarrow \Lambda' \colon
    \fmat{\zeta_1\\ c_1\\ z_1} * \fmat{\zeta_2\\ c_2\\ z_2} 
    = \fmatl{\zeta_1 * \zeta_2\\
             c_1 \wedge c_2\\
             \case(c_1, \zeta_1 \bullet z_2, z_1)}
\end{align*}
Then $(\Lambda', \lambda', *', \bullet')$ is a representation of function composition for $\bullet\colon A' \times X \rightarrow X$, and furthermore, if $*$ is associative then $*'$ is associative.
\end{lemma}

There are several recurrences of interest that have representations of function composition obtainable from Lemma~\ref{lemma:updating-with-case}

\subsubsection*{Example 38. Segmented Scans}

See Blelloch \cite{Blelloch1993}. Segmented scans have a recurrence of the form

\begin{equation*}
x_i = \ourcases{
    a_i           & \text{if $c_i$, else}\\
    a_i * x_{i-1} & }
\end{equation*}
where $*$ is an associative binary operation and $c_i$ is truth-valued. Clearly we can construct a representation of function composition for $\smat{a\\ c} \bullet x =\case(c, a, a * x) = \case(\neg c, a * x, a)$, as follows

\begin{equation*}
\lambda\fmat{a\\ c} = \fmat{a\\ \neg c\\ a}
, \quad
\fmat{a\\ c\\ z} \bullet x = \case(c, a * x, z)
, \quad
\fmat{a_1\\ c_1\\ z_1} * \fmat{a_2\\ c_2\\ z_2}
    = \fmatl{a_1 * a_2\\ c_1 \wedge c_2\\ \case(c_1, a_1 * z_2, z_1)}
\end{equation*}
where $\neg$ denotes logical negation. Note that Blelloch \cite{Blelloch1993} uses the following equivalent representation instead
\begin{equation*}
\lambda\fmat{a\\ c} = \fmat{a\\ c\\ a}
, \quad
\fmat{a\\ c\\ z} \bullet x = \case(c, z, a * x)
, \quad
\fmat{a_1\\ c_1\\ z_1} * \fmat{a_2\\ c_2\\ z_2}
    = \fmatl{a_1 * a_2\\ c_1 \vee c_2\\ \case(c_1, z_1, a_1 * z_2)}
\end{equation*}
This corresponds to using the function $\case_\tmat{c\\ x}^{(12)}(y) = \case(c, x, y)$ instead of $\case_\tmat{c\\ y}^{(13)}$. The companion operation $*$ is associative.

\subsubsection*{Example 39. Run Statistics}
The recurrence for run statistics is
\begin{equation*}
x_i = \ourcasesclosed{
    x_{i-1} + 1 & \text{if $a_i > 0$ else }\\
    0 }\\
= a_i \bullet x_{i-1}
\end{equation*}
where  $a \bullet x = \case(a > 0, x + 1, 0)$.
A representation of function composition is
\begin{equation*}
\lambda(a) = \fmatl{1\\ a > 0\\ 0}
, \quad
\fmat{r\\ c\\ z} \bullet x = \case(c, x + r, z)\\
, \quad
\fmat{r_1\\ c_1\\ z_1} * \fmat{r_2\\ c_2\\ z_2}
    = \fmatl{r_1 + r_2\\
             c_1 \wedge c_2\\
            \case(c_1, r_1 + z_2, z_1)}
\end{equation*}
The companion operation $*$ is associative.

\subsubsection*{Example 40. Case and If}

There are many associative, and semi-associative, operations related to the $\case$ or $\operatorname{if}$ function and these also have useful algebraic properties when composed with other functions. We begin with the basic definition

\begin{equation*}
\case(c, x, y)= (\text{$x$ if $c$ else $y$})    
\end{equation*}
Notationally it is also helpful to define the multi-case version

\begin{equation*}
\case(c_1, x_1, c_2, x_2, \ldots, c_k, x_k, y)
    = \ourcases{
    x_1    & \text{if $c_1$, else}\\
    x_2    & \text{if $c_2$, else}\\
    \vdots & \vdots\\
    x_k    & \text{if $c_k$, else}\\
    y      & }
\end{equation*}
In these definitions, $c, c_1, \ldots, c_k$ are truth-valued variables.
$\case$ has the following algebraic properties.

\begin{enumerate}[label=\alph*.]

\item 
$\case(\neg c, x, y)=\case(c, y, x)$

\item
$\case(c_1, \case(c_2, x_2, y_2), y_1)
    = \case(c_1 \wedge c_2, x_2, \case(c_1, y_2, y_1))$

\item
$\case(c_1, x_1, \case(c_2, x_2, y))
    = \case(c_1 \vee c_2, \case(c_1, x_1, x_2), y))$

\item
$\case(\case(c, c_1, c_2), x, y)= \case(c, \case(c_1, x, y), \case(c_2, x, y))$

\item 
$\case(c_1, x_1, \ldots, c_k, x_k, y)
    =\case(c_1, x_1, \case(c_2, x_2, \ldots, c_k, x_k, y))$

\item 
$f(\case(c, x, y)) = \case(c, f(x), f(y))$, where $f$ is an arbitrary function accepting $x$, $y$ as arguments.

\end{enumerate}
These properties suffice to prove the assertions of associativity and semi-associativity in the remainder of this example. There are $6$ ways to group the three variables in $\case(c, x, y)$ to form semi-associative set actions, and in each case the companion operations are also associative. In the table below we describe $7$ semi-associative set actions corresponding to these 6 cases, and also describe the companion operations and how both the set actions and companion operations behave under composition with functions. For each set action we also give a function name, so for example, when we say that $\case_\tmat{c\\ x}^{(12)}$ corresponds to $\smat{c\\ x} \bullet y = \case(c, x, y)$ we mean that $\case_\tmat{c\\ x}^{(12)}(y) = \smat{c\\ x} \bullet y = \case(c, x, y)$. At times it is helpful to have notation for the action of functions on pairs, and we use the notations
$\tilde{f}(\smat{c\\ x}) = \smat{c\\ f(x)}$, and 
$\bar{f}(\fmat{x\\ y})   = \fmat{f(x)\\ f(y)}$.

\medskip 
\begin{center}
{ 
\renewcommand{\arraystretch}{2.0}
{\bf{Semi-Associativity of the Case Function}}
\nopagebreak

\begin{tabular}{|m{0.72in}|m{1.30in}|m{1.72in}|m{2.06in}|}
\hline
function & action & companion & composition with functions\\
\hline
\nostretch{$\case_\tmat{c\\ x}^{(12)}(y)$} &  
\nostretch{$\smat{c\\ x} \bullet y = \case(c, x, y)$} &
\nostretch{$\smat{c_1\\ x_1} * \smat{c_2\\ x_2} = \smat{c_1 \vee c_2\\ \case(c_1, x_1, x_2)}$} &
\parbox[t]{2.06in}{\nostretch{
    $f(\smat{c\\ x} \bullet y) = \smat{c\\ f(x)} \bullet f(y)$\\
    $f \circ \case_\tmat{c\\ x}^{(12)} = \case_\tmat{c\\ f(x)}^{(12)} \circ f$\\
    $\tilde{f}(\smat{c_1\\ x_1} * \smat{c_2\\ x_2}) 
        = \tilde{f}(\smat{c_1\\ x_1}) * \tilde{f}(\smat{c_2\\ x_2})$
}}\\[0.6in]
\hline
\nostretch{$\case_\tmat{c\\ y}^{(13)}(x)$} &  
\nostretch{$\smat{c\\ y} \bullet x = \case(c, x, y)$} &
\nostretch{$\smat{c_1\\ y_1} * \smat{c_2\\ y_2} = \smat{c_1 \wedge c_2\\ \case(c_1, y_2, y_1)}$} &
\parbox[t]{2.06in}{\nostretch{
    $f(\smat{c\\ y} \bullet x) = \smat{c\\ f(y)} \bullet f(x)$\\
    $f \circ \case_\tmat{c\\ y}^{(13} = \case_\tmat{c\\ f(y)}^{(13)} \circ f$\\
    $\tilde{f}(\smat{c_1\\ y_1} * \smat{c_2\\ y_2}) 
        = \tilde{f}(\smat{c_1\\ y_1}) * \tilde{f}(\smat{c_2\\ y_2})$
}}\\[0.6in]
\hline
\nostretch{$\case_\tmat{x\\ y}^{(23)}(c)$} &  
\nostretch{$\smat{x\\ y} \bullet c = \case(c, x, y)$} &
\nostretch{$\smat{x_1\\ y_1} * \smat{x_2\\ y_2} 
                = \smatl{\case(x_2, x_1, y_1)\\ \case(y_2, x_1, y_1)}$
                } &
\parbox[t]{2.06in}{\nostretch{
    $f(\smat{x\\ y} \bullet c) = \smat{f(x)\\ f(y)} \bullet c$\\
    $f \circ \case_\tmat{x\\ y}^{(23)} = \case_\tmat{f(x)\\ f(y)}^{(23)}$\\
    $\bar{f}(\smat{x_1\\ y_1} * \smat{x_2\\ y_2}) 
        = \bar{f}(\smat{x_1\\ y_1}) * \smat{x_2\\ y_2}$
}}\\[0.6in]
\hline
\nostretch{$\case_c^{1\vee}(\smat{x\\ y})$} &
\nostretch{$c {\kern1.15pt}\bullet \smat{x\\ y} = \smat{x\\ \case(c, x, y)}$} &
$c_1 * c_2 = c_1 \vee c_2$ &
\parbox[t]{2.06in}{\nostretch{
$\bar{f}(c \bullet \smat{x\\ y}) = c \bullet \bar{f}(\smat{x\\ y})$\\
$\bar{f} \circ \case_c^{1\vee} = \case_c^{1\vee} \circ \bar{f}$\\
$\neg(c_1 \vee c_2) = \neg c_1 \wedge \neg c_2$, and for all\\
other Boolean functions, $f$,\\ $f(c_1\vee c_2) = f(c_1) \vee f(c_2)$
}}\\[0.7in]
\hline
\nostretch{$\case_c^{1\wedge}(\smat{x\\ y})$} &
\nostretch{$c {\kern1.15pt}\bullet \smat{x\\ y} = \smat{\case(c, x, y)\\ y}$} &
$c_1 * c_2 = c_1 \wedge c_2$ &
\parbox[t]{2.06in}{\nostretch{
$\bar{f}(c \bullet \smat{x\\ y}) = c \bullet \bar{f}(\smat{x\\ y})$\\
$\bar{f} \circ \case_c^{1\wedge} = \case_c^{1\wedge} \circ \bar{f}$\\
$\neg(c_1 \wedge c_2) = \neg c_1 \vee \neg c_2$, and for all\\
other Boolean functions, $f$,\\ $f(c_1\wedge c_2) = f(c_1) \wedge f(c_2)$
}}\\[0.7in]
\hline
\nostretch{$\case_x^2(\smat{c\\ y})$} &
\nostretch{$x {\kern1.15pt}\bullet \smat{c\\ y} = \smat{c\\ \case(c, x, y)}$} & 
$x_1 * x_2 = x_1$ &
\parbox[t]{2.06in}{\nostretch{
$\tilde{f}(x \bullet \smat{c\\ y}) = f(x) \bullet \smat{c\\ f(y)}$\\
$\tilde{f} \circ \case_x^2 = \case_{f(x)}^2 \circ \tilde{f}$\\
$f(x_1 * x_2) = f(x_1) * f(x_2)$
}}\\[0.45in]
\hline
\nostretch{$\case_y^3(\smat{c\\ x})$} &
\nostretch{$y {\kern1.15pt}\bullet \smat{c\\ x} = \smat{c\\ \case(c, x, y)}$} & 
$y_1 * y_2 = y_1$ &
\parbox[t]{2.06in}{\nostretch{
$\tilde{f}(y \bullet \smat{c\\ x}) = f(y) \bullet \smat{c\\ f(x)}$\\
$\tilde{f} \circ \case_y^3 = \case_{f(y)}^3 \circ \tilde{f}$\\
$f(y_1 * y_2) = f(y_1) * f(y_2)$
}}\\[0.45in]
\hline
\end{tabular}
} 
\end{center}

\noindent All the companion operations in the table are associative. Another associative operation associated with $\case$ is
\begin{equation*}
\mat{c_1\\ x_1} * \mat{c_2\\ x_2} = \mat{c_1 \vee c_2\\ \case(c_1, x_1, c_2, x_2, y)}
\end{equation*}
where $y$ is a fixed value (e.g. a `default' value for the problem in question or an $\undefined$ value). This satisfies $\tilde{f}(\smat{c_1\\ x_1} * \smat{c_2\\ x_2})=\tilde{f}(\smat{c_1\\ x_1}) * \tilde{f}(\smat{c_2\\ x_2})$ provided $f(y) = y$. However the companion operation to the action of $\case_\tmat{c\\ x}^{(12)}$ does not place requirements on $f$ in order for $\tilde{f}$ to distribute over the $*$ operation and hence seems preferable.

\subsubsection*{Example 41. List Composition and Function Composition on Finite Sets}

Suppose $k$ is a nonnegative integer, and $f\colon \theset{1, \ldots, k} \rightarrow\theset{1, \ldots, k}$ is a function in  $\Endop(\theset{1,\ldots,k})$. Let $\Lambda$ denote the set of all arrays of length $k$ with entries taken from $\theset{1, \ldots, k}$. Define $\lambda\colon \Endop(\theset{1, \ldots, k}) \rightarrow \Lambda$, and $\bullet$, $*$, by
\begin{equation*}
\lambda(f) = (f(1), \ldots, f(k))
, \quad
a \bullet i = a[i]
, \quad
a * b = (a[b[1]], \ldots, a[b[k]])
\end{equation*}
for $a, b \in \Lambda$, $i \in \theset{1, \ldots, k}$. Then $(\Lambda, \lambda, *, \bullet)$ is a representation of function composition for the functions in $\Endop(\theset{1, \ldots, k})$ acting on $\theset{1, \ldots, k}$.
Now suppose $X$ is a finite set of cardinality $k$. Then there is a function $h\colon X \rightarrow\theset{1, \ldots, k}$ with an inverse $h^{-1}\colon \theset{1, \ldots, k} \rightarrow X$. 
Define $\lambda_h\colon \Endop(X) \rightarrow \Lambda$, and $\bullet_h\colon \Lambda \times X \rightarrow X$, by
\begin{equation*}
\lambda_h(f) = \lambda(h \circ f \circ h^{-1})
    = ( h(f(h^{-1}(1))), \ldots, h(f(h^{-1}(k))) )
, \quad
\zeta \bullet_h x = h^{-1}(\zeta \bullet h(x))
\end{equation*}
Then $(\Lambda, \lambda_h, *, \bullet_h)$ is a representation of function composition for the action of $\Endop(X)$ acting on $X$. For this to be useful we must have a practical way to compute $h$ and $h^{-1}$. For `not too large' $k$ if we can enumerate the elements of $X$ then we can list them in an
array and use that array to compute $h^{-1}$. To compute $h$ we can use a dictionary data structure to store the mapping. For large $k$, this approach may not be feasible, however, and one must rely on efficient procedures to associate elements of $X$ with integers in $\theset{1, \ldots, k}$, when these exist.

\subsubsection*{Example 42. Inverse Functions}

The technique of Example 41 is easily generalized. Suppose $\bullet\colon A \times X \rightarrow X$ is a set action and $(\Lambda, \lambda, *, \bullet)$ is a representation of function composition for $\bullet\colon A \times X \rightarrow X$. Suppose $h\colon X \rightarrow Y$ is invertible. Define $\bullet_h\colon \Lambda \times Y \rightarrow Y$ by
\begin{equation*}
\zeta \bullet_h y = h(\zeta \bullet h^{-1}(y))
\end{equation*}
Then $(\Lambda, \lambda, *, \bullet_h)$ is a representation of function composition for the set action $\bullet\colon A \times Y \rightarrow Y\colon (a, y) \longmapsto h(a \bullet h^{-1}(y))$.

\subsubsection*{Example 43. Dictionary Composition}

Dictionary data structures are in direct correspondence with functions on finite sets. The semi-associative set action corresponding to function application is
\begin{equation*}
d \bullet x = d[x] = \text { The value at key } x
\end{equation*}
with companion operation
\begin{equation*}
d_1 * d_2= \text{The dictionary mapping $x$ to $d_1[d_2[x]]$ for each key $x$ of $d_2$}
\end{equation*}
The operation $*$ is defined provided the values of $d_2$ are contained in the set of keys of $d_1$. The companion operation $*$ is also associative.

\subsubsection*{Example 44.  Concatenation of Arrays}

Concatenation of arrays is associative, and the concatenation operator $*$ is the companion operation of
\begin{equation*}
\bullet\colon ((a_1, \ldots, a_k), x) 
    \longmapsto a_1 \bullet(a_2 \bullet (\ldots (a_k \bullet x) \ldots))  
\end{equation*}
for any set action $\bullet\colon A \times X\rightarrow X$, where the array elements are all in $A$. This example shows that any set action has a representation of function composition, as the lifting function $a \mapsto (a)$ embeds $A$ in the space of arrays, where $(a)$ denotes the one element array containing $a$.

\subsubsection*{Example 45. Merge of Sets: Union and Intersection}

The union of sets and intersection of sets are associative operations. Also, if $f$ is a function on $X$ and $A, B \subseteq X$, then
\begin{equation*}
    f(A \cup B)=f(A) \cup f(B)
\end{equation*}
Therefore `union of sets with updating' corresponds to the following semi-associative set action and companion operation.
\begin{displaymath}
\fmat{f\\ A} \bullet B = A \cup f(B)
, \quad 
\fmat{f_1\\ A_1} * \fmat{f_2\\ A_2} = \fmatl{f_1 \circ f_2\\ A_1 \cup f_1(A_2)}
\end{displaymath}
Computing a sliding window $\cup$-product may be used to compute the collection of distinct elements in a sliding window, by using the input sequence $\theset{a_1}, \theset{a_2}, \theset{a_3}, \ldots$.
For intersection of sets similar properties hold under the condition that the functions are 1:1. If we assume that $f\colon X \rightarrow X$ is 1:1, then it follows that for $A, B \subseteq X$ we have
\begin{equation*}
    f(A \cap B)=f(A) \cap f(B)
\end{equation*}
Thus, `intersection of sets with updating' using 1:1 functions corresponds to the following semi-associative set action and companion operation.
\begin{displaymath}
\fmat{f\\ A} \bullet B = A \cap f(B)
, \quad
\fmat{f_1\\ A_1} * \fmat{f_2\\ A_2} = \fmatl{f_1 \circ f_2\\ A_1 \cap f_1(A_2)}
\end{displaymath}

\subsubsection*{Example 46. Ordered Merge of Ordered Arrays}

Assume $\leq$ is a total order on a set $X$. Let $A$ be the set of arrays $(a_1,\ldots, a_k)$ of elements in $X$ with $a_1 \leq \ldots \leq a_k$, where $k \geq 0$.
For any array $(a_1,\ldots, a_k)$ of elements of $X$, let $\operatorname{sort}((a_1, \ldots, a_k))$ denote the rearrangement $(b_1,\ldots, b_k)$ of $(a_1,\ldots, a_k)$ with $b_1 \leq \ldots \leq b_k$. Define $*\colon A \times A \rightarrow A$ by
\begin{equation*}
(a_1, \ldots, a_k) *(b_1, \ldots, b_l)=\operatorname{sort}((a_1, \ldots, a_k, b_1, \ldots, b_l))
\end{equation*}
Then $*$ is associative. Furthermore, if $f\colon X \rightarrow X$ satisfies $x \leq y \Rightarrow f(x) \leq f(y)$ for $x, y \in X$, and we define
\begin{math}
f((a_1, \ldots, a_k)) = (f(a_1), \ldots, f(a_k))
\end{math}
for $(a_1, \ldots, a_k) = a \in A$, then $f(a * b) = f(a) * f(b)$ for all $a, b \in A$.

\subsubsection*{Example 47. Merge of Dictionaries}

Assume $X$ and $Y$ are sets, and $*$ is an associative operation on $Y$, and $d_1, d_2$ are dictionaries with keys in $X$ and values in $Y$. Let $d_1 * d_2$ be the dictionary defined by
\begin{equation*}
(d_1 * d_2)[x] = \ourcases{
d_1[x] * d_2[x] & \text{if $x$ is in the keys of both $d_1$ and $d_2$}\\
d_1[x]          & \text{if $x$ is in the keys of $d_1$ but not $d_2$}\\
d_2[x]          & \text{if $x$ is in the keys of $d_2$ but not $d_1$}
}
\end{equation*}
where the keys of $d_1 * d_2$ are the set $(\text{keys of $d_1$}) \cup (\text{keys of $d_2$})$.
Then $*$ is an associative operation on the set of dictionaries with keys in $X$ and values in $Y$.
Suppose $f\colon Y \rightarrow Z$ is a semigroup homomorphism from $Y$ to another semigroup $Z$. 
I.e., there is an associative operation $*$ on $Z$ and $f$ satisfies $f(y_1 * y_2) = f(y_1) * f(y_2)$ for all $y_1, y_2 \in Y$. 
Let $f(d)$ be the dictionary with $f(d)[x]=f(d[x])$ for any $x$ in the keys of $d$. Then
$f(d_1 * d_2)=f(d_1) * f(d_2)$.

\subsubsection*{Example 48. Histograms}

A sliding window histogram can be computed from an input sequence $a_1, a_2, \ldots, a_N$ as follows. The data contained in a histogram can be represented as a dictionary that maps `bins' to counts, where the bins are computed from the input values using a binning function which we denote $\operatorname{bin}$. Thus $\operatorname{bin}(a_i)$ denotes the histogram bin corresponding to $a_i$. Let us also denote the single entry dictionary that maps $b$ to $c$ by $\theset{b \rightarrow c}$. Let $*$ denote the operation of Example 47 on dictionaries corresponding to $+$ on $Y=\mathbb{Z}_{>0}$. Then the recurrence for computing histograms is
\begin{equation*}
    d_i = \theset{\operatorname{bin}(a_i) \rightarrow 1} * d_{i-1}    
\end{equation*}
where $d_i$ is the $i^\text{th}$ dictionary of bin counts. To compute the sliding window histograms, compute the sliding window $*$-product for the operation $*$ on the sequence $\theset{\operatorname{bin}(a_1) \rightarrow 1}, \theset{\operatorname{bin}(a_2) \rightarrow 2}, \ldots$.

\subsubsection*{Example 49. Continued Fractions}

These are also discussed in Example~\ref{example:sliding-window-continued-fractions}, Example~\ref{example:rofc-examples-5}, and Example~\ref{example:continued-fraction-algorithm}. Assume $F$ is a field, and extend the operations of $F$ to $F \cup \theset{\infty}$ by $a \cdot \infty = \infty \cdot a = \infty$, $b + \infty = \infty + b = \infty$, $b / \infty = 0$, where $a, b \in F$ and $a \neq 0$, and $\infty$ is an element not in $F$.%
\footnote{
Note that we have not defined $0 \cdot \infty$, $\infty / \infty$, $\infty \cdot \infty$, or $\infty + \infty$.
}
Then the recurrence for continued fractions is

\begin{equation*}
    x_i = a_i + \frac{1}{x_{i-1}}
\end{equation*}
where $a_i \in F$, and $x_i \in F \cup\theset{\infty}$. A representation of function composition for $F$ acting on $F \cup \theset{\infty}$ by $a \bullet x = a + \frac{1}{x}$ is given by
\begin{align*}
& \Lambda = GL_2(F) = \theset{\text{$2 \times 2$ matrices over $F$ with nonzero determinant}}\\
& \lambda(a) = \mat{a & 1\\ 1 & 0}\\
& * = \text{matrix multiplication}\\
& \mat{a & b\\ c & d} \bullet x = T_A(x) = \ourcasesclosed{
    \frac{ax + b}{cx + d} & \text{if $x \neq \infty$, $cx + d \neq 0$}\\
    \frac{a}{c}           & \text{if $x =\infty$}\\
    \infty                & \text{if $cx + d = 0$ and $x \neq \infty$}
} 
\end{align*}
Then $(\Lambda, \lambda, *, \bullet)$ is a representation of function composition for the action 
$a \bullet x = a + 1/x$ of $F$ on $F \cup\theset{\infty}$.%
\footnote{
Note there appear to be special cases associated with $\infty$ and zero denominators. These are easily handled however by noting that $F \cup \infty$ is the projective line $P^1(F)$ over $F$, and $T_A(x)=\operatorname{div}(A \smat{x\\ 1})$ for $x \neq \infty$ and $T_A(\infty)=\operatorname{div}(A\smat{0\\ 1})$, where $\operatorname{div}(\smat{x\\ y}) = (\text{$\frac{x}{y}$ if $y \neq 0$ else $\infty$})$, for $x, y \in F$ with $(x, y) \neq (0, 0)$.
}
As noted in Example~\ref{example:rofc-examples-5}, there are other companion operations to $\bullet$, some nonassociative, which are useful. If $F$ is a subfield of $\mathbb{C}$, then

\begin{equation*}
    A *_1 B = \frac{A B}{\|A B\|} \quad \text{ and } \quad A *_3 B = \frac{A B}{\|A\|}
\end{equation*}
are useful to prevent overflow in finite precision arithmetic, where $\|\ \|$ is a matrix norm.

\subsubsection*{Example 50. Linear Fractional Transformations}

Assume $F$ is a field and we extend $F$ to $F \cup\theset{\infty}$ as in Example 49.
%
%
The recurrence for iterated fractional linear transformations is
\begin{align*}
x_i & = \ourcases{
    \frac{a_i x_{i-1} + b_i}{c_i x_{i-1}+d_i} & \text{if $x_{i-1} \neq \infty$ and $c_i x_{i-1} + d_i \neq 0$ else}\\ 
    \frac{a_i}{c_i} & \text{if $x_{i-1} = \infty$ else}\\
    \infty &}
& = T_{A_i}(x_{i-1})
\end{align*}
where $A_i=\smat{a_i & b_i\\ c_i & d_i}$, and $a_i, b_i, c_i, d_i \in F$ with $a_i d_i - b_i c_i \neq 0$.
Let $A \bullet x = T_A(x)$ for $x \in F \cup\theset{\infty}$. Then $\bullet$ is semi-associative with companion operation matrix multiplication, or companion operation $*_1$, or $*_3$ of Example 49.

\subsubsection*{Example 51. Bayesian Filtering for Hidden Markov Models}

This example is adapted from S\"arkk\"a and Garc\'ia-Fern\'andez \cite{Sarkka2021} (see also \cite{Hassan2021}). We consider a hidden Markov model
\begin{center}
\begin{tikzpicture}[x=0.75in, y=0.75in]
\node (x0) at (0, 1) {$x_0$}; \node (x1) at (1, 1) {$x_1$}; \node (x2) at (2, 1) {$x_2$};
\node (x3) at (3, 1) {$x_3$}; \node (x4) at (4, 1) {$\ldots$};
\draw[->] (x0) -- (x1); \draw[->] (x1) -- (x2); \draw[->] (x2) -- (x3);
\draw[->] (x3) -- (x4);
\node (y1) at (1, 0) {$y_1$}; \node (y2) at (2, 0) {$y_2$}; \node (y3) at (3, 0) {$y_3$};
\draw[->] (x1) -- (y1); \draw[->] (x2) -- (y2); \draw[->] (x3) -- (y3);
\end{tikzpicture}
\end{center}
\noindent where $x_0$ is the initial hidden state (with a prior), the $x_i$ are hidden states and the $y_i$ are measurements. For background on hidden Markov models and Bayesian updating refer to \cite{Cappe2005} and \cite{Sarkka2023}. We work with conditional probability density functions. The unconditional density for $x_0$ is $p(x_0)$, the transition kernel is $p(x_i \mid x_{i-1})$, the measurement density given the $i^\text{th}$ hidden state is $p(y_i \mid x_i)$, and the posterior density 
given the measurements $y_1, \ldots, y_k$ is $p(x_k \mid y_1, \ldots, y_k)$.
To describe the recurrence for the posterior densities we define the following action on the space of densities
\begin{equation*}
\mat{f\\ g} \bullet h = x \longmapsto \frac{\int g(z) f(x, z) h(z) dz}{\int g(z) h(z) dz}
\end{equation*}
where $f\colon (x, z) \mapsto f(x, z)$ is a nonnegative measurable function of two variables and $z \mapsto g(z), z \mapsto h(z)$ are nonnegative functions of one variable. ($f, g, h$ should be such that the integrals are finite). A straightforward calculation shows that $\bullet$ is semi-associative with companion operation
\begin{equation*}
\mat{f_1\\ g_1} *\mat{f_2\\ g_2} = \matl{
    (x, z) \longmapsto \frac{\int g_1(u) f_1(x, u) f_2(u, z) du}{\int g_1(u) f_2(u, z) du}\\
    z \longmapsto g_2(z) \int g_1(u) f_2(u, z) du
}
\end{equation*}

S\"arkk\"a and Garc\'ia-Fern\'andez \cite{Sarkka2021} show that $*$ is in fact associative, though we do not need this fact as semi-associativity suffices for the computation of the posterior densities $p(x_k \mid y_1, \ldots, y_k)$. Note that we may multiply the second component of $\smat{f_1\\ g_1} * \smat{f_2\\ g_2}$ by a nonzero scalar and we will still obtain a companion operation because $g$ appears in both the numerator and denominator of the definition of $\bullet$ . Including this scalar factor can make the resulting operator nonassociative, but this is of no concern. On the other hand this means we must only keep track of $g$ up to a scalar multiple. Thus we assume
\begin{equation*}
\mat{f_1\\ g_1} * \mat{f_2\\ g_2} = \matl{
    (x, z) \longmapsto \frac{\int g_1(u) f_1(x, u) f_2(u, z) du}{\int g_1(u) f_2(u, z) du} \\[1ex]
    z \longmapsto c(f_1, f_2, g_1, g_2) \, g_2(z) \int g_1(u) f_2(u, z) dz
    }
\end{equation*}
where $c\left(f_1, f_2, g_1, g_2\right)$ is a strictly positive real number. Now let
\begin{equation}
a_i = \matl{
    (x_i, x_{i-1}) \longmapsto p(x_i \mid y_i, x_{i-1}) \\
    x_{i-1} \longmapsto c_i p(y_i \mid x_{i-1})
}   
\end{equation}
where the $c_i$ are strictly positive numbers. Then
\begin{align*}
& a_1 \bullet \left(x_0 \longmapsto p(x_0)\right)
    = \left(x_1 \longmapsto p(x_1 \mid y_1)\right), \quad \text{ and } \\
& a_i \bullet \left(x_{i-1} \longmapsto p(x_{i-1} \mid y_1, \ldots, y_{i-1})\right)
    =\left(x_i \longmapsto p(x_i \mid y_1, \ldots, y_i)\right)
\end{align*}
%
This is a recurrence for the posterior densities using the semi-associative set action $\bullet$ with companion operation $*$. Using these operators, we may therefore compute sliding window Bayesian filters which start from a sequence of initial priors. This is provided, of course, that we have a means to represent the densities using data (e.g., a formula with parameters), that the integrals can be computed, and the description of the functions does not increase in complexity too rapidly to be of practical use.

\subsubsection*{Example 52. Kalman Filters}

We continue Example 51, adapted from \cite{Sarkka2021} in the Gaussian case. For illustration we consider a simplified Kalman filter whose Gaussian state space model is
\begin{align*}
x_i & = A_i x_{i-1} + q_i \\
y_i & = H_i x_i + r_i
\end{align*}
where $A_i, H_i$ are known matrices and $q_i, r_i$ are Gaussian noise terms with zero mean and covariance matrices $Q_i$, $R_i$. Under this model we have
\begin{align*}
p(x_i \mid x_{i-1}) & = N(x_i ; A_i x_{i-1}, Q_i) \\
p(y_i \mid x_i)     & = N(y_i ; H_i x_i, R_i)
\end{align*}
%
where $N$ is the normal density. The functions $p(x_i \mid y_i, x_{i-1})$, $p(y_i \mid x_{i-1})$ appearing in the recurrence have the form
\begin{align*}
p(x_i \mid y_i, x_{i-1}) & = N(x_i ; (I - K_i H_i) A_i x_{i-1} + K_i y_i, (I - K_i H_i) Q_i) \\
p(y_i \mid x_{i-1}) & \propto N_I(x_{i-1} ; A_i^\top H_i^\top S_i^{-1} y_i, A_i^\top H_i^\top S_i^{-1} H_i A_i)
\end{align*}
where ${ }^\top$ is matrix transpose, $N_I(x ; \eta, J) = N(x; J^{-1} \eta, J^{-1})$, and
\begin{align*}
S_i & = H_i Q_i H_i^\top+R_i \\
K_i & = Q_i H_i^\top S_i^{-1}
\end{align*}
We now shift the representation of the functions to collections of vector and matrices.
\begin{enumerate}
\item 
The function $x \longmapsto N(x ; m, P)$ is represented by $\mat{m & P}$.

\item 
The function $(x, z) \longmapsto N(x ; Bz + b, C)$ is represented by $\mat{B & b & C}$.

\item 
The function $x \longmapsto N_I(x ; \eta, J)$ is represented by $\mat{\eta & J}$.

\end{enumerate}

\noindent In this notation the recurrence
\begin{equation*}
\left(x_i \longmapsto p(x_i \mid y_1,\ldots, y_i)\right)
    = a_i \bullet \left(x_{i-1} \longmapsto p(x_{i-1} \mid y_1, \ldots, y_{i-1})\right)    
\end{equation*}
becomes
\begin{equation*}
\mat{m_i & P_i} = a_i \bullet \mat{m_{i-1} & P_{i-1}}    
\end{equation*}
where
\begin{equation*}
a_i = \mat{
\begin{array}{ccc} B_i & b_i & C_i \end{array}\\
\begin{array}{cc} \eta_i & J_i \end{array}
} = \mat{
\begin{array}{ccc} (I - K_i H_i) A_i & K_i y_i & (I - K_i H_i) Q_i \end{array}\\
\begin{array}{cc} A_i^\top H_i^\top S_i^{-1} y_i & A_i^\top H_i^\top S_i^{-1} H_i A_i \end{array}
}
\end{equation*}
and the operations $\bullet$ and $*$ become
\begin{align*}
& \mat{ \begin{array}{ccc} B & b & C \end{array}\\
        \begin{array}{cc} \eta & J \end{array} }
  \bullet \mat{m & P}
    =  \mat{B (I + PJ)^{-1} (m + P\eta) + b & B (I +PJ)^{-1} P B^\top + C}\\
& \mat{ \begin{array}{ccc} B_1 & b_1 & C_1 \end{array}\\
        \begin{array}{cc} \eta_1 & J_1 \end{array} }
    * 
  \mat{ \begin{array}{ccc} B_2 & b_2 & C_2 \end{array}\\
        \begin{array}{cc} \eta_2 & J_2 \end{array} } \\
& \qquad = \mat{ 
    \begin{array}{ccc}
        B_1(I + C_2 J_1)^{-1} B_2 & 
        B_1(I + C_2 J_1)^{-1}(b_2 + C_2 \eta_1) + b_1 & 
        B_1(I + C_2 J_1)^{-1} C_2 B_1^\top + C_1 
    \end{array} \\
    \begin{array}{cc}
        B_2^\top (I + J_1 C_2)^{-1}(\eta_1 - J_1 b_2) + \eta_2 & 
        B_2^\top (I + J_1 C_2)^{-1} J_1 B_2 + J_2
    \end{array}
    }
\end{align*}
$m_i$ is the estimated (posterior) mean of $x_i$, and $P_i$ is the posterior covariance.
A proof of these formulae, up to notational differences and our use of set actions in addition to binary operations, is indicated in \cite{Sarkka2021}.

%% file: htcams-bibliography.bib
@article{Boyer2012,
    title={Sliding window analyses for optimal selection of mini-barcodes, and application to 454-pyrosequencing for specimen identification from degraded {DNA}},
    author={Boyer, Stephane and Brown, Samuel D. J. and Collins, Rupert A and Cruickshank, Robert H and Lefort, Marie-Caroline and Malumbres-Olarte, Jagoba and Wratten, Stephen D},
    journal={{PLoS} {ONE}},
    volume={7},
    number={5},
    pages={e38215},
    year={2012},
    publisher={Public Library of Science San Francisco, USA}
}

@article{Wang2015,
    title={{swDMR}: a sliding window approach to identify differentially methylated regions based on whole genome bisulfite sequencing},
    author={Wang, Zhen and Li, Xianfeng and Jiang, Yi and Shao, Qianzhi and Liu, Qi and Chen, BingYu and Huang, Dongsheng},
    journal={{PLoS} {ONE}},
    volume={10},
    number={7},
    pages={e0132866},  
    year={2015},
    publisher={Public Library of Science San Francisco, CA USA}
}

@misc{Bernstein2002,
    author = {Daniel J. Bernstein},
    title = {Pippenger's Exponentiation Algorithm},
    year = {2002},
    note = {Retrieved at semanticscholar.org CorpusID:116149978}
}

@article{Brauer1939,
    author = {Alfred Brauer},
    title = {On addition chains},
    volume = {45},
    journal = {Bull. Amer. Math. Soc.},
    number = {10},
    publisher = {American Mathematical Society},
    pages = {736--739},
    year = {1939},
}

@article{Clift2011,
    author = {Clift, Neill Michael},
    title = {Calculating optimal addition chains},
    year = {2011},
    issue_date = {March 2011},
    publisher = {Springer-Verlag},
    address = {Berlin, Heidelberg},
    volume = {91},
    number = {3},
    issn = {0010-485X},
    doi = {10.1007/s00607-010-0118-8},
    journal = {Computing},
    month = mar,
    pages = {265--284},
    numpages = {20},
}

@book{Cohen1996,
    author = {Henri Cohen},
    title = {A Course in Computational Algebraic Number Theory},
    publisher = {Springer-Verlag},
    year = {1996},
    edition = {{3rd}}
}

@article{deJonquieres1894,
    author = {E. de Jonqui\'eres },
    title = {Response 49},
    journal = {Interm\'ed. Math.},
    publisher = {Guathier-Villars et fils},
    volume = {I},
    year = {1894},
    pages = {162--164}
}

@article{Dellac1894,
    author = {H. Dellac},
    title = {Question 49},
    journal = {Interm\'ed. Math.},
    publisher = {Guathier-Villars et fils},
    volume = {I},
    year = {1894},
    pages = {20}
}

@incollection{Doche2005,
    author = {Christophe Doche},
    title = {Exponentiation},
    booktitle = {Handbook of Elliptic and Hyperelliptic Cryptography},
    publisher = {Chapman \& Hall/CRC},
    year = {2005},
    chapter = {9},
    isbn = {9781584885184},
    series = {Discrete Mathematics and Its Applications},
    volume = {34},
    pages = {145--168}
}

@article{Erdos1960,
    author = {Paul Erd\"os},
    title = {Remarks on number theory {III}. {On} addition chains},
    journal = {Acta Arith.},
    year = {1960},
    volume = {6},
    pages = {77--81}
}

@misc{Flammenkamp2022,
    author = {Achim Flammenkamp},
    title = {Shortest Addition Chains},
    howpublished = {\url{http://wwwhomes.uni-bielefeld.de/achim/addition_chain.html}},
    year = {2022},
    note = {Retrieved February 2025}
}

@article{Gordon1998,
    author = {Daniel M. Gordon},
    title = {A Survey of Fast Exponentiation Methods},
    journal = {J. Algorithms},
    volume = {27},
    number = {1},
    pages = {129--146},
    year = {1998},
    issn = {0196-6774},
}

@book{Knuth1998,
    author = {Knuth, Donald E.},
    title = {The Art of Computer Programming, Volume 2: Seminumerical Algorithms},
    edition = {Third},
    year = {1998},
    isbn = {0201896842},
    publisher = {Addison-Wesley},
    address = {USA}
}

@article{Scholz1937,
    author = {Arnold Scholz},
    title = {Aufgabe 253},
    Journal = {Jahresbericht der Deutschen Mathematiker-Vereinigung},
    publisher = {DMV},
    volume = {47},
    number = {II},
    year = {1937},
    pages = {41--42}
}

@article{Thurber1973b,
    author = {Edward G. Thurber},
    title = {{On addition chains $l(mn)\leq l(n)-b$ and lower bounds for $c(r)$}},
    volume = {40},
    journal = {Duke Math. J.},
    number = {4},
    publisher = {Duke University Press},
    pages = {907--913},
    year = {1973},
    month = dec,
    doi = {10.1215/S0012-7094-73-04085-4},
}

@article{Yao1976,
    author = {Yao, Andrew Chi-Chih},
    title = {On the Evaluation of Powers},
    journal = {SIAM J. Comput.},
    volume = {5},
    number = {1},
    pages = {100--103},
    year = {1976},
    month = mar,
    doi = {10.1137/0205008},
}

@book{Jacobson1985,
    author = {Nathan Jacobson},
    title = {Basic Algebra I},
    edition = {Second},
    year = {1985},
    publisher = {W. H. Freeman and Company},
}

@book{Jacobson1989,
    author = {Nathan Jacobson},
    title = {Basic Algebra II},
    edition = {Second},
    year = {1989},
    publisher = {W. H. Freeman and Company},
}

@book{Clifford1961,
    author = {A. H. Clifford and G. W. Preston},
    title = {The Algebraic Theory of Semigroups},
    volume = {I},
    year = {1961},
    publisher = {American Mathematical Society},
}

@article{Suschkewitsch1927,
  title={On a generalization of the associative law},
  author={Suschkewitsch, Anton},
  journal={Trans. Amer. Math. Soc.},
  volume={31},
  number={1},
  pages={204--214},
  year={1929},
}

@article{Marshall1967,
    author = {A. W. Marshall and D. W. Walkup and R. J.-B. Wets},
    title = {Order-preserving functions: Applications to majorization and order statistics},
    volume = {23},
    journal = {Pacific J. Math.},
    number = {3},
    publisher = {Mathematiucal Sciences Publishers},
    pages = {569--584},
    year = {1967},
}

@book{HardyLittlewoodPolya1934,
    author = {G. H. Hardy and J. E. Littlewood and G. P\'olya},
    title = {Inequalities},
    year = {1934},
    publisher = {Cambridge University Press}
}

@book{Olkin1979,
    author = {Albert W. Marshall and Ingram Olkin},
    title = {Inequalities: Theory of Majorization and Its Applications},
    publisher = {Academic Press},
    year = {1979},
    series = {Mathematics in Science and Engineering},
    volume = {143}
}

@book{Steele2008,
    author = {Steele, J. Michael},
    title = {The Cauchy-Schwarz Master Class},
    year = {2008},
    publisher = {Cambridge University Press}
}

@book{Basseville1993,
    title = {Detection of Abrupt Changes: Theory and Application},
    author = {Basseville, Mich\`{e}le and Nikiforov, Igor V.},
    year = {1993},
    publisher = {Prentice Hall, Inc.},
    address = {USA}
}

@book{Cappe2005,
    author = {Capp\'e, Olivier and Moulines, Eric and Ryd\'en, Tobias},
    title = {Inference in Hidden Markov Models},
    year = {2005},
    isbn = {9780387402642},
    publisher = {Springer Science+Business Media, Inc.},
}

@book{Sarkka2023, 
    author={Särkkä, Simo and Svensson, Lennart},
    title={Bayesian Filtering and Smoothing}, 
    edition={{2nd}}, 
    year={2023},
    place={Cambridge}, 
    series={Institute of Mathematical Statistics Textbooks}, 
    publisher={Cambridge University Press}, 
    collection={Institute of Mathematical Statistics Textbooks}
}

@book{Bentley1986,
    title = {Programming Pearls},
    author = {Jon Louis Bentley},
    year = {1986},
    publisher = {Addison-Wesley}
}

@article{Bentley1984,
    author = {Jon Louis Bentley},
    title = {Programming Pearls: Algorithm design techniques},
    journal = {Commun. ACM},
    volume={27},
    number={9},
    pages={865--871},
    year={1984},
    publisher = {Association for Computing Machinery},
    address = {New York, NY, USA},
}

@article{Landin1966,
    title={The Next 700 Programming Languages},
    author={Peter J. Landin},
    journal={Commun. ACM},
    year={1966},
    volume={9},
    number={3},
    pages={157--166},
    publisher = {Association for Computing Machinery},
    address = {New York, NY, USA},
}

@book{Blelloch1990,
    title = {Vector Models for Data-Parallel Computing},
    author = {Blelloch, Guy E},
    year = {1990},
    publisher = {The MIT Press},
    address = {Cambridge, Massachusetts}
}

@incollection{Blelloch1993,
    author = {Blelloch, Guy E.},
    title = {Prefix Sums and Their Applications},
    booktitle = {Synthesis of Parallel Algorithms},
    editor = {John H. Reif},
    publisher = {Morgan Kaufmann Publishers, Inc.},
    address = {San Mateo, CA, USA},
    year = {1993},
    chapter = {1},
    pages = {35--60},
}

@article{Blelloch1996,
    author = {Blelloch, Guy E.},
    title = {Programming Parallel Algorithms},
    year = {1996},
    issue_date = {March 1996},
    publisher = {Association for Computing Machinery},
    address = {New York, NY, USA},
    volume = {39},
    number = {3},
    issn = {0001-0782},
    doi = {10.1145/227234.227246},
    journal = {Commun. ACM},
    month = mar,
    pages = {85--97},
    numpages = {13}
}

@inproceedings{ChinTakano1998,
    author = {Chin, Wei-Ngan and Takano, Akihiko and Hu, Zhenjiang},
    title = {Parallelization via Context Preservation},
    year = {1998},
    isbn = {0818684542},
    publisher = {IEEE Computer Society},
    address = {USA},
    booktitle = {Proceedings of the 1998 International Conference on Computer Languages},
    pages = {153--162},
    series = {ICCL '98}
}

@inproceedings{ChinKhoo2004,
    author = {Chin, Wei Ngan and Khoo, Siau-Cheng and Hu, Zhenjiang and Takeichi, Masato},
    title = {Deriving Parallel Codes via Invariants},
    booktitle = {Proceedings of the 7th International Symposium on Static Analysis},
    year = {2000},
    pages = {75--94},
    series = {Lecture Notes in Computer Science},
    volume = {1824},
    isbn = {978-3-540-67668-3},
    doi = {10.1007/978-3-540-45099-3_5}
}

@inproceedings{FisherGhuloum1994,
    author = {Fisher, Allan L. and Ghuloum, Anwar M.},
    title = {Parallelizing complex scans and reductions},
    year = {1994},
    isbn = {089791662X},
    publisher = {Association for Computing Machinery},
    address = {New York, NY, USA},
    doi = {10.1145/178243.178255},
    booktitle = {Proceedings of the ACM SIGPLAN 1994 Conference on Programming Language Design and Implementation},
    pages = {135--146},
    numpages = {12},
    location = {Orlando, Florida, USA},
    series = {PLDI '94}
}

@article{Gibbons1996,
    title={The Third Homomorphism Theorem},
    author={Gibbons, Jeremy},
    journal={Journal of Functional Programming},
    volume={6},
    number={4},
    pages={657--665},
    year={1996},
    publisher={Journal of Functional Programming}
}

@article{HillisSteele1986,
    author = {Hillis, W. Daniel and Steele, Guy L.},
    title = {Data parallel algorithms},
    year = {1986},
    issue_date = {Dec. 1986},
    publisher = {Association for Computing Machinery},
    address = {New York, NY, USA},
    volume = {29},
    number = {12},
    issn = {0001-0782},
    doi = {10.1145/7902.7903},
    journal = {Commun. ACM},
    month = dec,
    pages = {1170--1183},
    numpages = {14}
}

@article{KoggeStone1973,
    author = {Kogge, Peter M. and Stone, Harold S.},
    journal = {IEEE Trans. Comput.}, 
    title = {A Parallel Algorithm for the Efficient Solution of a General Class of Recurrence Equations}, 
    year = {1973},
    volume = {C-22},
    number = {8},
    pages = {786--793},
    doi = {10.1109/TC.1973.5009159}
}

@article{LadnerFischer1980,
    author = {Ladner, Richard E. and Fischer, Michael J.},
    title = {Parallel Prefix Computation},
    year = {1980},
    issue_date = {Oct. 1980},
    publisher = {Association for Computing Machinery},
    address = {New York, NY, USA},
    volume = {27},
    number = {4},
    issn = {0004-5411},
    doi = {10.1145/322217.322232},
    journal = {J. ACM},
    month = oct,
    pages = {831--838},
    numpages = {8}
}

@inproceedings{Morita2007,
    author = {Morita, Kazutaka and Morihata, Akimasa and Matsuzaki, Kiminori and Hu, Zhenjiang and Takeichi, Masato},
    title = {Automatic Inversion Generates Divide-and-Conquer Parallel Programs},
    year = {2007},
    month = jun,
    isbn = {9781595936332},
    publisher = {Association for Computing Machinery},
    address = {New York, NY, USA},
    doi = {10.1145/1250734.1250752},
    booktitle = {Proceedings of the 28th {ACM} {SIGPLAN} Conference on Programming Language Design and Implementation},
    pages = {146--155},
    numpages = {10},
    location = {San Diego, California, USA},
    series = {PLDI '07}
}

@article{Ofman1962,
    author = {Yu. Ofman},
    title = {On the algorithmic complexity of discrete functions},
    journal = {Dokl. Akad. Nauk SSSR},
    year = {1962},
    volume = {145},
    number = {1},
    pages = {48--51},
}

@inproceedings{Steele2005,
    author={Steele, Guy L.},
    booktitle={14th International Conference on Parallel Architectures and Compilation Techniques (PACT'05)}, 
    title={Parallel Programming and Parallel Abstractions in {Fortress}}, 
    year={2005},
    doi={10.1109/PACT.2005.34}}

@inproceedings{Steele2009,
    author = {Steele, Guy L.},
    title = {Organizing functional code for parallel execution or, foldl and foldr considered slightly harmful},
    year = {2009},
    isbn = {9781605583327},
    publisher = {Association for Computing Machinery},
    address = {New York, NY, USA},
    doi = {10.1145/1596550.1596551},
    booktitle = {Proceedings of the 14th ACM SIGPLAN International Conference on Functional Programming},
    location = {Edinburgh, Scotland},
    series = {ICFP '09}
}

@phdthesis{Trout1972,
    author = {Trout, Harold Robert George},
    title = {Parallel Techniques},
    year = {1972},
    month = oct,
    school = {University of Illinois at Urbana-Champaign},
    doi = {10.5555/906816}
}

@misc{adamax2011,
    author = {adamax},
    title = {Re: Implement a queue in which push{\_}rear(), pop{\_}front() and get{\_}min() are all constant time operations},
    howpublished = {\url{https://stackoverflow.com/questions/4802038}},
    year = {2011},
    month = jan,
    note = {Retrieved June 2024},
}

@inproceedings{ArasuWidom2004,
    author = {Arasu, Arvind and Widom, Jennifer},
    title = {Resource Sharing in Continuous Sliding-Window Aggregates},
    year = {2004},
    isbn = {0120884690},
    publisher = {VLDB Endowment},
    booktitle = {Proceedings of the Thirtieth International Conference on Very Large Data Bases},
    pages = {336--347},
    numpages = {12},
    location = {Toronto, Canada},
    series = {VLDB '04}
}

@misc{Bou2019,
    author = {Savong Bou and Hiroyuki Kitagawa and Toshiyuki Amagasa},
    title = {{CBiX}: Incremental Sliding-Window Aggregation For Real-Time Analytics Over Out-of-Order Data Streams},
    howpublished = {DEIM Forum 2019 F7-9},
    year = {2019},
    month = mar,
}

@inproceedings{Carbone2016,
    author = {Carbone, Paris and Traub, Jonas and Katsifodimos, Asterios and Haridi, Seif and Markl, Volker},
    title = {Cutty: Aggregate Sharing for User-Defined Windows},
    year = {2016},
    isbn = {9781450340731},
    publisher = {Association for Computing Machinery},
    address = {New York, NY, USA},
    doi = {10.1145/2983323.2983807},
    booktitle = {Proceedings of the 25th ACM International on Conference on Information and Knowledge Management},
    pages = {1201--1210},
    numpages = {10},
    location = {Indianapolis, Indiana, USA},
    series = {CIKM '16}
}

@article{Hassan2021,
    author={Hassan, Syeda Sakira and S\"arkk\"a, Simo and  Garc\'ia-Fern\'andez, \'Angel F.},
    journal={IEEE Transactions on Signal Processing}, 
    title={Temporal Parallelization of Inference in Hidden {Markov} Models}, 
    year={2021},
    volume={69},
    pages={4875--4887},
    doi={10.1109/TSP.2021.3103338}
}

@inproceedings{Hirzel2017,
    author = {Hirzel, Martin and Schneider, Scott and Tangwongsan, Kanat},
    title = {Tutorial: Sliding-Window Aggregation Algorithms},
    year = {2017},
    isbn = {9781450350655},
    publisher = {Association for Computing Machinery},
    address = {New York, NY, USA},
    doi = {10.1145/3093742.3095107},
    booktitle = {Proceedings of the 11th ACM International Conference on Distributed and Event-Based Systems},
    pages = {11--14},
    numpages = {4},
    location = {Barcelona, Spain},
    series = {DEBS '17}
}

@inproceedings{Koliousis2016,
    author = {Koliousis, Alexandros and Weidlich, Matthias and Castro Fernandez, Raul and Wolf, Alexander L. and Costa, Paolo and Pietzuch, Peter},
    title = {{SABER}: Window-Based Hybrid Stream Processing for Heterogeneous Architectures},
    year = {2016},
    isbn = {9781450335317},
    publisher = {Association for Computing Machinery},
    address = {New York, NY, USA},
    doi = {10.1145/2882903.2882906},
    booktitle = {Proceedings of the 2016 International Conference on Management of Data},
    pages = {555--569},
    numpages = {15},
    location = {San Francisco, California, USA},
    series = {SIGMOD '16}
}

@inproceedings{Krishnamurthy2006,
    author = {Krishnamurthy, Sailesh and Wu, Chung and Franklin, Michael},
    title = {On-the-Fly Sharing for Streamed Aggregation},
    year = {2006},
    isbn = {1595934340},
    publisher = {Association for Computing Machinery},
    address = {New York, NY, USA},
    doi = {10.1145/1142473.1142543},
    booktitle = {Proceedings of the 2006 ACM SIGMOD International Conference on Management of Data},
    pages = {623--634},
    numpages = {12},
    location = {Chicago, IL, USA},
    series = {SIGMOD '06}
}

@article{LiMaier2005,
    author = {Li, Jin and Maier, David and Tufte, Kristin and Papadimos, Vassilis and Tucker, Peter A.},
    title = {No Pane, No Gain: Efficient Evaluation of Sliding-Window Aggregates over Data Streams},
    year = {2005},
    issue_date = {March 2005},
    publisher = {Association for Computing Machinery},
    address = {New York, NY, USA},
    volume = {34},
    number = {1},
    issn = {0163-5808},
    doi = {10.1145/1058150.1058158},
    journal = {SIGMOD Rec.},
    month = mar,
    pages = {39--44},
    numpages = {6}
}

@article{Sarkka2021,
  author={S\"arkk\"a, Simo and  Garc\'ia-Fern\'andez, \'Angel F.},
  journal={IEEE Transactions on Automatic Control}, 
  title={Temporal Parallelization of {Bayesian} Smoothers}, 
  year={2021},
  volume={66},
  number={1},
  pages={299--306},
  doi={10.1109/TAC.2020.2976316}
}

@inproceedings{Shein2017,
    author = {Shein, Anatoli U. and Chrysanthis, Panos K. and Labrinidis, Alexandros},
    title = {{FlatFIT}: Accelerated Incremental Sliding-Window Aggregation For Real-Time Analytics},
    year = {2017},
    isbn = {9781450352826},
    publisher = {Association for Computing Machinery},
    address = {New York, NY, USA},
    doi = {10.1145/3085504.3085509},
    booktitle = {Proceedings of the 29th International Conference on Scientific and Statistical Database Management},
    articleno = {5},
    numpages = {12},
    location = {Chicago, IL, USA},
    series = {SSDBM '17}
}

@inproceedings{Shein2018,
    author = {Anatoli U. Shein and Panos K. Chrysanthis and Alexandros Labrinidis},
    editor = {Michael H. B{\"{o}}hlen and Reinhard Pichler and Norman May and Erhard Rahm and Shan{-}Hung Wu and Katja Hose},
    title = {{SlickDeque}: High Throughput and Low Latency Incremental Sliding-Window Aggregation},
    booktitle = {Proceedings of the 21st International Conference on Extending Database Technology, {EDBT} 2018, Vienna, Austria, March 26-29, 2018},
    pages = {397--408},
    publisher = {OpenProceedings.org},
    year = {2018},
    doi = {10.5441/002/EDBT.2018.35},
}

@phdthesis{Shein2019,
    title = {Algorithms and Optimizations for Incremental Window-Based Aggregations},
    author = {Shein, Anatoli U},
    year = {2019},
    month = sep,
    school={University of Pittsburgh}
}

@article{Snytsar2023b,
    title = {Sliding Window Sum Algorithms for Deep Neural Networks},
    author = {Snytsar, Roman},
    year = {2023},
    month = {October},
    pages = {71--78},
    volume = {12},
    number = {5},
    journal = {International Journal on Cybernetics \& Informatics},
    doi = {10.5121/ijci.2023.120507}
}

@inproceedings{SnytsarTurakhia2019,
    title = {Parallel Approach to Sliding Window Sums},
    author = {Snytsar, Roman and Turakhia, Yatish},
    booktitle = {International Conference on Algorithms and Architectures for Parallel Processing, 19th International Conference, ICA3PP 2019, Melbource, VIC, Australia, December 9-11, 2019, Proceedings, Part II},
    pages = {19--26},
    year = {2020},
    month = dec,
    publisher={Springer}
}

@article{Tangwongsan2015a,
    author = {Tangwongsan, Kanat and Hirzel, Martin and Schneider, Scott and Wu, Kun-Lung},
    title = {General incremental sliding-window aggregation},
    year = {2015},
    issue_date = {February 2015},
    publisher = {VLDB Endowment},
    volume = {8},
    number = {7},
    issn = {2150-8097},
    doi = {10.14778/2752939.2752940},
    journal = {Proc. VLDB Endow.},
    month = feb,
    pages = {702--713},
    numpages = {12}
}

@techreport{Tangwongsan2015b,
    title = {Constant-Time Sliding Window Aggregation},
    author = {Kanat Tangwongsan and Martin Hirzel and Scott Schneider},
    institution = {IBM Research Division},
    number = {RC25574},
    series = {IBM Research Report},
    year={2015},
    month=nov,
}

@inproceedings{Tangwongsan2017,
    author = {Tangwongsan, Kanat and Hirzel, Martin and Schneider, Scott},
    title = {Low-Latency Sliding-Window Aggregation in Worst-Case Constant Time},
    year = {2017},
    isbn = {9781450350655},
    publisher = {Association for Computing Machinery},
    address = {New York, NY, USA},
    doi = {10.1145/3093742.3093925},
    booktitle = {Proceedings of the 11th ACM International Conference on Distributed and Event-Based Systems},
    pages = {66--77},
    numpages = {12},
    location = {Barcelona, Spain},
    series = {DEBS '17}
}

@article{Tangwongsan2019,
    author = {Tangwongsan, Kanat and Hirzel, Martin and Schneider, Scott},
    title = {Optimal and general out-of-order sliding-window aggregation},
    year = {2019},
    issue_date = {June 2019},
    publisher = {VLDB Endowment},
    volume = {12},
    number = {10},
    issn = {2150-8097},
    doi = {10.14778/3339490.3339499},
    journal = {Proc. VLDB Endow.},
    month = jun,
    pages = {1167--1180},
    numpages = {14}
}

@article{Tangwongsan2021,
    author = {Kanat Tangwongsan and Martin Hirzel and Scott Schneider},
    title = {In-order sliding-window aggregation in worst case constant time},
    journal = {VLDB J.},
    year = {2021},
    month = jun,
    volume = {30},
    number = {6},
    pages = {933--957},
    doi = {10.1007/s00778-021-00668-3}
}

@incollection{Tangwongsan2022,
    title = {Sliding-Window Aggregation Algorithms},
    author = {Kanat Tangwongsan and Martin Hirzel and Scott Schneider},
    booktitle = {Encyclopedia of Big Data Technologies},
    editor = {Albert Zomaya and Javid Taheri and Sheri Sakr},
    publisher = {Springer International Publishing},
    address = {Cham}, 
    year = {2022},
    month = mar,
    doi = {10.1007/978-3-319-63962-8_157-2},
    isbn={978-3-319-63962-8}
}

@inproceedings{Theodorakis2018,
    title = {{Hammer} {Slide}: Work- and {CPU}-efficient Streaming Window Aggregation},
    author = {Georgios Theodorakis and Alexandros Koliousis and Peter R. Pietzuch and Holger Pirk},
    booktitle = {Workshop on Accelerating Analystics and Data Management Systems using Modern Processor and Storage Architectures (ADMS)},
    year = {2018},
    month = aug,
    pages = {34--41},
    series = {ADMS '18}
}

@inproceedings{Theodorakis2020a,
    author = {Georgios Theodorakis and Peter R. Pietzuch and Holger Pirk},
    editor = {Angela Bonifati and Yongluan Zhou and Marcos Antonio Vaz Salles and Alexander B{\"{o}}hm and Dan Olteanu and George H. L. Fletcher and Arijit Khan and Bin Yang},
    title = {{SlideSide}: A fast Incremental Stream Processing Algorithm for Multiple Queries},
    booktitle = {Proceedings of the 23rd International Conference on Extending Database Technology, {EDBT} 2020, Copenhagen, Denmark, March 30 - April 02, 2020},
    pages = {435--438},
    publisher = {OpenProceedings.org},
    year = {2020},
    month = mar,
    doi = {10.5441/002/EDBT.2020.51},
}

@inproceedings{Theodorakis2020b,
    author = {Theodorakis, Georgios and Koliousis, Alexandros and Pietzuch, Peter and Pirk, Holger},
    title = {{LightSaber}: Efficient Window Aggregation on Multi-core Processors},
    year = {2020},
    isbn = {9781450367356},
    publisher = {Association for Computing Machinery},
    address = {New York, NY, USA},
    doi = {10.1145/3318464.3389753},
    booktitle = {Proceedings of the 2020 ACM SIGMOD International Conference on Management of Data},
    pages = {2505--2521},
    numpages = {17},
    location = {Portland, OR, USA},
    series = {SIGMOD '20}
}

@inproceedings{Traub2018,
    title={Scotty: Efficient window aggregation for out-of-order stream processing},
    author={Traub, Jonas and Grulich, Philipp Marian and Cu\'{e}llar, Alejandro Rodriguez and Bress, Sebastian and Katsifodimos, Asterios and Rabl, Tilmann and Markl, Volker},
    booktitle={2018 IEEE 34th International Conference on Data Engineering (ICDE)},
    pages={1300--1303},
    year={2018},
    organization={IEEE}
}

@article{Traub2021,
    author = {Traub, Jonas and Grulich, Philipp Marian and Cu\'{e}llar, Alejandro Rodr\'{\i}guez and Bress, Sebastian and Katsifodimos, Asterios and Rabl, Tilmann and Markl, Volker},
    title = {Scotty: General and Efficient Open-source Window Aggregation for Stream Processing Systems},
    year = {2021},
    issue_date = {March 2021},
    publisher = {Association for Computing Machinery},
    address = {New York, NY, USA},
    volume = {46},
    number = {1},
    issn = {0362-5915},
    doi = {10.1145/3433675},
    journal = {ACM Trans. Database Syst.},
    month = apr,
    articleno = {1},
    numpages = {46},
    pages = {1--46}
}

@article{Verweibe2023,
    author = {Verwiebe, Juliane and Grulich, Philipp M. and Traub, Jonas and Markl, Volker},
    title = {Survey of window types for aggregation in stream processing systems},
    year = {2023},
    issue_date = {Sep 2023},
    publisher = {Springer-Verlag},
    address = {Berlin, Heidelberg},
    volume = {32},
    number = {5},
    issn = {1066-8888},
    doi = {10.1007/s00778-022-00778-6},
    journal = {VLDB J.},
    month = feb,
    pages = {985--1011},
    numpages = {27},
}

@inproceedings{Zhang2021,
    author = {Zhang, Chao and Akbarinia, Reza and Toumani, Farouk},
    title = {Efficient Incremental Computation of Aggregations over Sliding Windows},
    year = {2021},
    isbn = {9781450383325},
    publisher = {Association for Computing Machinery},
    address = {New York, NY, USA},
    doi = {10.1145/3447548.3467360},
    booktitle = {Proceedings of the 27th ACM SIGKDD Conference on Knowledge Discovery \& Data Mining},
    pages = {2136--2144},
    numpages = {9},
    location = {Virtual Event, Singapore},
    series = {KDD '21}
}
